\documentclass[12pt]{article}
\usepackage{amsmath}
\usepackage{graphicx}
\usepackage{enumerate}
\usepackage{natbib}
\usepackage{smile}
\usepackage{algorithm}
\usepackage{algpseudocode}
\usepackage{setspace}
\usepackage{threeparttable}
\usepackage{multirow}
\usepackage{bbm}
\usepackage[dvipsnames,svgnames,table]{xcolor}
\usepackage[colorlinks=true,
linkcolor=blue,
urlcolor=blue,
citecolor=blue]{hyperref}
\usepackage{xr}
\usepackage{url} % not crucial - just used below for the URL
\usepackage{enumitem}
\usepackage{booktabs}
\usepackage{adjustbox}

\def\tr{{\mathop{\text{\rm Tr}}}}

\newcommand{\blind}{0}

\numberwithin{equation}{section}
\numberwithin{theorem}{section}
\begin{document}

\def\spacingset#1{\renewcommand{\baselinestretch}%
{#1}\small\normalsize} \spacingset{1.15}

%%%%%%%%%%%%%%%%%%%%%%%%%%%%%%%%%%%%%%%%%%%%%%%%%%%%%%%%%%%%%%%%%%%%%%%%%%%%%%

\if0\blind
{
  \title{\bf Expected Shortfall Panel Regression}
  \author{Yujie Hou\\
  Shandong University, Jinan 250100, China\\
  Xinbing Kong\\
	Southeast University, Nanjing 211189, China\\
    Yalin Wang\\
  Shandong University, Jinan 250100, China\\
	Bin Wu\thanks{Corresponding author: bin.w@ustc.edu.cn.\\Authors are listed in alphabetical order by surname.}\\
	University of Science and Technology of China, Hefei 230026, China
	}
  \maketitle
} \fi

\if1\blind
{
  \bigskip
  \bigskip
  \bigskip
  \begin{center}
    {\LARGE\bf Expected Shortfall Factor Models}
\end{center}
  \medskip
} \fi

%%\bigskip
\begin{abstract}

Expected Shortfall (ES) is a coherent measure of tail risk that captures the average loss beyond a quantile threshold. Despite the growing literature on ES regression conditional on covariates, no existing work considers ES modeling in panel data settings where both cross-sectional and temporal dependencies are present.
This paper introduces the  panel ES regression model with a latent factor structure to capture cross-sectional dependence. We develop a two-stage estimation procedure robust to heavy-tailed errors, recovering the conditional quantile in the first stage and iteratively estimating the ES factor model in the second stage. Theoretically, we establish the consistency and asymptotic normality of the proposed two-step ES estimators and derive non-asymptotic error bounds for both the panel quantile and ES estimators. We also provide a non-asymptotic normal approximation for the standardized ES regression estimator, bridging asymptotic theory and finite-sample practice. Simulation evidence shows that the proposed method delivers substantial gains in both parameter estimation and factor recovery, particularly in the presence of latent tail dependence. An empirical application further indicates that the extracted ES factors carry distinct pricing information that is not captured by conventional mean or quantile-based approaches.

\end{abstract}

\noindent%
\vfill

\newpage
\spacingset{1.2} % DON'T change the spacing! 1.9

\makeatletter
\setlength{\abovedisplayskip}{4pt}
\setlength{\belowdisplayskip}{4pt}
\setlength{\abovedisplayshortskip}{1pt}
\setlength{\belowdisplayshortskip}{1pt}
\setlength{\jot}{2pt}
\makeatother

\section{Introduction}\label{sect:introduction}

Understanding extreme downside risk is a central issue in financial econometrics and risk management. Expected Shortfall (ES), also known as superquantile or conditional Value-at-Risk (VaR), has been widely recognized as an important measure of tail risk with broad applications in finance \citep{Acerbi2002on,Rockafellar2002conditional}.
While VaR measures only a quantile of the loss distribution, ES captures the average loss beyond the quantile threshold and thus provides a more informative measure of tail risk \citep{Rockafellar2014Superquantile}. In addition, ES satisfies the axioms of coherent risk measures, including subadditivity, which VaR generally fails to satisfy \citep{artzner1999coherent,Acerbi2002on}. These advantages have led to the increasing adoption of ES in financial regulation and risk management frameworks.
A growing body of literature has studied methods for estimating and forecasting ES; see, e.g., \citet{Taylor2007using,Taylor2008Estimating,caiNonparametric2008,kato2012weighted,Taylor2019Forecasting}.

Recent studies have introduced regression-based frameworks that allow ES to depend on observable covariates, providing a flexible approach for analyzing the determinants and dynamics of tail risk. One key challenge in estimating ES in the regression setting is its lack of elicitability; that is, there does not exist a loss function such that the expected shortfall is the minimizer of the expected loss. \cite{fissler2016higher} show that the quantile and the expected shortfall are jointly elicitable, and construct a class of joint loss functions. Building on this result, \cite{dimitriadis2019joint} and \cite{Patton2019Dynamic} propose a regression framework which simultaneously models the quantile and the ES  of a response variable given a set of covariates. 
\cite{he2023robust} develop a statistically robust and computationally 
efficient two-step procedure for fitting joint quantile and ES regression models that can handle highly 
skewed and heavy-tailed data. \cite{Zhang2025High} propose
the lasso-penalized expected shortfall regression for high-dimensional settings. Additional related studies on ES regression include, among others, \cite{Barendse2020Efficiently}, \cite{Fissler2023deep}, \cite{Yu02025Estimation}, and \cite{chang2025predictive}.

In modern financial and macroeconomic studies, data are typically collected for a large number of assets over time, leading naturally to high-dimensional panel structures that have become fundamental in empirical asset pricing and macro-finance analysis.  This motivates the development of panel ES regression models that exploit both cross-sectional and temporal dependence. To the best of our knowledge, however, the literature has not yet considered ES regression in a panel data framework. 
To address this gap, we propose a panel ES regression model with latent factor structure, allowing for flexible cross-sectional dependence across units, a feature widely documented in large panel datasets (see, e.g., \citealp{Pesaran2006Estimation,bai2009panel,Ke2016Structure,Xiao2021homogeneity,Juodis02012022,jing2021community,hu2025aggregated}). 
\citet{bai2009panel} develop an interactive fixed-effects framework for panel data with a general multifactor error structure, while focusing on conditional mean regression and therefore not addressing tail risk.  Beyond the conditional mean, a growing literature investigates distributional features of panel data through quantile regression. For example, \cite{Kato2019Asymptotics} study panel quantile regression models with individual fixed effects. \cite{Lamarche2010robust}  investigates a class of penalized quantile regression estimators for panel data. To capture cross-sectional information, 
factor structures have also been incorporated into conditional quantile analysis through quantile factor models, as developed by \cite{Ma2021Estimation} and \cite{chen2021quantile}. 
\cite{ando2020quantile} introduce a panel quantile model to capture unobserved heterogeneity in financial time series, accounting for both sensitivity to explanatory variables and an underlying latent factor structure. 
While quantile regression does not fully capture extreme losses, ES overcomes this limitation by averaging losses beyond the chosen quantile, thereby providing a more complete measure of tail risk.
Motivated by this line of research, we incorporate the ES regression framework and its associated quantile component into a factor-augmented panel model, extending the existing joint quantile--ES regression methodology to panel settings and providing a flexible approach for modeling tail risk in high-dimensional financial panels.

In this paper, we propose a two-stage estimation method for the ES panel regression model with a factor structure (ESFM), as well as a modified information criterion (IC) to consistently estimate the number of factors. Numerical experiments demonstrate the superiority of our model and the effectiveness of our method. The main contributions of this work are summarised as follows:
\begin{itemize}
\item As far as we know, this is the first study to consider ES panel regression, which generalizes tail-specific conditional expected loss modeling to the panel data setting---the workhorse framework in modern empirical research---while accommodating latent cross-sectional dependence. To further capture interactive fixed effects, we incorporate a factor structure into the ES regression framework. We develop a two-step estimation procedure building upon recent advances in joint quantile and expected shortfall regression, where the conditional quantile is estimated in the first step and the ES factor model is iteratively estimated in the second step. This approach allows us to jointly model the conditional quantile and the tail expectation of the response variable while accounting for latent cross-sectional dependence.

\item %Theoretically, we first establish the asymptotic properties of the proposed two-step ES estimators, including consistency and asymptotic normality. We further derive non-asymptotic error bounds for both the panel quantile regression estimator and the ES estimator. The error of the ES estimator can be decomposed into three components: the inherent uncertainty of ES regression, the first-stage quantile estimation error, and the latent factor estimation error. Importantly, due to Neyman orthogonality, the first-stage quantile error contributes only at a higher order. Building on these results, we establish a finite-sample Gaussian approximation with an explicit bound on the Kolmogorov distance, providing a theoretical guarantee for inference in finite samples.
Theoretically, this paper provides the first analysis of ES regression in a panel setting with latent factors, extending existing panel factor models that focus only on conditional means or quantiles. This extension presents several methodological challenges, including accounting for cross-sectional correlation, temporal dependence, and potential high dimensionality when the number of assets or covariates is large. Unlike conventional panel factor models, ES is not elicitable alone, and its two-stage estimation procedure introduces dependence that does not arise in mean or quantile settings. Nevertheless, we show that the proposed ES estimator remains consistent and asymptotically normal under mild conditions. A key insight is that, due to Neyman orthogonality embedded in the second-stage objective, the first-stage quantile estimation error affects the ES estimator only at a second-order level. Consequently, the ES estimator achieves the same convergence rate as if the true quantile were known. Going beyond asymptotic theory, we derive non-asymptotic error bounds that explicitly decompose the ES estimation error into three sources, namely sampling variability, first-stage quantile error, and factor estimation error. This decomposition clarifies how the panel dimensions $N$ and $T$ trade off against each other. We further establish a finite-sample Gaussian approximation that justifies standard inference even in moderately sized panels. Collectively, these results demonstrate that ES regression can be extended to panel settings with latent factors without sacrificing its key properties: $\sqrt{T}$-consistency, asymptotic normality, and robustness to first-stage errors.

\item From an empirical perspective, we demonstrate that incorporating a latent factor structure into the ES framework substantially enhances its practical performance in asset pricing applications. While standard ES regression captures tail risk at the individual asset level, it neglects the pervasive cross-sectional dependence that is particularly pronounced during extreme market conditions. Our ESFM addresses this limitation by extracting common tail risk components, leading to economically meaningful return spreads and robust pricing implications. Compared with quantile-based factor constructions, which focus on threshold behavior, the ESFM captures the magnitude of tail losses and therefore delivers a sharper characterization of downside risk. The resulting ES-based factors exhibit stable and significant pricing effects that are not spanned by conventional factor models, highlighting the incremental information embedded in tail expectations.
\end{itemize}

Beyond financial applications, the proposed ESFM provides a general framework for modeling extreme outcomes in high-dimensional panel data. Potential applications include systemic risk monitoring in macroeconomics, tail-event analysis in climate and environmental studies, and extreme demand or failure modeling in operations and supply chain systems, where cross-sectional dependence and tail behavior jointly play a central role.

The rest of this article is organized as follows. We begin
with a brief introduction to the joint quantile and ES regression
framework and the ES factor model in Section \ref{sec:Model Setup}. Section \ref{subsec:Two-Stage Estimation} presents the two-step ES factor model estimation procedure, while Section \ref{subsec:determine} introduces a modified information criterion for determining the number of factors. In Section \ref{sec:Theoretical Results}, we investigate the theoretical properties of the
proposed estimators. Section \ref{sec:Simulation Study} and Section \ref{sec:Empirical Application} examine the finite-sample performance of the proposed estimator through simulation studies and a real data application.

Notations adopted throughout the paper are as follows. Let $Y$ be a real-valued random variable with finite first-order absolute moment ($E|Y|<\infty$), and let $F_Y$ be its cumulative distribution function. For any $\tau\in(0,1)$, the quantile and ES at level $\tau$ are defined as
	$Q_{\tau}(Y)=\inf\{y\in\mathbb{R}:F_Y(y)\geq\tau\}$ and
$\mathrm{ES}_{\tau}(Y)=E[Y|Y\leq Q_{\tau}(Y)]$,
respectively. When $F_Y$ is continuous, one also has the equivalent representation  $\mathrm{ES}_{\tau}(Y)=\tau^{-1}\int_{0}^{\tau}Q_{u}(Y)\mathrm{d}u$. For a vector or matrix $A$, denote $A^\top$ as the transpose of $A$, and its norm is defined as $\|A\| = \left(\tr(A^\top A)\right)^{1/2}$, where ``$\tr$" denotes the trace of a square matrix. 
$\lambda_{\min}(A)$ and $\lambda_{\max}(A)$ denote the smallest and largest 
eigenvalues of the matrix $A$, respectively.
For a symmetric matrix $A$, we write $A\succ 0$ to denote that $A$ is positive definite, and $A \succeq 0$ to denote that $A$ is positive semidefinite. Let $P_{A}=A(A^\top A)^{-1}A^\top$ denote the orthogonal projection matrix onto the space spanned by the columns of $A$.
$I_k$ denotes a $k$-order identity matrix and $\otimes$ denotes the Kronecker product. The notations $\stackrel{p}\longrightarrow,\stackrel{d}\longrightarrow,\stackrel{a.s.}\longrightarrow$ represent convergence in probability, in distribution and almost surely, respectively. The $o_p$ is for convergence to zero in probability and $O_p$ is for stochastic boundedness.  We denote the set $\{1,\ldots,T\}$ by $[T]$ for convenience.

\section{Model Setup and Methodologies}\label{sec:Model Setup} 
Suppose we observe, for each of $N$ units over $T$ periods, a response variable $Y_{it}$ along with a $(p+1)$-dimensional vector $X_{it}=(1,X_{it,1},...,X_{it,p})^\top$. We model the $\tau$-th conditional quantile of $Y_{it}$ given $X_{it}$ using a standard linear quantile regression framework, thereby characterizing the dependence of the conditional quantile on observable covariates. To capture
cross-sectional information, we  introduce a factor structure for the corresponding conditional expected shortfall (ES), which accommodates both observed covariate effects and latent common factors. In particular, at a given quantile level $\tau \in(0,1)$, assume that
\begin{equation}\label{eq:quantile models}
Q_{\tau}(Y_{it}|X_{it})=X_{it}^\top\alpha_{i,\tau}^0,
\end{equation}	
\begin{equation}\label{eq:ES models}
\mathrm{ES}_{\tau}(Y_{it}|X_{it},f_{t,\tau})=X_{it}^\top\beta_{i,\tau}^0+\lambda_{i,\tau}^{0\top} f_{t,\tau}^0,\quad i\in[N],\quad t\in[T], 
\end{equation}
where $\alpha_{i,\tau}^0=(\alpha_{i,0,\tau}^{0},\alpha_{i,1,\tau}^0,...,\alpha_{i,p,\tau}^0)^\top$ and $\beta_{i,\tau}^0=(\beta_{i,0,\tau}^0,\beta_{i,1,\tau}^0,...,\beta_{i,p,\tau}^0)^\top$ are the true quantile‐ and ES‐regression coefficients; $f_{t,\tau}^0$ is an $r_{\tau}^0$-dimensional vector  of unobservable factors, where $r_{\tau}^0$ denotes the true number of factors; $\lambda_{i,\tau}^{0}$ is the corresponding loading vector. Moreover, we include a leading constant equal to 1 in each $X_{it}$, so that the first elements of $\alpha_{i,\tau}^0$ and $\beta_{i,\tau}^0$ represent the intercept terms  in the quantile and ES models, respectively. Allowing $\alpha_{i,\tau}^0$, $\beta_{i,\tau}^0$, $f_{t,\tau}^0$, $\lambda_{i,\tau}^0$, and $r_{\tau}^0$ to vary with $\tau$ enables our ES factor model (ESFM) to flexibly capture quantile-specific heterogeneity and tail-specific dynamics.

We treat $f_{t,\tau}^0$ and $\lambda_{i,\tau}^0$ as  unknown parameters. This latent factor structure is commonly referred to as the ``interactive effects'' model in the mean‐regression literature (e.g., \citealt{bai2009panel}) and has also been extended to the quantile regression framework (e.g., \citealt{ando2020quantile}). By capturing common shocks through $f_{t,\tau}^0$ and unit‐specific loadings $\lambda_{i,\tau}^0$, it provides a flexible way to model cross‐sectional dependence in tail expectations.
Special cases include:
\begin{enumerate}[label=$\bullet$]
	\item If the number of factors is set to $r_{\tau}^0=0$, model \eqref{eq:ES models} reduces to a panel ES regression without latent factors (see standard ES regression models in \citealp{dimitriadis2019joint,he2023robust,Zhang2025High}).
	\item When $\beta_{i,\tau}^0=0_{p+1}$, model \eqref{eq:ES models} reduces to a pure factor model for ES, without observed covariates.
\end{enumerate}
One could also allow the quantile and ES equations to depend on different covariates, denoted by $X^q$ and $X^e$, respectively, in which case
\begin{equation}\nonumber
	Q_{\tau}\left(Y_{i t} \mid X_{i t}^q\right)=X_{i t}^{q \top} \alpha_{i, \tau}^0, \quad \mathrm{ES}_{\tau}\left(Y_{i t} \mid X_{i t}^e, f_{t, \tau}\right)=X_{i t}^{e \top} \beta_{i, \tau}^0+\lambda_{i, \tau}^{0\top} f_{t, \tau}^0.
\end{equation}
%In this case, the conditional $\tau$-quantile and $\tau$-ES of $Y-X^{q\top}\alpha_{\tau}$ and $Y-X^{e\top}\beta_{\tau}-\lambda_{\tau}^\top f_{\tau}$, respectively, given $X=(X^{q\top},X^{e\top})^\top$, are assumed to be zero. 
In this paper, for simplicity, we assume that both the quantile and ES equations share the same covariate vector.

\subsection{Two-Stage Estimation of the ES Factor Model}\label{subsec:Two-Stage Estimation}
Our objective is to estimate the  parameters
$B_\tau^0=(\beta_{1,\tau}^0,...,\beta_{N,\tau}^0)^\top$, $\Lambda_\tau^0=(\lambda_{1,\tau}^0,...,\lambda_{N,\tau}^0)^\top$ and $F_\tau^0=(f_{1,\tau}^0,...,f_{T,\tau}^0)^\top$.
A well-known result in the literature on factor models is that the factors $\{f_{t,\tau}^0\}$ and loadings $\{\lambda_{i,\tau}^0\}$ cannot be
separately identified without imposing normalizations, see \cite{bai2002determining}. Without loss of
generality, 
we impose the following normalization conditions:
 {
\begin{align}\label{eq:constraints for factor}
	F_\tau^\top F_\tau/T=I_{r_\tau^0},\quad\Lambda_\tau^\top\Lambda_\tau/N \longrightarrow \Sigma_{\Lambda_\tau} \succ 0 .
\end{align}}
 Let $\theta_\tau^0=(f_{1,\tau}^{0\top},...,f_{T,\tau}^{0\top},\lambda_{1,\tau}^{0\top},...,\lambda_{N,\tau}^{0\top})^\top$ denote the vector of true parameters. Define the admissible parameter space 
\[
\begin{aligned}
\Theta^{r_\tau^0}=\{\theta_\tau\in\mathbb{R}^{(N+T)\times r_{\tau}^0}:
&\ \lambda_{i,\tau},f_{t,\tau}\in\mathbb{R}^{r_{\tau}^0} \text{ for all } i, t, \\
&\ \{f_{t,\tau}\} \text{ and } \{\lambda_{i,\tau}\} \text{ satisfy the normalization in \eqref{eq:constraints for factor}} \bigr\}.
\end{aligned}
\]
% $$\Theta^{r_\tau^0}=\{\theta_\tau\in\mathbb{R}^{(N+T)\times r_{\tau}^0}:\lambda_{i,\tau},f_{t,\tau}\in\mathbb{R}^{r_{\tau}^0}~\text{for all}~i,t,\{f_{t,\tau}\}~\text{and}~\{\lambda_{i,\tau}\}~\text{satisfy the normalization in \eqref{eq:constraints for factor}}\}.$$

Since ES depends on the quantile but not vice versa, the quantile parameters $\alpha_{i,\tau}$ naturally act as nuisance parameters when ES is the primary object of interest. Inspired by the Neyman-orthogonality idea---which seeks score functions whose first‑order sensitivity to nuisance estimation vanishes \citep{neyman1979c,chernozhukov2018double,Barendse2020Efficiently}---one can adopt a two-stage approach that avoids non‑convex optimization altogether. Accordingly, to estimate the ES model \eqref{eq:ES models}, we adopt a two-stage procedure similar to that of \citet{Barendse2020Efficiently,he2023robust,Zhang2025High}. 
\begin{itemize}
    \item \textbf{Stage 1: Quantile Regression}
    
    Estimate the conditional quantile parameters $\alpha_{i,\tau}$ for each unit using standard quantile regression:
\begin{align}\label{eq:step 1 QR}
	\widehat{\alpha}_{i,\tau}=\argmin_{\alpha_{i,\tau}\in\mathbb{R}^{p+1}}\frac{1}{T}\sum_{t=1}^T\rho_{\tau}(Y_{it}-X_{it}^\top\alpha_{i,\tau}),\quad\text{for each $i$},
\end{align}
and let $\hat{A}_\tau=(\hat{\alpha}_{1,\tau},...,\hat{\alpha}_{N,\tau})^\top$, where $\rho_{\tau}(u)=(\tau-\mathbbm{1}(u<0))u$ is the check loss function (\citealt{koenker1978regression}).

\item \textbf{Stage 2: Orthogonalized ES Regression}

Define 
\begin{align*}
	S_0(\alpha_{i,\tau},\beta_{i,\tau},f_{t,\tau},\lambda_i;Y_{it},X_{it})=\tau \lambda_i^\top f_t+\tau X_{it}^\top(\beta_i-\alpha_i)-\left(Y_{it}-X_{it}^\top\alpha_i\right) \mathbbm{1}\left(Y_{it} \leq X_{it}^\top\alpha_i\right).
\end{align*}
Given $\widehat{A}_\tau$ obtained from the first step, the ES parameters are estimated as
\begin{align}\label{eq:step 2 ES}
	\begin{aligned}
	(\widehat{B}_\tau,\widehat{\theta}_\tau)=&\argmin_{B_\tau\in\mathbb{R}^{N\times (p+1)},\theta_\tau\in\Theta^{r_\tau^0}}\frac{1}{NT}\sum_{i=1}^N\sum_{t=1}^TS_0^2(\widehat{\alpha}_{i,\tau},\beta_{i,\tau},f_{t,\tau},\lambda_{i,\tau};Y_{it},X_{it})\\
	=&\argmin_{B_\tau\in\mathbb{R}^{N\times (p+1)},\theta_\tau\in\Theta^{r_\tau^0}}\frac{1}{NT}\sum_{i=1}^N\sum_{t=1}^T\left[Z_{it}(\widehat{\alpha}_{i,\tau})-\tau(X_{it}^\top\beta_{i,\tau}+\lambda_{i,\tau}^\top f_{t,\tau})\right]^2,
	\end{aligned}
\end{align}
where 
\begin{align}\label{eq:Z}
	Z_{it}(\alpha_{i,\tau})=\left(Y_{it}-X_{it}^\top\alpha_{i,\tau}\right) \mathbbm{1}\left(Y_{it} \leq X_{it}^\top\alpha_{i,\tau}\right)+\tau X_{it}^\top\alpha_{i,\tau},
\end{align}
for each $i\in[N]$ and $t\in[T]$.

\end{itemize}
The rationale for the estimation procedure in (\ref{eq:step 2 ES}) lies in the following considerations. 
Define \begin{equation*}\small
	\begin{aligned}
		\psi_0(\alpha_{i,\tau}, \beta_{i,\tau}, \lambda_{i,\tau}; X_{it},f_{t,\tau}) =E\left(Z_{it}(\alpha_{i,\tau}) \mid X_{it},f_{t,\tau}\right) -\tau \left(X_{it}^\top \beta_{i,\tau}+\lambda_{i,\tau}^\top f_{t,\tau}\right),
	\end{aligned}
\end{equation*}
which satisfies $\psi_0(\alpha_{i,\tau}^0, \beta_{i,\tau}^0, \lambda_{i,\tau}^0; X_{it},f_{t,\tau}^0) =0$. Let $F_{Y_{it}|X_{it}}$ be the conditional distribution function of $Y_{it}$ given $X_{it}$. Provided that $F_{Y_{it}|X_{it}}$ is continuously 
differentiable, it can be shown that for any $\beta_{i,\tau}$,
\begin{align*}
	\partial_{\alpha_{i,\tau}} \psi_0(\alpha_{i,\tau}, \beta_{i,\tau}, \lambda_{i,\tau}; X_{it},f_{t,\tau})\mid_{\alpha_{i,\tau}=\alpha_{i,\tau}^0}  =\left\{\tau-F_{Y_{it} \mid X_{it}}\left(X_{it}^\top \alpha_{i,\tau}^0\right)\right\} X_{it}=0,
\end{align*}
thereby satisfying the Neyman orthogonality condition.
Therefore, the quantile regression estimation error is first-order negligible in the ES regression estimation and thus does not affect the asymptotic distribution of the ES estimators.

Since both $\lambda_{i,\tau}$ and $f_{t,\tau}$ are unobservable in (\ref{eq:step 2 ES}), we adopt an iterative procedure similar to that of \cite{bai2009panel} to estimate $\theta_\tau$ and $B_\tau$. For a given $F_\tau$,
define the projection matrix
\begin{align*}
	M_{F_\tau}=I_T-F_\tau\left(F_\tau^\top F_\tau\right)^{-1}F_\tau^\top=I_T-F_\tau F_\tau^\top/T.
\end{align*}
For each cross-section unit $i$, stack its regressors and ``pseudo-response'' as 
\begin{align*}
	X_i=(X_{i1},...,X_{iT})^\top,\ \ Z_i^\ast(\widehat{\alpha}_{i,\tau})=(Z^\ast_{i1}(\widehat{\alpha}_{i,\tau}),...,Z^\ast_{iT}(\widehat{\alpha}_{i,\tau}))^\top,
\end{align*}
where $Z_{it}^\ast(\widehat{\alpha}_{i,\tau})=\tau^{-1}Z_{it}(\widehat{\alpha}_{i,\tau})$.
Given $\widehat{\alpha}_{i,\tau}$ and $F_\tau$, the least squares estimator for $\beta_{i,\tau}$ is
\begin{align*}
	\widehat{\beta}_{i,\tau}=\left(X_i^\top M_{F_\tau}X_i\right)^{-1}\left(X_i^\top M_{F_\tau}Z_i^\ast(\widehat{\alpha}_{i,\tau})\right).
\end{align*}
Having updated $\widehat{\beta}_{i,\tau}$, form the residuals
\begin{align*}
	\widehat{W}_{it,\tau}=Z_{it}^\ast(\widehat{\alpha}_{i,\tau})-X_{it}^\top\widehat{\beta}_{i,\tau},\quad\widehat{W}_\tau=(\widehat{W}_{it,\tau})_{N\times T}.
\end{align*}
We then obtain updated factor estimates by applying Principal Components Analysis (PCA) to $\widehat{W}_\tau^\top \widehat{W}_\tau/(TN)$. Let $\widehat{F}_\tau$ be $\sqrt{T}$ times 
the eigenvectors corresponding to the largest 
$r_\tau^0$ eigenvalues of 
$\widehat{W}_\tau^\top \widehat{W}_\tau/(TN)$, and the  corresponding loading matrix be
$\widehat{\Lambda}_\tau=\widehat{W}_\tau\widehat{F}_\tau/T$.
The final solution $(\hat{B}_\tau,\hat{\theta}_\tau)$
can be obtained by iteration. The two-stage estimation procedure is summarized in Algorithm \ref{alg:ES_factor}.

\begin{algorithm}[H]
\caption{Two-Stage Estimation of the ES Factor Model}
\label{alg:ES_factor}
\begin{algorithmic}[1]

\Require Quantile level $\tau$, number of factors $r_\tau^0$, 
initial factor estimator $\widehat{F}_\tau$.

\State Estimate $\widehat{A}_\tau$ via quantile regression in \eqref{eq:step 1 QR}.

\State Compute $Z_{it}(\widehat{\alpha}_{i,\tau})$ as in \eqref{eq:Z}$,$ and define
\[
Z_i^\ast(\widehat{\alpha}_{i,\tau})
=
\frac{1}{\tau}
\left(
Z_{i1}(\widehat{\alpha}_{i,\tau}),\ldots,
Z_{iT}(\widehat{\alpha}_{i,\tau})
\right)^\top.
\]

\State Update $\widehat{\beta}_{i,\tau}$ by
\[
\widehat{\beta}_{i,\tau}
=
\left(X_i^\top M_{\widehat{F}_\tau}X_i\right)^{-1}
X_i^\top M_{\widehat{F}_\tau}
Z_i^\ast(\widehat{\alpha}_{i,\tau}).
\]

\State Compute
\[
\widehat{W}_{it,\tau}
=
Z_i^\ast(\widehat{\alpha}_{i,\tau})
-
X_{it}^\top \widehat{\beta}_{i,\tau},
\]
form $\widehat{W}_\tau=(\widehat{W}_{it,\tau})_{N\times T}$,
and update $\widehat{F}_\tau$ as $\sqrt{T}$ times the eigenvectors associated with the largest $r_\tau^0$ eigenvalues of
$\widehat{W}_\tau^\top \widehat{W}_\tau/(TN)$.

\State Repeat Steps 3–4 until convergence. 
Compute the loading matrix
$
\widehat{\Lambda}_\tau
=
\widehat{W}_\tau \widehat{F}_\tau/T.
$

\Ensure Final estimators 
$\widehat{A}_\tau$, $\widehat{B}_\tau$, and $\widehat{\theta}_\tau$.

\end{algorithmic}
\end{algorithm}

To construct an initial factor estimator, we proceed as follows.
First, obtain $\widehat{\beta}_{i,\tau}^{(0)}$ by ordinary least squares from
$
Z_{it}^\ast(\widehat{\alpha}_{i,\tau})
=
X_{it}^\top \beta_{i,\tau}.
$
Let
$
\widehat{W}_{it,\tau}^{(0)}
=
Z_{it}^\ast(\widehat{\alpha}_{i,\tau})
-
X_{it}^\top \widehat{\beta}_{i,\tau}^{(0)},
$
and denote $\widehat{W}_\tau^{(0)}=(\widehat{W}_{it,\tau}^{(0)})_{N\times T}$.
The initial factor estimator $\widehat{F}_\tau$ is then given by
$\sqrt{T}$ times the eigenvectors corresponding to the largest
$r_\tau^0$ eigenvalues of
$\widehat{W}_\tau^{(0)\top} \widehat{W}_\tau^{(0)}/(TN)$.

%one can step the iteration based on $N^{-1}\sum_{i=1}^N\|\widehat{\beta}_{i,\tau}^{\text{new}}-\widehat{\beta}_{i,\tau}^{\text{old}}\|^2+(NT)^{-1}\sum_{i=1}^N\sum_{t=1}^T\left[(\widehat{\lambda}_{i,\tau}^\top\widehat{f}_{t,\tau})^{\text{new}}-(\widehat{\lambda}_{i,\tau}^\top\widehat{f}_{t,\tau})^{\text{old}}\right]^2<\omega$, where $\omega$ is small positive constant.
\subsection{Determining the Number of Factors}\label{subsec:determine}
In practice, the number of latent factors $r_{\tau}^0$ in model \eqref{eq:ES models} is unknown and must be determined before implementing the two-stage ES estimation procedure. Although several methods exist for selecting the number of factors in panel mean models (e.g., \citealp{bai2002determining,ahn2013eigenvalue}), they are designed for models where the response variable is directly observed and follows a mean regression structure. Within the ES factor model framework, 
the target variable $Z_{it}^{\ast}(\widehat{\alpha}_{i,\tau})$ 
is constructed based on the first-stage quantile estimates, 
while the factor structure is introduced only at the second stage. As a result, conventional Eigenvalue‑Ratio or Information‑Criterion methods are not directly applicable, since their theoretical validity depends on assumptions that may fail in our framework.

In this paper, we propose a modified Information Criterion (IC) tailored to the  ES factor model. Let $\widehat{V}_\tau(r)$ denote the mean squared residual 
from the ES model, defined as:
\begin{equation*}\label{eq:Vr}
\widehat{V}_\tau(r)=\frac{1}{NT}\sum_{i=1}^{N}\sum_{t=1}^{T}
\left[Z_{it}^{*}(\widehat{\alpha}_{i,\tau})-X_{it}^{\top}\widehat{\beta}_{i,\tau}(r)-\widehat{\lambda}_{i,\tau}(r)^{\top}\widehat{f}_{t,\tau}(r)\right]^{2},
\end{equation*}
where $\widehat{\beta}_{i,\tau}(r)$, 
$\widehat{\lambda}_{i,\tau}(r)$, 
and $\widehat{f}_{t,\tau}(r)$ 
are the estimators obtained from Algorithm \ref{alg:ES_factor} for a given number of factors $r$. Our information criterion is defined as
\begin{equation}\label{eq:IC}
\mathrm{IC}_{\tau}(r)=\log\widehat{V}_\tau(r)+r\cdot q(N,T),
\end{equation}
where $q(N,T)$ is a penalty function that depends on the panel dimensions. 
Following \cite{bai2002determining},
we specify
\begin{equation*}\label{eq:penalty}
q(N,T)=\log\left(\frac{NT}{N+T}\right)\left(\frac{N+T}{NT}\right).
\end{equation*}
The estimated number of factors is  selected as
\begin{equation*}\label{eq:rhat}
\widehat{r}_{\tau}= \underset{0\le r\le R_{\max}}{\arg\min}\; \mathrm{IC}_{\tau}(r),
\end{equation*}
where $R_{\max}$ is a prescribed
maximum number of factors.

\section{Theoretical Results}\label{sec:Theoretical Results}
In this section, we investigate the theoretical properties of the proposed estimators. We begin by establishing the asymptotic properties, including consistency, convergence rates, and asymptotic normality, under a set of regularity conditions. We then develop non‑asymptotic theory that provides explicit finite‑sample error bounds and a Gaussian approximation result, bridging the gap between asymptotic theory and practical inference. 

\subsection{Limiting Theory}\label{subsec:Limiting Theory}
%This section establishes the asymptotic properties of the proposed estimators, including the average consistency of the regression coefficients and factor loadings, the consistency of the estimated factor space, and the convergence rate of the ES regression coefficients. The results justify the statistical validity of the two-stage estimation procedure.
To analyze the asymptotic properties of the proposed estimators, we introduce the following concentrated objective function:
\begin{equation}\nonumber
    \begin{aligned}
        S_{NT}(B_{\tau}, F_{\tau}) =& \frac{1}{NT} \sum_{i=1}^{N} \left[Z_i^\ast(\widehat{\alpha}_{i,\tau})- X_i \beta_{i,\tau}\right]^\top M_{F_\tau}\left[Z_i^\ast(\widehat{\alpha}_{i,\tau})- X_i \beta_{i,\tau}\right] \\
        &-\frac{1}{NT} \sum_{i=1}^{N}\left[Z_i^\ast(\widehat{\alpha}_{i,\tau})- Z_i^\ast(\alpha^0_{i,\tau})\right]^\top M_{F_\tau^0}\left[Z_i^\ast(\widehat{\alpha}_{i,\tau})- Z_i^\ast(\alpha^0_{i,\tau})\right]\\
        &-\frac{2}{NT} \sum_{i=1}^{N}\left[Z_i^\ast(\widehat{\alpha}_{i,\tau})- Z_i^\ast(\alpha^0_{i,\tau})\right]^\top M_{F_\tau^0}\varepsilon_{i,\tau}- \frac{1}{NT} \sum_{i=1}^{N} \varepsilon_{i,\tau}^\top M_{F_{\tau}^0} \varepsilon_{i,\tau}.
    \end{aligned}
\end{equation}
%The function $S_{NT}(B_{\tau}, F_{\tau})$ serves as a pivotal theoretical construct for establishing the asymptotic properties of our estimators. 
The first term of function $S_{NT}(B_{\tau}, F_{\tau})$,
$$
\frac{1}{NT} \sum_{i=1}^{N} \left[Z_i^\ast(\widehat{\alpha}_{i,\tau})- X_i \beta_{i,\tau}\right]^\top M_{F_\tau}\left[Z_i^\ast(\widehat{\alpha}_{i,\tau})- X_i \beta_{i,\tau}\right],
$$
represents the sample analogue of the expected squared error after concentrating out the factor loadings $\Lambda_\tau$, where $M_{F_\tau} = I_T - P_{F_\tau}$ is the projection matrix that removes the factor space. If the true quantile coefficients $\alpha_{i,\tau}^0$ were known, this term alone would serve as the natural criterion for the joint estimation of $(B_\tau, F_\tau)$.

However, since $\alpha_{i,\tau}^0$ is unknown and replaced by the first-stage estimator $\widehat{\alpha}_{i,\tau}$, the remaining three terms are introduced as centering corrections. Importantly, these terms do not depend on $B_{\tau}$ or $F_{\tau}$ and therefore do not affect the minimization. Their role is to ensure that $S_{NT}(B_{\tau}, F_{\tau})$ is asymptotically equivalent to an infeasible objective function constructed with the true $\alpha_{i,\tau}^0$, which facilitates the derivation of the asymptotic properties of the estimators. The estimators of
 $ B^0_{\tau} $ and $ F_\tau^0 $ are then defined as
$$
(\widehat{B}_\tau, \widehat{F}_\tau) =\operatorname*{arg\,min}_{B_\tau, F_\tau} S_{NT}(B_\tau, F_\tau).
$$
%As explained in the previous section, $(\widehat{\beta}_{i,\tau}, \widehat{F}_\tau) $ satisfies$$\widehat{\beta}_{i,\tau} = \left( X_i^\top M_{\widehat{F}_\tau} X_i \right)^{-1}X_i^\top M_{\widehat{F}_\tau} Z_i^\ast(\widehat{\alpha}_{i,\tau}),$$$$\left[ \frac{1}{NT} \sum_{i=1}^{N} \left\{Z_i^\ast(\widehat{\alpha}_{i,\tau}) - X_i \widehat{\beta}_{i,\tau}\right\}\left\{Z_i^\ast(\widehat{\alpha}_{i,\tau}) - X_i \widehat{\beta}_{i,\tau}\right\}^\top \right] \widehat{F}_\tau = \widehat{F}_\tau V_{NT}^\tau,$$where $ \widehat{F}_\tau $ is the the matrix that consists of the first $ r^0_\tau $ eigenvectors (multiplied by $ \sqrt{T} $) of the matrix $ (NT)^{-1} \sum_{i=1}^{N} \{Z_i^\ast(\widehat{\alpha}_{i,\tau}) - X_i \widehat{\beta}_{i,\tau}\}\{Z_i^\ast(\widehat{\alpha}_{i,\tau}) - X_i \widehat{\beta}_{i,\tau}\}^\top $ and where $ V_{NT}^\tau $ is a diagonal matrix that consists of the first $ r^0_\tau $ largest eigenvalues of this matrix. 
%\subsection{Assumptions}\label{subsec:Assumptions}
With the objective function in place, we now impose a set of assumptions that are required for establishing the asymptotic properties of these estimators.

\begin{assumption}[Moment and Mixing Conditions]\label{assum:indentification}
\
\begin{enumerate}[label=(\roman*)]
    \item There exists a positive constant $M$ such that for all $i\in[N]$ and $t\in[T]$, 
$E\|X_{it}\|^4 \le M$.
    \item For each $i\in[N]$, the process $\{(X_{it}, Y_{it})\}_{t=1}^\infty$ is $\alpha$-mixing with mixing coefficients $\alpha(h)$ satisfying $\sum_{h=1}^\infty \alpha(h) < \infty$.
\end{enumerate}
\end{assumption}
Assumption \ref{assum:indentification} imposes standard moment and dependence conditions on the regressors. The bounded fourth moment ensures the validity of law of large numbers and central limit theorems in our asymptotic framework. The $\alpha$-mixing condition controls temporal dependence, guaranteeing that these limit theorems apply to the time series dimension. Together, these conditions ensure the convergence of sample moments to their population counterparts, which underlies consistent estimation.
\begin{assumption}[Factor and Loading Structure]\label{assum:factor&loading}
For a positive constant $M$,
\begin{enumerate}[label=(\roman*)]
    \item  $ E \| f_{t,\tau}^0 \|^4 \leq M $ and $ T^{-1} \sum_{t=1}^T f_{t,\tau}^0 f_{t,\tau}^{0\top} = I_{r_\tau^0} $.
    \item $  \| \lambda_{i,\tau}^0 \| \leq M $ and $ \Lambda_\tau^{0\top}   \Lambda_\tau^0 / N \longrightarrow \Sigma_{\Lambda_\tau} $ for some $ r^0_\tau \times r^0_\tau $ positive definite matrix $ \Sigma_{\Lambda_\tau} $, as $ N \to \infty $.
\end{enumerate}
\end{assumption}
Assumption \ref{assum:factor&loading} characterizes the properties of the latent factor structure. The moment conditions ensure that factors and loadings are well-behaved, while the convergence requirements guarantee that the factor space is non-degenerate and identifiable. These conditions are standard in factor model literature \citep{bai2002determining,bai2003inferential,bai2009panel,He2025huber} and ensure that the common components can be consistently estimated via PCA. 

Since $E[Z_{it}^\ast(\alpha_{i,\tau}^0) \mid X_{it}] = X_{it}^\top \beta_{i,\tau}^0 + \lambda_{i,\tau}^{0\top} f_{t,\tau}^0$, we define the regression error as $\varepsilon_{it,\tau} = Z_{it}^\ast(\alpha_{i,\tau}^0) - X_{it}^\top \beta_{i,\tau}^0 - \lambda_{i,\tau}^{0\top} f_{t,\tau}^0$. We then impose the following conditions on the error process $\{\varepsilon_{it,\tau}\}$.
\begin{assumption}[Error Structure]\label{assum:error}
For a positive constant $M$,
\begin{enumerate}[label=(\roman*)]
\item $ E(\varepsilon_{it,\tau}) = 0 $ and $ E |\varepsilon_{it,\tau}|^8 \leq M $.
    \item $ E(\varepsilon_{it,\tau} \varepsilon_{js,\tau}) = \sigma^\tau_{ij,ts}$, $|\sigma^\tau_{ij,ts}| \leq \delta_{ij}^\tau $ for all $ (t, s) $ and $ |\sigma^\tau_{ij,is}| \leq \eta^\tau_{ts} $ for all $ (i, j) $ such that
$$
\frac{1}{N} \sum_{i=1}^N\sum_{j=1}^{N} \delta^\tau_{ij} \leq M, \quad \frac{1}{T} \sum_{t=1}^T\sum_{s=1}^{T} \eta^\tau_{ts} \leq M, \quad \frac{1}{NT} \sum_{i=1}^N\sum_{j=1}^{N}\sum_{t=1}^T\sum_{s=1}^T |\sigma^\tau_{ij,ts}| \leq M.
$$
The largest eigenvalue of $ \Omega_{i,\tau} = E(\varepsilon_{i,\tau}\varepsilon_{i,\tau}^\top) $ is uniformly  bounded over $ i $ and $ T $, where $\varepsilon_{i,\tau}=(\varepsilon_{i1,\tau},\ldots,\varepsilon_{iT,\tau})^\top$.
\item For every $ (t, s) $, $ E\left|N^{-1/2} \sum_{i=1}^{N} [\varepsilon_{is,\tau} \varepsilon_{it,\tau} - E(\varepsilon_{is,\tau} \varepsilon_{it,\tau})]\right|^4 \leq M $.
\item Moreover,
$$
T^{-2} N^{-1} \sum_{t=1}^T\sum_{s=1}^T\sum_{u=1}^T\sum_{v=1}^T \sum_{i=1}^N\sum_{j=1}^N \left|\text{cov}(\varepsilon_{it,\tau} \varepsilon_{is,\tau}, \varepsilon_{ju,\tau} \varepsilon_{jv,\tau})\right| \leq M,
$$
$$
T^{-1} N^{-2} \sum_{t=1}^T\sum_{s=1}^T \sum_{i=1}^N\sum_{j=1}^N\sum_{k=1}^N\sum_{\ell=1}^N \left|\text{cov}(\varepsilon_{it,\tau} \varepsilon_{jt,\tau}, \varepsilon_{ks,\tau} \varepsilon_{\ell s,\tau})\right| \leq M.
$$
\end{enumerate}
\end{assumption}
Assumption \ref{assum:error} specifies conditions on the error process $\{\varepsilon_{it,\tau}\}$. Parts (i)-(ii) are standard in high-dimensional panel literature, allowing for weak serial and cross-sectional correlation while ensuring that the covariance structure remains well-behaved. Parts (iii) and (iv) impose higher-order moment restrictions that are consistent with those in \cite{bai2009panel}, which are needed to control the complex dependence arising in our two-stage estimators. Specifically, part (iii) facilitates a central limit theorem for the interaction between errors, while part (iv) ensures the stability of the fourth-moment matrices, which appear in the asymptotic variance of the ES estimators.
\begin{assumption}[Weak Dependence between Factors and Errors]\label{assum:weak dependence} For a positive constant $M$,
\begin{enumerate}[label=(\roman*)]
    \item $\displaystyle E\left\|\frac{1}{\sqrt{N}}\sum_{i=1}^N\lambda_{i,\tau}^{0}\varepsilon_{it,\tau}\right\|^2\leq M$, $t\in[T]$, $\displaystyle E\left\|\frac{1}{\sqrt{T}}\sum_{t=1}^{T}f_{t,\tau}^{0}\varepsilon_{it,\tau}\right\|^2\leq M$, $i\in[N]$.
    \item $\displaystyle E\left\|\frac{1}{\sqrt{NT}}\sum_{i=1}^N\sum_{t=1}^{T}\lambda_{i,\tau}^{0}f_{t,\tau}^{0\top}\varepsilon_{it,\tau}\right\|^2\leq M$.
\end{enumerate}
\end{assumption}
Assumption \ref{assum:weak dependence} controls the interaction between the latent factor structure and the error term, ensuring that factors, loadings, and errors are not excessively correlated in aggregate. %Similar conditions are standard in the large-dimensional factor model literature; see, for example, \citet{bai2002determining,bai2003inferential}. 
These conditions prevent the factor structure from being contaminated by the idiosyncratic errors and guarantee that PCA can consistently recover the true factor space.
\begin{assumption}[Rank and Identification Conditions]\label{assum:eigenvalue}
\ 
\begin{enumerate}[label=(\roman*)]

\item There exist positive constants $C_1$  and $C_2$ such that for each $i$,
    \begin{equation}\nonumber
        0 < C_1 < \lambda_{\min}(T^{-1}(X_i, F_{\tau}^0)^\top(X_i, F_{\tau}^0)) < \lambda_{\max}(T^{-1}(X_i, F_{\tau}^0)^\top(X_i, F^0_{\tau})) < C_2 < \infty,
    \end{equation}
almost surely as $T\to\infty$.
\item  Define $ A_{i,\tau} = T^{-1}X_i^\top M_{F_\tau}X_i $, $ B_{i,\tau} = (\lambda_{i,\tau}^0\lambda_{i,\tau}^{0\top}) \otimes I_T $, $ C_{i,\tau} = T^{-1/2}\lambda_{i,\tau}^{0\top}\otimes(M_{F_\tau}X_i) $. Let $
\mathcal{F}_\tau = \{ F_\tau : F_\tau^\top F_\tau / T = I_{r^0_\tau} \}
$ denote the set of all $F_\tau$ with orthonormal columns. We assume
\begin{equation}\nonumber
\inf_{F_\tau \in \mathcal{F}_\tau} \lambda_{\min}
\left[
\frac{1}{N} \sum_{i=1}^{N} E_{i,\tau}(F_\tau)
\right] > 0,    
\end{equation}
where 
$
E_{i,\tau}(F_\tau) = B_{i,\tau} - C_{i,\tau}^\top A_{i,\tau}^{-1} C_{i,\tau},
$ and the infimum is taken for the fixed $\tau$.% of interest.

\end{enumerate}
\end{assumption}
Assumption \ref{assum:eigenvalue} provides the key identification conditions. Part (i) requires that the regressors and factors jointly form a full-rank design matrix, ruling out multicollinearity. Part (ii) is a generalized identification condition that accommodates the interactive fixed effects. It ensures that, even after projecting out the latent factors, the remaining variation in the regressors is sufficient to identify the parameters of interest. This condition extends the identification framework of \citet{bai2009panel} to the expected shortfall context and is essential for the consistency of our estimators. Similar conditions are also used in \citet{song2013asymptotic}, \citet{ando2015asset}, and \citet{ando2020quantile}.
%These assumptions collectively provide a framework for establishing consistency and asymptotic normality of the proposed estimators while accommodating the key features of financial panel data: cross-sectional dependence and temporal persistence.

The following proposition establishes the average consistency of $\widehat{\gamma}_{i,\tau}=(\widehat{\beta}_{i,\tau}^\top,\widehat{\lambda}_{i,\tau}^\top)^\top$ and $\widehat{f}_{t,\tau}$. In factor models, the parameters $(f_{t,\tau}^0, \lambda_{i,\tau}^0)$ are only identifiable up to an orthogonal rotation. Consequently, $\widehat{f}_{t,\tau}$ and $\widehat{\lambda}_{i,\tau}$ estimate some rotated version of the true factors and loadings. For notational convenience, we omit the rotation matrix throughout.
\begin{proposition}\label{pro:average consistency}
Under Assumptions \ref{assum:indentification}-\ref{assum:eigenvalue}, as $ N, T \to \infty $, the following statements hold:
\begin{enumerate}[label=(\roman*)]
    \item Let ${\gamma}_{i,\tau}^0=({\beta}_{i,\tau}^{0\top},{\lambda}_{i,\tau}^{0\top})^\top$, we have 
  $$
    \frac{1}{N}\sum_{i=1}^N\left\|\widehat{\gamma}_{i,\tau}-{\gamma}_{i,\tau}^0\right\|^2=o_p(1).
    $$
    
    \item The matrix $ F^{0\top}_\tau\widehat{F}_\tau/T $ is invertible and 
  $$
    \left\|P_{\widehat{F}_\tau} - P_{F_\tau^0}\right\|\overset{p}{\longrightarrow} 0,\quad\frac{1}{T}\left\|\widehat{F}_{\tau}-F_\tau^0\right\|^2=\frac{1}{T}\sum_{t=1}^T\left\|\widehat{f}_{t,\tau}-f_{t,\tau}^0\right\|^2=o_p(1).
    $$
    
\end{enumerate}
\end{proposition}
Proposition \ref{pro:average consistency} establishes the fundamental consistency of the proposed estimators. Part (i) shows that the estimated parameters $\widehat{\gamma}_{i,\tau}$, which combine the ES regression coefficients and factor loadings, converge to their true values in a mean-squared sense averaged across all cross-sectional units. Part (ii) shows that the estimated common factors $\widehat{F}_\tau$ consistently recover the true factor space spanned by $F_\tau^0$, even though the factors themselves are only identified up to a rotation. These results align with those in \cite{ando2020quantile} for quantile factor models and \cite{bai2009panel} for mean factor models, indicating that the proposed two-stage procedure preserves the desirable large-sample properties despite the initial quantile regression step.

Proposition \ref{pro:average consistency} ensures that the estimation errors diminish as the sample size grows. To further characterize the speed of convergence, the following theorem establishes the convergence rate of the ES coefficient estimators.
\begin{theorem}\label{theo:consistency}
Under Assumptions \ref{assum:indentification}--\ref{assum:eigenvalue}, as $N,T \to \infty$ with $T/N \to \rho > 0$, we have
 $ \sqrt{T}(\widehat{\beta}_{i,\tau} - \beta^0_{i,\tau}) = O_p(1) $, $i\in[N]$.
\end{theorem}
Theorem \ref{theo:consistency} establishes that $\widehat{\beta}_{i,\tau}$ converges to $\beta^0_{i,\tau}$ at a $\sqrt{T}$ rate. Such a rate is standard in panel settings, where the time dimension $T$ determines the precision of unit-specific coefficient estimates. The assumption $T/N \to \rho > 0$ ensures that the time and cross-sectional dimensions grow at the same pace, a condition commonly met in macroeconomic and financial panel data. A noteworthy feature is that the first-stage quantile regression does not impair this convergence rate, thanks to the Neyman orthogonality property embedded in the second-stage loss function.

Regarding the selection of the number of factors, the information criterion \eqref{eq:IC} consistently estimates the true number $r_{\tau}^{0}$, as formalized in the following theorem.
\begin{theorem}\label{thm:selection}
Suppose that Assumptions \ref{assum:indentification}--\ref{assum:eigenvalue} hold. Under the information criterion $\mathrm{IC}_{\tau}(r)$ defined in \eqref{eq:IC} with a penalty function $q(N,T)$ satisfying
$$
q(N,T)\to 0\ \text{and}\ C_{NT}q(N,T)\to\infty,
$$
where $C_{NT} = \min{N, T}$, the estimator $\widehat{r}_{\tau}$ consistently selects the true number of factors. That is,
$$
\lim_{N,T\to\infty} P\left(\widehat{r}_{\tau}=r_{\tau}^{0}\right)=1.
$$
\end{theorem}
The intuition behind Theorem \ref{thm:selection} is as follows. When $r < r_{\tau}^{0}$, the model is underfitted and $\widehat{V}_{\tau}(r)$ exceeds $\widehat{V}_{\tau}(r_{\tau}^{0})$ by a non-negligible amount, making $\mathrm{IC}_{\tau}(r)$ asymptotically larger. When $r > r_{\tau}^{0}$, the model is overfitted. Although $\widehat{V}_{\tau}(r)$ is close to $\widehat{V}_{\tau}(r_{\tau}^{0})$, the additional penalty $r q(N,T)$ dominates and again inflates the criterion. The penalty function defined in \eqref{eq:penalty} satisfies the required rate conditions, guaranteeing that $\widehat{r}_{\tau}$ equals $r_{\tau}^{0}$ with probability approaching one.

\subsection{Asymptotic Normality}\label{subsec:Asymptotic Normality}
Building on the consistency and convergence rate results, we now establish the asymptotic normality of the ES regression coefficients. For each unit $i$, under appropriate relative rates for $T$ and $N$, the estimator admits the following representation:
\begin{equation}\nonumber
\begin{aligned}
\left( \frac{1}{T}X_i^\top M_{F^0_\tau} X_i \right)\sqrt{T}\left(\widehat{\beta}_{i,\tau}-\beta_{i,\tau}^0\right)
=&\frac{1}{\sqrt{T}N}\sum_{k=1}^N\left(X_i^\top M_{F^0_\tau}X_k a_{ik,\tau}\right)\sqrt{T}\left(\widehat{\beta}_{k,\tau}-\beta_{k,\tau}^0\right)\\
&+\frac{1}{\sqrt{T}} X_i^\top M_{{F}^0_\tau}\varepsilon_{i,\tau} - \frac{1}{\sqrt{T}N} \sum_{k=1}^{N} a_{ik,\tau} X_i^\top M_{{F}^0_\tau} \varepsilon_{k,\tau}+o_p(1),
\end{aligned}
\end{equation}
where $ a_{ik,\tau} = \lambda_{i,\tau}^\top (\Lambda^\top_\tau \Lambda_\tau / N)^{-1} \lambda_{k,\tau} $ captures cross-sectional dependence through the factor loadings. This representation has an intuitive interpretation. The left-hand side is the scaled estimation error for unit $i$, while the right-hand side consists of three components. The first term captures the \textit{cross-unit contamination effect}, describing how estimation errors from other units propagate to affect unit $i$'s estimator through the factor structure. The second and third terms together form the \textit{idiosyncratic score vector}. In the absence of cross-sectional dependence, these two terms would be the only sources of estimation uncertainty.

Multiplying both sides by the inverse of $T^{-1}X_i^\top M_{F^0_\tau} X_i$ and introducing compact notation
\begin{equation*}
\begin{gathered}
A_{i,\tau}=\left( \frac{1}{T}X_i^\top M_{F^0_\tau} X_i\right)^{-1},\quad 
H_{ik,\tau}=A_{i,\tau}\left(\frac{1}{T}X_i^\top M_{F^0_\tau}X_k a_{ik,\tau}\right), \\
U_{i,\tau}=\frac{1}{\sqrt{T}} X_i^\top M_{{F}^0_\tau}\varepsilon_{i,\tau} - \frac{1}{\sqrt{T}N} \sum_{k=1}^{N} a_{ik,\tau} X_i^\top M_{{F}^0_\tau} \varepsilon_{k,\tau},
\end{gathered}
\end{equation*}
we obtain the unit-level equation
\begin{equation}\label{eq:compact_form}
\sqrt{T}\left(\widehat{\beta}_{i,\tau}-\beta_{i,\tau}^0\right)=\frac{1}{N}\sum_{k=1}^N H_{ik,\tau}\sqrt{T}\left(\widehat{\beta}_{k,\tau}-\beta_{k,\tau}^0\right)+A_{i,\tau} U_{i,\tau}+o_p(1).
\end{equation}
This reveals a \textit{simultaneous equations system} where each unit's estimation error depends linearly on all other units' errors. To analyze this system globally, we stack the equations across all units. Defining
\begin{equation*}
\begin{gathered}
\sqrt{T}(\widehat{\beta}_\tau-\beta^0_\tau)=\left(\sqrt{T}(\widehat{\beta}_{1,\tau}-\beta_{1,\tau}^0)^\top,\ldots,\sqrt{T}(\widehat{\beta}_{N,\tau}-\beta_{N,\tau}^0)^\top\right)^\top, \\
H_\tau=[H_{ik,\tau}]_{i,k\in[N]},\quad A_\tau=\operatorname{diag}(A_{1,\tau},\ldots,A_{N,\tau}),\quad U_\tau=(U_{1,\tau}^\top,\ldots,U_{N,\tau}^\top)^\top,
\end{gathered}
\end{equation*}
the system of $N$ equations in (\ref{eq:compact_form}) can now be written compactly as
\begin{equation}\label{eq:system_matrix}
\sqrt{T}\left(\widehat{\beta}_\tau-\beta^0_\tau\right)=\frac{1}{N}H_\tau\sqrt{T}(\widehat{\beta}_\tau-\beta^0_\tau)+A_\tau U_\tau+o_p(1).
\end{equation}
Rearranging terms yields the key equation, that is
\begin{equation}\nonumber
\left(I-\frac{1}{N}H_\tau\right)\sqrt{T}(\widehat{\beta}_\tau-\beta^0_\tau)=A_\tau U_\tau+o_p(1).
\end{equation}
The matrix $I-N^{-1}H_\tau$ plays a crucial role in the asymptotic analysis. Its invertibility ensures that the propagation of estimation errors remains controlled, allowing us to solve the system explicitly. This motivates the following assumptions.
\begin{assumption}[Cross-Sectional Dependence Control]\label{assum:cross-sectional dependence}
\ 
\begin{enumerate}[label=(\roman*)]
\item There exists a constant $ c > 0 $ such that
$$
\lambda_{\min}\left( I - \frac{1}{N} H_\tau \right) \geq c > 0
$$
in probability approaching one.
\item For each $ i\in[N]$,
$$
\sum_{j=1}^N \| G_{ij,\tau} \| = O_p(1)
$$
where $G_{ij,\tau}$ is the $(i,j)$-th block of $ G_\tau = (I - N^{-1}H_\tau)^{-1} $.
\end{enumerate}
\end{assumption}
Assumption \ref{assum:cross-sectional dependence} ensures that the system of simultaneous equations linking the estimation errors across units is well-behaved. Part (i) guarantees that the matrix $I - N^{-1}H_\tau$ is invertible with eigenvalues bounded away from zero, preventing the uncontrolled propagation of estimation errors across cross-sectional units. Part (ii) controls the aggregate influence of other units' estimation errors on a given unit, ensuring that the asymptotic variance remains bounded as $N$ grows.

%To establish asymptotic normality, we need to replace sample-dependent weights by their probability limits. This requires uniform convergence across all cross-sectional units.

\begin{assumption}[Uniform Convergence of Sample Moments]\label{assum:uniform convergence}
    Assume that for all $i,\ j\in[N]$,
$$
\frac{1}{T} X_i^\top M_{F^0_\tau} X_j\overset{p}{\longrightarrow}\Sigma_{X_i,X_j},
$$
where $\Sigma_{X_i,X_j}=\lim_{T\to\infty}E(T^{-1}X_i^\top M_{F^0_\tau} X_j)$. Moreover, as $N,\ T \to \infty$, the following uniform convergence properties hold:
    \begin{enumerate}[label=(\roman*)]
        \item 
    $
    \displaystyle\max_{1 \le j \le N} \left\| \frac{1}{T} X_j^\top M_{F^0_\tau} X_j - \Sigma_{X_j,X_j} \right\| \overset{p}{\longrightarrow} 0.
    $
    \item 
    $\displaystyle
    \max_{1 \le j \le N} \left\| A_{j,\tau} - A_{j,\tau}^0 \right\| \overset{p}{\longrightarrow} 0.
    $
    where $A_{j,\tau}^0=\Sigma_{X_j,X_j}^{-1}$.
    \item 
    Let $H_\tau^0 = [H_{ij,\tau}^0]$ be the probability limit of $H_\tau = [H_{ij,\tau}]$, where 
    $$
    H_{ij,\tau}^0 = A_{i,\tau}^0 \Sigma_{X_i,X_j} a_{ij,\tau}^0,
    $$
    with $a_{ij,\tau}^0 = \lambda_{i,\tau}^{0\top} \Sigma_{\Lambda_\tau}^{-1} \lambda_{j,\tau}^0$. Define $G_\tau^0 = (I - N^{-1}H_\tau^0)^{-1}$ and let $G_{ij,\tau}^0$ be its $(i,j)$-th block. Then
    $$
    \max_{1 \le j \le N} \left\| G_{ij,\tau} A_{j,\tau} - G_{ij,\tau}^0 A_{j,\tau}^0 \right\| \overset{p}{\longrightarrow} 0.
    $$
    \end{enumerate}
\end{assumption}
Assumption \ref{assum:uniform convergence} strengthens the convergence requirements from pointwise to uniform across all cross-sectional units. This is necessary because the asymptotic distribution involves weighted sums over all units, and we need to ensure that estimation errors in the weights do not affect the limiting distribution. These conditions guarantee that the sample moments, their inverses, and the complex weighting matrices all converge uniformly to their population counterparts.

%The final component required for establishing asymptotic normality is the limiting behavior of the score vectors $U_{i,\tau}$.
\begin{assumption}[Asymptotic Normality of Score Vectors]\label{assum:Asymptotic Normality}
Define  
$$
U_{i,\tau}=\frac{1}{\sqrt{T}} \left( X_i^\top M_{F^0_\tau} \varepsilon_{i,\tau} - \frac{1}{N} \sum_{k=1}^{N} a_{ik,\tau} X_k^\top M_{F^0_\tau} \varepsilon_{k,\tau} \right).
$$
\begin{enumerate}[label=(\roman*)]
    \item For any fixed set of indices $i_1, \dots, i_m \in [N]$,
$$
\left( U_{i_1,\tau}^\top, \dots, U_{i_m,\tau}^\top \right)^\top \xrightarrow{d} \mathcal{N}\left(0, 
\begin{bmatrix}
\Sigma_{i_1 i_1,\tau} & \cdots & \Sigma_{i_1 i_m,\tau} \\
\vdots & \ddots & \vdots \\
\Sigma_{i_m i_1,\tau} & \cdots & \Sigma_{i_m i_m,\tau}
\end{bmatrix} \right),
$$
where $\Sigma_{ij,\tau} = \lim_{T\to\infty} \mathrm{Cov}(U_{i,\tau}, U_{j,\tau} \mid X_1,\ldots,X_N, F^0_\tau)$. In particular, for the full vector $U_\tau = (U_{1,\tau}^\top, \dots, U_{N,\tau}^\top)^\top$,
$$
U_\tau \xrightarrow{d} \mathcal{N}(0, \Sigma_\tau),
$$
where $\Sigma_\tau$ is an $N(p+1) \times N(p+1)$ block matrix with $(i,j)$-th block $\Sigma_{ij,\tau}$.
\item There exists a constant $ M < \infty $ such that
$$
\sup_{1 \le j \le N} \sum_{k=1}^N \left| \mathrm{Cov}(U_{j,\tau}, U_{k,\tau} \mid X_1,\ldots,X_N, F^0_\tau) \right| \leq M.
$$
\end{enumerate}
\end{assumption}
Assumption \ref{assum:Asymptotic Normality} provides the fundamental conditions for the central limit theorem. The vector $U_{i,\tau}$ incorporates both the direct estimation error and the indirect effect through other units. Part (i) assumes joint asymptotic normality of these scores, while part (ii) imposes a weak cross-sectional dependence condition, ensuring that the covariance structure is sufficiently sparse for the central limit theorem to hold as $N$ grows.

%With these additional assumptions, we now state the main asymptotic normality result.
Building on the assumptions above, the following theorem establishes the asymptotic normality of the proposed ES regression coefficient estimator.
\begin{theorem}\label{theo:asymptotic}
Under Assumptions \ref{assum:indentification}-\ref{assum:Asymptotic Normality}, for any cross-sectional unit $i\in[N]$, as $N, T \to \infty$,
$$
\sqrt{T}\left(\widehat{\beta}_{i,\tau}-\beta_{i,\tau}^0\right) \overset{d}{\longrightarrow} \mathcal{N}(0, \Omega_{i,\tau}),
$$
where the asymptotic variance matrix $\Omega_{i,\tau}$ is given by
$$
\Omega_{i,\tau} = \sum_{j=1}^N \sum_{k=1}^N (G_{ij,\tau}^0 A_{j,\tau}^0) \, \Sigma_{jk,\tau} \, (G_{ik,\tau}^0 A_{k,\tau}^0)^\top.
$$
\end{theorem}
Theorem \ref{theo:asymptotic} establishes the asymptotic normality of the ES regression coefficients. The asymptotic variance $\Omega_{i,\tau}$ admits a natural interpretation: $A_{j,\tau}^0$ is the unit-specific precision, $\Sigma_{jk,\tau}$ captures the covariance between score vectors, and $G_{ij,\tau}^0$ encodes how estimation errors propagate through the factor structure. This variance structure reflects two sources of uncertainty: sampling variability and cross-sectional dependence. Notably, the theorem shows that despite the two-stage estimation and interactive effects, the ES coefficients are $\sqrt{T}$-consistent and asymptotically normal, providing a theoretical foundation for valid inference.
\subsection{Non-Asymptotic Theory}\label{subsec:Non-asymptotic Theory}
Having established the asymptotic theory, we now develop explicit finite-sample theory for the two-step ES factor model estimator. We provide high-probability error bounds that clarify the interplay among the time dimension $T$, cross-section dimension $N$, covariate dimension $p$, and the distributional features of the data.

We first derive finite-sample error bounds for the first-stage panel quantile regression estimators, which serve as the foundation for the second-stage ES regression. Let $\Sigma_i = E[X_{it}X_{it}^\top]$ denote the population covariance matrix for unit $i\in[N]$, and let $\epsilon_{it} = Y_{it} - X_{it}^\top \alpha_{i,\tau}^0$ be the quantile regression error. We impose the following assumptions.
\begin{assumption}[Regularity of the Conditional Error Distribution]\label{assum:Regularity}
Given $X_{it}$, the conditional distribution function $F_{\epsilon|X}(\cdot \mid X_{it})$ of $\epsilon_{it}$ admits a Lipschitz‑continuous conditional density $f_{\epsilon|X}(\cdot \mid X_{it})$ in a neighbourhood of zero. There exist constants $\underline{f}, L_0 > 0$ such that almost surely
$$
f_{\epsilon|X}(0 \mid X_{it}) \geq \underline{f} \quad \text{and} \quad |f_{\epsilon|X}(x \mid X_{it}) - f_{\epsilon|X}(0 \mid X_{it})| \leq L_0 |x|, \quad \forall x \in \mathbb{R}.
$$
\end{assumption}
\begin{assumption}[Uniformly Bounded Covariance Matrices]\label{assum:UniformBound}
There exist constants $0 < \kappa_{\min} \leq \kappa_{\max} < \infty$ such that for all cross‑sectional units $i\in[N]$
$$
\kappa_{\min} I_{p+1} \preceq \Sigma_i \preceq \kappa_{\max} I_{p+1}.
$$
\end{assumption}
\begin{assumption}[Sub‑Gaussianity of Standardized Covariates]\label{assum:Sub-Gaussianity}
Define the standardized covariate vector $W_{it} = \Sigma_i^{-1/2} X_{it} \in \mathbb{R}^{p+1}$, so that $E[W_{it}W_{it}^\top] = I_{p+1}$. There exists a constant $\upsilon_1 \geq 1$ such that for any unit vector $u \in \mathbb{S}^{p}$ (the unit sphere in $\mathbb{R}^{p+1}$),
$$
P\left( |u^\top W_{it}| \geq \upsilon_1 x \right) \leq 2e^{-x^2/2}, \quad \forall x \geq 0.
$$
\end{assumption}
Assumption \ref{assum:Regularity} is standard in quantile regression. It ensures that the conditional density of the error is bounded away from zero at the origin and does not change too rapidly, which guarantees sufficient curvature of the population objective function. Assumption \ref{assum:UniformBound} requires the eigenvalues of each unit's covariate covariance matrix to be uniformly bounded across units. This accommodates cross‑sectional heterogeneity while preventing ill‑conditioned designs. Assumption \ref{assum:Sub-Gaussianity} controls the tail behavior of the covariates. It is a common condition in high‑dimensional statistics that allows the covariate dimension $p$ to grow with sample size and facilitates concentration inequalities (e.g., \citealt{guo2016spline}).

For a parameter vector $u \in \mathbb{R}^{p+1}$, define the unit‑specific $\Sigma_i$-norm: $\|u\|_{\Sigma_i} = \|\Sigma_i^{1/2} u\|_2$. This norm naturally reflects the heterogeneity in covariate distributions across units.

\begin{proposition}\label{pro:quantile}
Under Assumptions \ref{assum:Regularity}--\ref{assum:Sub-Gaussianity}, for any confidence level $\delta \in (0,1)$, there exist constants $C_1, C_2 > 0$ depending only on $\upsilon_1$ such that for each cross‑sectional unit $i \in [N]$, the estimator $\widehat{\alpha}_{i,\tau}$ satisfies
$$
\|\widehat{\alpha}_{i,\tau} - \alpha_{i,\tau}^0\|_{\Sigma_i} \leq C_1 \underline{f}^{-1} \sqrt{\frac{p + \log(1/\delta)}{T}},
$$
with probability at least $1 - \delta$, provided $T \geq C_2 L_0^2 \underline{f}^{-4} (p + \log(1/\delta))$.
\end{proposition}
Proposition \ref{pro:quantile} provides a finite-sample error bound for the first-stage quantile regression estimator. With high probability, the estimation error for each unit $i\in[N]$ decays at rate $\sqrt{(p + \log(1/\delta))/T}$. The $\Sigma_i$-norm on the left-hand side accounts for heterogeneity in covariate distributions across units. Importantly, the constants are independent of $i$, so the bound holds uniformly for all cross-sectional units.
%Proposition \ref{pro:quantile} provides an explicit finite‑sample error bound for panel quantile regression. The estimation error decays at the rate $O_p\bigl(\sqrt{(p + \log(1/\delta))/T}\,\bigr)$, matching the optimal rate for cross‑sectional quantile regression and reflecting the complexity cost incurred by the parameter dimension in high‑dimensional settings. The bound holds with high probability $1-\delta$, offering a stronger foundation for subsequent hypothesis testing and confidence‑interval construction. The error is measured in the unit‑specific $\Sigma_i$-norm, naturally accommodating heterogeneity in covariate distributions; the constants $C_1$ and $C_2$ are independent of the unit index, indicating uniformity over the entire cross‑section. When $N=1$, the result reduces to the non‑asymptotic bound for a single cross‑sectional quantile regression; the main challenge in the panel setting lies in obtaining such uniform bounds across all units.

Next, we establish non‑asymptotic error bounds for the two‑step ES factor model estimators. We introduce two additional assumptions that build on the previous assumptions.
\begin{assumption}[Bounded Conditional Tail Variance]\label{assum:TailBehavior}
Define the negative part of the quantile regression error as $\epsilon_{it,-} = \min\{\epsilon_{it},\ 0\}$. There exists a constant $\overline{\sigma} > 0$ such that
$
\operatorname{Var}_{X_{it}}(\epsilon_{it,-}) \leq \overline{\sigma}^2
$ almost surely.
\end{assumption}
\begin{assumption}[Smoothness of the Conditional Distribution]\label{assum:Smoothness}
The conditional distribution function $F_{\epsilon|X}(\cdot \mid X_{it})$ of $\epsilon_{it}$ satisfies Lipschitz continuity, i.e., there exists a constant $\overline{f} > 0$ such that
$$
|F_{\epsilon|X}(x \mid X_{it}) - F_{\epsilon|X}(0 \mid X_{it})| \leq \overline{f} |x|, \quad \forall x \in \mathbb{R} \quad \text{almost surely}.
$$
\end{assumption}
Assumption \ref{assum:TailBehavior} controls the volatility of the left tail in the quantile regression and is crucial for the robustness of ES estimation. Assumption \ref{assum:Smoothness} ensures the validity of the second‑order expansion for the first‑stage estimation error and is key to the Neyman orthogonality property.
\begin{theorem}\label{theo:nonasymES}
Under Assumptions \ref{assum:Regularity}--\ref{assum:Smoothness}, for any confidence level $\delta \in (0, 1)$, there exist constants $C_1, C_2, C_3, C_4 > 0$ depending only on $(\upsilon_1, \kappa_{\min}, \kappa_{\max}, \overline{\sigma}, \overline{f})$ such that for each cross‑sectional unit $i\in[N]$, the two‑step estimator $\widehat{\beta}_{i,\tau}$ satisfies
\begin{equation}\nonumber
\begin{aligned}
    \tau \|\widehat{\beta}_{i,\tau} - \beta_{i,\tau}^0\|_{\Sigma_i} \leq& C_1 \overline{\sigma} \sqrt{\frac{p + \log(1/\delta)}{T}} + C_2 \left( \sqrt{\frac{p + \log(1/\delta)}{T}} \cdot c_0 + \overline{f} c_0^2 \right) \\
    &+ C_3 \left( \frac{1}{\sqrt{T}} + \frac{1}{\sqrt{N}} \right)(p + \log(1/\delta)),
\end{aligned}
\end{equation}
with probability at least $1 - 5\delta$, where $c_0 = \max_{1 \le i \le N} \|\widehat{\alpha}_{i,\tau} - \alpha_{i,\tau}^0\|_{\Sigma_i}$, and $T \ge C_4 (p + \log(1/\delta))$.
\end{theorem}
Theorem \ref{theo:nonasymES} provides a non-asymptotic error bound for the two-step ES estimator. The three terms on the right-hand side reflect three distinct sources of error: sampling variability in ES regression (first term), first-stage quantile estimation error (second term), and latent factor estimation error (third term). The first term decays at the parametric rate $1/\sqrt{T}$. The second term decays faster because $c_0 = O(\sqrt{(p+\log(1/\delta))/T})$; this faster decay is a consequence of Neyman orthogonality, which makes the first-stage error only a higher-order concern. %The third term captures the cost of estimating the factors. When there are no latent factors, that is $r_\tau^0 = 0$, this term vanishes and the bound reduces to the ES regression result of \cite{he2023robust}.
When there are no latent factors (i.e., $r_\tau^0 = 0$), the third term vanishes. In this case, the bound resembles the ES regression result for i.i.d. data established in \cite{he2023robust}, but adapted to a panel data setting with potential cross-sectional and temporal dependence.

Finally, we establish a finite-sample Gaussian approximation for the standardized ES regression coefficient estimator, which provides a theoretical foundation for inference. Define the standardized statistic
$$
T_{i,\tau} = \frac{\sqrt{T} \, a^\top (\widehat{\beta}_{i,\tau} - \beta_{i,\tau}^0)}{\sqrt{a^\top \Omega_{i,\tau} a}},
$$
where $a \in \mathbb{R}^{p+1}$ is an arbitrary unit vector.
\begin{theorem}\label{theo:ESnon-asy}
Under Assumptions \ref{assum:Regularity}--\ref{assum:Smoothness}, for any $i \in [N]$ and any unit vector $a$, we have
$$
\sup_{t\in\mathbb{R}} \Bigl| P\!\bigl( T_{i,\tau}\le t\mid X,F^0 \bigr)-{\Phi(t)} \Bigr|
\lesssim \frac{p+\log N+\log T}{\sqrt{T}} + \sqrt{\frac{\log N + \log T}{N}}.
$$
\end{theorem}
Theorem \ref{theo:ESnon-asy} provides a non-asymptotic bound for the normal approximation of the standardized ES estimator. The first term $(p + \log N + \log T)/\sqrt{T}$ captures the approximation error from time-series estimation, covariate dimension, and the complexity of controlling uniform deviations across $N$ units. The second term $\sqrt{\log (NT)}/\sqrt{N}$ comes from estimating the latent factors. The bound shrinks as $T$ and $N$ increase, making the normal approximation valid in large panels.

\section{Simulation Study}\label{sec:Simulation Study}

We conduct a Monte Carlo study to assess the finite-sample performance of the proposed ESFM under a variety of economically and statistically relevant environments. The design is organized to isolate (i) heavy-tailed innovations, (ii) cross-sectional heterogeneity in slope coefficients, and (iii) latent factors that primarily affect downside risk through time-varying tail dispersion. Throughout, we compare ESFM to a baseline ES regression that ignores latent factors, with particular emphasis on whether incorporating tail factors improves the accuracy of slope estimation and factor-space recovery.

\subsection{Simulation Design}\label{subsec:Simulation Design}

For $i=1,\ldots,N$ and $t=1,\ldots,T$, we generate a covariate vector $X_{it}\in\mathbb{R}^{p+1}$ that includes an intercept and $p=3$ observed regressors. The latent tail component is driven by a factor structure with true dimension $r_0=2$. To align with the two-stage ESFM estimation strategy, we construct the data so that the conditional $\tau$-quantile satisfies a standard panel quantile regression,
$Q_{\tau}(Y_{it}\mid X_{it}) = X_{it}^\top\alpha_i$, and the conditional expected shortfall follows a location--scale form in which latent factors enter through the tail scale:
\begin{equation}
Y_{it} = \mu_{it} + \sigma_{it}\varepsilon_{it},\qquad \mu_{it}=X_{it}^\top\alpha_i.
\label{eq:DGP_new}
\end{equation}
The innovation $\varepsilon_{it}$ is generated from a standardized Student-$t$ distribution with $\nu=5$ degrees of freedom and then shifted so that its unconditional $\tau$-quantile is approximately zero. This construction ensures that $Q_{\tau}(Y_{it}\mid X_{it})=\mu_{it}$ holds by design, while the magnitude of tail losses is governed by $\sigma_{it}$. The conditional ES is therefore $ES_{\tau}(Y_{it}\mid X_{it},\sigma_{it})=\mu_{it}+\sigma_{it}\,ES_{\tau}(\varepsilon_{it})$, so that latent factors affect tail risk through the time variation in $\sigma_{it}$, without mechanically entering the quantile equation.

We consider tail probability levels $\tau\in\{0.01,0.05,0.10\}$. Panel dimensions vary over $N\in\{100,200,300\}$ and $T\in\{100,200,300\}$. For each $(N,T,\tau)$ configuration and each scenario described below, we run $MC=100$ replications with independent random seeds. In estimation, we set the number of latent ES factors to match the data-generating value ($r_0=2$) when reporting the main results.

For each simulated dataset, we estimate:
(i) an ES regression that ignores latent factors (equivalently, ESFM with $r=0$), and
(ii) the proposed ESFM with $r=r_0$ factors.
Performance is summarized by (a) mean squared errors of the slope coefficients $\beta_i$ (averaged over $i$), (b) factor-space recovery measured by the projection-matrix error $\|P_{\widehat F}-P_F\|_F^2/T$, (c) ES estimators, and (d) number of factors (included in the appendix). %The key comparison is whether ESFM delivers more accurate $\beta_i$ estimates than ES regression when tail factors are present.

To assess the performance of the proposed method under different economic environments, we consider seven data-generating scenarios. Scenario 1 introduces a baseline tail-factor structure; Scenario 2 strengthens tail-factor dependence; Scenario 3 incorporates heterogeneous slope coefficients; Scenario 4 allows for endogenous covariates; Scenario 5 emphasizes volatility-driven tail comovement; Scenario 6 introduces rare jump shocks; and Scenario 7 features asymmetric tail behavior. Together, these scenarios cover a range of empirically relevant mechanisms, including latent tail dependence, heterogeneity, endogeneity, and extreme risk dynamics. Detailed specifications are provided in the appendix.

\subsection{Simulation Results}\label{Simulation Results}

Table~\ref{tab:Estimation error of beta} reports the root mean squared errors (RMSEs) of the estimated covariate
coefficients $\beta_i$ under the expected shortfall regression (ESR) and the proposed ESFM. Results are reported for seven data-generating scenarios, three cross-sectional dimensions ($N=100,200,300$),
three time dimensions ($T=100,200,300$), and three tail probability levels $\tau\in\{0.10,0.05,0.01\}$.

\begin{table}[htbp]
\centering
\caption{Estimation error of $\beta_\tau$ in expected shortfall (ES) models. {\footnotesize{\textit{Notes}. This table reports the root mean squared error (RMSE) of the estimated covariate coefficients $\beta_\tau$ across Monte Carlo replications. Results are presented for seven data-generating scenarios, comparing the benchmark expected shortfall regression (ESR) and the proposed expected shortfall factor model (ESFM).}}}
\label{tab:Estimation error of beta}

\setlength{\tabcolsep}{2.5pt}       % Reduce column spacing
\renewcommand{\arraystretch}{0.95} % Reduce row spacing
\small                       % Font size
\begin{adjustbox}{max totalsize={\textwidth}{0.85\textheight},center}
\begin{tabular}{lcccccccccccccc}
\toprule
      & \multicolumn{2}{c}{Scenario 1} & \multicolumn{2}{c}{Scenario 2} & \multicolumn{2}{c}{Scenario 3} & \multicolumn{2}{c}{Scenario 4} & \multicolumn{2}{c}{Scenario 5} & \multicolumn{2}{c}{Scenario 6} & \multicolumn{2}{c}{Scenario 7} \\
      & ESR & ESFM & ESR & ESFM & ESR & ESFM & ESR & ESFM & ESR & ESFM & ESR & ESFM & ESR & ESFM \\
\midrule
\multicolumn{15}{c}{Panel A: $\tau = 0.10$} \\
\midrule
\multicolumn{15}{c}{$N$ = 100} \\
\midrule
$T=100$ & 0.4130 & 0.3683 & 0.4179 & 0.3722 & 0.4143 & 0.3740 & 0.4146 & 0.3685 & 0.4146 & 0.3706 & 0.4166 & 0.3705 & 0.4147 & 0.3716 \\
$T=200$ & 0.3559 & 0.3232 & 0.3536 & 0.3221 & 0.3557 & 0.3243 & 0.3545 & 0.3228 & 0.3525 & 0.3223 & 0.3523 & 0.3194 & 0.3556 & 0.3226 \\
$T=300$ & 0.3306 & 0.3057 & 0.3311 & 0.3065 & 0.3286 & 0.3032 & 0.3296 & 0.3098 & 0.3297 & 0.3028 & 0.3297 & 0.3052 & 0.3285 & 0.3028 \\
\midrule
\multicolumn{15}{c}{$N$ = 200} \\
\midrule
$T=100$ & 0.4179 & 0.3662 & 0.4141 & 0.3643 & 0.4143 & 0.3631 & 0.4175 & 0.3664 & 0.4142 & 0.3631 & 0.4115 & 0.3616 & 0.4148 & 0.3639 \\
$T=200$ & 0.3551 & 0.3223 & 0.3532 & 0.3193 & 0.3530 & 0.3191 & 0.3520 & 0.3185 & 0.3554 & 0.3206 & 0.3529 & 0.3184 & 0.3547 & 0.3211 \\
$T=300$ & 0.3295 & 0.3060 & 0.3299 & 0.3071 & 0.3273 & 0.3030 & 0.3307 & 0.3056 & 0.3294 & 0.3064 & 0.3284 & 0.3067 & 0.3296 & 0.3075 \\
\midrule
\multicolumn{15}{c}{$N$ = 300} \\
\midrule
$T=100$ & 0.4148 & 0.3640 & 0.4173 & 0.3650 & 0.4151 & 0.3645 & 0.4170 & 0.3626 & 0.4170 & 0.3634 & 0.4162 & 0.3635 & 0.4174 & 0.3641 \\
$T=200$ & 0.3533 & 0.3216 & 0.3530 & 0.3202 & 0.3529 & 0.3191 & 0.3550 & 0.3232 & 0.3529 & 0.3190 & 0.3537 & 0.3206 & 0.3541 & 0.3212 \\
$T=300$ & 0.3292 & 0.3144 & 0.3297 & 0.3099 & 0.3288 & 0.3123 & 0.3289 & 0.3089 & 0.3296 & 0.3128 & 0.3287 & 0.3111 & 0.3285 & 0.3122 \\
\midrule
\multicolumn{15}{c}{Panel B: $\tau = 0.05$} \\
\midrule
\multicolumn{15}{c}{$N$ = 100} \\
\midrule
$T=100$ & 0.4809 & 0.3603 & 0.4826 & 0.3622 & 0.4840 & 0.3632 & 0.4855 & 0.3604 & 0.4813 & 0.3612 & 0.4836 & 0.3596 & 0.4826 & 0.3621 \\
$T=200$ & 0.4160 & 0.3277 & 0.4131 & 0.3227 & 0.4133 & 0.3249 & 0.4129 & 0.3242 & 0.4090 & 0.3228 & 0.4134 & 0.3265 & 0.4163 & 0.3276 \\
$T=300$ & 0.3806 & 0.3087 & 0.3803 & 0.3088 & 0.3801 & 0.3087 & 0.3779 & 0.3061 & 0.3816 & 0.3068 & 0.3810 & 0.3084 & 0.3805 & 0.3071 \\
\midrule
\multicolumn{15}{c}{$N$ = 200} \\
\midrule
$T=100$ & 0.4828 & 0.3551 & 0.4817 & 0.3553 & 0.4855 & 0.3550 & 0.4865 & 0.3585 & 0.4784 & 0.3515 & 0.4827 & 0.3536 & 0.4848 & 0.3568 \\
$T=200$ & 0.4149 & 0.3227 & 0.4106 & 0.3202 & 0.4112 & 0.3193 & 0.4130 & 0.3197 & 0.4151 & 0.3212 & 0.4137 & 0.3207 & 0.4134 & 0.3208 \\
$T=300$ & 0.3791 & 0.3053 & 0.3789 & 0.3050 & 0.3768 & 0.3028 & 0.3805 & 0.3058 & 0.3797 & 0.3049 & 0.3788 & 0.3050 & 0.3800 & 0.3057 \\
\midrule
\multicolumn{15}{c}{$N$ = 300} \\
\midrule
$T=100$ & 0.4813 & 0.3536 & 0.4833 & 0.3528 & 0.4806 & 0.3517 & 0.4871 & 0.3523 & 0.4844 & 0.3526 & 0.4833 & 0.3535 & 0.4838 & 0.3516 \\
$T=200$ & 0.4137 & 0.3203 & 0.4127 & 0.3189 & 0.4133 & 0.3203 & 0.4152 & 0.3207 & 0.4136 & 0.3190 & 0.4126 & 0.3188 & 0.4134 & 0.3200 \\
$T=300$ & 0.3786 & 0.3050 & 0.3797 & 0.3044 & 0.3795 & 0.3041 & 0.3791 & 0.3037 & 0.3795 & 0.3046 & 0.3793 & 0.3041 & 0.3786 & 0.3050 \\
\midrule
\multicolumn{15}{c}{Panel C: $\tau = 0.01$} \\
\midrule
\multicolumn{15}{c}{$N$ = 100} \\
\midrule
$T=100$ & 0.4048 & 0.3789 & 0.4010 & 0.3747 & 0.4041 & 0.3797 & 0.4054 & 0.3800 & 0.4094 & 0.3825 & 0.4045 & 0.3792 & 0.4070 & 0.3804 \\
$T=200$ & 0.4895 & 0.3404 & 0.5002 & 0.3399 & 0.4952 & 0.3389 & 0.4958 & 0.3418 & 0.4904 & 0.3351 & 0.4954 & 0.3400 & 0.4994 & 0.3391 \\
$T=300$ & 0.5224 & 0.3134 & 0.5212 & 0.3130 & 0.5210 & 0.3124 & 0.5211 & 0.3124 & 0.5359 & 0.3128 & 0.5215 & 0.3106 & 0.5212 & 0.3128 \\
\midrule
\multicolumn{15}{c}{$N$ = 200} \\
\midrule
$T=100$ & 0.4061 & 0.3803 & 0.4088 & 0.3809 & 0.4078 & 0.3808 & 0.4080 & 0.3819 & 0.4049 & 0.3774 & 0.4066 & 0.3798 & 0.4104 & 0.3820 \\
$T=200$ & 0.4932 & 0.3385 & 0.4986 & 0.3365 & 0.4880 & 0.3370 & 0.4982 & 0.3347 & 0.4985 & 0.3351 & 0.5002 & 0.3386 & 0.4976 & 0.3387 \\
$T=300$ & 0.5257 & 0.3100 & 0.5113 & 0.3108 & 0.5175 & 0.3123 & 0.5234 & 0.3097 & 0.5236 & 0.3117 & 0.5202 & 0.3115 & 0.5189 & 0.3114 \\
\midrule
\multicolumn{15}{c}{$N$ = 300} \\
\midrule
$T=100$ & 0.4083 & 0.3797 & 0.4074 & 0.3793 & 0.4059 & 0.3786 & 0.4059 & 0.3785 & 0.4099 & 0.3807 & 0.4050 & 0.3772 & 0.4048 & 0.3779 \\
$T=200$ & 0.4998 & 0.3346 & 0.4957 & 0.3365 & 0.4960 & 0.3364 & 0.4989 & 0.3343 & 0.4956 & 0.3352 & 0.4974 & 0.3351 & 0.4944 & 0.3371 \\
$T=300$ & 0.5236 & 0.3122 & 0.5186 & 0.3091 & 0.5231 & 0.3111 & 0.5222 & 0.3102 & 0.5180 & 0.3117 & 0.5228 & 0.3104 & 0.5188 & 0.3102 \\
\bottomrule
\end{tabular}
\end{adjustbox}
\end{table}

Several patterns stand out. First, ESFM consistently delivers lower RMSEs than ESR across all scenarios and sample sizes. The improvement is particularly clear in settings where latent factors drive tail risk (Scenarios 2, 4, 5, and 6), where ignoring such structure leads to noticeable bias. Second, the advantage of ESFM becomes stronger as we move further into the tail. The gap between the two methods widens from $\tau=0.10$ to $\tau=0.01$, indicating that latent factor effects play a larger role in more extreme outcomes. Third, both methods improve with larger samples, but ESFM benefits more. In particular, increasing $T$ leads to faster reductions in RMSEs for ESFM, especially when $N$ is moderate or large, suggesting that accounting for latent factors helps mitigate bias in panel settings. Overall, the performance gain of ESFM is stable across all scenarios. These results provide clear finite-sample evidence that incorporating latent tail factors improves the estimation of covariate effects in expected shortfall models.

Table~\ref{tab:Estimation error of the factor space} reports the estimation error of the factor space associated with the expected shortfall factors in ESFM. The error is measured by the normalized
Frobenius norm $\|P_{\hat F_\tau}-P_{F_\tau}\|_F^2 / T$, where $P_F = F(F^\top F)^{-1}F^\top$ denotes the projection matrix onto the true factor space. Results are reported for all seven scenarios and the same set of sample sizes and tail probabilities as in Table~\ref{tab:Estimation error of beta}.

\begin{table}[htbp]
  \centering
  \caption{Estimation error of the factor space $P_{F_\tau}$. {\footnotesize{\textit{Notes}. This table reports the estimation error of the factor space associated with the expected shortfall factors. The error is measured by the normalized Frobenius norm $\|P_{\hat{F}_\tau}-P_{F_\tau}\|_F^2/T$, where $P_F=F^\top(F^\top F)^{-1}F^\top$ denotes the projection matrix.}}}
  \label{tab:Estimation error of the factor space}

  \setlength{\tabcolsep}{4pt}      
  \renewcommand{\arraystretch}{0.95} 
  \small                  

  \begin{adjustbox}{max totalsize={\textwidth}{0.85\textheight},center}
  \begin{tabular}{lccccccc}
    \toprule
          & Scenario 1 & Scenario 2 & Scenario 3 & Scenario 4 & Scenario 5 & Scenario 6 & Scenario 7 \\
    \midrule
    \multicolumn{8}{c}{Panel A: $\tau = 0.10$} \\
    \midrule
    \multicolumn{8}{c}{$N$ = 100} \\
    \midrule
    $T=100$ & 0.1938  & 0.1939  & 0.1936  & 0.1939  & 0.1932  & 0.1942  & 0.1940  \\
    $T=200$ & 0.1373  & 0.1372  & 0.1368  & 0.1372  & 0.1366  & 0.1375  & 0.1376  \\
    $T=300$ & 0.1115  & 0.1113  & 0.1114  & 0.1102  & 0.1117  & 0.1112  & 0.1118  \\
    \midrule
    \multicolumn{8}{c}{$N$ = 200} \\
    \midrule
    $T=100$ & 0.1921  & 0.1926  & 0.1922  & 0.1921  & 0.1932  & 0.1930  & 0.1927  \\
    $T=200$ & 0.1336  & 0.1346  & 0.1348  & 0.1345  & 0.1350  & 0.1353  & 0.1344  \\
    $T=300$ & 0.1073  & 0.1071  & 0.1081  & 0.1082  & 0.1069  & 0.1072  & 0.1065  \\
    \midrule
    \multicolumn{8}{c}{$N$ = 300} \\
    \midrule
    $T=100$ & 0.1900  & 0.1902  & 0.1896  & 0.1918  & 0.1902  & 0.1904  & 0.1907  \\
    $T=200$ & 0.1307  & 0.1318  & 0.1332  & 0.1300  & 0.1326  & 0.1319  & 0.1319  \\
    $T=300$ & 0.0981  & 0.1014  & 0.0995  & 0.1018  & 0.0996  & 0.0998  & 0.1002  \\
    \midrule
    \multicolumn{8}{c}{Panel B: $\tau = 0.05$} \\
    \midrule
    \multicolumn{8}{c}{$N$ = 100} \\
    \midrule
    $T=100$ & 0.0385  & 0.0384  & 0.0385  & 0.0385  & 0.0384  & 0.0387  & 0.0387  \\
    $T=200$ & 0.0194  & 0.0195  & 0.0194  & 0.0195  & 0.0194  & 0.0194  & 0.0194  \\
    $T=300$ & 0.0130  & 0.0130  & 0.0130  & 0.0130  & 0.0130  & 0.0130  & 0.0131  \\
    \midrule
    \multicolumn{8}{c}{$N$ = 200} \\
    \midrule
    $T=100$ & 0.0384  & 0.0383  & 0.0384  & 0.0381  & 0.0387  & 0.0385  & 0.0383  \\
    $T=200$ & 0.0193  & 0.0194  & 0.0194  & 0.0193  & 0.0194  & 0.0193  & 0.0194  \\
    $T=300$ & 0.0129  & 0.0129  & 0.0130  & 0.0129  & 0.0130  & 0.0129  & 0.0129  \\
    \midrule
    \multicolumn{8}{c}{$N$ = 300} \\
    \midrule
    $T=100$ & 0.0382  & 0.0384  & 0.0381  & 0.0385  & 0.0384  & 0.0382  & 0.0383  \\
    $T=200$ & 0.0193  & 0.0194  & 0.0194  & 0.0193  & 0.0193  & 0.0193  & 0.0193  \\
    $T=300$ & 0.0128  & 0.0128  & 0.0129  & 0.0129  & 0.0128  & 0.0128  & 0.0128  \\
    \midrule
    \multicolumn{8}{c}{Panel C: $\tau = 0.01$} \\
    \midrule
    \multicolumn{8}{c}{$N$ = 100} \\
    \midrule
    $T=100$ & 0.0374  & 0.0373  & 0.0376  & 0.0374  & 0.0374  & 0.0375  & 0.0375  \\
    $T=200$ & 0.0188  & 0.0186  & 0.0185  & 0.0187  & 0.0184  & 0.0186  & 0.0185  \\
    $T=300$ & 0.0125  & 0.0125  & 0.0126  & 0.0126  & 0.0124  & 0.0124  & 0.0125  \\
    \midrule
    \multicolumn{8}{c}{$N$ = 200} \\
    \midrule
    $T=100$ & 0.0375  & 0.0378  & 0.0376  & 0.0372  & 0.0379  & 0.0376  & 0.0373  \\
    $T=200$ & 0.0189  & 0.0191  & 0.0189  & 0.0189  & 0.0191  & 0.0188  & 0.0188  \\
    $T=300$ & 0.0128  & 0.0129  & 0.0127  & 0.0129  & 0.0127  & 0.0128  & 0.0128  \\
    \midrule
    \multicolumn{8}{c}{$N$ = 300} \\
    \midrule
    $T=100$ & 0.0374  & 0.0376  & 0.0372  & 0.0374  & 0.0376  & 0.0376  & 0.0377  \\
    $T=200$ & 0.0191  & 0.0191  & 0.0190  & 0.0191  & 0.0192  & 0.0191  & 0.0191  \\
    $T=300$ & 0.0129  & 0.0129  & 0.0129  & 0.0129  & 0.0128  & 0.0127  & 0.0128  \\
    \bottomrule
  \end{tabular}
  \end{adjustbox}
\end{table}
%Overall, ESFM exhibits accurate recovery of the latent factor space across a wide range of data-generating environments. For a fixed $N$, the factor-space error decreases monotonically as $T$ increases, indicating consistent recovery of the common components. For example, when $N=100$ and $\tau=0.10$, the error declines from approximately $0.19$ at $T=100$ to around $0.11$ at $T=300$ across all scenarios. Similar convergence patterns are observed for larger cross sections.

%The estimation accuracy also improves as the cross-sectional dimension grows. Holding $T$ fixed, the factor-space error is uniformly smaller for $N=300$ than for $N=100$, reflecting the benefit of increased cross-sectional information in identifying common tail-risk factors. In addition, the factor-space recovery remains stable across different tail probability levels. Comparing Panels~A--C, the magnitude of $\|P_{\hat{F}_\tau}-P_{F_\tau}\|_F^2/T$ is largely insensitive to the choice of $\tau$, even for extreme tail probabilities such as $\tau=0.01$. This robustness suggests that ESFM can reliably extract latent factors governing tail risk even in very low-probability regions of the outcome distribution. Finally, the similarity of factor-space errors across scenarios indicates that the proposed estimation procedure is robust to heterogeneous slopes, endogeneity, volatility-driven tails, jump components, and asymmetric distributions. Taken together, these results confirm that ESFM consistently recovers the latent tail-risk structure underlying the data.

The results indicate that ESFM recovers the latent factor space with good accuracy. For a given $N$, the error declines steadily as $T$ increases, showing improved estimation of the common components. For instance, when $N=100$ and $\tau=0.10$, the error decreases from about $0.19$ at $T=100$ to around $0.11$ at $T=300$, with similar patterns observed for larger $N$. A larger cross section also improves accuracy. Holding $T$ fixed, the error is consistently smaller when $N=300$ than when $N=100$, reflecting the role of cross-sectional information in identifying the factors. The results are stable across different tail levels. The magnitude of the error changes little across $\tau=0.10$, $0.05$, and $0.01$, suggesting that factor recovery remains reliable even in the far tail. In addition, the error levels are similar across scenarios, indicating that the procedure is not sensitive to the specific features of the data-generating process. Overall, these findings support that ESFM is able to recover the underlying tail-risk factor structure in finite samples.

%To assess the finite-sample accuracy of tail-risk estimation, we examine the bias of the conditional ES under varying cross-sectional and temporal dimensions. In the presence of latent common components, misspecification of cross-sectional dependence can distort tail expectations even when first-stage quantile estimation remains accurate. We therefore compare  the results of ESR and ESFM. For each design, the true conditional ES is computed directly from the data-generating process via Monte Carlo integration.
To evaluate the finite-sample accuracy of tail-risk estimation, we examine the bias in conditional ES across different cross-sectional and temporal dimensions. When latent common factors are present, misspecifying cross-sectional dependence can distort tail expectations, even if first-stage quantile estimation is accurate. Thus, we compare the results of ESR and ESFM. The true conditional ES is computed using Monte Carlo integration from the data-generating process.
\begin{table}[htbp]
\centering
\caption{Finite-sample bias of conditional expected shortfall (ES) estimators. {\footnotesize{\textit{Notes}. This table reports the bias of conditional expected shortfall estimators across Monte Carlo replications. Bias is defined as the difference between the estimated and true ES. Results are presented for seven data-generating scenarios, comparing the benchmark expected shortfall regression (ESR) and the proposed expected shortfall factor model (ESFM).}}}
\label{tab:Estimation error of ES}
\setlength{\tabcolsep}{2.5pt}       % Reduce column spacing
\renewcommand{\arraystretch}{0.95} % Reduce row spacing
\small                       % Font size
\begin{adjustbox}{max totalsize={\textwidth}{0.85\textheight},center}
\begin{tabular}{lcccccccccccccc}
\toprule
      & \multicolumn{2}{c}{Scenario 1} & \multicolumn{2}{c}{Scenario 2} & \multicolumn{2}{c}{Scenario 3} & \multicolumn{2}{c}{Scenario 4} & \multicolumn{2}{c}{Scenario 5} & \multicolumn{2}{c}{Scenario 6} & \multicolumn{2}{c}{Scenario 7} \\
      & ESR & ESFM & ESR & ESFM & ESR & ESFM & ESR & ESFM & ESR & ESFM & ESR & ESFM & ESR & ESFM \\
\midrule
    \multicolumn{15}{c}{Panel A: $\tau$=0.10} \\
    \midrule
    \multicolumn{15}{c}{$N=100$} \\
    \midrule
    $T=100$ & 1.0207  & 0.9671  & 1.0336  & 0.9797  & 1.0286  & 0.9749  & 1.0274  & 0.9735  & 1.0274  & 0.9735  & 1.0333  & 0.9789  & 1.0375  & 0.9826  \\
    $T=200$ & 1.1538  & 1.0977  & 1.1577  & 1.1022  & 1.1644  & 1.1088  & 1.1536  & 1.0983  & 1.1619  & 1.1066  & 1.1592  & 1.1044  & 1.1512  & 1.0947  \\
    $T=300$ & 1.2128  & 1.1563  & 1.2327  & 1.1764  & 1.1977  & 1.1413  & 1.2117  & 1.1561  & 1.2119  & 1.1550  & 1.2156  & 1.1597  & 1.2110  & 1.1555  \\
    \midrule
    \multicolumn{15}{c}{$N=200$} \\
    \midrule
    $T=100$ & 1.0265  & 0.9727  & 1.0317  & 0.9771  & 1.0244  & 0.9709  & 1.0311  & 0.9773  & 1.0220  & 0.9685  & 1.0298  & 0.9754  & 1.0230  & 0.9683  \\
    $T=200$ & 1.1474  & 1.0921  & 1.1464  & 1.0911  & 1.1558  & 1.1000  & 1.1550  & 1.0993  & 1.1539  & 1.0984  & 1.1623  & 1.1067  & 1.1509  & 1.0949  \\
    $T=300$ & 1.2063  & 1.1501  & 1.2112  & 1.1552  & 1.2119  & 1.1558  & 1.2078  & 1.1518  & 1.2051  & 1.1493  & 1.2103  & 1.1540  & 1.2052  & 1.1494  \\
    \midrule
    \multicolumn{15}{c}{$N=300$} \\
    \midrule
    $T=100$ & 1.0356  & 0.9815  & 1.0337  & 0.9793  & 1.0285  & 0.9745  & 1.0309  & 0.9764  & 1.0434  & 0.9888  & 1.0278  & 0.9738  & 1.0276  & 0.9732  \\
    $T=200$ & 1.1603  & 1.1049  & 1.1497  & 1.0945  & 1.1520  & 1.0967  & 1.1535  & 1.0982  & 1.1574  & 1.1017  & 1.1486  & 1.0933  & 1.1556  & 1.1003  \\
    $T=300$ & 1.2173  & 1.1615  & 1.2087  & 1.1526  & 1.1989  & 1.1428  & 1.1943  & 1.1377  & 1.2090  & 1.1529  & 1.2104  & 1.1539  & 1.2163  & 1.1604  \\
    \midrule
    \multicolumn{15}{c}{Panel B: $\tau$=0.05} \\
    \midrule
    \multicolumn{15}{c}{$N=100$} \\
    \midrule
    $T=100$ & 0.8184  & 0.6970  & 0.8400  & 0.7189  & 0.8412  & 0.7188  & 0.8429  & 0.7212  & 0.8370  & 0.7150  & 0.8445  & 0.7215  & 0.8457  & 0.7232  \\
    $T=200$ & 1.0214  & 0.8966  & 1.0278  & 0.9041  & 1.0300  & 0.9077  & 1.0129  & 0.8895  & 1.0276  & 0.9050  & 1.0328  & 0.9100  & 1.0086  & 0.8833  \\
    $T=300$ & 1.1105  & 0.9859  & 1.1301  & 1.0049  & 1.0971  & 0.9720  & 1.1054  & 0.9812  & 1.1096  & 0.9836  & 1.1157  & 0.9906  & 1.1113  & 0.9865  \\
    \midrule
    \multicolumn{15}{c}{$N=200$} \\
    \midrule
    $T=100$ & 0.8310  & 0.7093  & 0.8464  & 0.7229  & 0.8291  & 0.7067  & 0.8372  & 0.7156  & 0.8371  & 0.7162  & 0.8395  & 0.7176  & 0.8241  & 0.7013  \\
    $T=200$ & 1.0109  & 0.8874  & 1.0067  & 0.8834  & 1.0224  & 0.8981  & 1.0197  & 0.8957  & 1.0229  & 0.8990  & 1.0317  & 0.9072  & 1.0146  & 0.8906  \\
    $T=300$ & 1.0927  & 0.9683  & 1.1123  & 0.9887  & 1.1076  & 0.9829  & 1.1009  & 0.9764  & 1.1021  & 0.9779  & 1.1027  & 0.9777  & 1.1091  & 0.9852  \\
    \midrule
    \multicolumn{15}{c}{$N=300$} \\
    \midrule
    $T=100$ & 0.8407  & 0.7181  & 0.8421  & 0.7189  & 0.8303  & 0.7085  & 0.8401  & 0.7171  & 0.8560  & 0.7334  & 0.8362  & 0.7140  & 0.8285  & 0.7058  \\
    $T=200$ & 1.0272  & 0.9038  & 1.0112  & 0.8886  & 1.0164  & 0.8933  & 1.0183  & 0.8950  & 1.0220  & 0.8983  & 1.0110  & 0.8875  & 1.0268  & 0.9031  \\
    $T=300$ & 1.1195  & 0.9952  & 1.1009  & 0.9760  & 1.0914  & 0.9668  & 1.0838  & 0.9589  & 1.1020  & 0.9777  & 1.1067  & 0.9809  & 1.1169  & 0.9928  \\
    \midrule
    \multicolumn{15}{c}{Panel C: $\tau$=0.01} \\
    \midrule
    \multicolumn{15}{c}{$N=100$} \\
    \midrule
    $T=100$ & 0.2899  & 0.2932  & 0.3666  & 0.3699  & 0.3314  & 0.3347  & 0.3711  & 0.3743  & 0.3236  & 0.3270  & 0.3251  & 0.3288  & 0.3539  & 0.3572  \\
    $T=200$ & 0.5127  & 0.3323  & 0.5344  & 0.3491  & 0.5467  & 0.3627  & 0.4970  & 0.3164  & 0.5315  & 0.3486  & 0.5493  & 0.3696  & 0.5409  & 0.3551  \\
    $T=300$ & 0.6798  & 0.3985  & 0.7094  & 0.4315  & 0.6411  & 0.3639  & 0.6617  & 0.3851  & 0.6949  & 0.4137  & 0.7022  & 0.4262  & 0.6835  & 0.4064  \\
    \midrule
    \multicolumn{15}{c}{$N=200$} \\
    \midrule
    $T=100$ & 0.3153  & 0.3187  & 0.3611  & 0.3645  & 0.3338  & 0.3374  & 0.3469  & 0.3500  & 0.3583  & 0.3617  & 0.3348  & 0.3384  & 0.3451  & 0.3486  \\
    $T=200$ & 0.5094  & 0.3272  & 0.4858  & 0.3077  & 0.5359  & 0.3558  & 0.5245  & 0.3414  & 0.5465  & 0.3636  & 0.5638  & 0.3786  & 0.5177  & 0.3368  \\
    $T=300$ & 0.6379  & 0.3612  & 0.6857  & 0.4120  & 0.6756  & 0.3985  & 0.6658  & 0.3905  & 0.6867  & 0.4076  & 0.6496  & 0.3739  & 0.6921  & 0.4120  \\
    \midrule
    \multicolumn{15}{c}{$N=300$} \\
    \midrule
    $T=100$ & 0.3440  & 0.3476  & 0.3375  & 0.3409  & 0.3236  & 0.3272  & 0.3418  & 0.3454  & 0.3295  & 0.3331  & 0.3071  & 0.3107  & 0.3290  & 0.3324  \\
    $T=200$ & 0.5573  & 0.3744  & 0.5050  & 0.3253  & 0.5385  & 0.3570  & 0.5263  & 0.3441  & 0.5343  & 0.3538  & 0.5301  & 0.3478  & 0.5568  & 0.3750  \\
    T=300 & 0.6992  & 0.4214  & 0.6543  & 0.3773  & 0.6502  & 0.3727  & 0.6326  & 0.3528  & 0.6348  & 0.3587  & 0.6488  & 0.3703  & 0.7159  & 0.4398  \\
    \bottomrule
\end{tabular}
\end{adjustbox}
\end{table}

%Table \ref{tab:Estimation error of ES} reports the Monte Carlo bias of the ES estimators across tail levels and sample sizes. Several patterns emerge. First, for moderate tail levels ($\tau = 0.10$ and $\tau = 0.05$), ESFM consistently exhibits smaller bias than ESR across all designs. This improvement becomes more pronounced as the cross-sectional dimension increases, reflecting the benefit of exploiting latent factor structure when estimating tail expectations. Second, at the extreme tail level ($\tau = 0.01$), the relative performance becomes less uniform. While ESFM remains competitive, its bias advantage narrows in smaller samples. This behavior is consistent with the amplification of estimation noise inherent in extreme tail extrapolation, where the pseudo-response used in ES construction magnifies lower-tail variability. Overall, the results demonstrate that incorporating cross-sectional dependence materially improves tail-risk estimation in finite samples, particularly when the effective tail probability is not vanishingly small.
Table \ref{tab:Estimation error of ES} reports the Monte Carlo bias of the ES estimators across various tail levels and sample sizes. A few key patterns emerge. First, for moderate tail levels ($\tau = 0.10$ and $\tau = 0.05$), ESFM consistently shows smaller bias than ESR across all scenarios. This improvement is more pronounced as the cross-sectional dimension increases, highlighting the advantage of latent factor modeling in estimating tail risk. Second, at the extreme tail level ($\tau = 0.01$), the performance difference narrows, especially in smaller samples. This is expected, as extreme tail extrapolation amplifies estimation noise, and the pseudo-response in ES construction increases lower-tail variability. Overall, these results confirm that modeling cross-sectional dependence significantly improves tail-risk estimation, especially when the tail probability is not too small.

\section{Empirical Application}\label{sec:Empirical Application}
\subsection{Data}\label{subsec:Data}
Our empirical analysis is conducted using a large cross-section of Chinese equities. We focus on the constituents of the CSI 300 index over the period from January 2011 to December 2023. This sample covers the most liquid and largest-cap stocks in the Chinese market, providing a representative laboratory for studying cross-sectional risk heterogeneity. To ensure data quality, we exclude assets with more than 5\% missing observations as well as stocks under special treatment (ST). The resulting panel consists of approximately 250 equities with relatively stable time-series coverage. Daily closing prices are obtained from the Wind database, from which we compute log returns.

For asset pricing analysis, we augment the dataset with standard risk factors. Specifically, we employ the Fama–French five-factor returns—including the market, size, value, profitability, and investment factors—along with the risk-free rate, obtained from publicly available data sources. Both equal-weighted (EW) and value-weighted (VW) Fama–French five-factor returns are considered. These factors enter our ESFM framework as observable covariates, capturing systematic components of asset returns.

The sample period spans several major market episodes, including the 2015 stock market crash, the 2018 trade tensions, and the COVID-19 pandemic. These episodes provide substantial time variation in tail risk and are particularly informative for identifying systematic components in downside risk. During such periods, return distributions exhibit pronounced skewness and heavy tails, which motivates the construction of tail-based risk factors.

\subsection{Estimation Results}\label{subsec:Estimation Results}
This subsection describes the estimation procedures and the presentation design for the latent factors and their associated loadings. We consider three classes of models---ESFM, the mean factor model (\citealt{bai2009panel}), and the quantile factor model (\citealt{ando2020quantile})---and estimate both the factor realizations and the cross-sectional exposure structure.

On the estimation side, for each model, asset returns are modeled conditional on observable covariates (the Fama–French five factors), and a low-dimensional set of latent factors is extracted together with asset-specific loadings. The ESFM captures tail-risk-driven components via a two-stage procedure, the mean factor model characterizes conditional mean dynamics, and the quantile factor model focuses on distributional asymmetries. The unified framework ensures comparability across models, facilitating a coherent evaluation of the extracted factor structures. On the presentation side, we report two key objects: (1) the cross-sectional loadings on Fama–French five factors, and (ii) the time series of latent factors. The factor paths describe the evolution of systematic risk, while the loadings capture heterogeneity in asset exposures. Presenting both components allows us to disentangle variation arising from common shocks and that arising from cross-sectional sensitivity.

For clarity and interpretability, we selectively report results across weighting schemes and parameter choices. In the main text, we present factor paths based on VW data with $\tau = 0.1$, while results for alternative $\tau$ values and EW specifications are relegated to the appendix. Similarly, for the loading estimates, value-weighted results are reported in the main text, with equal-weighted counterparts provided as robustness checks in the appendix.

\begin{figure}[htb]
	\vspace{0pt}
	\centering
	\includegraphics[scale=0.45]{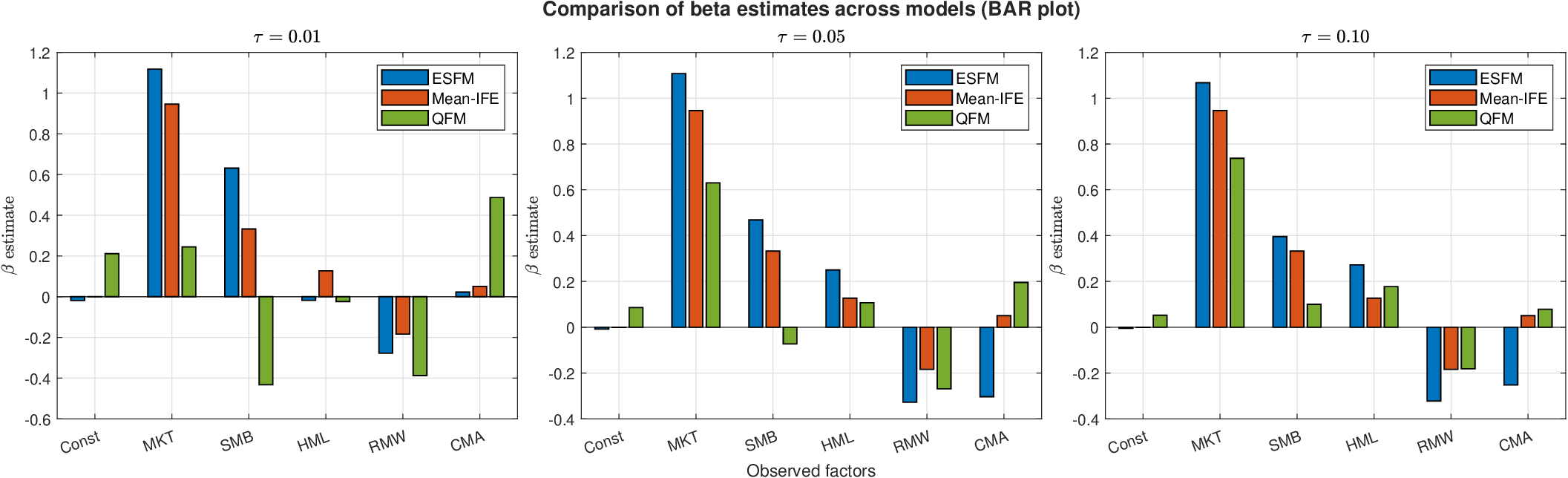}
	\caption{\footnotesize Estimated coefficients ($\beta$) on observable risk factors. \textit{Notes}. The figure reports the estimated coefficients on observable factors (MKT, SMB, HML, RMW, CMA, and a constant) from three models: ESFM, the mean factor model (Mean-IFE), and the quantile factor model (QFM). The estimates are obtained using VW observable factor returns, and results are shown for three tail levels ($\tau = 0.01, 0.05, 0.10$). Each panel corresponds to a different $\tau$, and bars represent the average coefficients across assets.}\label{fig:beta_VW}
\end{figure}
Figure \ref{fig:beta_VW} reports the estimated coefficients on observable factors across the ESFM, mean factor model, and quantile factor model for different tail levels ($\tau = 0.01, 0.05, 0.10$). The figure compares how each model loads on the Fama–French factors, providing a unified view of their sensitivity to systematic risk.

Several patterns emerge. First, the market factor is consistently the dominant driver across all models, with significant and stable loadings. However, the contribution of the market factor is notably weaker in the quantile model at extreme tail levels, suggesting that market-wide variation explains a smaller fraction of extreme downside risk under quantile specifications. Second, the ESFM generally exhibits larger coefficients across most factors, indicating a stronger sensitivity to systematic risk components, particularly in the tails. This is consistent with its construction, which aggregates tail realizations and thus amplifies exposure to extreme events. Third, the quantile model performs relatively poorly in the constant term, pointing to potential limitations in capturing unconditional tail behavior. In addition, its loadings on several factors are less stable and occasionally attenuated, reflecting weaker identification of systematic structure in extreme quantiles. Finally, factor-specific heterogeneity is evident across models. For example, size and profitability factors display noticeable variation in magnitude and even sign across specifications, highlighting that different models extract distinct aspects of risk. Overall, these results suggest that ES-based estimation provides a more pronounced and economically interpretable linkage between latent factors and observable sources of risk.

\begin{figure}[htb]
	\vspace{0pt}
	\centering
	\includegraphics[scale=0.55]{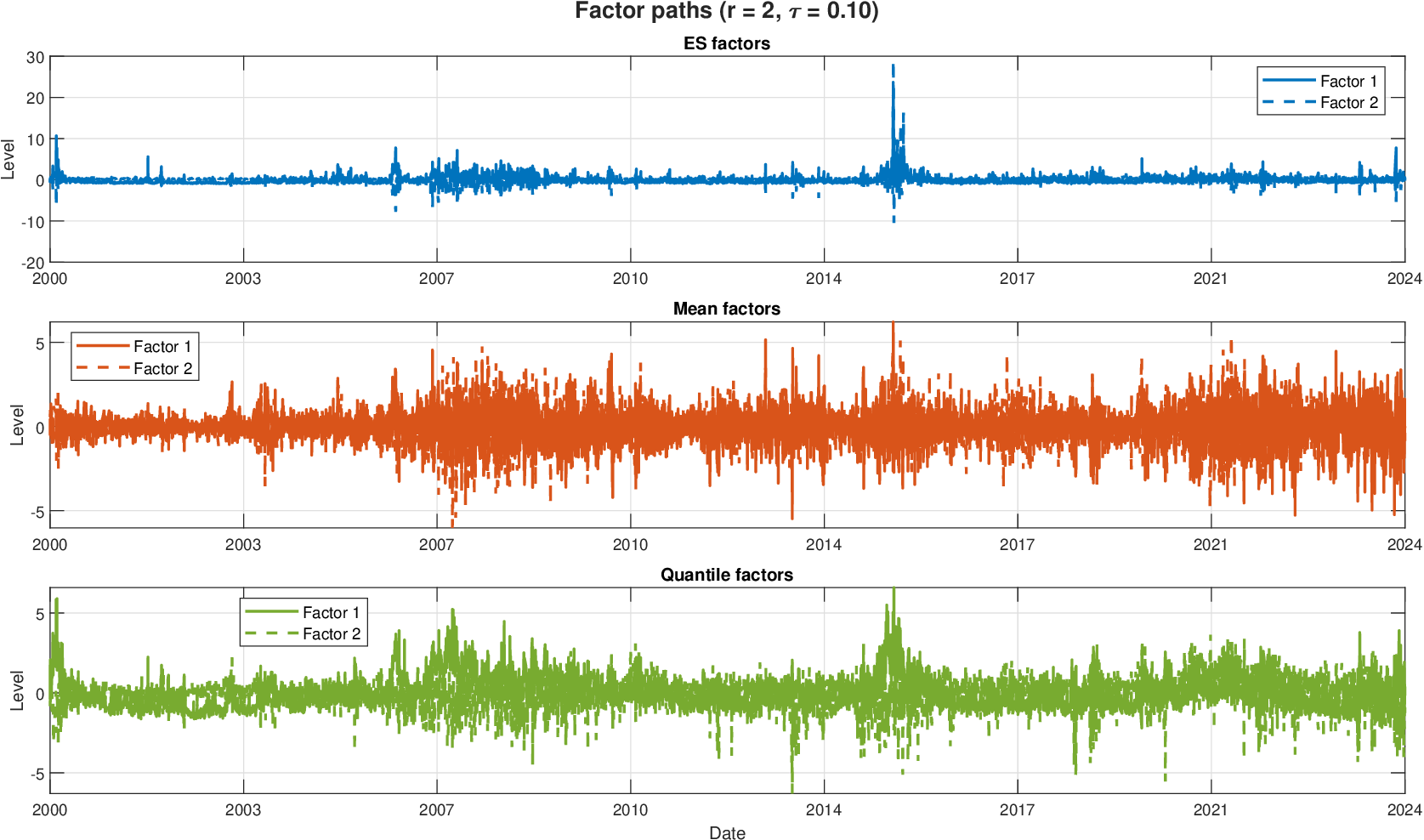}
	\caption{\footnotesize Estimated factor paths across models ($r=2,\tau=0.10$). \textit{Notes}. The figure plots the estimated latent factor paths from three models: ESFM (top panel), the mean factor model (middle panel), and the quantile factor model (bottom panel). Each model extracts two factors ($r = 2$). The covariates in these models include the value-weighted Fama–French five factors. The quantile and ES factors are constructed using tail information at level $\tau = 0.10$.}\label{fig:factor_VW_0p10}
\end{figure}
Figure \ref{fig:factor_VW_0p10} shows the factor path estimates for the three models. The estimated factor paths reveal clear differences across models. Both the ES and quantile factors exhibit pronounced time variation, reflecting their sensitivity to downside risk. This is consistent with their construction based on the lower tail of the return distribution. In contrast, the mean factors display more symmetric fluctuations and are less informative about tail events. 

A key distinction emerges when comparing ES and quantile factors. The ES factors remain relatively subdued during normal market conditions but exhibit sharp spikes during extreme market downturns, most notably around the 2015 market crash. By contrast, the quantile factors, while capturing general downside movements, do not display a similarly pronounced response at these extreme episodes. Their magnitudes during the crash period are comparable to those observed during other high-volatility periods, suggesting limited sensitivity to the severity of tail realizations. This difference is economically intuitive and directly linked to the definitions of the two measures. Quantile-based factors (VaR-type) capture the location of the tail but are insensitive to the magnitude of losses beyond the threshold, whereas ES explicitly aggregates tail realizations and thus responds strongly to extreme events. As a result, the ES factors provide a sharper characterization of crash risk and tail thickness, while quantile factors primarily reflect shifts in tail probability. This observation is in line with the theoretical distinction between VaR and ES emphasized in the risk management literature (e.g., \citealt{artzner1999coherent}), where ES is shown to be more sensitive to tail risk and better suited for capturing extreme downside exposure. Finally, during relatively tranquil periods, the ES factors remain close to zero, indicating limited tail risk, whereas the quantile factors continue to exhibit noticeable fluctuations. This further suggests that ES provides a more selective measure of extreme risk, while quantile-based factors may overreact to moderate tail movements.

Finally, following the quantile factor literature (e.g., \citealt{ando2020quantile}), we examine the correlations across factors estimated from different models. The purpose of this exercise is to assess whether these models capture similar underlying sources of systematic risk or uncover distinct dimensions. This comparison provides a structural benchmark for the subsequent asset pricing analysis. The results are presented in Table \ref{tab:GC}.

\begin{table}[htbp]
  \centering
  \caption{Generalized correlations of estimated factors across models. {\footnotesize{\textit{Notes}. This table reports the generalized correlations (GC) between estimated latent factors across the ESFM, mean factor model (Mean), and quantile factor model (QFM). Results are presented for three tail levels ($\tau = 0.01, 0.05, 0.10$)}}}
  \label{tab:GC}
    \begin{tabular}{lcccccccc}
    \toprule
          & \multicolumn{2}{c}{$\tau=0.01$} &       & \multicolumn{2}{c}{$\tau=0.05$} &       & \multicolumn{2}{c}{$\tau=0.10$} \\
\cmidrule{2-3}\cmidrule{5-6}\cmidrule{8-9}          & 1.GC  & 2.GC  &       & 1.GC  & 2.GC  &       & 1.GC  & 2.GC \\
    \midrule
    \multicolumn{9}{c}{Panel A: based on the value--weighted Fama–French five-factors} \\
    \midrule
    ESFM vs Mean & 0.1903 & 0.0632 &       & 0.3026 & 0.0802 &       & 0.3670 & 0.0819 \\
    ESFM vs QFM & 0.1991 & 0.0025 &       & 0.4756 & 0.006 &       & 0.6187 & 0.0401 \\
    Mean vs QFM & 0.1023 & 0.0401 &       & 0.5918 & 0.0443 &       & 0.6853 & 0.0377 \\
    \midrule
    \multicolumn{9}{c}{Panel B: based on the equal-weighted Fama–French five-factors} \\
    \midrule
    ESFM vs Mean & 0.2863 & 0.0342 &       & 0.3781 & 0.1084 &       & 0.4229 & 0.1575 \\
    ESFM vs QFM & 0.1904 & 0.0033 &       & 0.4417 & 0.0029 &       & 0.5462 & 0.0338 \\
    Mean vs QFM & 0.1304 & 0.0391 &       & 0.6277 & 0.1862 &       & 0.7545 & 0.2471 \\
    \bottomrule
    \end{tabular}%
  %\label{tab:addlabel}%
\end{table}%
Lower generalized correlation indicates weaker overlap in factor space and hence more model-specific information, whereas higher generalized correlation suggests that the corresponding factors reflect more similar underlying risk components. In our results, the ES factors are generally the least correlated with the other factors, especially at higher tail levels, implying that they contain additional information not captured by mean- or quantile-based specifications. By contrast, the mean and quantile factors are the most strongly related, particularly when $\tau$ is relatively high. This pattern changes when $\tau$ is very small, suggesting that the dependence structure across models becomes materially different in the extreme tail.

\subsection{Asset Pricing}\label{subsec:Asset Pricing}
\subsubsection{Excess Returns}\label{subsubsec:Excess Returns}
We begin by examining whether exposure to the estimated tail factors is priced in the cross-section of stock returns. To this end, we follow a standard portfolio-sorting approach. At each month $t$, we sort stocks into five portfolios based on their estimated factor exposures obtained from the previous 60-month rolling window. The portfolios are then held for one period ahead, and returns are computed using an equal-weighted scheme. We focus on the high-minus-low (H–L) portfolio, defined as the return difference between the highest- and lowest-exposure portfolios, which captures the return premium associated with the corresponding factor exposure. We focus on tail levels around $\tau\approx 0.2$, which correspond to economically meaningful downside events while retaining a sufficient number of observations for reliable estimation. This choice reflects a balance between capturing systematic lower-tail risk and avoiding the instability associated with extreme quantiles.\footnote{A growing literature links downside risk to moderately adverse states rather than extreme tail realizations. For instance, \cite{giglio2016systemic} and \cite{delikouras2019single} associate investors’ disappointment with approximately the worst 20% of outcomes.
}

\begin{figure}[htb]
	\vspace{0pt}
	\centering
	\includegraphics[scale=0.45]{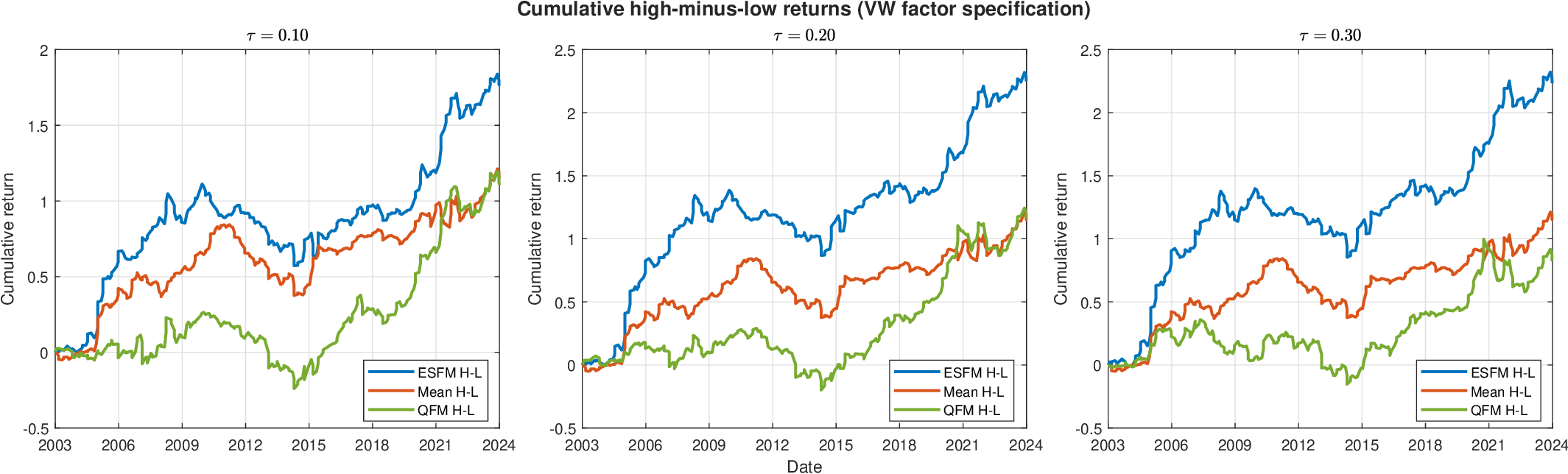}
	\caption{\footnotesize Cumulative high-minus-low returns across models. \textit{Notes}. The figure plots cumulative returns of high-minus-low (H–L) portfolios formed on factor exposures estimated from three models: ESFM, the mean factor model (Mean), and the quantile factor model (QFM). Stocks are sorted into five portfolios based on estimated exposures, and the H–L portfolio is constructed as the difference between the highest- and lowest-exposure portfolios. Portfolio returns are computed using equal-weighted (EW) returns, while the underlying Fama–French five factors are constructed using value-weighted (VW) returns. Each panel corresponds to a different tail level ($\tau=0.10,0.20,0.30$).}\label{fig:cumu_returns_VW}
\end{figure}
Figure \ref{fig:cumu_returns_VW} reports the cumulative returns of the H–L portfolios across the three models. A clear pattern emerges. The portfolio based on the ESFM exposures delivers a persistent and monotonic increase in cumulative returns across all values of $\tau$. In contrast, the portfolios constructed from the mean factor model and the quantile factor model exhibit substantially weaker performance, with periods of stagnation and reversals. This difference is particularly pronounced during episodes of market stress, where the ES-based strategy shows sharper increases, indicating that exposures to expected shortfall risks are systematically rewarded.

\begin{table}[H]
\centering
\caption{Portfolio returns and alphas sorted on factor exposures across models. {\footnotesize{\textit{Notes}. The table reports annualized portfolio returns and alphas (in \%) sorted on factor exposures estimated from different models (ESFM, Mean, and QFM) across three tail levels ($\tau=0.10,0.20,0.30$). We report estimated intercept (alphas) from regressing the returns on various sets of asset pricing factors: market (CAPM), three factor factors of \cite{fama1993common} (FF3), and five factors of \cite{fama2015five}. Stocks are sorted into No.G $=5$ or $10$ portfolios based on estimated exposures, and portfolio returns are computed using equal-weighted (EW) scheme.  The underlying Fama–French
three and five factors are constructed using value-weighted (VW) returns. Reported numbers correspond to time-series averages of one-period-ahead returns, with Newey-West $t$-statistics (six lags) shown in parentheses. The ``High–Low'' spread captures the return difference between the highest- and lowest-exposure portfolios (the average is placed in column ``Average'').}}}
\label{tab:Alphas}
\setlength{\tabcolsep}{2.5pt}       % Reduce column spacing
\renewcommand{\arraystretch}{0.95} % Reduce row spacing
\small                       % Font size
\begin{adjustbox}{max totalsize={\textwidth}{0.85\textheight},center}
\begin{tabular}{rrccccrccccrcccc}
\toprule
          &       & \multicolumn{4}{c}{$\tau=0.10$} &       & \multicolumn{4}{c}{$\tau=0.20$} &       & \multicolumn{4}{c}{$\tau=0.30$} \\
\cmidrule{3-6}\cmidrule{8-11}\cmidrule{13-16}       No.G   &       & Average & CAPM  & FF3   & FF5   &       & Average & CAPM  & FF3   & FF5   &       & Average & CAPM  & FF3   & FF5 \\
    \midrule
    \multicolumn{1}{l}{5} & \multicolumn{1}{l}{ESFM} & 8.04\% & 9.01\% & 9.40\% & 10.31\% &       & 10.27\% & 11.26\% & 11.75\% & 12.25\% &       & 10.20\% & 11.28\% & 11.81\% & 12.34\% \\
          &       & (2.62) & (3.09) & (3.38) & (3.33) &       & (3.08) & (3.49) & (3.84) & (3.61) &       & (2.96) & (3.41) & (3.78) & (3.48) \\
    \multicolumn{1}{l}{5} & \multicolumn{1}{l}{Mean} & 5.29\% & 5.20\% & 5.33\% & 5.18\% &       & 5.29\% & 5.20\% & 5.33\% & 5.18\% &       & 5.29\% & 5.20\% & 5.33\% & 5.18\% \\
          &       & (2.18) & (2.14) & (2.19) & (2.18) &       & (2.19) & (2.14) & (2.19) & (2.18) &       & (2.19) & (2.14) & (2.19) & (2.18) \\
    \multicolumn{1}{l}{5} & \multicolumn{1}{l}{QFM} & 5.05\% & 5.92\% & 5.71\% & 5.73\% &       & 5.29\% & 5.98\% & 6.13\% & 8.38\% &       & 3.77\% & 3.87\% & 4.43\% & 5.25\% \\
          &       & (1.80) & (2.13) & (1.97) & (1.87) &       & (1.87) & (2.07) & (2.13) & (2.51) &       & (1.22) & (1.24) & (1.58) & (1.48) \\
    \multicolumn{1}{l}{10} & \multicolumn{1}{l}{ESFM} & 9.34\% & 10.55\% & 11.02\% & 13.09\% &       & 10.78\% & 11.97\% & 12.54\% & 14.04\% &       & 11.68\% & 13.01\% & 13.69\% & 14.96\% \\
          &       & (2.48) & (2.99) & (3.28) & (3.33) &       & (2.73) & (3.13) & (3.46) & (3.28) &       & (2.84) & (3.27) & (3.69) & (3.39) \\
    \multicolumn{1}{l}{10} & \multicolumn{1}{l}{Mean} & 8.27\% & 8.23\% & 8.52\% & 8.14\% &       & 8.27\% & 8.23\% & 8.52\% & 8.14\% &       & 8.27\% & 8.23\% & 8.52\% & 8.14\% \\
          &       & (2.54) & (2.52) & (2.60) & (2.59) &       & (2.54) & (2.52) & (2.60) & (2.59) &       & (2.54) & (2.52) & (2.60) & (2.59) \\
    \multicolumn{1}{l}{10} & \multicolumn{1}{l}{QFM} & 7.76\% & 8.74\% & 8.42\% & 8.26\% &       & 6.88\% & 7.79\% & 8.01\% & 10.56\% &       & 4.28\% & 4.49\% & 5.10\% & 7.28\% \\
          &       & (2.23) & (2.53) & (2.33) & (2.11) &       & (1.90) & (2.09) & (2.18) & (2.47) &       & (1.19) & (1.24) & (1.54) & (1.92) \\
    \bottomrule
\end{tabular}
\end{adjustbox}
\end{table}
Table \ref{tab:Alphas} complements this evidence by reporting the average H-L portfolio returns and the corresponding pricing alphas. Consistent with the graphical evidence, the ESFM portfolios deliver the largest return spreads within each sorting scheme, both economically and statistically. To assess whether these spreads can be explained by standard risk factors, we regress the H–L portfolio returns on CAPM, FF3, and FF5. The resulting alphas remain sizable and statistically significant across specifications. By construction, a larger alpha indicates that the return spread is less spanned by these benchmark models. The consistently higher alphas for ESFM therefore imply that the ES-based portfolios capture a dimension of risk that is not accounted for by conventional factors.

In contrast, portfolios formed using the mean factor model and the quantile factor model exhibit smaller return spreads and substantially weaker alphas, suggesting that their pricing content is more limited. Moreover, the differences across models become more pronounced at higher values of $\tau$, where the ESFM portfolios show a clearer separation in both returns and alphas. Taken together, these results indicate that ES-based exposures provide incremental pricing information beyond both mean-based and quantile-based factor structures.

\subsubsection{Fama-MacBeth Regressions}\label{subsubsec:Fama-MacBeth Regressions}
We complement the portfolio-sorting evidence with two-pass Fama–MacBeth regressions following \cite{fama1973risk}. This approach allows us to assess whether the pricing effects associated with ES-based exposures persist after controlling for other sources of risk in a unified cross-sectional framework. In the first step, we estimate factor loadings by regressing individual stock returns on the factor-mimicking portfolio returns constructed from ESFM, Mean, and QFM specifications. In the second step, we run cross-sectional regressions of next-period returns on these estimated loadings to obtain the corresponding prices of risk. We consider specifications with individual factors, jointly estimated factors, and augmentations that include the Fama–French five factors.

\begin{table}[htbp]
\centering
\caption{Fama–MacBeth regressions for factor risk premia. {\footnotesize{\textit{Notes}. The table reports estimated prices of risk from two-pass Fama–MacBeth regressions. In the first stage, factor loadings are obtained from time-series regressions of individual stock returns on factor-mimicking portfolio returns constructed from ESFM, Mean, and QFM specifications. In the second stage, cross-sectional regressions are performed each month to estimate factor risk premia. We report specifications with individual factors, jointly estimated factors, and augmentations with the Fama–French five factors (MKT, SMB, HML, RMW, CMA), constructed using value-weighted returns. The reported coefficients are time-series averages of monthly estimates. Adjusted $R^2$ is reported for the cross-sectional regressions. The coefficients are multiplied by 100 to ensure the clarity of the presentation.}}}
\label{tab:Fama–MacBeth}
\setlength{\tabcolsep}{2.5pt}       % Reduce column spacing
\renewcommand{\arraystretch}{0.95} % Reduce row spacing
\small                       % Font size
\begin{adjustbox}{max totalsize={\textwidth}{0.85\textheight},center}
\begin{tabular}{lcccccccccc}
    \toprule
          & \multicolumn{1}{l}{ESFM} & \multicolumn{1}{l}{Mean} & \multicolumn{1}{l}{QFM} & \multicolumn{1}{l}{Joint} & \multicolumn{1}{l}{ESFM\_MKT} & \multicolumn{1}{l}{ESFM\_SMB} & \multicolumn{1}{l}{ESFM\_HML} & \multicolumn{1}{l}{ESFM\_RMW} & \multicolumn{1}{l}{ESFM\_CMA} & \multicolumn{1}{l}{ESFM\_FF5} \\
    \midrule
    \multicolumn{11}{c}{Panel A: $\tau=0.10$} \\
    \midrule
    ESFM  & 1.3647  &       &       & 0.9168  & 0.9857  & 0.9196  & 0.9438  & 1.1717  & 1.1575  & 0.9540  \\
    Mean  &       & 0.5329  &       & 0.7383  &       &       &       &       &       &  \\
    QFM   &       &       & 1.2761  & 0.8392  &       &       &       &       &       &  \\
    MTK   &       &       &       &       & -3.7597  &       &       &       &       & -4.1021  \\
    SMB   &       &       &       &       &       & 2.4672  &       &       &       & 1.9328  \\
    HML   &       &       &       &       &       &       & -1.3910  &       &       & -1.4622  \\
    RMW   &       &       &       &       &       &       &       & -1.3169  &       & -0.3998  \\
    CMA   &       &       &       &       &       &       &       &       & 1.2188  & 0.4814  \\
    Intercept & -4.9495  & -6.5308  & -6.0740  & -6.9415  & -3.5136  & -8.2988  & -6.8780  & -8.4304  & -7.6352  & -4.0660  \\
    $R^2_{\rm adj}$ & 1.6926  & 1.2676  & 1.6981  & 3.9438  & 2.9411  & 4.1018  & 3.4559  & 3.7234  & 3.4245  & 7.5766  \\
    \midrule
    \multicolumn{11}{c}{Panel B: $\tau=0.20$} \\
    \midrule
    ESFM  & 1.5763  &       &       & 1.5114  & 1.3221  & 1.0540  & 1.2372  & 1.3216  & 1.1529  & 1.1411  \\
    Mean  &       & 0.5329  &       & 0.8828  &       &       &       &       &       &  \\
    QFM   &       &       & 0.9383  & 0.7903  &       &       &       &       &       &  \\
    MTK   &       &       &       &       & -3.5932  &       &       &       &       & -4.1551  \\
    SMB   &       &       &       &       &       & 2.5241  &       &       &       & 1.9809  \\
    HML   &       &       &       &       &       &       & -1.3113  &       &       & -1.5855  \\
    RMW   &       &       &       &       &       &       &       & -1.4412  &       & -0.4665  \\
    CMA   &       &       &       &       &       &       &       &       & 1.2467  & 0.4924  \\
    Intercept & -4.8764  & -6.5308  & -6.4467  & -5.1027  & -3.4591  & -8.4723  & -6.8684  & -8.6122  & -8.0020  & -4.0429  \\
    $R^2_{\rm adj}$ & 1.5654  & 1.2676  & 1.5754  & 4.1850  & 2.9794  & 4.0088  & 3.4352  & 3.6315  & 3.2959  & 7.6445  \\
    \midrule
    \multicolumn{11}{c}{Panel C: $\tau=0.30$} \\
    \midrule
    ESFM  & 1.6291  &       &       & 1.4751  & 1.4691  & 0.9541  & 1.3822  & 1.2611  & 1.1664  & 1.0543  \\
    Mean  &       & 0.5329  &       & 0.9036  &       &       &       &       &       &  \\
    QFM   &       &       & -0.7204  & -0.5199  &       &       &       &       &       &  \\
    MTK   &       &       &       &       & -3.5169  &       &       &       &       & -4.1316  \\
    SMB   &       &       &       &       &       & 2.5074  &       &       &       & 1.8856  \\
    HML   &       &       &       &       &       &       & -1.3335  &       &       & -1.5727  \\
    RMW   &       &       &       &       &       &       &       & -1.4713  &       & 0.4522  \\
    CMA   &       &       &       &       &       &       &       &       & 1.2274  & 0.4904  \\
    Intercept & -4.7123  & -6.5308  & -7.0182  & -6.5134  & -3.5416  & -8.3365  & -6.6911  & -8.4056  & -7.8404  & -4.0318  \\
    $R^2_{\rm adj}$ & 1.6064  & 1.2676  & 1.5518  & 4.3523  & 3.0341  & 4.0294  & 3.4728  & 3.6972  & 3.3020  & 7.7286  \\
    \bottomrule
\end{tabular}
\end{adjustbox}
\end{table}

The results are reported in Table \ref{tab:Fama–MacBeth}. Several findings emerge. First, the ESFM factor carries a positive and economically meaningful price of risk across all specifications. The magnitude of the ESFM premium remains stable when estimated jointly with the Mean and QFM factors, indicating that it captures a distinct dimension of risk rather than a reparameterization of existing factor structures. Second, in contrast, the Mean and QFM factors exhibit weaker and less stable pricing patterns, and the QFM factors' contributions diminish once ESFM is included. Importantly, the ESFM premium is not materially affected by the inclusion of the Fama–French five factors. The estimated price of risk remains largely unchanged when augmenting the specification with MKT, SMB, HML, RMW, and CMA, suggesting that the ES-based factor is not spanned by standard linear factor models. This robustness is consistent across different tail levels, with the pricing effect becoming more pronounced at higher $\tau$, pointing to stronger cross-sectional variation associated with more extreme tail risks. Taken together, these results provide further evidence that ES-based factors contain pricing information beyond both mean-based and quantile-based specifications, and that this information is not subsumed by conventional risk factors.

\section{Conclusion}\label{sec:Conclusion}
This paper studies expected shortfall (ES) in panel data through a factor-augmented regression framework that accounts for cross-sectional dependence in tail risk. While existing work has largely focused on mean or quantile dynamics, our approach targets tail expectations and accommodates both observed covariates and latent common components. The proposed two-step procedure builds on the joint structure of quantile and ES, and the orthogonal construction avoids the numerical difficulties typically associated with non-convex formulations.

On the theoretical side, we establish consistency and asymptotic normality of the estimators, together with non-asymptotic bounds that make explicit how the time and cross-sectional dimensions enter the problem. The results highlight a decomposition of estimation error into three parts: the intrinsic variability of ES, the first-stage quantile error, and the factor estimation error. The orthogonality of the construction implies that the first-stage effect is of higher order, which helps explain the stability of the procedure in heavy-tailed settings. Empirically, the ES factor structure provides a direct way to capture comovement in downside risk. Compared with mean- and quantile-based specifications, the ES factors isolate variation that is more closely tied to extreme losses. The evidence suggests that these factors contain pricing information not accounted for by standard models, while the estimation remains straightforward in large panels.

Several directions for future research appear promising. First, while the present work focuses on financial applications, the proposed framework is broadly applicable to other domains where tail behavior is central, such as macroeconomic stress analysis, climate extremes, and insurance risk modeling. Second, it would be of interest to extend the current formulation to matrix-valued (e.g., \citealt{yuan2023two}) or operator-based ES factor structures, allowing for richer dependence patterns across both units and distributions. Third, incorporating structural changes or regime shifts into the ES factor model---through, for example, change-point detection in tail factors---may further enhance its ability to capture episodic systemic risk (e.g., \citealt{wang2025efficient}).

%\section*{Acknowledgments} xxx

\bibliographystyle{chicago}
\bibliography{ref}

\end{document}

\if0\blind
{
  \title{\LARGE\bf  Supplementary Materials for ``Expected Shortfall Panel Regression''}
  \author{}
  \maketitle
} \fi

\if1\blind
{
  \bigskip
  \bigskip
  \bigskip
  \begin{center}
    {\LARGE\bf  Supplementary Materials for ``Expected Shortfall Panel Regression''}
\end{center}
  \medskip
} \fi

\bigskip

In this supplement, we provide the detailed proofs and additional results for the main text. The proofs of Proposition \ref{pro:average consistency} and Theorem \ref{theo:consistency} are given in Sections \ref{secA} and \ref{secB}, respectively. Sections \ref{secC} and \ref{secD} establish Theorems \ref{thm:selection} and \ref{theo:asymptotic}, while Proposition \ref{pro:quantile} is proved in Section \ref{secE}. The proofs of Theorems \ref{theo:nonasymES} and \ref{theo:ESnon-asy} are deferred to Sections \ref{secF} and \ref{secG}. Additional simulation results, including data-generating scenarios and simulation outcomes, are reported in Section \ref{sec:Additional Simulation Results}. Section \ref{sec:Additional Empirical Results} provides further empirical evidence on estimation and asset pricing.

For notational simplicity, we suppress the dependency of $\tau$ such that $\alpha_{i,\tau}=\alpha_i$, $\beta_{i,\tau}=\beta_i$, $f_{t,\tau}=f_t$, $\lambda_{i,\tau}=\lambda_i$, $\varepsilon_{it,\tau}=\varepsilon_{it}$, $B_\tau=B$, $F_\tau=F$, $\Lambda_\tau=\Lambda$, $r^0_\tau=r^0$, etc throughout the proof. We denote the estimated parameters as $\widehat{\alpha}_i$, $\widehat{\beta}_i$, $\widehat{f}_t$, $\widehat{\lambda}_i$, $\widehat{B}$, $\widehat{F}$, $\widehat{\Lambda}$ and $\widehat{r}^0$, etc. Also, we denote the true parameters as $\alpha_i^0$, $\beta_i^0$, $f_t^0$, $\lambda_i^0$, $B^0$, $F^0$, $\Lambda^0$ and $r^0$, etc.

%\newpage

\tableofcontents

\newpage
We use the following facts throughout: $T^{-1} \|X_i\|^2 = T^{-1} \sum_{i=1}^T \|X_{it}\|^2 = O_p(1)$ or $T^{-1/2} \|X_i\| = O_p(1)$. Averaging over $i$, $(TN)^{-1} \sum_{i=1}^N \|X_i\|^2 = O_p(1)$. Similarly, $T^{-1/2} \|F^0\| = O_p(1)$, $T^{-1} \|\widehat{F}\| = r$, $T^{-1/2} \|\widehat{F}\| = \sqrt{r}$, $T^{-1} \|X_i^\top F^0\| = O_p(1)$, and so forth. Throughout, we define $\delta_{NT} = \min[\sqrt{N}, \sqrt{T}]$ so that $\delta_{NT}^2 = \min[N, T]$.
\section{Proof of Proposition \ref{pro:average consistency}}\label{secA}
\begin{proof}
Without loss of generality, assume $ \beta_i^0 = 0 $ for $i\in[N]$ (purely for notational simplicity). From $Z_i^\ast(\alpha_i^0)= X_i \beta_i^0 + F^0 \lambda_i^0 + \varepsilon_i = F^0 \lambda_i^0 + \varepsilon_i $, expanding $ S_{NT}(B, F) $, we obtain
\begin{equation}\nonumber
    \begin{aligned}
        S_{NT}(B, F) =& \frac{1}{NT} \sum_{i=1}^{N} \left\{Z_i^\ast(\widehat{\alpha}_{i})- X_i \beta_{i}\right\}^\top M_{F}\left\{Z_i^\ast(\widehat{\alpha}_{i})- X_i \beta_{i}\right\}\\
        &-\frac{1}{NT} \sum_{i=1}^{N}\left\{Z_i^\ast(\widehat{\alpha}_{i})- Z_i^\ast(\alpha^0_{i})\right\}^\top M_{F^0}\left\{Z_i^\ast(\widehat{\alpha}_{i})- Z_i^\ast(\alpha^0_{i})\right\}\\
        &-\frac{2}{NT} \sum_{i=1}^{N}\left\{Z_i^\ast(\widehat{\alpha}_{i})- Z_i^\ast(\alpha^0_{i})\right\}^\top M_{F^0}\varepsilon_{i}- \frac{1}{NT} \sum_{i=1}^{N} \varepsilon_{i}^\top M_{F^0} \varepsilon_{i}\\
        =&\frac{1}{NT} \sum_{i=1}^{N} \left[\left\{Z_i^\ast(\widehat{\alpha}_{i})-Z_i^\ast(\alpha^0_{i})\right\}+Z_i^\ast(\alpha^0_{i})- X_i \beta_{i}\right]^\top M_{F}\left[\left\{Z_i^\ast(\widehat{\alpha}_{i})- Z_i^\ast(\alpha^0_{i})\right\}+Z_i^\ast(\alpha^0_{i})-X_i \beta_{i}\right]\\
        &-\frac{1}{NT} \sum_{i=1}^{N}\left\{Z_i^\ast(\widehat{\alpha}_{i})- Z_i^\ast(\alpha^0_{i})\right\}^\top M_{F^0}\left\{Z_i^\ast(\widehat{\alpha}_{i})- Z_i^\ast(\alpha^0_{i})\right\}\\
        &-\frac{2}{NT} \sum_{i=1}^{N}\left\{Z_i^\ast(\widehat{\alpha}_{i})- Z_i^\ast(\alpha^0_{i})\right\}^\top M_{F^0}\varepsilon_{i}- \frac{1}{NT} \sum_{i=1}^{N} \varepsilon_{i}^\top M_{F^0} \varepsilon_{i}\\
        =&\widetilde{S}_{NT}(B, F)+  \frac{2}{NT} \sum_{i=1}^N \beta_i^\top X_i^\top  M_F^\prime \varepsilon_i +  \frac{2}{NT} \sum_{i=1}^N \lambda_i^{0\top}  F^{0\top} M_F \varepsilon_i \\
        &+ \frac{1}{NT} \sum_{i=1}^N \varepsilon_i^\top (P_{F_0} - P_{F}) \varepsilon_i+\frac{2}{NT} \sum_{i=1}^N\beta_i^\top X_i^\top M_F\left\{Z_i^\ast({\alpha}^0_{i})-Z_i^\ast(\widehat{\alpha}_{i})\right\}\\
        &+\frac{2}{NT} \sum_{i=1}^N\lambda_i^{0\top}F^{0\top}M_F\left\{Z_i^\ast(\widehat{\alpha}_{i})-Z_i^\ast(\alpha^0_{i})\right\}\\
        &+\frac{2}{NT} \sum_{i=1}^N\varepsilon_i^\top(P_{F^0}-P_F)\left\{Z_i^\ast(\widehat{\alpha}_{i})-Z_i^\ast(\alpha^0_{i})\right\}\\
        &+\frac{1}{NT} \sum_{i=1}^N\left\{Z_i^\ast(\widehat{\alpha}_{i})-Z_i^\ast(\alpha^0_{i})\right\}^\top(P_{F^0}-P_F)\left\{Z_i^\ast(\widehat{\alpha}_{i})-Z_i^\ast(\alpha^0_{i})\right\},
    \end{aligned}
\end{equation}
where
\begin{equation}\label{eq:A.1}
\widetilde{S}_{NT}(B, F) =  \frac{1}{NT} \sum_{i=1}^N \beta_i^\top X_i^\top  M_F X_i \beta_i+\frac{1}{NT}\sum_{i=1}^N\lambda_i^\top F^{0\top}M_F F^0\lambda_i^0 +  \frac{2}{NT} \sum_{i=1}^N \beta_i^\top X_i^\top  M_F^\prime F^0 \lambda_i^0
\end{equation}
and $M_F^\prime=-M_F=FF^\top/T-I_T$. By Lemma \ref{lemmaA.2},
\begin{equation}\label{eq:A.2}
S_{NT}(\beta, F) = \widetilde{S}_{NT}(\beta, F) + o_p(1)
\end{equation}
uniformly over bounded $ \beta_i $ and over $ F $ such that $ F^\top F/T = I $. Bounded $ \beta_i $ is in fact not necessary because the objective function is quadratic in $ \beta_i $ (that is, it is easy to argue that the objective function cannot achieve its minimum for very large $ \beta_i $).

Clearly, $ \widetilde{S}_{NT}(B^0, F^0 H) = 0 $ for $B^0=0$ and any $ r \times r $ invertible $ H $, because $ M_{F^0 H} = M_{F^0} $ and $ M_{F^0} F^0 = 0 $. The identification restrictions implicitly fix an $ H $. Define
$$
A_i = \frac{1}{T} X_i^\top  M_{\widehat{F}} X_i, \quad B_i = (\lambda_i^0\lambda_i^{0\top})\otimes I_T,\quad C_i=\frac{1}{T}\lambda_i^{0}  \otimes (M_{\widehat{F}} X_i),\quad\eta = \frac{1}{\sqrt{T}}\text{vec}(M_{\widehat{F}} F^0)
$$
Then
$$
\widetilde{S}_{NT}(\widehat{B}, \widehat{F}) = \frac{1}{N}\sum_{i=1}^N\widehat{\beta}_i^\top A_i \widehat{\beta}_i + \eta^\top \left(\frac{1}{N}\sum_{i=1}^NB_i\right) \eta + \frac{2}{N}\sum_{i=1}^N\widehat{\beta}_i^\top C_i^\top \eta.
$$
Completing the square, we have
$$
\widetilde{S}_{NT}(\widehat{B}, \widehat{F}) = \eta^\top \left(\frac{1}{N}\sum_{i=1}^NE_i\right) \eta+\frac{1}{N}\sum_{i=1}^N\left(\widehat{\beta}_i+A_i^{-1}C_i\eta\right)^\top A_i\left(\widehat{\beta}_i+A_i^{-1}C_i\eta\right),
$$
where $ E_i=B_i-C_i^\top A_i^{-1}C_i$. Because each of the two terms is non-negative and the centered objective function satisfies $ S_{NT}(B^0, F^0) = 0 $ and, by definition, $ S_{NT}(\widehat{B}, \widehat{F}) \leq 0 $. Therefore, in view of \eqref{eq:A.2},
$$
0 \geq S_{NT}(\widehat{B}, \widehat{F}) = \widetilde{S}_{NT}(\widehat{B}, \widehat{F}) + o_p(1).
$$
Combined with $ \widetilde{S}_{NT}(\widehat{B}, \widehat{F}) \geq 0 $, it must be true that
$$
\widetilde{S}_{NT}(\widehat{B}, \widehat{F}) = o_p(1).
$$
Then,
\begin{equation}\label{eq:A.3}
\eta^\top \left(\frac{1}{N}\sum_{i=1}^NE_i\right) \eta=o_p(1),
\end{equation}
\begin{equation}\label{eq:A.4}
\frac{1}{N}\sum_{i=1}^N\left(\widehat{\beta}_i+A_i^{-1}C_i\eta\right)^\top A_i\left(\widehat{\beta}_i+A_i^{-1}C_i\eta\right)=o_p(1).
\end{equation}
From Assumption $\ref{assum:eigenvalue}$ 2, the matrix $N^{-1}\sum_{i=1}^N E_i$ is positive definite, and thus equation \eqref{eq:A.3} implies that $\|\eta\|^2=o_p(1)$. That is,
\begin{equation}\label{eq:A.5}
\frac{F^{0\top}M_{\widehat{F}}F^0}{T} = \frac{F^{0\top}F^0}{T} - \frac{F^{0\top}\widehat{F}}{T} \frac{\widehat{F}^\top F^0}{T} = o_p(1).
\end{equation}
By Assumption \ref{assum:factor&loading}, $ F^{0\top}F^0/T $ is invertible, so it follows that $ F^{0\top}\widehat{F}/T $ is invertible. Next,
$$
\|P_{\widehat{F}} - P_{F^0}\|^2 = \tr[(P_{\widehat{F}} - P_{F^0})^2] = 2\tr(I_r - {F}^{0\top} P_{\widehat{F}}{F}^0/T).
$$
But \eqref{eq:A.5} implies $ {F}^{0\top} P_{\widehat{F}}{F}^0/T \overset{p}{\longrightarrow} I_r $, which is equivalent to $ \|P_{\widehat{F}} - P_{F^0}\| \overset{p}{\longrightarrow} 0 $. That is, the space spanned by $ F^0 $ and the space spanned by the estimated factors $ \widehat{F} $ are asymptotically the same. We then have
\begin{equation}\nonumber
\begin{aligned}
\frac{1}{\sqrt{T}}\left\|M_{F^0}\widehat{F}\right\| &= \frac{1}{\sqrt{T}}\left\|(M_{F^0} - M_{\widehat{F}})\widehat{F}\right\| \\
&= \frac{1}{\sqrt{T}}\left\|(P_{F^0} - P_{\widehat{F}})\hat{F}\right\| \\
&\leq \left\|P_{F^0} - P_{\widehat{F}}\right\| \times \left(\frac{1}{\sqrt{T}}\left\|\widehat{F}\right\|\right) \\
&= o_p(1) \times O_p(1),
\end{aligned}
\end{equation}
where we used $\|\widehat{F}\|/T^{1/2} = O_p(1)$. This implies that
$$
\frac{1}{\sqrt{T}}\left\|\widehat{F} - F^0(F^{0\top}F^0)^{-1}F^{0\top}\widehat{F}\right\| = \frac{1}{\sqrt{T}}\left\|\widehat{F} - F^0G\right\| = o_p(1),
$$
where $G=(F^{0\top}F^0)^{-1}F^{0\top}\widehat{F}$ is the rotation matrix. Because we omit this rotation matrix $G$ in Proposition \ref{pro:average consistency}, as explained in the main text, we obtain
$$
\frac{1}{\sqrt{T}}\left\|\widehat{F} - F^0\right\| = o_p(1).
$$

From $\|\eta\|^2=o_p(1)$, equation \eqref{eq:A.4} implies that
$$
o_p(1) = \frac{1}{N} \sum_{i=1}^N \widehat{\beta}_i^\top A_{i}^0 \widehat{\beta}_i + \frac{1}{N} \sum_{i=1}^N \widehat{\beta}_i^\top(A_i - A_{i}^0) \widehat{\beta}_i \geq (\rho_A + o_p(1)) \frac{1}{N} \sum_{i=1}^N \widehat{\beta}_i^\top \widehat{\beta}_i,
$$
where $0 < \rho_A$ is the lower bound of the eigenvalues of $A_{i}^0 = \frac{1}{T} X_i^\top M_{F^0} X_i$, $i\in[N]$. Because of Assumption $\ref{assum:eigenvalue}$ 2, $\rho_A > 0$ exists. We also used the fact that 
$$
\|A_i - A_{i,0}\|=\frac{1}{T}\left\|X_i^\top(P_{F^0}-P_{\widehat{F}})X_i\right\|\leq\frac{1}{T}\|X_i\|^2\left\|P_{F^0}-P_{\widehat{F}}\right\| = o_p(1).
$$
The average consistency of $\widehat{\beta}_i$ follows from $\frac{1}{N} \sum_{i=1}^N  \widehat{\beta}_i^\top\widehat{\beta}_i = o_p(1)$ (recall we normalize ${\beta}_i^0 = 0$). The average consistency of $\widehat{f}_t$ and the average consistency of $\widehat{{\beta}}_i$ further imply the average consistency of $\widehat{{\lambda}}_i$ (see \cite{ando2015asset}). That is, $N^{-1} \sum_{i=1}^N \|\widehat{{\gamma}}_i - {\gamma}_{i}^0\|^2 = o_p(1)$. This completes the proof of Proposition 1.
\end{proof}
\begin{lemma}\label{lemmaA.1}
Under Assumptions \ref{assum:indentification}-\ref{assum:weak dependence}, as $ N, T \to \infty $, for $i\in[N]$, we have 
$$
\left\|Z_i^\ast(\widehat{\alpha}_{i}) - Z_i^\ast(\alpha_{i}^0)\right\| = O_p\left(1\right),
$$
\end{lemma}

\begin{proof}
By definition,  
$$
Z^\ast_i(\alpha_i)=\frac{1}{\tau}(Z_{i1}(\alpha_i),\ldots,Z_{iT}(\alpha_i))^\top,\quad Z_{it}(\alpha_i) = (Y_{it} - X_{it}^\top \alpha_i) \mathbbm{1}(Y_{it} \le X_{it}^\top \alpha_i) + \tau X_{it}^\top \alpha_i.
$$
Let $q_t^0 = X_{it}^\top \alpha_{i}^0$, $\widehat{q}_t = X_{it}^\top \widehat{\alpha}_{i}$ and $\Delta q_t = \widehat{q}_t - q_t^0$. Consider the difference  
$$
\Delta Z_t := Z_{it}^\ast(\widehat{\alpha}_{i}) - Z_{it}^\ast(\alpha_{i}^0) 
= \frac{1}{\tau} \left[ (Y_{it} - \widehat{q}_t) \mathbbm{1}(Y_{it} \le \widehat{q}_t) - (Y_{it} - q_t^0) \mathbbm{1}(Y_{it} \le q_t^0) + \tau \Delta q_t \right].
$$
Let $D_t = (Y_{it} - \widehat{q}_t) \mathbbm{1}(Y_{it} \le \widehat{q}_t) - (Y_{it} - q_t^0) \mathbbm{1}(Y_{it} \le q_t^0)$.  
By analyzing the relative position of $Y_{it}$ with respect to $q_t^0$ and $\widehat{q}_t$, we consider four cases:
\begin{enumerate}
    \item $Y_{it} \le \min(q_t^0, \widehat{q}_t)$: $D_t = -\Delta q_t$, so $|D_t| = |\Delta q_t|$.
    \item $Y_{it} > \max(q_t^0, \widehat{q}_t)$: $D_t = 0$, so $|D_t| = 0$.
    \item $q_t^0 < Y_{it} \le \widehat{q}_t$ ($\widehat{q}_t > q_t^0$): $|D_t| = |Y_{it} - \widehat{q}_t| \le \widehat{q}_t - q_t^0 = \Delta q_t = |\Delta q_t|$.
    \item $\widehat{q}_t < Y_{it} \le q_t^0$ ($q_t^0 > \widehat{q}_t$): $|D_t| = q_t^0 - Y_{it} \le q_t^0 - \widehat{q}_t = -\Delta q_t = |\Delta q_t|$.
\end{enumerate}
In all cases, $|D_t| \le |\Delta q_t|$. Therefore,  
$$
|\Delta Z_t| \le \frac{1}{\tau} \left( |D_t| + \tau |\Delta q_t| \right) \le \frac{1+\tau}{\tau} |\Delta q_t| = \frac{1+\tau}{\tau} \left| X_{it}^\top (\widehat{\alpha}_{i} - \alpha_{i}^0) \right|.
$$
From quantile regression theory, $\widehat{\alpha}_{i} - \alpha_{i}^0 = O_p(T^{-1/2})$, and by Assumption \ref{assum:indentification}, $E\|X_{it}\|^4 \le M$, so  
$$
|\Delta Z_t| = O_p(T^{-1/2}).
$$
Now consider the squared Euclidean norm:  
$$
\left\| Z_i^\ast(\widehat{\alpha}_{i}) - Z_i^\ast(\alpha_{i}^0) \right\|^2 
= \sum_{t=1}^T (\Delta Z_t)^2=O_p(T \cdot \|\widehat{\alpha}_{i} - \alpha_{i}^0\|^2) = O_p(T \cdot T^{-1}) = O_p(1).
$$
Hence,  
$$
\left\| Z_i^\ast(\widehat{\alpha}_{i}) - Z_i^\ast(\alpha_{i}^0) \right\| = O_p(1).
$$
\end{proof}

\begin{lemma}\label{lemmaA.2}
Under Assumptions \ref{assum:indentification}-\ref{assum:weak dependence}, as $ N, T \to \infty $, we have
\begin{enumerate}[label=(\roman*)]
\item $\displaystyle\sup_{F\in\mathcal{F}}\left\|\frac{1}{NT} \sum_{i=1}^N \beta_i^\top X_i^\top  M_F^\prime \varepsilon_i\right\|=o_p(1)$,
\item $\displaystyle\sup_{F\in\mathcal{F}}\left\|\frac{1}{NT} \sum_{i=1}^N \lambda_i^{0\top}  F^{0\top} M_F \varepsilon_i\right\|=o_p(1)$,
\item $\displaystyle\sup_{F\in\mathcal{F}}\left\|\frac{1}{NT} \sum_{i=1}^N \varepsilon_i^\top (P_{F_0} - P_{F}) \varepsilon_i\right\|=o_p(1)$,
\item $\displaystyle\sup_{F\in\mathcal{F}}\left\|\frac{1}{NT} \sum_{i=1}^N\beta_i^\top X_i^\top M_F\left\{Z_i^\ast({\alpha}^0_{i})-Z_i^\ast(\widehat{\alpha}_{i})\right\}\right\|=o_p(1)$,
\item $\displaystyle\sup_{F\in\mathcal{F}}\left\|\frac{1}{NT} \sum_{i=1}^N\lambda_i^{0\top}F^{0\top}M_F\left\{Z_i^\ast(\widehat{\alpha}_{i})-Z_i^\ast(\alpha^0_{i})\right\}\right\|=o_p(1)$,
\item $\displaystyle\sup_{F\in\mathcal{F}}\left\|\frac{1}{NT} \sum_{i=1}^N\varepsilon_i^\top(P_{F^0}-P_F)\left\{Z_i^\ast(\widehat{\alpha}_{i})-Z_i^\ast(\alpha^0_{i})\right\}\right\|=o_p(1)$,
\item $\displaystyle\sup_{F\in\mathcal{F}}\left\|\frac{1}{NT} \sum_{i=1}^N\left\{Z_i^\ast(\widehat{\alpha}_{i})-Z_i^\ast(\alpha^0_{i})\right\}^\top(P_{F^0}-P_F)\left\{Z_i^\ast(\widehat{\alpha}_{i})-Z_i^\ast(\alpha^0_{i})\right\}\right\|=o_p(1)$,
\end{enumerate}
where $\mathcal{F}=\{F:F^\top F/T=I_r\}$.
\end{lemma}
\begin{proof}
(i) According to $M_F^\prime=-M_F=P_F-I_T$, we have
$$
\frac{1}{NT} \sum_{i=1}^N \beta_i^\top X_i^\top  M_F^\prime \varepsilon_i=\frac{1}{NT} \sum_{i=1}^N \beta_i^\top X_i^\top  P_F \varepsilon_i-\frac{1}{NT} \sum_{i=1}^N \beta_i^\top X_i^\top \varepsilon_i.
$$
From $\frac{1}{NT}\sum_{i=1}^{N}X_{i}^{\top}\varepsilon_{i}=o_{p}(1)$, it is sufficient to show $\sup_{F\in\mathcal{F}}\frac{1}{NT}\sum_{i=1}^{N}X_{i}^{\top}P_{F}\varepsilon_{i}=o_{p}(1)$. Using $P_{F}=FF^{\top}/T$,
\begin{equation}\nonumber
\begin{aligned}
\frac{1}{NT}\left\|\sum_{i=1}^{N}X_{i}P_{F}\varepsilon_{i}\right\| &=\left\|\frac{1}{N}\sum_{i=1}^{N}\left(\frac{X_{i}^{\top}F}{T} \right)\frac{1}{T}\sum_{i=1}^{T}f_{t}\varepsilon_{it}\right\|\\
&\leq\frac{1}{N}\sum_{i=1}^{N}\left\|\frac{X_{i}^{\top}F}{T}\right\|\times\left\|\frac{1}{T}\sum_{t=1}^{T}f_{t}\varepsilon_{it}\right\|.
\end{aligned}
\end{equation}
Note that $T^{-1}\|X_{i}^{\top}F\|\leq T^{-1}\|X_{i}\|\|F\|=\sqrt{r}T^{-1/2}\|X_{i}\| \leq\sqrt{r}(\frac{1}{T}\sum_{t=1}^{T}\|X_{it}\|^{2})^{1/2}$ because $T^{-1/2}\|F\|=\sqrt{r}$. Thus, using the Cauchy-Schwarz inequality, the above is bounded by
$$
\sqrt{r}\Bigg{(}\frac{1}{N}\sum_{i=1}^{N}\frac{1}{T}\sum_{t=1}^{T}\|X_{it}\|^{2 }\Bigg{)}^{1/2}\Bigg{(}\frac{1}{N}\sum_{i=1}^{N}\left\|\frac{1}{T}\sum_{t=1}^{T }f_{t}\varepsilon_{it}\right\|^{2}\Bigg{)}^{1/2}.
$$
The first expression is $O_{p}(1)$. It suffices to show that the second term is $o_{p}(1)$ uniformly in $F$. Now
\begin{equation}\nonumber
\begin{aligned}
\frac{1}{N}\sum_{i=1}^{N}\left\|\frac{1}{T}\sum_{t=1}^{T}f_{t} \varepsilon_{it}\right\|^{2} =&\tr\left(\frac{1}{N}\sum_{i=1}^{N}\frac{1}{T^{2}} \sum_{t=1}^{T}\sum_{s=1}^{T}f_{t}f_{s}^{\top}\varepsilon_{it}\varepsilon_{is}\right)\\
=&\tr\left(\frac{1}{T^{2}}\sum_{t=1}^{T}\sum_{s=1}^{T}f_{t}f_{s}^{\top}\frac{1}{N}\sum_{i=1}^{N}[\varepsilon_{it}\varepsilon_{is}- E(\varepsilon_{it}\varepsilon_{is})]\right)\\
&+\tr\left(\frac{1}{T^{2}}\sum_{t=1}^{T}\sum_{s=1}^{T}f_{t}f_{s}^{\top}\frac{1}{N}\sum_{i=1}^{N}\sigma_{ii,ts}\right),
\end{aligned}
\end{equation}
where $\sigma_{ii,ts}=E(\varepsilon_{it}\varepsilon_{is})$. The first expression is bounded by the Cauchy-Schwarz inequality:
$$
\frac{1}{\sqrt{N}}\left(\frac{1}{T^{2}}\sum_{t=1}^{T}\sum_{s=1}^{T}\|f_{t}\|^{2}\| f_{s}\|^{2}\right)^{1/2}\left(\frac{1}{T^{2}}\sum_{t=1}^{T}\sum_{s=1}^{T}\left[\frac{1}{\sqrt{N}}\sum_{i=1}^{N}[\varepsilon_{it}\varepsilon_{is}-E(\varepsilon_{it}\varepsilon_{is})]\right]^{2}\right)^{1/2}.
$$
But $T^{-1}\sum_{i=1}^{T}\|f_{t}\|^{2}=\|F^{\top}F/T\|=r$. Thus the above expression is equal to $rN^{-1/2}O_{p}(1)$. Next, $|\frac{1}{N}\sum_{i=1}^{N}\sigma_{ii,ts}|\leq \tau_{ts}$ by Assumption \ref{assum:error} 2. Again by the Cauchy-Schwarz inequality,
\begin{equation}\nonumber
\begin{aligned}
\left\|\frac{1}{T^{2}}\sum_{t=1}^{T}\sum_{s=1}^{T}f_{t}f_{s}^{\top}\frac{1}{N}\sum_{i=1}^{N}\sigma_{ii,ts}\right\| &\leq\left(\frac{1}{T^{2}}\sum_{t=1}^{T}\sum_{s=1}^{T}\|f_{t}\|^{2} \|f_{s}\|^{2}\right)^{1/2}\left(\frac{1}{T^{2}}\sum_{t=1}^{T}\sum_{s=1}^{T}\tau_{ts}^{ 2}\right)^{1/2}\\
&=rT^{-1/2}\left(\frac{1}{T}\sum_{t=1}^{T}\sum_{s=1}^{T}\tau_{ts}^{2 }\right)^{1/2}\\
&=rO_p\left(T^{-1/2}\right).
\end{aligned}
\end{equation}
Hence, we have
$$
\sup_{F\in\mathcal{F}}\left\|\frac{1}{NT} \sum_{i=1}^N \beta_i^\top X_i^\top  M_F^\prime \varepsilon_i\right\|=o_p(1).
$$

(ii) According to $M_F=I_T-P_F$, we have
$$
\frac{1}{NT} \sum_{i=1}^N \lambda_i^{0\top}  F^{0\top} P_F \varepsilon_i=\frac{1}{NT} \sum_{i=1}^N \lambda_i^{0\top}  F^{0\top} \varepsilon_i-\frac{1}{NT} \sum_{i=1}^N \lambda_i^{0\top}  F^{0\top} P_F \varepsilon_i.
$$
We first prove $\sup_{F\in\mathcal{F}}\frac{1}{NT} \sum_{i=1}^N \lambda_i^{0\top}  F^{0\top}  \varepsilon_i=o_{p}(1)$.
\begin{equation}\nonumber
\begin{aligned}
\frac{1}{NT}\left\|\sum_{i=1}^N \lambda_i^{0\top}  F^{0\top}  \varepsilon_i\right\|&\leq\frac{1}{NT}\sum_{i=1}^N\left\| \lambda_i^{0\top}  F^{0\top}  \varepsilon_i\right\|\\
&\leq\frac{1}{N}\sum_{i=1}^N\|\lambda_i\|\left\|\frac{1}{T}\sum_{t=1}^Tf_t^0\varepsilon_{it}\right\|\\
&\leq\left(\frac{1}{N}\sum_{i=1}^N\|\lambda_i\|^2\right)^{1/2}\left(\frac{1}{N}\sum_{i=1}^N\left\|\frac{1}{T}\sum_{t=1}^Tf_t^0\varepsilon_{it}\right\|^2\right)^{1/2}.
\end{aligned}
\end{equation}
Same as the proof in (i), we have $\sup_{F\in\mathcal{F}}\frac{1}{NT} \sum_{i=1}^N \lambda_i^{0\top}  F^{0\top}  \varepsilon_i=o_{p}(1)$.

For the second term, using $P_{F}=FF^{\top}/T$, we have
\begin{equation}\nonumber
\begin{aligned}
\frac{1}{NT}\left\|\sum_{i=1}^N \lambda_i^{0\top}  F^{0\top} P_F \varepsilon_i\right\|&=\frac{1}{NT}\left\|\sum_{i=1}^N \lambda_i^{0\top}  F^{0\top} F\left(\frac{F^\top\varepsilon_i}{T}\right) \right\|\\
&\leq\frac{1}{NT}\sum_{i=1}^N\left\|F^{0\top}F\right\|\left\|\frac{1}{T}\sum_{t=1}^Tf_t\varepsilon_{it}\right\|\\
&\leq\left(\frac{1}{N}\sum_{i=1}^N\frac{1}{T^2}\|F^0\|^2\|F\|^2\right)^{1/2}\left(\frac{1}{N}\sum_{i=1}^N\left\|\frac{1}{T}\sum_{t=1}^Tf_t\varepsilon_{it}\right\|^2\right)^{1/2}.
\end{aligned}
\end{equation}
Note that $T^{-1/2}\|F\|=\sqrt{r}$, $T^{-1/2}\|F^0\|=\sqrt{r}$ and according to the proof of (i), we have $\sup_{F\in\mathcal{F}}\frac{1}{NT} \sum_{i=1}^N \lambda_i^{0\top}  F^{0\top} P_F \varepsilon_i=o_{p}(1)$.

Hence, we have
$$
\sup_{F\in\mathcal{F}}\left\|\frac{1}{NT} \sum_{i=1}^N \lambda_i^{0\top}  F^{0\top} M_F \varepsilon_i\right\|=o_p(1).
$$

(iii) Using $P_{F}=FF^{\top}/T$, we have
$$
\frac{1}{NT}\left\|\sum_{i=1}^N \varepsilon_i^\top P_{F} \varepsilon_i\right\|=\frac{1}{N}\left\|\sum_{i=1}^N \left(\frac{\varepsilon_i^\top F}{T}\right)\left(\frac{F^\top\varepsilon_i}{T}\right)\right\|\leq\frac{1}{N}\sum_{i=1}^N \left\|\frac{1}{T}\sum_{t=1}^Tf_t\varepsilon_{it}\right\|^2=o_p(1).
$$
Similarly,
$$
\frac{1}{NT}\left\|\sum_{i=1}^N \varepsilon_i^\top P_{F^0} \varepsilon_i\right\|=\frac{1}{N}\left\|\sum_{i=1}^N \left(\frac{\varepsilon_i^\top F^0}{T}\right)\left(\frac{F^{0\top}\varepsilon_i}{T}\right)\right\|\leq\frac{1}{N}\sum_{i=1}^N \left\|\frac{1}{T}\sum_{t=1}^Tf_t^0\varepsilon_{it}\right\|^2=o_p(1).
$$
Hence, we have
$$
\sup_{F\in\mathcal{F}}\left\|\frac{1}{NT} \sum_{i=1}^N \varepsilon_i^\top (P_{F_0} - P_{F}) \varepsilon_i\right\|=o_p(1).
$$

(iv) According to $M_F=I_T-P_F$, we have
$$
\frac{1}{NT} \sum_{i=1}^N\beta_i^\top X_i^\top M_F\left\{Z_i^\ast({\alpha}^0_{i})-Z_i^\ast(\widehat{\alpha}_{i})\right\}=\frac{1}{NT} \sum_{i=1}^N\beta_i^\top X_i^\top \left\{Z_i^\ast({\alpha}^0_{i})-Z_i^\ast(\widehat{\alpha}_{i})\right\}-\frac{1}{NT} \sum_{i=1}^N\beta_i^\top X_i^\top P_F\left\{Z_i^\ast({\alpha}^0_{i})-Z_i^\ast(\widehat{\alpha}_{i})\right\}.
$$
For the first term, by Lemma \ref{lemmaA.1},
\begin{equation}\nonumber
\frac{1}{NT}\left\|\sum_{i=1}^N\beta_i^\top X_i^\top \left\{Z_i^\ast({\alpha}^0_{i})-Z_i^\ast(\widehat{\alpha}_{i})\right\}\right\|\leq\frac{1}{NT}\sum_{i=1}^N\left\|\beta_i\right\| \left\|X_i\right\| \left\|Z_i^\ast({\alpha}^0_{i})-Z_i^\ast(\widehat{\alpha}_{i})\right\|
=O_p\left(\frac{1}{\sqrt{T}}\right).
\end{equation}
Similarly, for the second term,
\begin{equation}\nonumber
\frac{1}{NT}\left\|\sum_{i=1}^N\beta_i^\top X_i^\top P_F \left\{Z_i^\ast({\alpha}^0_{i})-Z_i^\ast(\widehat{\alpha}_{i})\right\}\right\|\leq\frac{1}{NT}\sum_{i=1}^N\left\|\beta_i\right\| \left\|X_i\right\| \left\|P_F\right\|\left\|Z_i^\ast({\alpha}^0_{i})-Z_i^\ast(\widehat{\alpha}_{i})\right\|=O_p\left(\frac{1}{\sqrt{T}}\right).
\end{equation}
Hence, we have
$$
\sup_{F\in\mathcal{F}}\left\|\frac{1}{NT} \sum_{i=1}^N\beta_i^\top X_i^\top M_F\left\{Z_i^\ast({\alpha}^0_{i})-Z_i^\ast(\widehat{\alpha}_{i})\right\}\right\|=o_p(1).
$$

(v) According to $M_F=I_T-P_F$, we have
$$
\frac{1}{NT} \sum_{i=1}^N\lambda_i^{0\top}F^{0\top}M_F\left\{Z_i^\ast(\widehat{\alpha}_{i})-Z_i^\ast(\alpha^0_{i})\right\}=\frac{1}{NT} \sum_{i=1}^N\lambda_i^{0\top}F^{0\top}\left\{Z_i^\ast(\widehat{\alpha}_{i})-Z_i^\ast(\alpha^0_{i})\right\}-\frac{1}{NT} \sum_{i=1}^N\lambda_i^{0\top}F^{0\top}P_F\left\{Z_i^\ast(\widehat{\alpha}_{i})-Z_i^\ast(\alpha^0_{i})\right\}.
$$
For the first term, by Lemma \ref{lemmaA.1},
$$
\frac{1}{NT}\left\|\sum_{i=1}^N\lambda_i^{0\top}F^{0\top}\left\{Z_i^\ast(\widehat{\alpha}_{i})-Z_i^\ast(\alpha^0_{i})\right\}\right\|\leq\frac{1}{NT}\sum_{i=1}^N\left\|\lambda_i^{0}\right\|\left\|F^{0}\right\|\left\|Z_i^\ast(\widehat{\alpha}_{i})-Z_i^\ast(\alpha^0_{i})\right\|=O_p\left(\frac{1}{\sqrt{T}}\right).
$$
Similarly, for the second term,
$$
\frac{1}{NT}\left\|\sum_{i=1}^N\lambda_i^{0\top}F^{0\top}P_F\left\{Z_i^\ast(\widehat{\alpha}_{i})-Z_i^\ast(\alpha^0_{i})\right\}\right\|\leq\frac{1}{NT}\sum_{i=1}^N\left\|\lambda_i^{0}\right\|\left\|F^{0}\right\|\left\|P_F\right\|\left\|Z_i^\ast(\widehat{\alpha}_{i})-Z_i^\ast(\alpha^0_{i})\right\|=O_p\left(\frac{1}{\sqrt{T}}\right).
$$
Hence, we have
$$
\sup_{F\in\mathcal{F}}\left\|\frac{1}{NT} \sum_{i=1}^N\lambda_i^{0\top}F^{0\top}M_F\left\{Z_i^\ast(\widehat{\alpha}_{i})-Z_i^\ast(\alpha^0_{i})\right\}\right\|=o_p(1).
$$

(vi) Using $P_{F^0}=F^0F^{0\top}/T$, by Lemma \ref{lemmaA.1},
\begin{equation}\nonumber
\begin{aligned}
\frac{1}{NT}\left\|\sum_{i=1}^N\varepsilon_i^\top P_{F^0}\left\{Z_i^\ast(\widehat{\alpha}_{i})-Z_i^\ast(\alpha^0_{i})\right\}\right\|&=\frac{1}{NT}\left\|\sum_{i=1}^N\varepsilon_i^\top \left(\frac{F^0F^{0\top}}{T}\right)\left\{Z_i^\ast(\widehat{\alpha}_{i})-Z_i^\ast(\alpha^0_{i})\right\}\right\|\\
&\leq\frac{1}{NT}\sum_{i=1}^N\left\|\frac{1}{T}\sum_{t=1}^Tf_t^0\varepsilon_{it}\right\|\left\|F^0\right\|\left\|Z_i^\ast(\widehat{\alpha}_{i})-Z_i^\ast(\alpha^0_{i})\right\|\\
&\leq\left(\frac{1}{N}\sum_{i=1}^N\left\|\frac{1}{T}\sum_{t=1}^Tf_t^0\varepsilon_{it}\right\|^2\right)^{1/2}\left(\frac{1}{N}\sum_{i=1}^N\frac{1}{T^2}\left\|F^0\right\|^2\left\|Z_i^\ast(\widehat{\alpha}_{i})-Z_i^\ast(\alpha^0_{i})\right\|^2\right)^{1/2}.
\end{aligned}
\end{equation}
Hence, we have
$$
\sup_{F\in\mathcal{F}}\left\|\frac{1}{NT} \sum_{i=1}^N\varepsilon_i^\top(P_{F^0}-P_F)\left\{Z_i^\ast(\widehat{\alpha}_{i})-Z_i^\ast(\alpha^0_{i})\right\}\right\|=o_p(1).
$$

(vii) We first prove $\sup_{F\in\mathcal{F}}\frac{1}{NT} \sum_{i=1}^N\left\{Z_i^\ast(\widehat{\alpha}_{i})-Z_i^\ast(\alpha^0_{i})\right\}^\top P_{F^0}\left\{Z_i^\ast(\widehat{\alpha}_{i})-Z_i^\ast(\alpha^0_{i})\right\}=o_p(1)$.
$$
\frac{1}{NT}\left\|\sum_{i=1}^N\left\{Z_i^\ast(\widehat{\alpha}_{i})-Z_i^\ast(\alpha^0_{i})\right\}^\top P_{F^0}\left\{Z_i^\ast(\widehat{\alpha}_{i})-Z_i^\ast(\alpha^0_{i})\right\}\right\|\leq\frac{1}{NT}\sum_{i=1}^N\left\|Z_i^\ast(\widehat{\alpha}_{i})-Z_i^\ast(\alpha^0_{i})\right\|^2\left\|P_{F^0}\right\|=O_p\left(\frac{1}{T}\right).
$$
Similarly,
$$
\sum_{i=1}^N\left\|\sum_{i=1}^N\left\{Z_i^\ast(\widehat{\alpha}_{i})-Z_i^\ast(\alpha^0_{i})\right\}^\top P_F\left\{Z_i^\ast(\widehat{\alpha}_{i})-Z_i^\ast(\alpha^0_{i})\right\}\right\|\leq\frac{1}{NT}\sum_{i=1}^N\left\|Z_i^\ast(\widehat{\alpha}_{i})-Z_i^\ast(\alpha^0_{i})\right\|^2\left\|P_F\right\|=O_p\left(\frac{1}{T}\right).
$$
Hence, we have
$$
\sup_{F\in\mathcal{F}}\left\|\frac{1}{NT} \sum_{i=1}^N\left\{Z_i^\ast(\widehat{\alpha}_{i})-Z_i^\ast(\alpha^0_{i})\right\}^\top(P_{F^0}-P_F)\left\{Z_i^\ast(\widehat{\alpha}_{i})-Z_i^\ast(\alpha^0_{i})\right\}\right\|=o_p(1).
$$
\end{proof}
%\clearpage

\section{Proof of Theorem \ref{theo:consistency}}\label{secB}
\begin{proof}
From $ Z_i^\ast(\widehat{\alpha}_{i})  =Z_i^\ast(\widehat{\alpha}_{i})-Z_i^\ast({\alpha}_{i}^0)+ X_i \beta_i^0  + F^0 \lambda_i^0 + \varepsilon_i $, we have
\begin{equation}\nonumber
\begin{aligned}
\widehat{\beta}_{i} =& \left( X_i^\top M_{\widehat{F}} X_i \right)^{-1}X_i^\top M_{\widehat{F}} Z_i^\ast(\widehat{\alpha}_{i})\\
=&\left( X_i^\top M_{\widehat{F}} X_i \right)^{-1}X_i^\top M_{\widehat{F}}\left\{Z_i^\ast(\widehat{\alpha}_{i})-Z_i^\ast({\alpha}_{i}^0)+ X_i \beta_i^0  + F^0 \lambda_i^0 + \varepsilon_i\right\}\\
=&\left( X_i^\top M_{\widehat{F}} X_i \right)^{-1}X_i^\top M_{\widehat{F}}\left\{Z_i^\ast(\widehat{\alpha}_{i})-Z_i^\ast({\alpha}_{i}^0)\right\}+\beta_i^0\\
&+\left( X_i^\top M_{\widehat{F}} X_i \right)^{-1}X_i^\top M_{\widehat{F}}F^0 \lambda_i^0+\left( X_i^\top M_{\widehat{F}} X_i \right)^{-1}X_i^\top M_{\widehat{F}}\varepsilon_i.
\end{aligned}
\end{equation}
Thus,
\begin{equation}\nonumber
\begin{aligned}
\widehat{\beta}_{i}-\beta_i^0=&\left( X_i^\top M_{\widehat{F}} X_i \right)^{-1}X_i^\top M_{\widehat{F}}F^0 \lambda_i^0+\left( X_i^\top M_{\widehat{F}} X_i \right)^{-1}X_i^\top M_{\widehat{F}}\varepsilon_i\\
&+\left( X_i^\top M_{\widehat{F}} X_i \right)^{-1}X_i^\top M_{\widehat{F}}\left\{Z_i^\ast(\widehat{\alpha}_{i})-Z_i^\ast({\alpha}_{i}^0)\right\}
\end{aligned}
\end{equation}
or
\begin{equation}\label{eq:beta-beta0}
\left( \frac{1}{T}X_i^\top M_{\widehat{F}} X_i \right)\left(\widehat{\beta}_{i}-\beta_i^0\right)=\frac{1}{T}X_i^\top M_{\widehat{F}}F^0 \lambda_i^0+\frac{1}{T}X_i^\top M_{\widehat{F}}\varepsilon_i+\frac{1}{T}X_i^\top M_{\widehat{F}}\left\{Z_i^\ast(\widehat{\alpha}_{i})-Z_i^\ast({\alpha}_{i}^0)\right\}.
\end{equation}
In view of $ M_{\widehat{F}} \widehat{F} = 0 $, we have $ M_{\widehat{F}} F^0 = M_{\widehat{F}}(F^0 - \widehat{F} H^{-1}) $. From \eqref{eq:B.2}, we get
$$
F^0 - \widehat{F} H^{-1} = -\left[I1 + \cdots + I15\right]\left(\frac{F^{0\top} \widehat{F}}{T}\right)^{-1}\left(\frac{\Lambda^{0\top} \Lambda^0}{N}\right)^{-1}.
$$
It follows that
\begin{equation}\nonumber
\begin{aligned}
\frac{1}{T}X_i^\top M_{\widehat{F}}F^0 \lambda_i^0=&\frac{1}{T}X_i^\top M_{\widehat{F}}\left(F^0 - \widehat{F} H^{-1}\right) \lambda_i^0\\
=&-\frac{1}{T}X_i^\top M_{\widehat{F}}\left[I1 + \cdots + I15\right]\left(\frac{F^{0\top} \widehat{F}}{T}\right)^{-1}\left(\frac{\Lambda^{0\top} \Lambda^0}{N}\right)^{-1}\lambda_i^0\\
:=&J1+\cdots+J15.
\end{aligned}
\end{equation}
For the first term,
\begin{equation}\nonumber
\begin{aligned}
J1=-\frac{1}{T}X_i^\top M_{\widehat{F}}\left(I1\right)\left(\frac{F^{0\top} \widehat{F}}{T}\right)^{-1}\left(\frac{\Lambda^{0\top} \Lambda^0}{N}\right)^{-1}\lambda_i^0,
\end{aligned}
\end{equation}
By the proof of Proposition \ref{pro:B.1},
\begin{equation}\nonumber
\begin{aligned}
\left\|J1\right\|=&\left\|-\frac{1}{T}X_i^\top M_{\widehat{F}}\left(I1\right)\left(\frac{F^{0\top} \widehat{F}}{T}\right)^{-1}\left(\frac{\Lambda^{0\top} \Lambda^0}{N}\right)^{-1}\lambda_i^0\right\|\\
\leq&\frac{1}{\sqrt{T}}\left\|I1\right\|\\
=&O_p\left(\frac{1}{N}\sum_{i=1}^N\left\|\widehat{\beta}_i-\beta_i^0\right\|^2\right).
\end{aligned}
\end{equation}
Consider
\begin{equation}\nonumber
\begin{aligned}
J2=&-\frac{1}{T}X_i^\top M_{\widehat{F}}\left(I2\right)\left(\frac{F^{0\top} \widehat{F}}{T}\right)^{-1}\left(\frac{\Lambda^{0\top} \Lambda^0}{N}\right)^{-1}\lambda_i^0\\
=&\frac{1}{T}X_i^\top M_{\widehat{F}}\left[\frac{1}{N} \sum_{k=1}^N X_k \left(\widehat{\beta}_k-\beta_k^0\right) \lambda_k^{0\top}  \right]\left(\frac{\Lambda^{0\top} \Lambda^0}{N}\right)^{-1}\lambda_i^0\\
=&\frac{1}{TN}\sum_{k=1}^N\left(X_i^\top M_{\widehat{F}}X_k\right)\left[\lambda_k^{0\top}\left(\frac{\Lambda^{0\top} \Lambda^0}{N}\right)^{-1}\lambda_i^0\right]\left(\widehat{\beta}_k-\beta_k^0\right)\\
=&\frac{1}{TN}\sum_{k=1}^N\left(X_i^\top M_{\widehat{F}}X_ka_{ik}\right)\left(\widehat{\beta}_k-\beta_k^0\right),
\end{aligned}
\end{equation}
where $ a_{ik} = \lambda_i^{0\top}  (\Lambda^{0\top} \Lambda^0/N)^{-1} \lambda_k^0 $ is a scalar and thus commutable with $ \widehat{\beta}_k - \beta_k^0 $. Now consider
\begin{equation}\nonumber
\begin{aligned}
J3=&-\frac{1}{T}X_i^\top M_{\widehat{F}}\left(I3\right)\left(\frac{F^{0\top} \widehat{F}}{T}\right)^{-1}\left(\frac{\Lambda^{0\top} \Lambda^0}{N}\right)^{-1}\lambda_i^0\\
=&\frac{1}{T}X_i^\top M_{\widehat{F}}\left[\frac{1}{NT} \sum_{k=1}^N X_k (\widehat{\beta}_k-\beta_k^0) \varepsilon_k^\top \widehat{F}\right]\left(\frac{F^{0\top} \widehat{F}}{T}\right)^{-1}\left(\frac{\Lambda^{0\top} \Lambda^0}{N}\right)^{-1}\lambda_i^0\\
=&\frac{1}{NT}\sum_{k=1}^NX_i^\top M_{\widehat{F}}  X_k (\widehat{\beta}_k-\beta_k^0) \left(\frac{\varepsilon_k^\top \widehat{F}}{T}\right)\left(\frac{F^{0\top} \widehat{F}}{T}\right)^{-1}\left(\frac{\Lambda^{0\top} \Lambda^0}{N}\right)^{-1}\lambda_i^0
\end{aligned}
\end{equation}
By Assumption \ref{assum:weak dependence} and Proposition \ref{pro:B.1}, writing 
\begin{equation}\nonumber
\begin{aligned}
\frac{1}{T}\varepsilon_k^\top \widehat{F}=&\frac{1}{T}\varepsilon_k^\top F^0 H+\frac{1}{T}\varepsilon_k^\top (\widehat{F} - F^0 H)\\
=&O_p\left(\frac{1}{\sqrt{T}}\right)+O_p\left(\left(\frac{1}{N}\sum_{i=1}^N\left\|\widehat{\beta}_i-\beta_i^0\right\|^2\right)^{1/2}\right) + O_p\left(\frac{1}{\min[\sqrt{N}, \sqrt{T}]}\right),
\end{aligned}
\end{equation}
it is easy to see that $ \left\|J3\right\| = o_p(1)\cdot O_p((\frac{1}{N}\sum_{i=1}^N\|\widehat{\beta}_i-\beta_i^0\|^2)^{1/2})$. Next
\begin{equation}\nonumber
\begin{aligned}
J4&=-\frac{1}{T}X_i^\top M_{\widehat{F}}\left(I4\right)\left(\frac{F^{0\top} \widehat{F}}{T}\right)^{-1}\left(\frac{\Lambda^{0\top} \Lambda^0}{N}\right)^{-1}\lambda_i^0\\
&=\frac{1}{T}X_i^\top M_{\widehat{F}}\left[\frac{1}{NT} \sum_{k=1}^N F^0 \lambda_k^0 (\widehat{\beta}_k-\beta_k^0)^\top  X_k^\top  \widehat{F}\right]\left(\frac{F^{0\top} \widehat{F}}{T}\right)^{-1}\left(\frac{\Lambda^{0\top} \Lambda^0}{N}\right)^{-1}\lambda_i^0\\
&=\frac{1}{NT}\sum_{k=1}^NX_i^\top M_{\widehat{F}}F^0 \lambda_k^0 (\widehat{\beta}_k-\beta_k^0)^\top\left(\frac{X_k^\top  \widehat{F}}{T}\right)\left(\frac{F^{0\top} \widehat{F}}{T}\right)^{-1}\left(\frac{\Lambda^{0\top} \Lambda^0}{N}\right)^{-1}\lambda_i^0.
\end{aligned}
\end{equation}
Writing $ M_{\widehat{F}} F^0 = M_{\widehat{F}} (F^0 - \widehat{F} H^{-1}) $ and using that $ T^{-1/2} \| F^0 - \widehat{F} H^{-1} \| $ is small, then $ \left\|J4\right\| = o_p(1)\cdot O_p((\frac{1}{N}\sum_{i=1}^N\|\widehat{\beta}_i-\beta_i^0\|^2)^{1/2})$. It is easy to show $ \left\|J5\right\| = o_p(1)\cdot O_p((\frac{1}{N}\sum_{i=1}^N\|\widehat{\beta}_i-\beta_i^0\|^2)^{1/2})$ and thus it is omitted.

The terms $ J6-J8 $ do not explicitly depend on $ \widehat{\beta}_i - \beta_i^0 $. Consider
\begin{equation}\nonumber
\begin{aligned}
J6&=-\frac{1}{T}X_i^\top M_{\widehat{F}}\left(I6\right)\left(\frac{F^{0\top} \widehat{F}}{T}\right)^{-1}\left(\frac{\Lambda^{0\top} \Lambda^0}{N}\right)^{-1}\lambda_i^0\\
&=-\frac{1}{T}X_i^\top M_{\widehat{F}}\left(\frac{1}{NT} \sum_{k=1}^N F^0 \lambda_k^0 \varepsilon_k^\top \widehat{F}\right)\left(\frac{F^{0\top} \widehat{F}}{T}\right)^{-1}\left(\frac{\Lambda^{0\top} \Lambda^0}{N}\right)^{-1}\lambda_i^0\\
&=-\frac{1}{NT}\sum_{k=1}^N X_i^\top M_{\widehat{F}}F^0 \lambda_k^0\left(\frac{\varepsilon_k^\top \widehat{F}}{T}\right)\left(\frac{F^{0\top} \widehat{F}}{T}\right)^{-1}\left(\frac{\Lambda^{0\top} \Lambda^0}{N}\right)^{-1}\lambda_i^0.
\end{aligned}
\end{equation}
Denote $ G = (F^{0\top} \widehat{F} / T)^{-1} (\Lambda^{0\top} \Lambda^0 / N)^{-1} $ for the moment: it is a matrix of fixed dimension and does not vary with $ i $ and $k$. Using $ M_{\widehat{F}} F^0 = M_{\widehat{F}} (F^0 - \widehat{F} H^{-1}) $, we can write
$$
J6 = -\frac{1}{T} X_i^\top  M_{\widehat{F}} \left(F^0 - \widehat{F} H^{-1}\right) \left[ \frac{1}{N} \sum_{k=1}^{N} \lambda_k^0 \left( \frac{\varepsilon_k^\top \widehat{F}}{T} \right) \right] G \lambda_i^0.
$$
Now, by Assumption \ref{assum:weak dependence} and Proposition \ref{pro:B.1}, 
\begin{equation}\nonumber
\begin{aligned}
\left\|\frac{1}{NT} \sum_{k=1}^{N} \lambda_k^0 \varepsilon_k^\top \widehat{F}\right\| &=\left\|\frac{1}{NT} \sum_{k=1}^{N} \lambda_k^0 \varepsilon_k^\top F^0 H + \frac{1}{NT} \sum_{k=1}^{N} \lambda_k^0 \varepsilon_k^\top (\widehat{F} - F^0 H)\right\| \\
&\leq \left\|\frac{1}{NT} \sum_{k=1}^{N} \lambda_k^0 \varepsilon_k^\top F^0 H\right\| + \left\|\frac{1}{NT} \sum_{k=1}^{N} \lambda_k^0 \varepsilon_k^\top (\widehat{F} - F^0 H)\right\|\\
&= O_p \left( \frac{1}{\sqrt{NT}} \right) + O_p\left(\left(\frac{1}{N}\sum_{i=1}^N\left\|\widehat{\beta}_i-\beta_i^0\right\|^2\right)^{1/2}\right) + O_p\left(\frac{1}{\min[\sqrt{N}, \sqrt{T}]}\right)\\
&= O_p\left(\left(\frac{1}{N}\sum_{i=1}^N\left\|\widehat{\beta}_i-\beta_i^0\right\|^2\right)^{1/2}\right) + O_p\left(\frac{1}{\min[\sqrt{N}, \sqrt{T}]}\right).
\end{aligned}
\end{equation}

Furthermore,
$$
\left\|\frac{1}{T}  X_i^\top  M_{\widehat{F}} (\widehat{F} - F^0 H)\right\|
  = O_p\left(\left(\frac{1}{N}\sum_{i=1}^N\left\|\widehat{\beta}_i-\beta_i^0\right\|^2\right)^{1/2}\right) + O_p\left(\frac{1}{\min[\sqrt{N}, \sqrt{T}]}\right)
$$
and noting $ G $ does not depend on $ i $ and $k$, and $ \|G\| = O_p(1) $, we have
\begin{equation}\nonumber
\begin{aligned}
\left\|J6\right\|&=\left[O_p\left(\left(\frac{1}{N}\sum_{i=1}^N\left\|\widehat{\beta}_i-\beta_i^0\right\|^2\right)^{1/2}\right) + O_p\left(\frac{1}{\min[\sqrt{N}, \sqrt{T}]}\right)\right]^2\\
&=o_p(1)\cdot O_p\left(\left(\frac{1}{N}\sum_{i=1}^N\left\|\widehat{\beta}_i-\beta_i^0\right\|^2\right)^{1/2}\right)+O_p\left(\frac{1}{\min[{N}, {T}]}\right).
\end{aligned}
\end{equation}
The term $ J7 $ is simply
\begin{equation}\nonumber
\begin{aligned}
J7&=-\frac{1}{T}X_i^\top M_{\widehat{F}}\left(I7\right)\left(\frac{F^{0\top} \widehat{F}}{T}\right)^{-1}\left(\frac{\Lambda^{0\top} \Lambda^0}{N}\right)^{-1}\lambda_i^0\\
&=-\frac{1}{T}X_i^\top M_{\widehat{F}}\left(\frac{1}{NT} \sum_{k=1}^N \varepsilon_k \lambda_k^{0\top}  F^0 \widehat{F}\right)\left(\frac{F^{0\top} \widehat{F}}{T}\right)^{-1}\left(\frac{\Lambda^{0\top} \Lambda^0}{N}\right)^{-1}\lambda_i^0\\
&=-\frac{1}{NT}\sum_{k=1}^NX_i^\top M_{\widehat{F}}\varepsilon_k \lambda_k^{0\top}\left(\frac{\Lambda^{0\top} \Lambda^0}{N}\right)^{-1}\lambda_i^0\\
&=-\frac{1}{NT}\sum_{k=1}^Na_{ik}X_i^\top M_{\widehat{F}}\varepsilon_k
\end{aligned}
\end{equation}
Next consider $ J8 $, which has the expression
\begin{equation}\nonumber
\begin{aligned}
J8&=-\frac{1}{T}X_i^\top M_{\widehat{F}}\left(I8\right)\left(\frac{F^{0\top} \widehat{F}}{T}\right)^{-1}\left(\frac{\Lambda^{0\top} \Lambda^0}{N}\right)^{-1}\lambda_i^0\\
&=-\frac{1}{T}X_i^\top M_{\widehat{F}}\left(\frac{1}{NT} \sum_{k=1}^N \varepsilon_k \varepsilon_k^\top \widehat{F}\right)\left(\frac{F^{0\top} \widehat{F}}{T}\right)^{-1}\left(\frac{\Lambda^{0\top} \Lambda^0}{N}\right)^{-1}\lambda_i^0\\
&=-\frac{1}{NT^2}\sum_{k=1}^NX_i^\top M_{\widehat{F}}\varepsilon_k \varepsilon_k^\top \widehat{F}\left(\frac{F^{0\top} \widehat{F}}{T}\right)^{-1}\left(\frac{\Lambda^{0\top} \Lambda^0}{N}\right)^{-1}\lambda_i^0.
\end{aligned}
\end{equation}
Let $ E(\varepsilon_k \varepsilon_k^\top) = \Omega_k$. Note that $ G = (F^{0\top} \widehat{F}/T)^{-1}(\Lambda^{0\top} \Lambda^{0}/N)^{-1} $ and $ \|G\| = O_p(1) $, and rewriting gives
\begin{equation}\label{eq:J8}
\begin{aligned}
J8=&-\frac{1}{NT^2} \sum_{k=1}^N X_i^\top M_{\widehat{F}} \Omega_k \widehat{F} G \lambda_i^0\\
&-\frac{1}{NT^2} \sum_{k=1}^N X_i^\top M_{\widehat{F}} (\varepsilon_k \varepsilon_k^\top - \Omega_k) \widehat{F} G \lambda_i^0.
\end{aligned}
\end{equation}
Denote the first term on the right by $ A_{NT} $. Let $\Omega=\frac{1}{N}\sum_{i=1}\Omega_k$, then
\begin{equation}\nonumber
\begin{aligned}
\left\|A_{NT}\right\|=&\left\|-\frac{1}{NT^2} \sum_{k=1}^N X_i^\top M_{\widehat{F}} \Omega_k \widehat{F} G \lambda_i^0\right\|\\
=&\frac{1}{T^2}\left\|X_i^\top M_{\widehat{F}} \Omega\widehat{F} G \lambda_i^0\right\|\\
=&\leq\frac{1}{T^2}\left\|X_i^\top M_{\widehat{F}}\right\|\left\|\Omega\widehat{F}\right\|\\
=&O_p\left(\frac{1}{T}\right)
\end{aligned}
\end{equation}
because $\|X_{i}^{\top}M_{\widehat{F}}\|\leq\|X_{i}\|$, and $\|\Omega\widehat{F}\|\leq\lambda_{\max}(\Omega)\times\|\widehat{F}\|=\lambda_{\max}( \Omega)\sqrt{rT}$, where $\lambda_{\max}(\Omega)$ is the largest eigenvalue of $\Omega$ and is bounded by assumption.
By Lemma \ref{lemma:B.1}, we have
\begin{equation}\nonumber
\begin{aligned}
J8 &= O_p\left(\frac{1}{T}\right)+O_p \left( \frac{1}{T\sqrt{N}} \right) +o_p(1)\cdot O_p\left(\left(\frac{1}{N}\sum_{i=1}^N\left\|\widehat{\beta}_i-\beta_i^0\right\|^2\right)^{1/2}\right)+\frac{1}{\sqrt{NT}} O_p(\delta_{NT}^{-1})+\frac{1}{\sqrt{N}} O_p (\delta_{NT}^{-2})\\
&=O_p\left(\frac{1}{T}\right)+o_p(1)\cdot O_p\left(\left(\frac{1}{N}\sum_{i=1}^N\left\|\widehat{\beta}_i-\beta_i^0\right\|^2\right)^{1/2}\right).
\end{aligned}
\end{equation}
The remaining terms $J9$-$J15$ are related to $\left\{Z_k^\ast(\widehat{\alpha}_{k})-Z_k^\ast({\alpha}_{k}^0)\right\}$, and we can use Lemma \ref{lemma:B.5} to analyze. For terms $J9$ and $J15$,
\begin{equation}\nonumber
\begin{aligned}
\left\|J9\right\|&=\left\|-\frac{1}{T}X_i^\top M_{\widehat{F}}\left(I9\right)\left(\frac{F^{0\top} \widehat{F}}{T}\right)^{-1}\left(\frac{\Lambda^{0\top} \Lambda^0}{N}\right)^{-1}\lambda_i^0\right\|\\
&=\left\|-\frac{1}{T}X_i^\top M_{\widehat{F}}\left(\frac{1}{NT} \sum_{k=1}^N\left\{Z_k^\ast(\widehat{\alpha}_{k})-Z_k^\ast({\alpha}_{k}^0)\right\}\left\{Z_k^\ast(\widehat{\alpha}_{k})-Z_k^\ast({\alpha}_{k}^0)\right\}^\top\widehat{F}\right)\left(\frac{F^{0\top} \widehat{F}}{T}\right)^{-1}\left(\frac{\Lambda^{0\top} \Lambda^0}{N}\right)^{-1}\lambda_i^0\right\|\\
&=O_p\left(\frac{1}{T}\right),
\end{aligned}
\end{equation}
\begin{equation}\nonumber
\begin{aligned}
\left\|J10\right\|&=\left\|-\frac{1}{T}X_i^\top M_{\widehat{F}}\left(I10\right)\left(\frac{F^{0\top} \widehat{F}}{T}\right)^{-1}\left(\frac{\Lambda^{0\top} \Lambda^0}{N}\right)^{-1}\lambda_i^0\right\|\\
&=\left\|-\frac{1}{T}X_i^\top M_{\widehat{F}}\left(\frac{1}{NT} \sum_{k=1}^N\left\{Z_k^\ast(\widehat{\alpha}_{k})-Z_k^\ast({\alpha}_{k}^0)\right\}(\beta_k^0 - \widehat{\beta}_k)^\top X_k^\top\widehat{F}\right)\left(\frac{F^{0\top} \widehat{F}}{T}\right)^{-1}\left(\frac{\Lambda^{0\top} \Lambda^0}{N}\right)^{-1}\lambda_i^0\right\|\\
&\leq\frac{1}{\sqrt{T}}\frac{1}{N}\sum_{k=1}^N\left\|\frac{1}{\sqrt{T}}X_i^\top M_{\widehat{F}}\right\|\left\|\widehat{\beta}_k-\beta_k^0\right\|\left\|\frac{1}{T}X_k^\top\widehat{F}\right\|\\
&=\frac{1}{\sqrt{T}}O_p\left(\left(\frac{1}{N}\sum_{i=1}^N\left\|\widehat{\beta}_i-\beta_i^0\right\|^2\right)^{1/2}\right).
\end{aligned}
\end{equation}
For terms $J11$ and $J12$, by Lemma \ref{lemma:B.5}, we have
\begin{equation}\nonumber
\begin{aligned}
\left\|J11\right\|&=\left\|-\frac{1}{T}X_i^\top M_{\widehat{F}}\left(I11\right)\left(\frac{F^{0\top} \widehat{F}}{T}\right)^{-1}\left(\frac{\Lambda^{0\top} \Lambda^0}{N}\right)^{-1}\lambda_i^0\right\|\\
&=\left\|-\frac{1}{T}X_i^\top M_{\widehat{F}}\left(\frac{1}{NT} \sum_{k=1}^N\left\{Z_k^\ast(\widehat{\alpha}_{k})-Z_k^\ast({\alpha}_{k}^0)\right\}\lambda_k^{0\top}  F^{0\top}\widehat{F}\right)\left(\frac{F^{0\top} \widehat{F}}{T}\right)^{-1}\left(\frac{\Lambda^{0\top} \Lambda^0}{N}\right)^{-1}\lambda_i^0\right\|\\
&\leq\frac{1}{\sqrt{T}}\frac{1}{N}\sum_{k=1}^N\left\|\frac{1}{\sqrt{T}}X_i^\top M_{\widehat{F}}\left\{Z_k^\ast(\widehat{\alpha}_{k})-Z_k^\ast({\alpha}_{k}^0)\right\}\right\|\left\|\frac{1}{T}F^{0\top}\widehat{F}\right\|\\
&=o_p\left(\frac{1}{\sqrt{T}}\right),
\end{aligned}
\end{equation}
\begin{equation}\nonumber
\begin{aligned}
\left\|J12\right\|&=\left\|-\frac{1}{T}X_i^\top M_{\widehat{F}}\left(I12\right)\left(\frac{F^{0\top} \widehat{F}}{T}\right)^{-1}\left(\frac{\Lambda^{0\top} \Lambda^0}{N}\right)^{-1}\lambda_i^0\right\|\\
&=\left\|-\frac{1}{T}X_i^\top M_{\widehat{F}}\left(\frac{1}{NT} \sum_{k=1}^N\left\{Z_k^\ast(\widehat{\alpha}_{k})-Z_k^\ast({\alpha}_{k}^0)\right\}\varepsilon_k^\top\widehat{F}\right)\left(\frac{F^{0\top} \widehat{F}}{T}\right)^{-1}\left(\frac{\Lambda^{0\top} \Lambda^0}{N}\right)^{-1}\lambda_i^0\right\|\\
&\leq\frac{1}{\sqrt{T}}\frac{1}{N}\sum_{k=1}^N\left\|\frac{1}{\sqrt{T}}X_i^\top M_{\widehat{F}}\left\{Z_k^\ast(\widehat{\alpha}_{k})-Z_k^\ast({\alpha}_{k}^0)\right\}\right\|\left\|\frac{1}{T}\varepsilon_k^\top\widehat{F}\right\|\\
&=o_p\left(\frac{1}{\sqrt{T}}\right).
\end{aligned}
\end{equation}
Consider $J13$, we have
\begin{equation}\nonumber
\begin{aligned}
\left\|J13\right\|&=\left\|-\frac{1}{T}X_i^\top M_{\widehat{F}}\left(I13\right)\left(\frac{F^{0\top} \widehat{F}}{T}\right)^{-1}\left(\frac{\Lambda^{0\top} \Lambda^0}{N}\right)^{-1}\lambda_i^0\right\|\\
&=\left\|-\frac{1}{T}X_i^\top M_{\widehat{F}}\left(\frac{1}{NT} \sum_{k=1}^NX_k (\beta_k^0 - \widehat{\beta}_k)\left\{Z_k^\ast(\widehat{\alpha}_{k})-Z_k^\ast({\alpha}_{k}^0)\right\}^\top\widehat{F}\right)\left(\frac{F^{0\top} \widehat{F}}{T}\right)^{-1}\left(\frac{\Lambda^{0\top} \Lambda^0}{N}\right)^{-1}\lambda_i^0\right\|\\
&\leq\frac{1}{\sqrt{T}}\frac{1}{N}\sum_{k=1}^N\left\|\frac{1}{T}X_i^\top M_{\widehat{F}}X_k\right\|\left\|\widehat{\beta}_k-\beta_k^0\right\|\left\|\frac{1}{\sqrt{T}}\widehat{F}\right\|\\
&=\frac{1}{\sqrt{T}}O_p\left(\left(\frac{1}{N}\sum_{i=1}^N\left\|\widehat{\beta}_i-\beta_i^0\right\|^2\right)^{1/2}\right).
\end{aligned}
\end{equation}
For term $J14$, by $M_{\widehat{F}}-M_{F^0}=o_p(1)$ and $M_{F^0}F^0=0$,
\begin{equation}\nonumber
\begin{aligned}
\left\|J14\right\|&=\left\|-\frac{1}{T}X_i^\top M_{\widehat{F}}\left(I14\right)\left(\frac{F^{0\top} \widehat{F}}{T}\right)^{-1}\left(\frac{\Lambda^{0\top} \Lambda^0}{N}\right)^{-1}\lambda_i^0\right\|\\
&=\left\|-\frac{1}{T}X_i^\top M_{\widehat{F}}\left(\frac{1}{NT} \sum_{k=1}^NF^0 \lambda_k^0\left\{Z_k^\ast(\widehat{\alpha}_{k})-Z_k^\ast({\alpha}_{k}^0)\right\}^\top\widehat{F}\right)\left(\frac{F^{0\top} \widehat{F}}{T}\right)^{-1}\left(\frac{\Lambda^{0\top} \Lambda^0}{N}\right)^{-1}\lambda_i^0\right\|\\
&\leq\frac{1}{\sqrt{T}}\frac{1}{N}\sum_{k=1}^N\left\|\frac{1}{T}X_i^\top M_{\widehat{F}}F^0 \lambda_k^0\left\{Z_k^\ast(\widehat{\alpha}_{k})-Z_k^\ast({\alpha}_{k}^0)\right\}^\top\right\|\left\|\frac{1}{\sqrt{T}}\widehat{F}\right\|\\
&\leq\frac{1}{\sqrt{T}}\frac{1}{N}\sum_{k=1}^N\left\|\frac{1}{T}X_i^\top (M_{\widehat{F}}-M_{F^0})F^0 \lambda_k^0\left\{Z_k^\ast(\widehat{\alpha}_{k})-Z_k^\ast({\alpha}_{k}^0)\right\}^\top\right\|\left\|\frac{1}{\sqrt{T}}\widehat{F}\right\|\\
&=o_p\left(\frac{1}{\sqrt{T}}\right),
\end{aligned}
\end{equation}
For term $J15$, by Lemma \ref{lemma:B.5}, we have
\begin{equation}\nonumber
\begin{aligned}
\left\|J15\right\|&=\left\|-\frac{1}{T}X_i^\top M_{\widehat{F}}\left(I15\right)\left(\frac{F^{0\top} \widehat{F}}{T}\right)^{-1}\left(\frac{\Lambda^{0\top} \Lambda^0}{N}\right)^{-1}\lambda_i^0\right\|\\
&=\left\|-\frac{1}{T}X_i^\top M_{\widehat{F}}\left(\frac{1}{NT} \sum_{k=1}^N\varepsilon_k\left\{Z_k^\ast(\widehat{\alpha}_{k})-Z_k^\ast({\alpha}_{k}^0)\right\}^\top\widehat{F}\right)\left(\frac{F^{0\top} \widehat{F}}{T}\right)^{-1}\left(\frac{\Lambda^{0\top} \Lambda^0}{N}\right)^{-1}\lambda_i^0\right\|\\
&\leq\frac{1}{\sqrt{T}}\frac{1}{N}\sum_{k=1}^N\left\|\frac{1}{T}X_i^\top M_{\widehat{F}}\varepsilon_k\right\|\left\|\frac{1}{\sqrt{T}}\left\{Z_k^\ast(\widehat{\alpha}_{k})-Z_k^\ast({\alpha}_{k}^0)\right\}^\top\widehat{F}\right\|\\
&=o_p\left(\frac{1}{\sqrt{T}}\right).
\end{aligned}
\end{equation}
Collecting terms from $J1$ to $J15$ with dominated terms ignored gives
\begin{equation}\nonumber
\begin{aligned}
\frac{1}{T}X_i^\top M_{\widehat{F}}F^0 \lambda_i^0=&J2+J7+O_p\left(\frac{1}{\min[{N}, {T}]}\right)+o_p\left(\frac{1}{\sqrt{T}}\right)\\
&+\frac{1}{\sqrt{T}}O_p\left(\left(\frac{1}{N}\sum_{i=1}^N\left\|\widehat{\beta}_i-\beta_i^0\right\|^2\right)^{1/2}\right)+o_p(1)\cdot O_p\left(\left(\frac{1}{N}\sum_{i=1}^N\left\|\widehat{\beta}_i-\beta_i^0\right\|^2\right)^{1/2}\right).
\end{aligned}
\end{equation}
By Lemma \ref{lemma:B.5}, we have
$$
\frac{1}{T}X_i^\top M_{\widehat{F}}\left\{Z_i^\ast(\widehat{\alpha}_{i})-Z_i^\ast({\alpha}_{i}^0)\right\}=o_p\left(\frac{1}{\sqrt{T}}\right).
$$
According to $M_{\widehat{F}}-M_{F^0}=o_p(1)$, then,
$$
\frac{1}{T}X_i^\top M_{\widehat{F}} X_i=\frac{1}{T}X_i^\top M_{F^0} X_i+o_p(1)
$$
and
$$
\frac{1}{T}X_i^\top M_{\widehat{F}}X_ka_{ik}=\frac{1}{T}X_i^\top M_{F^0}X_ka_{ik}+o_p(1).
$$
Thus, by Lemma \ref{lemma:B.3}, we have
\begin{equation}\nonumber
\begin{aligned}
\left( \frac{1}{T}X_i^\top M_{F^0} X_i \right)\left(\widehat{\beta}_{i}-\beta_i^0\right)=&J2+\frac{1}{T}X_i^\top M_{\widehat{F}}\varepsilon_i+J7+O_p\left(\frac{1}{\min[{N}, {T}]}\right)+o_p\left(\frac{1}{\sqrt{T}}\right)\\
&+\frac{1}{\sqrt{T}}O_p\left(\left(\frac{1}{N}\sum_{i=1}^N\left\|\widehat{\beta}_i-\beta_i^0\right\|^2\right)^{1/2}\right)+o_p(1)\cdot O_p\left(\left(\frac{1}{N}\sum_{i=1}^N\left\|\widehat{\beta}_i-\beta_i^0\right\|^2\right)^{1/2}\right)\\
=&\frac{1}{TN}\sum_{k=1}^N\left(X_i^\top M_{F^0}X_ka_{ik}\right)\left(\widehat{\beta}_k-\beta_k^0\right)+\frac{1}{T} X_i^\top M_{{F}^0}\varepsilon_i - \frac{1}{NT} \sum_{k=1}^{N} a_{ik} X_i^\top M_{{F}^0} \varepsilon_k\\
&+\left(\frac{1}{N}+\frac{1}{\sqrt{T}}\right)\left[O_p\left(\left(\frac{1}{N}\sum_{i=1}^N\left\|\widehat{\beta}_i-\beta_i^0\right\|^2\right)^{1/2}\right) + O_p\left(\frac{1}{\min[\sqrt{N}, \sqrt{T}]}\right)\right]\\
&+O_p\left(\frac{1}{\min[{N}, {T}]}\right)+o_p(1)\cdot O_p\left(\left(\frac{1}{N}\sum_{i=1}^N\left\|\widehat{\beta}_i-\beta_i^0\right\|^2\right)^{1/2}\right)+o_p\left(\frac{1}{\sqrt{T}}\right).
\end{aligned}
\end{equation}
By left-multiplying both sides of the equation by $(T^{-1}X_i^\top M_{F^0} X_i)^{-1}$, taking the square of the Frobenius norm, and then averaging the sum over $i$ from 1 to $N$, we can obtain
$$
\frac{1}{N}\sum_{i=1}^N\left\|\widehat{\beta}_i-\beta_i^0\right\|^2=O_p\left(\frac{1}{T}\right).
$$
Thus, we have
$$
\widehat{\beta}_i-\beta_i^0=O_p\left(\frac{1}{\sqrt{T}}\right).
$$
\end{proof}
\begin{proposition}\label{pro:B.1}
Under Assumptions \ref{assum:indentification}-\ref{assum:weak dependence}, we can make the following statements:
\begin{enumerate}[label=(\roman*)]
    \item $ V_{NT} $ is invertible and $ V_{NT} \xrightarrow{p} V $, where $V_{NT} $ is a diagonal matrix that consists of the first $ r $ largest eigenvalues of the matrix $\frac{1}{NT} \sum_{i=1}^{N} \{Z_i^\ast(\widehat{\alpha}_{i}) - X_i \widehat{\beta}_{i}\}\{Z_i^\ast(\widehat{\alpha}_{i}) - X_i \widehat{\beta}_{i}\}^\top$ and $ V  $ is a diagonal matrix consisting of the eigenvalues of $ \Sigma_F\Sigma_{\Lambda}$.
    \item Let $ H = (\Lambda^\top  \Lambda / N)(F^{0\top} \widehat{F} / T)V^{-1}_{NT} $. Then $ H $ is an $ r \times r $ invertible matrix and
$$
\frac{1}{T} \left\| \widehat{F} - F^0 H \right\|^2 = \frac{1}{T} \sum_{i=1}^T \left\| \widehat{f}_t - H^\top f^0_t \right\|^2
= O_p\left(\frac{1}{N}\sum_{t=1}^N\left\|\widehat{\beta}_i-\beta_i^0\right\|^2\right) + O_p\left(\frac{1}{\min[N, T]}\right).
$$
\end{enumerate}
\end{proposition}
\begin{proof}
From
$$
\left[ \frac{1}{NT} \sum_{i=1}^{N} \left\{Z_i^\ast(\widehat{\alpha}_{i}) - X_i \widehat{\beta}_{i}\right\}\left\{Z_i^\ast(\widehat{\alpha}_{i}) - X_i \widehat{\beta}_{i}\right\}^\top \right] \widehat{F} = \widehat{F} V_{NT},
$$
and $ Z_i^\ast(\widehat{\alpha}_{i}) - X_i \widehat{\beta}_{i} =Z_i^\ast(\widehat{\alpha}_{i})-Z_i^\ast({\alpha}_{i}^0)+ X_i (\beta_i^0 - \widehat{\beta}_i) + F^0 \lambda_i^0 + \varepsilon_i $, by expanding terms, we obtain
\begin{equation}\nonumber
\begin{aligned}
\widehat{F} V_{NT}=&\frac{1}{NT} \sum_{i=1}^N X_i (\beta_i^0 - \widehat{\beta}_i) (\beta_i^0 - \widehat{\beta}_i)^\top  X_i^\top  \widehat{F}+\frac{1}{NT} \sum_{i=1}^N X_i (\beta_i^0 - \widehat{\beta}_i) \lambda_i^{0\top}  F^{0\top} \widehat{F}\\
&+\frac{1}{NT} \sum_{i=1}^N X_i (\beta_i^0 - \widehat{\beta}_i) \varepsilon_i^\top \widehat{F}+\frac{1}{NT} \sum_{i=1}^N F^0 \lambda_i^0 (\beta_i^0 - \widehat{\beta}_i)^\top  X_i^\top  \widehat{F}\\
&+\frac{1}{NT} \sum_{i=1}^N \varepsilon_i (\beta_i^0 - \widehat{\beta}_i)^\top  X_i^\top  \widehat{F}\\
&+\frac{1}{NT} \sum_{i=1}^N F^0 \lambda_i^0 \varepsilon_i^\top \widehat{F} + \frac{1}{NT} \sum_{i=1}^N \varepsilon_i \lambda_i^{0\top}  F^0 \widehat{F}+\frac{1}{NT} \sum_{i=1}^N \varepsilon_i \varepsilon_i^\top \widehat{F}\\
&+\frac{1}{NT} \sum_{i=1}^N\left\{Z_i^\ast(\widehat{\alpha}_{i})-Z_i^\ast({\alpha}_{i}^0)\right\}\left\{Z_i^\ast(\widehat{\alpha}_{i})-Z_i^\ast({\alpha}_{i}^0)\right\}^\top\widehat{F}\\
&+\frac{1}{NT} \sum_{i=1}^N\left\{Z_i^\ast(\widehat{\alpha}_{i})-Z_i^\ast({\alpha}_{i}^0)\right\}(\beta_i^0 - \widehat{\beta}_i)^\top X_i^\top\widehat{F}+\frac{1}{NT} \sum_{i=1}^N\left\{Z_i^\ast(\widehat{\alpha}_{i})-Z_i^\ast({\alpha}_{i}^0)\right\}\lambda_i^{0\top}  F^{0\top}\widehat{F}\\
&+\frac{1}{NT} \sum_{i=1}^N\left\{Z_i^\ast(\widehat{\alpha}_{i})-Z_i^\ast({\alpha}_{i}^0)\right\}\varepsilon_i^\top\widehat{F}+\frac{1}{NT} \sum_{i=1}^NX_i (\beta_i^0 - \widehat{\beta}_i)\left\{Z_i^\ast(\widehat{\alpha}_{i})-Z_i^\ast({\alpha}_{i}^0)\right\}^\top\widehat{F}\\
&+\frac{1}{NT} \sum_{i=1}^NF^0 \lambda_i^0\left\{Z_i^\ast(\widehat{\alpha}_{i})-Z_i^\ast({\alpha}_{i}^0)\right\}^\top\widehat{F}+\frac{1}{NT} \sum_{i=1}^N\varepsilon_i\left\{Z_i^\ast(\widehat{\alpha}_{i})-Z_i^\ast({\alpha}_{i}^0)\right\}^\top\widehat{F}\\
&+\frac{1}{NT} \sum_{i=1}^N F^0 \lambda_i^0 \lambda_i^{0\top}  F^{0\top} \widehat{F}\\
:=&I1+\cdots+I16.
\end{aligned}
\end{equation}
The last term $I16$ on the right is equal to $ F^0 (\Lambda^{0\top}  \Lambda^0 / N)(F^{0\top} \widehat{F} / T) $. Then, the above can be rewritten as
\begin{equation}\label{eq:B.1}
\widehat{F} V_{NT} - F^0 (\Lambda^{0\top}  \Lambda^0 / N)(F^{0\top} \widehat{F} / T) = I 1 + \cdots + I 15.
\end{equation}
Multiplying $ (F^{0\top} \widehat{F}/T)^{-1} (\Lambda^{0\top}  \Lambda^0 /N)^{-1} $ on each side of \eqref{eq:B.1}, we obtain
\begin{equation}\label{eq:B.2}
\widehat{F}[V_{NT}(F^{0\top} \widehat{F}/T)^{-1} (\Lambda^{0\top}  \Lambda^0 /N)^{-1}] - F^0 = (I1 + \cdots + I15)(F^{0\top} \widehat{F}/T)^{-1} (\Lambda^{0\top}  \Lambda^0 /N)^{-1}.
\end{equation}
Note that the matrix $ V_{NT}(F^{0\top} \widehat{F}/T)^{-1} (\Lambda^{0\top}  \Lambda^0 /N)^{-1} $ is equal to $ H^{-1} $, but the invertibility of $ V_{NT} $ is not proved yet. We have
$$
\frac{1}{\sqrt{T}}\left\| \widehat{F}[V_{NT}(F^{0\top} \widehat{F}/T)^{-1} (\Lambda^{0\top}  \Lambda^0 /N)^{-1}] - F^0 \right\| \leq \frac{1}{\sqrt{T}}\left(\|I1\| + \cdots + \|I15\|\right) \cdot \left\| (F^{0\top} \widehat{F}/T)^{-1} (\Lambda^{0\top}  \Lambda^0 /N)^{-1} \right\|.
$$
Consider each term on the right. For the first term, note that $ T^{-1/2}\|\widehat{F}\| = \sqrt{r} $, we have
\begin{equation}\nonumber
\begin{aligned}
\frac{1}{\sqrt{T}}\left\|I1\right\|&=\frac{1}{\sqrt{T}}\left\|\frac{1}{NT} \sum_{i=1}^N X_i (\beta_i^0 - \widehat{\beta}_i) (\beta_i^0 - \widehat{\beta}_i)^\top  X_i^\top  \widehat{F}\right\|\\
&\leq\frac{1}{NT}\sum_{i=1}^N\left(\frac{\left\|X_i\right\|^2}{T}\right)\left\|\beta_i^0 - \widehat{\beta}_i\right\|^2\sqrt{r}\\
&=O_p\left(\frac{1}{N}\sum_{i=1}^N\left\|\widehat{\beta}_i-\beta_i^0\right\|^2\right).
\end{aligned}
\end{equation}
For the second term, note that $T^{-1/2}\|F\|=\sqrt{r}$ and $ T^{-1/2}\|\widehat{F}\| = \sqrt{r} $, we have
\begin{equation}\nonumber
\begin{aligned}
\frac{1}{\sqrt{T}}\left\|I2\right\|&=\frac{1}{\sqrt{T}}\left\|\frac{1}{NT} \sum_{i=1}^N X_i (\beta_i^0 - \widehat{\beta}_i) \lambda_i^{0\top}  F^{0\top} \widehat{F}\right\|\\
&\leq\frac{1}{\sqrt{T}N}\sum_{i=1}^N\left\|X_i\right\|\left\| \beta_i^0 - \widehat{\beta}_i\right\|r\\
&\leq r\left(\frac{1}{N}\sum_{i=1}^N\frac{\left\|X_i\right\|^2}{T}\right)^{1/2}\left(\frac{1}{N}\sum_{i=1}^N\left\|\widehat{\beta}_i-\beta_i^0\right\|^2\right)^{1/2}\\
&=O_p\left(\left(\frac{1}{N}\sum_{i=1}^N\left\|\widehat{\beta}_i-\beta_i^0\right\|^2\right)^{1/2}\right).
\end{aligned}
\end{equation}
Using the same argument, it is easy to prove that next three terms (I3-I5) are each $O_p((\frac{1}{N}\sum_{i=1}^N\|\widehat{\beta}_i-\beta_i^0\|^2)^{1/2})$. The terms (I6-I8) do not explicitly depend on $ \widehat{\beta}_i - \beta_i $ and they have the same expressions as those in \cite{bai2002determining}. Each of these terms is $ O_p(1/\min[\sqrt{N}, \sqrt{T}]) $, which was proved in Theorem 1 of \cite{bai2002determining}. The proof there only uses the property that $ \widehat{F}^\top  \widehat{F}/T = I_r $ and the assumptions on $ \varepsilon_i $; thus the proof needs no modification.
For terms $I9-I15$, we have
\begin{equation}\nonumber
\begin{aligned}
\frac{1}{\sqrt{T}}\left\|I9\right\|&=\frac{1}{\sqrt{T}}\left\|\frac{1}{NT} \sum_{i=1}^N\left\{Z_i^\ast(\widehat{\alpha}_{i})-Z_i^\ast({\alpha}_{i}^0)\right\}\left\{Z_i^\ast(\widehat{\alpha}_{i})-Z_i^\ast({\alpha}_{i}^0)\right\}^\top\widehat{F}\right\|\\
&\leq\frac{1}{NT}\sum_{i=1}^N\left\|Z_i^\ast(\widehat{\alpha}_{i})-Z_i^\ast({\alpha}_{i}^0)\right\|^2\sqrt{r}\\
&=O_p\left(\frac{1}{T}\right),
\end{aligned}
\end{equation}
\begin{equation}\nonumber
\begin{aligned}
\frac{1}{\sqrt{T}}\left\|I10\right\|&=\frac{1}{\sqrt{T}}\left\|\frac{1}{NT} \sum_{i=1}^N\left\{Z_i^\ast(\widehat{\alpha}_{i})-Z_i^\ast({\alpha}_{i}^0)\right\}(\beta_i^0 - \widehat{\beta}_i)^\top X_i^\top\widehat{F}\right\|\\
&\leq\frac{1}{NT} \sum_{i=1}^N\left\|\beta_i^0 - \widehat{\beta}_i\right\|\left\|X_i\right\|\sqrt{r}\\
&\leq\frac{1}{\sqrt{T}}\left(\frac{1}{N} \sum_{i=1}^N\left\|\widehat{\beta}_i-\beta_i^0\right\|^2\right)^{1/2}\left(\frac{1}{N} \sum_{i=1}^N\frac{\left\|X_i\right\|^2}{T}\right)^{1/2}\sqrt{r}\\
&=\frac{1}{\sqrt{T}}O_p\left(\left(\frac{1}{N}\sum_{i=1}^N\left\|\widehat{\beta}_i-\beta_i^0\right\|^2\right)^{1/2}\right),
\end{aligned}
\end{equation}
\begin{equation}\nonumber
\begin{aligned}
\frac{1}{\sqrt{T}}\left\|I11\right\|&=\frac{1}{\sqrt{T}}\left\|\frac{1}{NT} \sum_{i=1}^N\left\{Z_i^\ast(\widehat{\alpha}_{i})-Z_i^\ast({\alpha}_{i}^0)\right\}\lambda_i^{0\top}  F^{0\top}\widehat{F}\right\|\\
&\leq\frac{1}{NT} \sum_{i=1}^N\left\|Z_i^\ast(\widehat{\alpha}_{i})-Z_i^\ast({\alpha}_{i}^0)\right\|\left\|F^{0}\right\|\sqrt{r}\\
&=O_p\left(\frac{1}{\sqrt{T}}\right),
\end{aligned}
\end{equation}
\begin{equation}\nonumber
\begin{aligned}
\frac{1}{\sqrt{T}}\left\|I12\right\|&=\frac{1}{\sqrt{T}}\left\|\frac{1}{NT} \sum_{i=1}^N\left\{Z_i^\ast(\widehat{\alpha}_{i})-Z_i^\ast({\alpha}_{i}^0)\right\}\varepsilon_i^\top\widehat{F}\right\|\\
&\leq\frac{1}{NT} \sum_{i=1}^N\left\|Z_i^\ast(\widehat{\alpha}_{i})-Z_i^\ast({\alpha}_{i}^0)\right\|\left\|\varepsilon_i\right\|\sqrt{r}\\
&=O_p\left(\frac{1}{\sqrt{T}}\right),
\end{aligned}
\end{equation}
\begin{equation}\nonumber
\begin{aligned}
\frac{1}{\sqrt{T}}\left\|I13\right\|&=\frac{1}{\sqrt{T}}\left\|\frac{1}{NT} \sum_{i=1}^NX_i (\beta_i^0 - \widehat{\beta}_i)\left\{Z_i^\ast(\widehat{\alpha}_{i})-Z_i^\ast({\alpha}_{i}^0)\right\}^\top\widehat{F}\right\|\\
&\leq\frac{1}{NT} \sum_{i=1}^N\left\|X_i\right\|\left\|\beta_i^0 - \widehat{\beta}_i\right\|\sqrt{r}\\
&\leq\frac{1}{\sqrt{T}}\left(\frac{1}{N} \sum_{i=1}^N\frac{\left\|X_i\right\|^2}{T}\right)^{1/2}\left(\frac{1}{N} \sum_{i=1}^N\left\|\widehat{\beta}_i-\beta_i^0\right\|^2\right)^{1/2}\sqrt{r}\\
&=\frac{1}{\sqrt{T}}O_p\left(\left(\frac{1}{N}\sum_{i=1}^N\left\|\widehat{\beta}_i-\beta_i^0\right\|^2\right)^{1/2}\right),
\end{aligned}
\end{equation}
\begin{equation}\nonumber
\begin{aligned}
\frac{1}{\sqrt{T}}\left\|I14\right\|&=\frac{1}{\sqrt{T}}\left\|\frac{1}{NT} \sum_{i=1}^NF^0 \lambda_i^0\left\{Z_i^\ast(\widehat{\alpha}_{i})-Z_i^\ast({\alpha}_{i}^0)\right\}^\top\widehat{F}\right\|\\
&\leq\frac{1}{NT} \sum_{i=1}^N\left\|F^0\right\|\left\|Z_i^\ast(\widehat{\alpha}_{i})-Z_i^\ast({\alpha}_{i}^0)\right\|\sqrt{r}\\
&=O_p\left(\frac{1}{\sqrt{T}}\right),
\end{aligned}
\end{equation}
\begin{equation}\nonumber
\begin{aligned}
\frac{1}{\sqrt{T}}\left\|I15\right\|&=\frac{1}{\sqrt{T}}\left\|\frac{1}{NT} \sum_{i=1}^N\varepsilon_i\left\{Z_i^\ast(\widehat{\alpha}_{i})-Z_i^\ast({\alpha}_{i}^0)\right\}^\top\widehat{F}\right\|\\
&\leq\frac{1}{NT} \sum_{i=1}^N\left\|\varepsilon_i\right\|\left\|Z_i^\ast(\widehat{\alpha}_{i})-Z_i^\ast({\alpha}_{i}^0)\right\|\sqrt{r}\\
&=O_p\left(\frac{1}{\sqrt{T}}\right).
\end{aligned}
\end{equation}
In summary, we have
\begin{equation}\label{eq:B.3}
\frac{1}{\sqrt{T}}\left\|\widehat{F}\left[V_{NT}(F^{0\top} \widehat{F}/T)^{-1} (\Lambda^{0\top}  \Lambda^0 /N)^{-1}\right] - F^0 \right\| = O_p\left(\left(\frac{1}{N}\sum_{i=1}^N\left\|\widehat{\beta}_i-\beta_i^0\right\|^2\right)^{1/2}\right) + O_p\left(\frac{1}{\min[\sqrt{N}, \sqrt{T}]}\right).
\end{equation}
(i) Left multiplying \eqref{eq:B.1} by $ \widehat{F}^\top  $ and using $ T^{-1}\widehat{F}^\top  \widehat{F} = I_T $, we have
$$
V_{NT} - (\widehat{F}^\top F^0/T)(\Lambda^{0\top}  \Lambda^0 /N)(F^{0\top} \widehat{F}/T) = T^{-1} \widehat{F}^\top  (I1 + \cdots + I15) = o_p(1)
$$
because $ T^{-1/2}\|\widehat{F}\| = \sqrt{r} $ and $ T^{-1/2}\| (I1 + \cdots + I15)\| = o_p(1) $. Thus
$$
V_{NT} = (\widehat{F}^\top F^0/T)(\Lambda^{0\top}  \Lambda^0 /N)(F^{0\top} \widehat{F}/T) + o_p(1).
$$
Proposition \ref{pro:average consistency} shows that $ \widehat{F}^\top F^0/T $ is invertible; thus $ V_{NT} $ is invertible. To obtain the limit of $ V_{NT} $, left multiply \eqref{eq:B.1} by $ F^{0\top} $ and then divide by $ T $ to yield
$$
(F^{0\top} F^0/T)(\Lambda^{0\top}  \Lambda^0 /N)(F^{0\top} \widehat{F}/T) + o_p(1) = (F^{0\top} \widehat{F}/T)V_{NT}
$$
because $ T^{-1}F^{0\top}(I1 + \cdots + I15) = o_p(1) $. The above equality shows that the columns of $ F^{0\top}\widehat{F}/T $ are the (nonnormalized) eigenvectors of the matrix $ (F^{0\top}F^0/T)(\Lambda^{0\top} \Lambda^0/N) $, and $ V_{NT} $ consists of the eigenvalues of the same matrix (in the limit). Thus $ V_{NT} \xrightarrow{p} V $, where $ V $ is $ r \times r $, consisting of the $ r $ eigenvalues of the matrix $ \Sigma_F\Sigma_{\Lambda} $.

(ii) Since $ V_{NT} $ is invertible, the left-hand side of \eqref{eq:B.3} can be written as $ T^{-1/2}\|\widehat{F}H^{-1}-F^0\| $; thus \eqref{eq:B.3} is equivalent to
$$
T^{-1/2}\|\widehat{F}-F^0H\|=O_p\left(\left(\frac{1}{N}\sum_{i=1}^N\left\|\widehat{\beta}_i-\beta_i^0\right\|^2\right)^{1/2}\right) + O_p\left(\frac{1}{\min[\sqrt{N}, \sqrt{T}]}\right).
$$
Taking squares on each side gives part (ii). Note that the cross-product term from expanding the square has the same bound.
\end{proof}
\begin{lemma}\label{lemma:B.1}
Let $ G = (F^{0\top} \widehat{F}/T)^{-1}(\Lambda^{0\top} \Lambda^{0}/N)^{-1} $. Under Assumptions \ref{assum:indentification}-\ref{assum:weak dependence}, we have
\begin{equation}\nonumber
\begin{aligned}
&\frac{1}{NT^2} \sum_{k=1}^N X_i^\top M_{\widehat{F}} (\varepsilon_k \varepsilon_k^\top - \Omega_k) \widehat{F} G \lambda_i^0\\
&\quad=O_p \left( \frac{1}{T\sqrt{N}} \right) + (NT)^{-1/2}\left[O_p\left(\left(\frac{1}{N}\sum_{i=1}^N\left\|\widehat{\beta}_i-\beta_i^0\right\|^2\right)^{1/2}\right) + O_p(\delta_{NT}^{-1})\right]\\
&\quad\quad+\frac{1}{\sqrt{N}} O_p\left(\frac{1}{N}\sum_{i=1}^N\left\|\widehat{\beta}_i-\beta_i^0\right\|^2\right) + \frac{1}{\sqrt{N}} O_p(\delta_{NT}^{-2}).
\end{aligned}
\end{equation}
\end{lemma}
\begin{proof}
According to $M_{\widehat{F}}=I_T-P_{\widehat{F}}$ and $P_{\widehat{F}}=\widehat{F}\widehat{F}^{\top}/T$, rewrite the left-hand side as
\begin{equation}\nonumber
\begin{aligned}
\frac{1}{NT^2} \sum_{k=1}^N X_i^\top M_{\widehat{F}} (\varepsilon_k \varepsilon_k^\top - \Omega_k) \widehat{F} G \lambda_i^0=&\frac{1}{N T^2} \sum_{k=1}^{N} X^\top_i (\varepsilon_k \varepsilon^\top_k - \Omega_k) \widehat{F} G \lambda_i^0\\
&-\left( \frac{X^\top_i \widehat{F}}{T} \right) \frac{1}{NT^2} \sum_{k=1}^{N} \widehat{F}^\top  (\varepsilon_k \varepsilon^\top_k - \Omega_k) \widehat{F} G \lambda_i^0\\
:=&I + II.
\end{aligned}
\end{equation}
Adding and subtracting terms yields
\begin{equation}\nonumber
\begin{aligned}
I=&\frac{1}{N T^2} \sum_{k=1}^{N} X_i^\top (\varepsilon_k \varepsilon_k^\top - \Omega_k) F^0 H G \lambda_i^0\\
&+\frac{1}{N T^2} \sum_{k=1}^{N} X_i^\top (\varepsilon_k \varepsilon_k^\top - \Omega_k) (\widehat{F} - F^0 H) G \lambda_i^0.
\end{aligned}
\end{equation}
The first term on the right is equal to
\begin{equation}\nonumber
\begin{aligned}
&\left( \frac{1}{N T^2} \right)  \sum_{k=1}^{N} \left\{ \sum_{t=1}^{T} \sum_{s=1}^{T} X_{it} [\varepsilon_{kt} \varepsilon_{ks} - E(\varepsilon_{kt} \varepsilon_{ks})] f_s^{0\top} H G \lambda_i \right\}\\
&\quad=\frac{1}{T \sqrt{N}} \left[ N^{-1/2} \sum_{k=1}^{N} \frac{1}{T} \sum_{t=1}^{T} \sum_{s=1}^{T} X_{it} [\varepsilon_{kt} \varepsilon_{ks} - E(\varepsilon_{kt} \varepsilon_{ks})] f_s^{0\top} \right] H G \lambda_i^0\\
&\quad =O_p \left( \frac{1}{T \sqrt{N}} \right)
\end{aligned}
\end{equation}
by Lemma \ref{lemma:B.2} (ii). Denote
$$
a_s = \left( \frac{1}{\sqrt{NT}} \sum_{k=1}^{N} \sum_{t=1}^{T} X_{it} [\varepsilon_{kt} \varepsilon_{ks} - E(\varepsilon_{kt} \varepsilon_{ks})] \right) = O_p (1).
$$
Then the second term of $ I $ is
$$
\frac{1}{\sqrt{NT}}  \frac{1}{T} \sum_{s=1}^{T} a_s (\widehat{f}_s - f_s^0 H)^\top  G \lambda_i^0.
$$
Notice
\begin{equation}\nonumber
\begin{aligned}
\left\| \frac{1}{T} \sum_{s=1}^{T} a_s (\widehat{f}_s - f_s^0 H) \right\| &\leq \left( \frac{1}{T} \sum_{s=1}^{T} \| a_s \|^2 \right)^{1/2} \left( \frac{1}{T} \sum_{s=1}^{T} \| \widehat{f}_s - f_s^0 H \|^2 \right)^{1/2}\\
&=O_p\left(\left(\frac{1}{N}\sum_{i=1}^N\left\|\widehat{\beta}_i-\beta_i^0\right\|^2\right)^{1/2}\right) + O_p(\delta_{NT}^{-1}).
\end{aligned}
\end{equation}
Thus the second term of $ I $ is $(NT)^{-1/2} [O_p((\frac{1}{N}\sum_{i=1}^N\|\widehat{\beta}_i-\beta_i^0\|^2)^{1/2}) + O_p (\delta_{NT}^{-1})]$. 

Consider $ II $:
\begin{equation}\nonumber
\begin{aligned}
\left\| II \right\|&=\left\| \left( \frac{X^\top_i \widehat{F}}{T} \right) \frac{1}{NT^2} \sum_{k=1}^{N} \widehat{F}^\top  (\varepsilon_k \varepsilon^\top_k - \Omega_k) \widehat{F} G \lambda_i^0 \right\|\\
&\leq\left\| \frac{X_i \widehat{F}}{T} \right\| \| G \lambda_i^0 \| \cdot \left\| \frac{1}{NT^2} \sum_{k=1}^{N} \widehat{F}^\top  (\varepsilon_k \varepsilon_k^\top - \Omega_k) \widehat{F} \right\|\\
&= O_{p}(1)\left\|\frac{1}{NT^{2}}\sum_{k=1}^{N}\widehat{F}^{\top}(\varepsilon _{k}\varepsilon^{\top}_{k}-\Omega_{k})\widehat{F}\right\|.
\end{aligned}
\end{equation}
But
\begin{equation}\nonumber
\begin{aligned}
\frac{1}{NT^{2}}\sum_{k=1}^{N}\widehat{F}^{\top}(\varepsilon_{k} \varepsilon^{\top}_{k}-\Omega_{k})\widehat{F}=&H\frac{1}{NT^{2}}\sum_{k=1}^{N}F^{0\top}(\varepsilon_{k} \varepsilon^{\top}_{k}-\Omega_{k})F^{0}H\\
&+ H\frac{1}{NT^{2}}\sum_{k=1}^{N}F^{0\top}(\varepsilon_{k} \varepsilon^{\top}_{k}-\Omega_{k})(\widehat{F}-F^{0}H)\\
&+\frac{1}{NT^{2}}\sum_{k=1}^{N}(\widehat{F}-F^{0}H)^{\top}(\varepsilon _{k}\varepsilon^{\top}_{k}-\Omega_{k})F^{0}H\\
&+\frac{1}{NT^{2}}\sum_{k=1}^{N}(\widehat{F}-F^{0}H)^{\top}(\varepsilon _{k}\varepsilon^{\top}_{k}-\Omega_{k})(\widehat{F}-F^{0}H)\\
:=&b1+b2+b3+b4.
\end{aligned}
\end{equation}
Now
\begin{equation}\nonumber
\begin{aligned}
b1&=H\frac{1}{NT^{2}}\sum_{k=1}^{N}F^{0\top}(\varepsilon_{k} \varepsilon^{\top}_{k}-\Omega_{k})F^{0}H\\
&=H\left(\frac{1}{T^{2}N}\right)\sum_{k=1}^{N}\sum_{t=1}^{T}\sum_ {s=1}^{T}f_{s}f_{t}^{\top}[\varepsilon_{kt}\varepsilon_{ks}-E(\varepsilon_{kt}\varepsilon_{ks})]H\\
&=O_{p}\left(\frac{1}{T\sqrt{N}}\right)
\end{aligned}
\end{equation}
by Lemma \ref{lemma:B.2} (i). Next
\begin{equation}\nonumber
\begin{aligned}
b2&=H\frac{1}{NT^{2}}\sum_{k=1}^{N}F^{0\top}(\varepsilon_{k} \varepsilon^{\top}_{k}-\Omega_{k})(\widehat{F}-F^{0}H)\\
&=H\frac{1}{\sqrt{NT}}\frac{1}{T}\sum_{s=1}^{T}\left[\frac{1}{ \sqrt{NT}}\sum_{t=1}^{T}\sum_{k=1}^{N}f_{t}^{0}[\varepsilon_{kt}\varepsilon_{ks }-E(\varepsilon_{kt}\varepsilon_{ks})]\right](\widehat{f}_{s}-H^\top f_{s}^{0}).
\end{aligned}
\end{equation}
Thus if we let $A_{s}=\frac{1}{\sqrt{NT}}\sum_{t=1}^{T}\sum_{k=1}^{N}f_{t}^{0}[\varepsilon_{kt}\varepsilon_{ks}-E(\varepsilon_{kt}\varepsilon_{ks})]$, then
\begin{equation}\nonumber
\begin{aligned}
\|b2\| &\leq \|H\|\frac{1}{\sqrt{NT}}\left(\frac{1}{T}\sum_{s=1}^{T}\|A_{s}\| ^{2}\right)^{1/2}\left(\frac{1}{T}\sum_{s=1}^{T}\|\widehat{f}_{s}-H^\top f_{s}^{0}\|^ {2}\right)^{1/2}\\
&= \frac{1}{\sqrt{NT}}\left[O_p\left(\left(\frac{1}{N}\sum_{i=1}^N\left\|\widehat{\beta}_i-\beta_i^0\right\|^2\right)^{1/2}\right) + O_p(\delta_{NT}^{-1})\right].
\end{aligned}
\end{equation}
The term $b3$ has the same upper bound because it is the transpose of $b2$. The last term is
\begin{equation}\nonumber
\begin{aligned}
b4&=\frac{1}{NT^{2}}\sum_{k=1}^{N}(\widehat{F}-F^{0}H)^{\top}(\varepsilon _{k}\varepsilon^{\top}_{k}-\Omega_{k})(\widehat{F}-F^{0}H)\\
&=\frac{1}{\sqrt{N}}\frac{1}{T^{2}}\sum_{t=1}^{T}\sum_{s=1}^{T}( \widehat{f}_{t}-H^{\top}f_{t}^{0})(\widehat{f}_{s}-H^{\top}f_{s}^{0})^{\top}\left[\frac{1}{\sqrt{N}}\sum_{k=1}^{N}\left[\varepsilon _{kt}\varepsilon_{ks}-E(\varepsilon_{kt}\varepsilon_{ks})\right]\right].
\end{aligned}
\end{equation}
Thus by the Cauchy-Schwarz inequality,
\begin{equation}\nonumber
\begin{aligned}
\|b4\| &\leq\frac{1}{\sqrt{N}}\left(\frac{1}{T}\sum_{t=1}^{T}\|f_{t}-H^{ \top}f_{t}^{0}\|^{2}\right)\left(\frac{1}{T^{2}}\sum_{t=1}^{T}\sum_{s=1}^{T}\left[ \frac{1}{\sqrt{N}}\sum_{k=1}^{N}\left[\varepsilon_{kt}\varepsilon_{ks}-E( \varepsilon_{kt}\varepsilon_{ks})\right]\right]^{2}\right)^{1/2}\\
&=\frac{1}{\sqrt{N}}O_p\left(\frac{1}{N}\sum_{i=1}^N\left\|\widehat{\beta}_i-\beta_i^0\right\|^2\right)+\frac{1}{ \sqrt{N}}O_{p}(\delta_{NT}^{-2}).
\end{aligned}
\end{equation}
Now collecting terms yields the lemma. 
\end{proof}

\begin{lemma}\label{lemma:B.2}
Under Assumptions \ref{assum:indentification}-\ref{assum:error}, there exists an $ M < \infty $, such that statements (i) and (ii) hold:
\begin{enumerate}[label=(\roman*)]
\item We have
$$
E\left\| N^{-1/2} \sum_{k=1}^{N} \frac{1}{T} \sum_{t=1}^{T} \sum_{s=1}^{T} f_s f_t^\top [\varepsilon_{kt} \varepsilon_{ks} - E(\varepsilon_{kt} \varepsilon_{ks})] \right\|^2 \leq M.
$$
\item For all $ i \in[N] $ and $ h \in[r] $, we have
$$
E\left\| N^{-1/2} \sum_{k=1}^{N} \frac{1}{T} \left\{ \sum_{t=1}^{T} \sum_{s=1}^{T} X_{it} [\varepsilon_{kt} \varepsilon_{ks} - E(\varepsilon_{kt} \varepsilon_{ks})] f_{hs} \right\} \right\|^2 \leq M.
$$
\end{enumerate}
\end{lemma}
\begin{proof}
(i) Let 
$$
A = N^{-1/2} \sum_{k=1}^{N} \frac{1}{T} \sum_{t=1}^{T} \sum_{s=1}^{T} f_s f_t^\top [\varepsilon_{kt} \varepsilon_{ks} - {E}(\varepsilon_{kt} \varepsilon_{ks})].
$$
We need to show ${E}\|A\|^2 \leq M$. Note that for any matrix $A$, $\|A\|^2 = \tr(AA^\top)$. Therefore,
\begin{equation}\nonumber
\begin{aligned}
{E}\|A\|^2 &= {E}[\tr(AA^\top)] \\
&= {E}\left[ \tr\left( \frac{1}{N} \sum_{k=1}^{N} \sum_{\ell=1}^{N} \frac{1}{T^2} \sum_{t,s,u,v} f_s f_t^\top [\varepsilon_{kt} \varepsilon_{ks} - {E}(\varepsilon_{kt} \varepsilon_{ks})]   [\varepsilon_{\ell u} \varepsilon_{\ell v} - {E}(\varepsilon_{\ell u} \varepsilon_{\ell v})] f_u f_v^\top \right) \right].
\end{aligned}
\end{equation}
Using the cyclic property of the trace, we have
\begin{equation}\nonumber
\begin{aligned}
\tr(AA^\top) &= \frac{1}{N} \sum_{k,\ell} \frac{1}{T^2} \sum_{t,s,u,v} \tr\left(f_s f_t^\top [\varepsilon_{kt} \varepsilon_{ks} - {E}(\varepsilon_{kt} \varepsilon_{ks})] [\varepsilon_{\ell u} \varepsilon_{\ell v} - {E}(\varepsilon_{\ell u} \varepsilon_{\ell v})] f_u f_v^\top\right) \\
&= \frac{1}{N} \sum_{k,\ell} \frac{1}{T^2} \sum_{t,s,u,v} f_v^\top f_s \cdot f_t^\top f_u \cdot [\varepsilon_{kt} \varepsilon_{ks} - {E}(\varepsilon_{kt} \varepsilon_{ks})] \cdot [\varepsilon_{\ell u} \varepsilon_{\ell v} - {E}(\varepsilon_{\ell u} \varepsilon_{\ell v})].
\end{aligned}
\end{equation}
Taking expectations, we have
$$
{E}[\tr(AA^\top)]= \frac{1}{N} \sum_{k,\ell} \frac{1}{T^2} \sum_{t,s,u,v} f_v^\top f_s \cdot f_t^\top f_u \cdot \mathrm{cov}(\varepsilon_{kt} \varepsilon_{ks}, \varepsilon_{\ell u} \varepsilon_{\ell v}).
$$
By Assumption \ref{assum:error}, specifically,
$$
T^{-2} N^{-1} \sum_{t,s,u,v} \sum_{k,\ell} |\mathrm{cov}(\varepsilon_{kt} \varepsilon_{ks}, \varepsilon_{\ell u} \varepsilon_{\ell v})| \leq M,
$$
Combined with Assumption \ref{assum:factor&loading}, we have
\begin{equation}\nonumber
\begin{aligned}
{E}\|A\|^2 &= {E}[\tr(AA^\top)]\\
&\leq \frac{1}{N} \sum_{k,\ell} \frac{1}{T^2} \sum_{t,s,u,v} \|f_v\| \|f_s\| \|f_t\| \|f_u\| \cdot |\mathrm{cov}(\varepsilon_{kt} \varepsilon_{ks}, \varepsilon_{\ell u} \varepsilon_{\ell v})|\\
&\leq M.
\end{aligned}
\end{equation}

(ii) For fixed $i$ and $h$, let
$$
B = N^{-1/2} \sum_{k=1}^{N} \frac{1}{T} \left\{ \sum_{t=1}^{T} \sum_{s=1}^{T} X_{it} [\varepsilon_{kt} \varepsilon_{ks} - {E}(\varepsilon_{kt} \varepsilon_{ks})] f_{hs} \right\}.
$$
We need to show ${E}\|B\|^2 \leq M$. Again, using $\|B\|^2 = \tr(BB^\top)$, we have
$$
\tr(BB^\top)= \frac{1}{N} \sum_{k,\ell} \frac{1}{T^2} \sum_{t,s,u,v} X_{it} X_{iu} f_{hs} f_{hv} [\varepsilon_{kt} \varepsilon_{ks} - {E}(\varepsilon_{kt} \varepsilon_{ks})] [\varepsilon_{\ell u} \varepsilon_{\ell v} - {E}(\varepsilon_{\ell u} \varepsilon_{\ell v})].
$$
Taking expectations,
$$
{E}[\tr(BB^\top)] = \frac{1}{N} \sum_{k,\ell} \frac{1}{T^2} \sum_{t,s,u,v} X_{it} X_{iu} f_{hs} f_{hv} \cdot \mathrm{cov}(\varepsilon_{kt} \varepsilon_{ks}, \varepsilon_{\ell u} \varepsilon_{\ell v}).
$$
By Assumption \ref{assum:error}, specifically,
$$
T^{-1} N^{-2} \sum_{t,s} \sum_{i,j,k,\ell} |\mathrm{cov}(\varepsilon_{it} \varepsilon_{jt}, \varepsilon_{ks} \varepsilon_{\ell s})| \leq M.
$$
Combined with Assumptions \ref{assum:indentification} and \ref{assum:factor&loading}, we obtain
\begin{equation}\nonumber
\begin{aligned}
{E}\|B\|^2 &= {E}[\tr(BB^\top)]\\
&\leq \frac{1}{N} \sum_{k,\ell} \frac{1}{T^2} \sum_{t,s,u,v} |X_{it}| |X_{iu}| |f_{hs}| |f_{hv}| \cdot |\mathrm{cov}(\varepsilon_{kt} \varepsilon_{ks}, \varepsilon_{\ell u} \varepsilon_{\ell v})|\\
&\leq M.
\end{aligned}
\end{equation}
This completes the proof.
\end{proof}

\begin{lemma}\label{lemma:B.3}
Under Assumptions \ref{assum:indentification}-\ref{assum:weak dependence},
\begin{equation}\nonumber
\begin{aligned}
\frac{1}{\sqrt{T}} X_i^\top M_{\widehat{F}}\varepsilon_i - \frac{1}{\sqrt{T}}\frac{1}{N} \sum_{k=1}^{N} a_{ik} X_i^\top M_{\widehat{F}} \varepsilon_k=&\frac{1}{\sqrt{T}} X_i^\top M_{{F}^0}\varepsilon_i - \frac{1}{\sqrt{T}}\frac{1}{N} \sum_{k=1}^{N} a_{ik} X_i^\top M_{{F}^0} \varepsilon_k\\
&+\left(\frac{\sqrt{T}}{N}+1\right)\left[O_p\left(\left(\frac{1}{N}\sum_{i=1}^N\left\|\widehat{\beta}_i-\beta_i^0\right\|^2\right)^{1/2}\right) + O_p\left(\delta_{NT}^{-1}\right)\right]\\
&+\sqrt{T}\left[O_p\left(\frac{1}{N}\sum_{i=1}^N\left\|\widehat{\beta}_i-\beta_i^0\right\|^2\right)+O_p(\delta_{NT}^{-2})\right]+o_p(1).
\end{aligned}
\end{equation}
\end{lemma}
\begin{proof}
First consider $\frac{1}{\sqrt{T}}X_{i}^\top(M_{F^{0}}-M_{\widehat{F}})\varepsilon_{i}$. Note that $M_{F^{0}}-M_{\widehat{F}}=P_{\widehat{F}}-P_{F^{0}}$ and $P_{\widehat{F}}=\widehat{F}\widehat{F}^{\top}/T$. By adding and subtracting terms,
\begin{equation}\nonumber
\begin{aligned}
\frac{1}{\sqrt{T}}\frac{X^{\top}_{i}\widehat{F}}{T}\,\widehat{F}^{ \top}\varepsilon_{i}-\frac{1}{\sqrt{T}}X^{\top}_{i}P_{F^{0}} \varepsilon_{i}=&\frac{1}{\sqrt{T}}\frac{X^{\top}_{i}(\widehat{F}-F^{0}H)}{T}H^{ \top}F^{0\top}\varepsilon_{i}\\
&+\frac{1}{\sqrt{T}}\frac{X^{\top}_{i}(\widehat{F}-F^{0}H)}{T}( \widehat{F}-F^{0}H)^{\top}\varepsilon_{i}\\
&+\frac{1}{\sqrt{T}}\frac{X^{\top}_{i}F^{0}H}{T}\,(\widehat{F}-F^{0 }H)^{\top}\varepsilon_{i}\\
&+\frac{1}{\sqrt{T}}\frac{X^{\top}_{i}F^{0}}{T}\left[HH^{ \top}-\left(\frac{F^{0\top}F^{0}}{T}\right)^{-1}\right]F^{0\top}\varepsilon _{i}\\
:=&a+b+c+d.
\end{aligned}
\end{equation}

Consider $a$. Note that $(\widehat{f}_{s}-H^{\top}f_{s}^{0})^{\top}H^{\top}f_{t}^{0}$ is scalar and thus commutable with $X_{it}$:
\begin{equation}\nonumber
\begin{aligned}
a=&\frac{1}{\sqrt{T}}\frac{X^{\top}_{i}(\widehat{F}-F^{0}H)}{T}H^{ \top}F^{0\top}\varepsilon_{i}\\
=&\frac{1}{T}\sum_{s=1}^{T}(\widehat{f}_{s}-H^{\top}f_{s}^{0})^{\top}H^{\top }\left(\frac{1}{\sqrt{T}}\sum_{t=1}^{T}f_{t}^{0}X_{is} \varepsilon_{it}\right).
\end{aligned}
\end{equation}
Thus
\begin{equation}\nonumber
\begin{aligned}
\left\|a\right\|=&\left\|\frac{1}{T}\sum_{s=1}^{T}(\widehat{f}_{s}-H^{\top}f_{s}^{0})^{\top}H^{\top }\left(\frac{1}{\sqrt{T}}\sum_{t=1}^{T}f_{t}^{0}X_{is} \varepsilon_{it}\right)\right\|\\
\leq&\left[\frac{1}{T}\sum_{s=1}^{T}\left\|\widehat{f}_{s}-H^{\top}f_{s}^{0}\right\| ^{2}\right]^{1/2}\left\|H\right\|\left[\frac{1}{T}\sum_{s=1}^{T}\left\|\left(\frac{1}{\sqrt{T} }\sum_{t=1}^{T}f_{t}^{0}X_{is}\varepsilon_{it}\right)\right\|^2\right]^{1/2}\\
=&O_p\left(\left(\frac{1}{N}\sum_{i=1}^N\left\|\widehat{\beta}_i-\beta_i^0\right\|^2\right)^{1/2}\right) + O_p\left(\delta_{NT}^{-1 }\right).\\
\end{aligned}
\end{equation}
Similarly,
\begin{equation}\nonumber
\begin{aligned}
b=&\frac{1}{\sqrt{T}}\frac{X^{\top}_{i}(\widehat{F}-F^{0}H)}{T}( \widehat{F}-F^{0}H)^{\top}\varepsilon_{i}\\
=&\sqrt{T}\frac{1}{T^2}\sum_{s=1}^T\sum_{t=1}^T(\widehat{f}_s - H^\top f_s^0)^\top(\widehat{f}_t - H^\top f_t^0)\left( X_{is} \varepsilon_{it}\right)
\end{aligned}
\end{equation}
and
\begin{equation}\nonumber
\begin{aligned}
\left\|b\right\|=&\left\|\sqrt{T}\frac{1}{T^2}\sum_{s=1}^T\sum_{t=1}^T(\widehat{f}_s - H^\top f_s^0)^\top(\widehat{f}_t - H^\top f_t^0)\left( X_{is} \varepsilon_{it}\right)\right\|\\
\leq&\sqrt{T}\left(\frac{1}{T}\sum_{t=1}^T\|\widehat{f}_t - H^\top f_t^0\|^2\right)\left(\frac{1}{T^2}\sum_{s=1}^T\sum_{t=1}^T\left\|  X_{it}\varepsilon_{it}\right\|^2\right)^{1/2}\\
=&\sqrt{T}\left[O_p\left(\frac{1}{N}\sum_{i=1}^N\left\|\widehat{\beta}_i-\beta_i^0\right\|^2\right)+O_p(\delta_{NT}^{-2})\right].
\end{aligned}
\end{equation}
Consider $c$:
\begin{equation}\nonumber
\begin{aligned}
c=&\frac{1}{\sqrt{T}}\frac{X^{\top}_{i}F^{0}H}{T}\,(\widehat{F}-F^{0 }H)^{\top}\varepsilon_{i}\\
=&\frac{1}{\sqrt{T}}\frac{X_i^\top F^0}{T}HH^\top (\widehat{F} H^{-1} - F^{0})^\top \varepsilon_i\\
=&\frac{1}{\sqrt{T}}\frac{X_i^\top F^0}{T}\left(\frac{F^{0\top}F^0}{T}\right)^{-1}(\widehat{F} H^{-1} - F^{0})^\top \varepsilon_i\\
&+\frac{1}{\sqrt{T}}\frac{X_i^\top F^0}{T}\left[HH^\top  - \left(\frac{F^{0\top}F^0}{T}\right)^{-1}\right](\widehat{F} H^{-1} - F^{0})^\top \varepsilon_i\\
:=&c1+c2.
\end{aligned}
\end{equation}
Denote $Q = HH^\top  - (F^{0\top}F^0/T)^{-1}$ for the moment. For the term $c2$,
\begin{equation}\nonumber
\begin{aligned}
c2=&\frac{1}{\sqrt{T}}\frac{X_i^\top F^0}{T}\left[HH^\top  - \left(\frac{F^{0\top}F^0}{T}\right)^{-1}\right](\widehat{F} H^{-1} - F^{0})^\top \varepsilon_i\\
=&\sqrt{T}\left(\frac{1}{T}\left[\varepsilon_i^\top (\widehat{F} H^{-1} - F_0)\otimes\left(\frac{X_i^\top F^0}{T}\right)\right]\right)\text{vec}(Q)
\end{aligned}
\end{equation}
We first analyze $\text{vec}(Q)$. From the Proposition \ref{pro:B.1}, we have
$$
T^{-1/2}\|\widehat{F}-F^0H\|=O_p\left(\left(\frac{1}{N}\sum_{i=1}^N\left\|\widehat{\beta}_i-\beta_i^0\right\|^2\right)^{1/2}\right) + O_p\left(\delta_{NT}^{-1}\right).
$$
Then, 
\begin{equation}\nonumber
\begin{aligned}
\frac{1}{T}\left\|F^{0\top}(\widehat{F}-F^0H)\right\|\leq&\frac{1}{\sqrt{T}}\left\|F^{0}\right\|\cdot\frac{1}{\sqrt{T}}\left\|\widehat{F}-F^0H\right\|\\
=&O_p\left(\left(\frac{1}{N}\sum_{i=1}^N\left\|\widehat{\beta}_i-\beta_i^0\right\|^2\right)^{1/2}\right) + O_p\left(\delta_{NT}^{-1}\right)
\end{aligned}
\end{equation}
and
\begin{equation}\nonumber
\begin{aligned}
\frac{1}{T}\left\|\widehat{F}^{\top}(\widehat{F}-F^0H)\right\|\leq&\frac{1}{\sqrt{T}}\left\|\widehat{F}\right\|\cdot\frac{1}{\sqrt{T}}\left\|\widehat{F}-F^0H\right\|\\
=&O_p\left(\left(\frac{1}{N}\sum_{i=1}^N\left\|\widehat{\beta}_i-\beta_i^0\right\|^2\right)^{1/2}\right) + O_p\left(\delta_{NT}^{-1}\right).
\end{aligned}
\end{equation}
The two results can be rewritten as
$$
\left\|\frac{F^{0\top}\widehat{F}}{T}-\frac{F^{0\top}F^{0}}{T}H\right\|=O_p\left(\left(\frac{1}{N}\sum_{i=1}^N\left\|\widehat{\beta}_i-\beta_i^0\right\|^2\right)^{1/2}\right) + O_p\left(\delta_{NT}^{-1}\right)
$$
and
$$
\left\|I_T-\frac{\widehat{F}^{\top}F^{0}}{T}H\right\|=O_p\left(\left(\frac{1}{N}\sum_{i=1}^N\left\|\widehat{\beta}_i-\beta_i^0\right\|^2\right)^{1/2}\right) + O_p\left(\delta_{NT}^{-1}\right).
$$
Left multiply the first equation by $H^{\top}$ and use the transpose of the second equation to obtain
\begin{equation}\nonumber
\begin{aligned}
\left\|H^{\top}\left(\frac{F^{0\top}\widehat{F}}{T}-\frac{F^{0\top}F^{0}}{T}H\right)+\left(I_T-\frac{\widehat{F}^{\top}F^{0}}{T}H\right)^\top\right\|\leq&\left\|\frac{F^{0\top}\widehat{F}}{T}-\frac{F^{0\top}F^{0}}{T}H\right\|+\left\|I_T-\frac{\widehat{F}^{\top}F^{0}}{T}H\right\|\\
=&O_p\left(\left(\frac{1}{N}\sum_{i=1}^N\left\|\widehat{\beta}_i-\beta_i^0\right\|^2\right)^{1/2}\right) + O_p\left(\delta_{NT}^{-1}\right).
\end{aligned}
\end{equation}
That is,
$$
\left\|I_T-H^{\top}\frac{F^{0\top}F^{0}}{T}H\right\|=O_p\left(\left(\frac{1}{N}\sum_{i=1}^N\left\|\widehat{\beta}_i-\beta_i^0\right\|^2\right)^{1/2}\right) + O_p\left(\delta_{NT}^{-1}\right).
$$
Right multiplying by $H^{\top}$ and left multiplying by $H^{\top-1}$, we obtain
$$
\left\|I_T-\frac{F^{0\top}F^{0}}{T}HH^\top\right\|=O_p\left(\left(\frac{1}{N}\sum_{i=1}^N\left\|\widehat{\beta}_i-\beta_i^0\right\|^2\right)^{1/2}\right) + O_p\left(\delta_{NT}^{-1}\right).
$$
Then,
$$
\left\|\text{vec}(Q)\right\|=O_p\left(\left(\frac{1}{N}\sum_{i=1}^N\left\|\widehat{\beta}_i-\beta_i^0\right\|^2\right)^{1/2}\right) + O_p\left(\delta_{NT}^{-1}\right).
$$
Moreover,
\begin{equation}\nonumber
\begin{aligned}
\left\|\frac{1}{T}\varepsilon_i^\top (\widehat{F} H^{-1} - F_0)\right\|\leq&\frac{1}{T}\left\|\varepsilon_i\right\|\cdot\frac{1}{T}\left\|\widehat{F} H^{-1} - F_0\right\|\\
=&O_p\left(\left(\frac{1}{N}\sum_{i=1}^N\left\|\widehat{\beta}_i-\beta_i^0\right\|^2\right)^{1/2}\right) + O_p\left(\delta_{NT}^{-1}\right).
\end{aligned}
\end{equation}
Therefore,
\begin{equation}\nonumber
\begin{aligned}
\left\|c2\right\|&=\sqrt{T}\left[O_p\left(\left(\frac{1}{N}\sum_{i=1}^N\left\|\widehat{\beta}_i-\beta_i^0\right\|^2\right)^{1/2}\right) + O_p\left(\delta_{NT}^{-1}\right)\right]^2\\
&=\sqrt{T} O_p\left(\frac{1}{N}\sum_{i=1}^N\left\|\widehat{\beta}_i-\beta_i^0\right\|^2\right)+ \sqrt{T} O_p (\delta_{NT}^{-2}).
\end{aligned}
\end{equation}
For the term $c1$, by Lemma \ref{lemma:B.4}, 
\begin{equation}\nonumber
\begin{aligned}
\left\|c1\right\|=&\left\|\frac{1}{\sqrt{T}}\frac{X_i^\top F^0}{T}\left(\frac{F^{0\top}F^0}{T}\right)^{-1}(\widehat{F} H^{-1} - F^{0})^\top \varepsilon_i\right\|\\
\leq&\left\|\frac{1}{\sqrt{T}}\varepsilon_i^\top(\widehat{F}-F^0H)\right\|\\
=&O_p\left(\left(\frac{1}{N}\sum_{i=1}^N\left\|\widehat{\beta}_i-\beta_i^0\right\|^2\right)^{1/2}\right) + O_p\left(\delta_{NT}^{-1}\right)\\
&+\frac{\sqrt{T}}{N}\left[O_p\left(\left(\frac{1}{N}\sum_{i=1}^N\left\|\widehat{\beta}_i-\beta_i^0\right\|^2\right)^{1/2}\right) + O_p\left(\delta_{NT}^{-1}\right)\right]+o_p(1).
\end{aligned}
\end{equation}
For $d$, again let $Q = HH^\top  - (F^0 F^0 / T)^{-1}$. Then
\begin{equation}\nonumber
\begin{aligned}
d=&\frac{1}{\sqrt{T}}\frac{X^{\top}_{i}F^{0}}{T}\left[HH^{ \top}-\left(\frac{F^{0\top}F^{0}}{T}\right)^{-1}\right]F^{0\top}\varepsilon _{i}\\
=&\frac{1}{\sqrt{T}}\left[\varepsilon_i^\top F^0\otimes\left( \frac{X^\top_i F^0}{T} \right)\right]\text{vec}(Q)\\
=&O_p(1) \text{vec}(Q)\\
=&O_p\left(\left(\frac{1}{N}\sum_{i=1}^N\left\|\widehat{\beta}_i-\beta_i^0\right\|^2\right)^{1/2}\right) + O_p\left(\delta_{NT}^{-1}\right).
\end{aligned}
\end{equation}
In summary,
\begin{equation}\label{eq:B.6}
\begin{aligned}
\frac{1}{\sqrt{T}}X_{i}^\top(M_{F^{0}}-M_{\widehat{F}})\varepsilon_{i}=&\left(\frac{\sqrt{T}}{N}+1\right)\left[O_p\left(\left(\frac{1}{N}\sum_{i=1}^N\left\|\widehat{\beta}_i-\beta_i^0\right\|^2\right)^{1/2}\right) + O_p\left(\delta_{NT}^{-1}\right)\right]\\
&+\sqrt{T}\left[O_p\left(\frac{1}{N}\sum_{i=1}^N\left\|\widehat{\beta}_i-\beta_i^0\right\|^2\right)+O_p(\delta_{NT}^{-2})\right]+o_p(1).
\end{aligned}
\end{equation}
Let $V_{ik} =  a_{ik} X_i^\top$. Then replacing $X_i^\top$ with $V_{ik}$, the same argument leads to
\begin{equation}\label{eq:B.7}
\begin{aligned}
\frac{1}{\sqrt{T}}\frac{1}{N} \sum_{k=1}^{N}  V_{ik}(M_{F^0}-M_{\widehat{F}}) \varepsilon_k=&\left(\frac{\sqrt{T}}{N}+1\right)\left[O_p\left(\left(\frac{1}{N}\sum_{i=1}^N\left\|\widehat{\beta}_i-\beta_i^0\right\|^2\right)^{1/2}\right) + O_p\left(\delta_{NT}^{-1}\right)\right]\\
&+\sqrt{T}\left[O_p\left(\frac{1}{N}\sum_{i=1}^N\left\|\widehat{\beta}_i-\beta_i^0\right\|^2\right)+O_p(\delta_{NT}^{-2})\right]+o_p(1).
\end{aligned}
\end{equation}
Combining \eqref{eq:B.6} and \eqref{eq:B.7}, we obtain the lemma:

$$\frac{1}{\sqrt{NT}} \sum_{i=1}^N \left[ X^\top_i M_{\widehat{F}} - \frac{1}{N} \sum_{k=1}^N a_{ik} X^\top_k M_{\widehat{F}} \right] \varepsilon_i$$

$$= \frac{1}{\sqrt{NT}} \sum_{i=1}^N \left[ X^\top_i M_{F^0} - \frac{1}{N} \sum_{k=1}^N a_{ik} X^\top_k M_{F^0} \right] \varepsilon_i.$$

$$-\left(\frac{\sqrt{NT}}{N}\right)(\psi_{NT}-\psi^{\ast}_{NT})+\sqrt{T}O_p(\| \widehat{\beta}-\beta^{0}\|^{2})$$

$$+O_p(\|\widehat{\beta}-\beta^{0}\|)+\sqrt{T}O_p(\delta^{-2}_{NT}). \quad \text{Q.E.D.}$$

\end{proof}

\begin{lemma}\label{lemma:B.4}
Under Assumptions \ref{assum:indentification}-\ref{assum:weak dependence}, we have
\begin{equation}\nonumber
\begin{aligned}
\frac{1}{\sqrt{T}}\varepsilon_i^\top(\widehat{F}H^{-1}-F^0)=&O_p\left(\left(\frac{1}{N}\sum_{i=1}^N\left\|\widehat{\beta}_i-\beta_i^0\right\|^2\right)^{1/2}\right) + O_p\left(\delta_{NT}^{-1}\right)\\
&+\frac{\sqrt{T}}{N}\left[O_p\left(\left(\frac{1}{N}\sum_{i=1}^N\left\|\widehat{\beta}_i-\beta_i^0\right\|^2\right)^{1/2}\right) + O_p\left(\delta_{NT}^{-1}\right)\right]+o_p(1).
\end{aligned}
\end{equation}
\end{lemma}
\begin{proof}
From \eqref{eq:B.1} and denoting $G=(F^{0\top}\widehat{F}/T)^{-1}(\Lambda^{0\top}\Lambda^0/N)^{-1}$ for the moment,
\begin{equation}\nonumber
\begin{aligned}
\frac{1}{\sqrt{T}}\varepsilon^{\top}_{i}(\widehat{F} H^{-1}-F^{0}) =&\frac{1}{\sqrt{T}}\varepsilon^{\top}_{k}(I1+\cdots+ I8)G\\
:=a1+\cdots+a8.
\end{aligned}
\end{equation}
For the first four terms, we have
\begin{equation}\nonumber
\begin{aligned}
\left\|a1\right\|=&\left\|\frac{1}{\sqrt{T}}\varepsilon^{\top}_{k}\left(I1\right)G\right\|\\
=&\left\|\frac{1}{\sqrt{T}}\varepsilon^{\top}_{k}\left(\frac{1}{NT} \sum_{i=1}^N X_i (\beta_i^0 - \widehat{\beta}_i) (\beta_i^0 - \widehat{\beta}_i)^\top  X_i^\top  \widehat{F}\right)G\right\|\\
\leq&\frac{1}{N}\sum_{i=1}^N\left\|\frac{\varepsilon^{\top}_{k}X_i}{\sqrt{T}}\right\|\left\|\widehat{\beta}_i-\beta_i^0\right\|^2\\
=&\frac{1}{N}\sum_{i=1}^N\left\|\frac{1}{\sqrt{T}}\sum_{t=1}^TX_{it}\varepsilon_{kt}\right\|\left\|\widehat{\beta}_i-\beta_i^0\right\|^2\\
=&O_p\left(\frac{1}{N}\sum_{i=1}^N\left\|\widehat{\beta}_i-\beta_i^0\right\|^2\right),
\end{aligned}
\end{equation}
\begin{equation}\nonumber
\begin{aligned}
\left\|a2\right\|=&\left\|\frac{1}{\sqrt{T}}\varepsilon^{\top}_{k}\left(I2\right)G\right\|\\
=&\left\|\frac{1}{\sqrt{T}}\varepsilon^{\top}_{k}\left(\frac{1}{NT} \sum_{i=1}^N X_i (\beta_i^0 - \widehat{\beta}_i) \lambda_i^{0\top}  F^{0\top} \widehat{F}\right)G\right\|\\
=&\left\|\frac{1}{N\sqrt{T}}\sum_{i=1}^N\varepsilon^{\top}_{k}X_i (\beta_i^0 - \widehat{\beta}_i) \lambda_i^{0\top}\left(\frac{\Lambda ^{0\top}\Lambda^0}{N}\right)\right\|\\
\leq&\frac{1}{N}\sum_{i=1}^N\left\|\frac{\varepsilon^{\top}_{k}X_i}{\sqrt{T}}\right\|\left\|\widehat{\beta}_i-\beta_i^0\right\|\\
=&O_p\left(\left(\frac{1}{N}\sum_{i=1}^N\left\|\widehat{\beta}_i-\beta_i^0\right\|^2\right)^{1/2}\right),
\end{aligned}
\end{equation}
\begin{equation}\nonumber
\begin{aligned}
\left\|a3\right\|=&\left\|\frac{1}{\sqrt{T}}\varepsilon^{\top}_{k}\left(I3\right)G\right\|\\
=&\left\|\frac{1}{\sqrt{T}}\varepsilon^{\top}_{k}\left(\frac{1}{NT} \sum_{i=1}^N X_i (\beta_i^0 - \widehat{\beta}_i) \varepsilon_i^\top \widehat{F}\right)G\right\|\\
\leq&\frac{1}{N}\sum_{i=1}^N\left\|\frac{\varepsilon^{\top}_{k}X_i}{\sqrt{T}}\right\|\left\|\widehat{\beta}_i-\beta_i^0\right\|\left\|\frac{\varepsilon_{i}}{\sqrt{T}}\right\|\\
=&O_p\left(\left(\frac{1}{N}\sum_{i=1}^N\left\|\widehat{\beta}_i-\beta_i^0\right\|^2\right)^{1/2}\right),
\end{aligned}
\end{equation}
\begin{equation}\nonumber
\begin{aligned}
\left\|a4\right\|=&\left\|\frac{1}{\sqrt{T}}\varepsilon^{\top}_{k}\left(I4\right)G\right\|\\
=&\left\|\frac{1}{\sqrt{T}}\varepsilon^{\top}_{k}\left(\frac{1}{NT} \sum_{i=1}^N F^0 \lambda_i^0 (\beta_i^0 - \widehat{\beta}_i)^\top  X_i^\top  \widehat{F}\right)G\right\|\\
\leq&\frac{1}{N}\sum_{i=1}^N\left\|\frac{\varepsilon^{\top}_{k}F^0}{\sqrt{T}}\right\|\left\|\widehat{\beta}_i-\beta_i^0\right\|\left\|\frac{X_i}{\sqrt{T}}\right\|\\
=&\frac{1}{N}\sum_{i=1}^N\left\|\frac{1}{\sqrt{T}}\sum_{t=1}^Tf_t^0\varepsilon_{kt}\right\|\left\|\widehat{\beta}_i-\beta_i^0\right\|\left\|\frac{X_i}{\sqrt{T}}\right\|\\
=&O_p\left(\left(\frac{1}{N}\sum_{i=1}^N\left\|\widehat{\beta}_i-\beta_i^0\right\|^2\right)^{1/2}\right).
\end{aligned}
\end{equation}
For $ a5 $, let $ W_i = X_i^\top  \widehat{F}/T $ and note that $\|W_i\|^2 \leq \|X_i\|^2/T$:
\begin{equation}\nonumber
\begin{aligned}
\left\|a5\right\|=&\left\|\frac{1}{\sqrt{T}}\varepsilon^{\top}_{k}\left(I5\right)G\right\|\\
=&\left\|\frac{1}{\sqrt{T}}\varepsilon^{\top}_{k}\left(\frac{1}{NT} \sum_{i=1}^N \varepsilon_i (\beta_i^0 - \widehat{\beta}_i)^\top  X_i^\top  \widehat{F}\right)G\right\|\\
=&\left\|\frac{1}{N\sqrt{T}}\sum_{i=1}^N\varepsilon^{\top}_{k}\varepsilon_i (\beta_i^0 - \widehat{\beta}_i)^\top W_iG\right\|\\
\leq&\frac{1}{N}\sum_{i=1}^N\left\|\frac{1}{\sqrt{T}}\sum_{t=1}^T\varepsilon_{kt}\varepsilon_{it}\right\|\left\|\widehat{\beta}_i-\beta_i^0\right\|\\
=&O_p\left(\left(\frac{1}{N}\sum_{i=1}^N\left\|\widehat{\beta}_i-\beta_i^0\right\|^2\right)^{1/2}\right).
\end{aligned}
\end{equation}
For $ a6 $,
\begin{equation}\nonumber
\begin{aligned}
a6=&\frac{1}{\sqrt{T}}\varepsilon^{\top}_{k}\left(I6\right)G\\
=&\frac{1}{\sqrt{T}}\varepsilon^{\top}_{k}\left(\frac{1}{NT} \sum_{i=1}^N F^0 \lambda_i^0 \varepsilon_i^\top \widehat{F}\right)G\\
=&\left(\frac{1}{\sqrt{T}}\varepsilon^{\top}_{k}F^0\right)\left(\frac{1}{NT} \sum_{i=1}^N  \lambda_i^0 \varepsilon_i^\top F^0HG\right)+\left(\frac{1}{\sqrt{T}}\varepsilon^{\top}_{k}F^0\right)\frac{1}{NT} \sum_{i=1}^N  \lambda_i^0 \varepsilon_i^\top (\widehat{F}-F^0H)G\\
:=&a6.1 + a6.2.
\end{aligned}
\end{equation}
For terms $a6.1$ and $a6.2$, we have
\begin{equation}\nonumber
\begin{aligned}
\left\|a6.1\right\|=&\left\|\left(\frac{1}{\sqrt{T}}\varepsilon^{\top}_{k}F^0\right)\left(\frac{1}{NT} \sum_{i=1}^N  \lambda_i^0 \varepsilon_i^\top F^0HG\right)\right\|\\
\leq&\frac{1}{\sqrt{NT}}\left\|\frac{1}{\sqrt{T}}\sum_{t=1}^Tf^0_t\varepsilon_{kt}\right\|\left\|\frac{1}{\sqrt{NT}}\sum_{i=1}^N\sum_{t=1}^T\lambda_i^0f_t^{0\top}\varepsilon_{it}\right\|\\
=&O_p\left(\frac{1}{\sqrt{NT}}\right)
\end{aligned}
\end{equation}
and
\begin{equation}\nonumber
\begin{aligned}
\left\|a6.2\right\|=&\left\|\left(\frac{1}{\sqrt{T}}\varepsilon^{\top}_{k}F^0\right)\frac{1}{NT} \sum_{i=1}^N  \lambda_i^0 \varepsilon_i^\top (\widehat{F}-F^0H)G\right\|\\
\leq&\frac{1}{\sqrt{N}}\left\|\frac{1}{\sqrt{T}}\sum_{t=1}^Tf^0_t\varepsilon_{kt}\right\|\left\|\frac{1}{\sqrt{NT}}\sum_{i=1}^N  \lambda_i^0 \varepsilon_i^\top\right\|\left\|\frac{\widehat{F}-F^0H}{\sqrt{T}}\right\|\\
=&\frac{1}{\sqrt{N}}\left[O_p\left(\left(\frac{1}{N}\sum_{i=1}^N\left\|\widehat{\beta}_i-\beta_i^0\right\|^2\right)^{1/2}\right) + O_p\left(\delta_{NT}^{-1}\right)\right].
\end{aligned}
\end{equation}
Next consider $a7$ and $a8$:
\begin{equation}\nonumber
\begin{aligned}
a7=&\frac{1}{\sqrt{T}}\varepsilon^{\top}_{k}\left(I7\right)G\\
=&\frac{1}{\sqrt{T}}\varepsilon^{\top}_{k}\left( \frac{1}{NT} \sum_{i=1}^N \varepsilon_i \lambda_i^{0\top}  F^0 \widehat{F}\right)G\\
=&\frac{1}{\sqrt{T}}\frac{1}{NT} \sum_{i=1}^N \varepsilon^{\top}_{k}\varepsilon_i \lambda_i^{0\top}\left(\frac{\Lambda^{0\top}\Lambda^0}{N}\right)^{-1}\\
=&\frac{1}{\sqrt{NT}}\frac{1}{T}\sum_{t=1}^T \left[\varepsilon_{kt}\left(\frac{1}{\sqrt{N}}\sum_{i=1}^N \varepsilon_{it}\lambda_i^{0\top}\right)\right]\left(\frac{\Lambda^{0\top}\Lambda^0}{N}\right)^{-1}\\
=&O_p\left(\frac{1}{\sqrt{NT}}\right),
\end{aligned}
\end{equation}
\begin{equation}\nonumber
\begin{aligned}
a8=&\frac{1}{\sqrt{T}}\varepsilon^{\top}_{k}\left(I8\right)G\\
=&\frac{1}{\sqrt{T}}\varepsilon^{\top}_{k}\left(\frac{1}{NT} \sum_{i=1}^N \varepsilon_i \varepsilon_i^\top \widehat{F}\right)G\\
=&\frac{1}{\sqrt{T}}\frac{1}{NT}\sum_{i=1}^N\varepsilon^{\top}_{k}\varepsilon_i (\varepsilon_i^\top \widehat{F})G\\
=&\frac{1}{\sqrt{T}}\frac{1}{NT}\sum_{i=1}^N\varepsilon^{\top}_{k}\varepsilon_i (\varepsilon_i^\top F^0)HG+\frac{1}{\sqrt{T}}\frac{1}{NT}\sum_{i=1}^N\varepsilon^{\top}_{k}\varepsilon_i (\varepsilon_i^\top (\widehat{F}-F^0H))G\\
:=&b8+c8,
\end{aligned}
\end{equation}
\begin{equation}\nonumber
b8=\frac{1}{\sqrt{T}}\frac{1}{NT}\sum_{i=1}^N\varepsilon^{\top}_{k}\varepsilon_i (\varepsilon_i^\top F^0)HG=O_p\left(\frac{1}{N}+\frac{1}{\sqrt{T}}\right)
\end{equation}
because $\sum_{i=1}^N\varepsilon^{\top}_{k}\varepsilon_i=O_p(T+N\sqrt{T})$.
And 
\begin{equation}\nonumber
\begin{aligned}
\left\|c8\right\|=&\left\|\frac{1}{\sqrt{T}}\frac{1}{NT}\sum_{i=1}^N\varepsilon^{\top}_{k}\varepsilon_i (\varepsilon_i^\top (\widehat{F}-F^0H))G\right\|\\
\leq&\frac{1}{N\sqrt{T}}\sum_{i=1}^N\left\|\varepsilon^{\top}_{k}\varepsilon_i\right\|\left\|\frac{\widehat{F}-F^0H}{\sqrt{T}}\right\|\\
=&\left(\frac{\sqrt{T}}{N}+1\right)\left[O_p\left(\left(\frac{1}{N}\sum_{i=1}^N\left\|\widehat{\beta}_i-\beta_i^0\right\|^2\right)^{1/2}\right) + O_p\left(\delta_{NT}^{-1}\right)\right].
\end{aligned}
\end{equation}
For $a9$ and $a10$,
\begin{equation}\nonumber
\begin{aligned}
a9=&\frac{1}{\sqrt{T}}\varepsilon^{\top}_{k}\left(I9\right)G\\
=&\frac{1}{\sqrt{T}}\varepsilon^{\top}_{k}\left(\frac{1}{NT} \sum_{i=1}^N\left\{Z_i^\ast(\widehat{\alpha}_{i})-Z_i^\ast({\alpha}_{i}^0)\right\}\left\{Z_i^\ast(\widehat{\alpha}_{i})-Z_i^\ast({\alpha}_{i}^0)\right\}^\top\widehat{F}\right)G\\
=&O_p\left(\frac{1}{\sqrt{T}}\right),
\end{aligned}
\end{equation}
\begin{equation}\nonumber
\begin{aligned}
\left\|a10\right\|=&\left\|\frac{1}{\sqrt{T}}\varepsilon^{\top}_{k}\left(I10\right)G\right\|\\
=&\left\|\frac{1}{\sqrt{T}}\varepsilon^{\top}_{k}\left(\frac{1}{NT} \sum_{i=1}^N\left\{Z_i^\ast(\widehat{\alpha}_{i})-Z_i^\ast({\alpha}_{i}^0)\right\}(\beta_i^0 - \widehat{\beta}_i)^\top X_i^\top\widehat{F}\right)G\right\|\\
=&O_p\left(\left(\frac{1}{N}\sum_{i=1}^N\left\|\widehat{\beta}_i-\beta_i^0\right\|^2\right)^{1/2}\right).
\end{aligned}
\end{equation}
By Lemma \ref{lemma:B.5},
\begin{equation}\nonumber
\begin{aligned}
a11=&\frac{1}{\sqrt{T}}\varepsilon^{\top}_{k}\left(I11\right)G\\
=&\frac{1}{\sqrt{T}}\varepsilon^{\top}_{k}\left(\frac{1}{NT} \sum_{i=1}^N\left\{Z_i^\ast(\widehat{\alpha}_{i})-Z_i^\ast({\alpha}_{i}^0)\right\}\lambda_i^{0\top}  F^{0\top}\widehat{F}\right)G\\
=&\frac{1}{N} \sum_{i=1}^N\left(\frac{1}{\sqrt{T}}\varepsilon^{\top}_{k}\right)\left(\frac{1}{\sqrt{T}}\left\{Z_i^\ast(\widehat{\alpha}_{i})-Z_i^\ast({\alpha}_{i}^0)\right\}\lambda_i^{0\top}  F^{0\top}\right)\left(\frac{1}{\sqrt{T}}\widehat{F}\right)G\\
=&o_p(1).
\end{aligned}
\end{equation}
For terms $a12$-$a15$,
\begin{equation}\nonumber
\begin{aligned}
\left\|a12\right\|=&\left\|\frac{1}{\sqrt{T}}\varepsilon^{\top}_{k}\left(I12\right)G\right\|\\
=&\left\|\frac{1}{\sqrt{T}}\varepsilon^{\top}_{k}\left(\frac{1}{NT} \sum_{i=1}^N\left\{Z_i^\ast(\widehat{\alpha}_{i})-Z_i^\ast({\alpha}_{i}^0)\right\}\varepsilon_i^\top\widehat{F}\right)G\right\|\\
\leq&\frac{1}{NT} \frac{1}{\sqrt{T}}\sum_{i=1}^N\left\|\varepsilon_{k}\varepsilon_i^\top\widehat{F}G\right\|\\
=&O_p\left(\frac{1}{N}+\frac{1}{\sqrt{T}}\right),
\end{aligned}
\end{equation}
\begin{equation}\nonumber
\begin{aligned}
\left\|a13\right\|=&\left\|\frac{1}{\sqrt{T}}\varepsilon^{\top}_{k}\left(I13\right)G\right\|\\
=&\left\|\frac{1}{\sqrt{T}}\varepsilon^{\top}_{k}\left(\frac{1}{NT} \sum_{i=1}^NX_i (\beta_i^0 - \widehat{\beta}_i)\left\{Z_i^\ast(\widehat{\alpha}_{i})-Z_i^\ast({\alpha}_{i}^0)\right\}^\top\widehat{F}\right)G\right\|\\
=&O_p\left(\left(\frac{1}{N}\sum_{i=1}^N\left\|\widehat{\beta}_i-\beta_i^0\right\|^2\right)^{1/2}\right),
\end{aligned}
\end{equation}
\begin{equation}\nonumber
\begin{aligned}
\left\|a14\right\|=&\left\|\frac{1}{\sqrt{T}}\varepsilon^{\top}_{k}\left(I14\right)G\right\|\\
=&\left\|\frac{1}{\sqrt{T}}\varepsilon^{\top}_{k}\left(\frac{1}{NT} \sum_{i=1}^NF^0 \lambda_i^0\left\{Z_i^\ast(\widehat{\alpha}_{i})-Z_i^\ast({\alpha}_{i}^0)\right\}^\top\widehat{F}\right)G\right\|\\
=&O_p\left(\frac{1}{\sqrt{T}}\right),
\end{aligned}
\end{equation}
\begin{equation}\nonumber
\begin{aligned}
\left\|a15\right\|=&\left\|\frac{1}{\sqrt{T}}\varepsilon^{\top}_{k}\left(I15\right)G\right\|\\
=&\left\|\frac{1}{\sqrt{T}}\varepsilon^{\top}_{k}\left(\frac{1}{NT} \sum_{i=1}^N\varepsilon_i\left\{Z_i^\ast(\widehat{\alpha}_{i})-Z_i^\ast({\alpha}_{i}^0)\right\}^\top\widehat{F}\right)G\right\|\\
\leq&\frac{1}{NT} \sum_{i=1}^N\left\|\varepsilon^{\top}_{k}\varepsilon_i\right\|\\
=&O_p\left(\frac{1}{N}+\frac{1}{\sqrt{T}}\right).
\end{aligned}
\end{equation}
Hence, we have
\begin{equation}\nonumber
\begin{aligned}
\frac{1}{\sqrt{T}}\varepsilon_i^\top(\widehat{F}H^{-1}-F^0)=&O_p\left(\left(\frac{1}{N}\sum_{i=1}^N\left\|\widehat{\beta}_i-\beta_i^0\right\|^2\right)^{1/2}\right) + O_p\left(\delta_{NT}^{-1}\right)\\
&+\frac{\sqrt{T}}{N}\left[O_p\left(\left(\frac{1}{N}\sum_{i=1}^N\left\|\widehat{\beta}_i-\beta_i^0\right\|^2\right)^{1/2}\right) + O_p\left(\delta_{NT}^{-1}\right)\right]+o_p(1).
\end{aligned}
\end{equation}
\end{proof}
\begin{lemma}\label{lemma:B.5}
Under Assumptions \ref{assum:indentification}-\ref{assum:weak dependence}, we have
\begin{enumerate}[label=(\roman*)]
\item $\displaystyle\frac{1}{\sqrt{T}}X_i^\top M_{\widehat{F}}\left\{Z_k^\ast(\widehat{\alpha}_{k})-Z_k^\ast({\alpha}_{k}^0)\right\}=o_p(1)$, for $i,k\in[N]$.
\item $\displaystyle\frac{1}{\sqrt{T}}\left\{Z_k^\ast(\widehat{\alpha}_{k})-Z_k^\ast({\alpha}_{k}^0)\right\}^\top\widehat{F}=o_p(1)$, for $k\in[N]$.
\item $\displaystyle\frac{1}{\sqrt{T}}F^{0}\lambda_i^{0}  \left\{Z_i^\ast(\widehat{\alpha}_{i})-Z_i^\ast({\alpha}_{i}^0)\right\}^\top=o_p(1)$, for $i\in[N]$.
\end{enumerate}
\end{lemma}
\begin{proof}
(i) From Proposition \ref{pro:average consistency}, we have $M_{\widehat{F}}-M_{F^0}=o_p(1)$, then,
\begin{equation}\nonumber
\begin{aligned}
\frac{1}{\sqrt{T}}X_i^\top M_{\widehat{F}}\left\{Z_k^\ast(\widehat{\alpha}_{k})-Z_k^\ast({\alpha}_{k}^0)\right\}=&\frac{1}{\sqrt{T}}X_i^\top M_{F^0}\left\{Z_k^\ast(\widehat{\alpha}_{k})-Z_k^\ast({\alpha}_{k}^0)\right\}\\
&+\frac{1}{\sqrt{T}}X_i^\top (M_{\widehat{F}}-M_{F^0})\left\{Z_k^\ast(\widehat{\alpha}_{k})-Z_k^\ast({\alpha}_{k}^0)\right\}\\
=&\frac{1}{\sqrt{T}}X_i^\top M_{F^0}\left\{Z_k^\ast(\widehat{\alpha}_{k})-Z_k^\ast({\alpha}_{k}^0)\right\}+o_p(1).
\end{aligned}
\end{equation}
It is sufficient to show $\frac{1}{\sqrt{T}}X_i^\top M_{F^0}\left\{Z_k^\ast(\widehat{\alpha}_{k})-Z_k^\ast({\alpha}_{k}^0)\right\}=o_p(1)$. Expand $Z_k^\ast(\widehat{\alpha}_k)$ around $\alpha_k^0$, we have
$$
Z_k^\ast(\widehat{\alpha}_k) - Z_k^\ast(\alpha_k^0) = \partial_{\alpha_k} Z_k^\ast(\alpha_k^0)^\top (\widehat{\alpha}_k - \alpha_k^0) + R_k(\widehat{\alpha}_k),
$$
where the remainder term $R_k(\widehat{\alpha}_k)$ satisfies $R_k(\widehat{\alpha}_k)=(R_{k1}(\widehat{\alpha}_k),\ldots,R_{kT}(\widehat{\alpha}_k))^\top$ and $\|R_{kt}(\widehat{\alpha}_k)\| = O_p(\|\widehat{\alpha}_k - \alpha_k^0\|^2)$. Then,
\begin{equation}\nonumber
\begin{aligned}
\frac{1}{\sqrt{T}}X_i^\top M_{F^0}\left\{Z_k^\ast(\widehat{\alpha}_{k})-Z_k^\ast({\alpha}_{k}^0)\right\}=&\frac{1}{\sqrt{T}}X_i^\top M_{F^0} \partial_{\alpha_k} Z_k^\ast(\alpha_k^0)^\top (\widehat{\alpha}_k - \alpha_k^0)+\frac{1}{\sqrt{T}}X_i^\top M_{F^0} R_k(\widehat{\alpha}_k)\\
:=&A_T + B_T.
\end{aligned}
\end{equation}
For the term $A_T$, let $C = X_i^\top M_{F^0} \partial_{\alpha_k} Z_k^\ast(\alpha_k^0)^\top$, then $A_T = \frac{1}{\sqrt{T}} C (\widehat{\alpha}_k - \alpha_k^0)$. $C$ is a $(p+1) \times (p+1)$ matrix, whose the $(j,\ell)$-th element is:
$$
C_{j\ell} = \sum_{t=1}^T A_t, \ A_t = (M_{F^0}^\top X_i)^{(j)}_t \cdot [X_{kt} m_{kt}(\alpha_k^0)]^{(\ell)},
$$
where $m_{kt}(\alpha_k^0) = 1 - \frac{1}{\tau}\mathbbm{1}(Y_{kt} \le X_{kt}^\top\alpha_k^0)$. Compute the variance of $C_{j\ell}$,
$$
\mathrm{Var}(C_{j\ell}) = \sum_{t=1}^T \mathrm{Var}(A_t) + 2\sum_{1 \le t < s \le T} \mathrm{Cov}(A_t, A_s).
$$
By the Cauchy-Schwarz and Hölder inequalities, we have
$$
\mathrm{Var}(A_t) \le E[A_t^2] \le \left( E|(M_{F^0}^\top X_i)^{(j)}_t|^4 \right)^{1/2} \cdot \left( E|X_{kt}^{(\ell)} m_{kt}|^4 \right)^{1/2}.
$$
By $E[\|X_{it}\|^4] \le M$, and $|m_{kt}| \le 1 + \frac{1}{\tau}$, so $\mathrm{Var}(A_t) \le C \quad \text{for some constant } C > 0.$ Thus, $\sum_{t=1}^T \mathrm{Var}(A_t) = O(T)$. For $\alpha$-mixing sequences $A_t$ and $A_s$ ($s > t$), by Davydov's inequality, we have
$$
|\mathrm{Cov}(A_t, A_s)| \le 8 \alpha(s-t)^{1/2} (E[A_t^4])^{1/4} (E[A_s^4])^{1/4} \le C \cdot \alpha(|s-t|)^{1/2}.
$$
Therefore,
$$
\sum_{1 \le t < s \le T} |\mathrm{Cov}(A_t, A_s)| \le C \sum_{h=1}^{T-1} (T-h) \alpha(h)^{1/2} \le CT \sum_{h=1}^\infty \alpha(h)^{1/2}.
$$
By the mixing condition $\sum_{h=1}^\infty \alpha(h) < \infty$, it follows that $\sum_{h=1}^\infty \alpha(h)^{1/2} < \infty$, so
$$
\sum_{1 \le t < s \le T} |\mathrm{Cov}(A_t, A_s)| = O(T).
$$
Then,
$$
\mathrm{Var}(C_{j\ell}) = O(T) + O(T) = O(T).
$$
By Chebyshev's inequality, for any $\epsilon > 0$, there exists $M_\epsilon$ such that:
$$
P\left( \left| \frac{1}{\sqrt{T}} C_{j\ell} \right| > M_\epsilon \right) \le \frac{\mathrm{Var}(C_{j\ell}/ \sqrt{T})}{M_\epsilon^2} = \frac{O(1)}{M_\epsilon^2} < \epsilon \quad \text{for sufficiently large $M_\epsilon$}.
$$
Hence, $\frac{1}{\sqrt{T}} C_{j\ell} = O_p(1)$.
Since $p$ is fixed, the matrix norm satisfies
$$
\left\| \frac{1}{\sqrt{T}} C \right\|  = O_p(1).
$$
Thus:
$$
A_T = \frac{1}{\sqrt{T}} C (\widehat{\alpha}_k - \alpha_k^0) = O_p(1) \cdot O_p(T^{-1/2}) = O_p(T^{-1/2}) = o_p(1).
$$
For the term $B_T$, each component $R_{kt}(\widehat{\alpha}_k) = O_p(T^{-1})$, then
$$
\|R_k(\widehat{\alpha}_k)\|^2 = \sum_{t=1}^T R_{kt}^2(\widehat{\alpha}_k) = T \cdot O_p(T^{-2}) = O_p(T^{-1}),
$$
Also, by $\|X_i^\top M_{F^0}\| = O_p(\sqrt{T})$, we have
$$
\|B_T\| = \left\| \frac{1}{\sqrt{T}} X_i^\top M_{F^0} R_k(\widehat{\alpha}_k) \right\| \le \frac{1}{\sqrt{T}} \cdot O_p(\sqrt{T}) \cdot O_p(T^{-1/2}) = O_p(T^{-1/2}) = o_p(1).
$$
Therefore,
$$
\frac{1}{\sqrt{T}}X_i^\top M_{F^0}\left\{Z_k^\ast(\widehat{\alpha}_{k})-Z_k^\ast({\alpha}_{k}^0)\right\} = A_T + B_T = o_p(1) + o_p(1) = o_p(1).
$$
The proof for the remaining statements are the same, so are omitted.
\end{proof}
%\clearpage
\section{Proof of Theorem \ref{thm:selection}}\label{secC}
\begin{proof}
For any fixed $r < r^0$, consider the difference in information criteria,
$$
\mathrm{IC}(r) - \mathrm{IC}(r^0) = \left( \log \widehat{V}(r) - \log \widehat{V}(r^0) \right) + (r - r^0) q(N,T).
$$
By Lemma \ref{lemma:Vr}, for $r < r^0$,
\begin{equation}\nonumber
\widehat{V}(r) = \sigma_{\varepsilon}^2 + \Delta(r) + o_p(1),
\end{equation}
and for $r \ge r^0$,
$$
\widehat{V}(r^0) = \sigma_{\varepsilon}^2 + O_p(1/\min(N,T)).
$$
Therefore,
$$
\frac{\widehat{V}(r)}{\widehat{V}(r^0)} = \frac{\sigma_{\varepsilon}^2 + \Delta(r) + o_p(1)}{\sigma_{\varepsilon}^2 + o_p(1)} = 1 + \frac{\Delta(r)}{\sigma_{\varepsilon}^2} + o_p(1).
$$
Let
$$
c = \log\left( 1 + \frac{\Delta(r)}{\sigma_{\varepsilon}^2} \right) > 0,
$$
then,
$$
\log\left( \frac{\widehat{V}(r)}{\widehat{V}(r^0)} \right) = c + o_p(1).
$$
and
$$
\mathrm{IC}(r) - \mathrm{IC}(r^0) = c + (r - r^0) q(N,T)+o_p(1).
$$
Since $r - r^0 \le -1$ and $q(N,T) > 0$,
$$
(r - r^0) q(N,T) \le -q(N,T).
$$
Let $R_{NT}=o_p(1)$, we need to prove:
$$
\lim_{N,T \to \infty} P\left( \mathrm{IC}(r) > \mathrm{IC}(r^0) \right) = 1,
$$
which is equivalent to:
$$
\lim_{N,T \to \infty} P\left( c + R_{NT} + (r - r^0) q(N,T) > 0 \right) = 1.
$$
Fix an arbitrary small $\epsilon > 0$. Take $\delta = c/4 > 0$. Since $R_{NT} = o_p(1)$, there exist $N_1, T_1$ such that for $N > N_1, T > T_1$,
$$
P(|R_{NT}| < \delta) > 1 - \epsilon/2.
$$
Since $q(N,T) \to 0$, there exist $N_2, T_2$ such that for $N > N_2, T > T_2$,
$$
q(N,T) < \delta/2.
$$
Let $N_0 = \max(N_1, N_2)$, $T_0 = \max(T_1, T_2)$. Then for $N > N_0, T > T_0$, with probability at least $1 - \epsilon/2$,
\begin{equation}\nonumber
\begin{aligned}
\mathrm{IC}(r) - \mathrm{IC}(r^0) &= c + R_{NT} + (r - r^0) q(N,T) \\
&\ge c - |R_{NT}| - q(N,T) \\
&> c - \delta - \delta/2 \\
&= 5c/8 > 0.
\end{aligned}
\end{equation}
Therefore,
$$
P\left( \mathrm{IC}(r) > \mathrm{IC}(r^0) \right) > 1 - \epsilon\ \text{for } N > N_0, T > T_0.
$$
Since $\epsilon$ is arbitrary,
$$
\lim_{N,T \to \infty} P\left( \mathrm{IC}(r) > \mathrm{IC}(r^0) \right) = 1,\ \text{for all } r < r^0.
$$
For any fixed $r > r^0$,
$$
\mathrm{IC}(r) - \mathrm{IC}(r^0) = \left( \log \widehat{V}(r) - \log \widehat{V}(r^0) \right) + (r - r^0) q(N,T).
$$
By Lemma \ref{lemma:Vr}, for $r \ge r^0$,
$$
\widehat{V}(r) = \sigma_{\varepsilon}^2 + O_p(1/\min(N,T)), \quad
\widehat{V}(r^0) = \sigma_{\varepsilon}^2 + O_p(1/\min(N,T)).
$$
Thus,
$$
\frac{\widehat{V}(r)}{\widehat{V}(r^0)} = 1+O_p\left( \frac{1}{\min(N,T)} \right).
$$
Taking logarithms and using $\log(1+x) = x + O(x^2)$, we obtain
$$
\log \widehat{V}(r) - \log \widehat{V}(r^0) = O_p\left( \frac{1}{\min(N,T)} \right).
$$
Then,
$$
\mathrm{IC}(r) - \mathrm{IC}(r^0) = O_p\left( \frac{1}{\min(N,T)} \right) + (r - r^0) q(N,T).
$$
From the condition $\min(N,T) \cdot q(N,T) \to \infty$, we have
$$
\frac{1}{\min(N,T)} = o( q(N,T) ) \ \text{and }O_p\left( \frac{1}{\min(N,T)} \right) = o_p( q(N,T)).
$$
Therefore,
$$
\mathrm{IC}(r) - \mathrm{IC}(r^0) = (r - r^0) q(N,T) + o_p( q(N,T)).
$$
Since $r - r^0 \ge 1$ and $q(N,T) > 0$ for large $N,T$, it follows that
$$
\lim_{N,T \to \infty} P( \mathrm{IC}(r) > \mathrm{IC}(r^0)) = 1\ \text{for all }r> r^0.
$$
Let the candidate factor number set be $\mathcal{R} = \{0, 1, \dots, R_{\max}\}$, where $R_{\max}$ is a predetermined upper bound. From the above analysis, for any $r \neq r^0$:
$$
\lim_{N,T \to \infty} P\left( \mathrm{IC}(r) > \mathrm{IC}(r^0) \right) = 1.
$$
Therefore,
\begin{equation}\nonumber
\begin{aligned}
\lim_{N,T \to \infty} P\left( \widehat{r} = r^0 \right) 
&= \lim_{N,T \to \infty} P\left( \bigcap_{r \neq r^0} \{ \mathrm{IC}(r) > \mathrm{IC}(r^0) \} \right) \\
&\ge 1 - \sum_{r \neq r^0} \lim_{N,T \to \infty} P\left( \mathrm{IC}(r) \le \mathrm{IC}(r^0) \right) \\
&= 1.
\end{aligned}
\end{equation}
This completes the detailed proof of the theorem.
\end{proof}

\begin{lemma}\label{lemma:Vr}
Under Assumptions \ref{assum:indentification}-\ref{assum:factor&loading}, for each fixed number of factors $r$, define
\begin{equation}\nonumber
\sigma_{\varepsilon}^2 = \lim_{N,T \to \infty} \frac{1}{NT} \sum_{i=1}^N \sum_{t=1}^T E(\varepsilon_{it}^2),
\end{equation}
when $r < r^0$, we have
\begin{equation}\nonumber
\widehat{V}(r) = \sigma_{\varepsilon}^2 + \Delta(r) + o_p(1),
\end{equation}
and when $r \ge r^0$,
\begin{equation}\nonumber
\widehat{V}(r) = \sigma_{\varepsilon}^2 + O_p\left( \frac{1}{\min(N,T)} \right),
\end{equation}
where $\widehat{V}(r)$ is defined in equation \eqref{eq:Vr}, and $\Delta(r)$ is a positive constant independent of $N,T$.
\end{lemma}
\begin{proof}
For any $r \ge 0$, define the true parameters $(\beta_{i}^0(r), \lambda_{i}^0(r), f_{t}^0(r))$ as the solution to the following minimization problem:
\begin{equation}\label{eq:lemmini}
\argmin_{B\in\mathbb{R}^{N\times (p+1)},\theta\in\Theta^{r}}\frac{1}{NT}\sum_{i=1}^N\sum_{t=1}^T\left[ \left( Z_{it}^\ast(\alpha_{i}^0) - X_{it}^\top \beta_i - \lambda_i^\top f_t \right)^2 \right].
\end{equation}
The corresponding residual is
\begin{equation}\nonumber
W_{it}^0(r) = Z_{it}^\ast(\alpha_{i}^0) - X_{it}^\top \beta_{i}^0(r) - \lambda_{i}^0(r)^\top f_{t}^0(r).
\end{equation}
From $Z_{it}^\ast({\alpha}_{i}^0)=X_{it}^\top\beta_i^0+\lambda_i^{0\top}f_t^0+\varepsilon_{it}$, we have
\begin{equation}\label{eq:lemWit}
W_{it}^0(r)=X_{it}^\top (\beta_{i}^0-\beta_{i}^0(r))+[\lambda_{i}^{0\top} f_{t}^0-\lambda_{i}^0(r)^\top f_{t}^0(r)] + \varepsilon_{it}.
\end{equation}
From the first-order condition of \eqref{eq:lemmini} with respect to $\beta_i$, we have
\begin{equation}\label{eq:lemfirst}
E\left( X_{it} W_{it}^0(r) \right) = 0.
\end{equation}
Substituting \eqref{eq:lemWit} into \eqref{eq:lemfirst}, we have
$$
E(X_{it} X_{it}^\top) (\beta_{i}^0 - \beta_{i}^0(r)) + E[X_{it} (\lambda_{i}^{0\top} f_{t}^0 - \lambda_{i}^0(r)^\top f_{t}^0(r))] + E(X_{it} \varepsilon_{it}) = 0.
$$
Because errors and common components are independent of covariates, then $E[X_{it} \varepsilon_{it}] = 0$ and $E[X_{it} (\lambda_{i}^{0\top} f_{t}^0 - \lambda_{i}^0(r)^\top f_{t}^0(r))] = 0$. Furthermore, $E[X_{it} X_{it}^\top]$ is non-singular, hence,
$$
\beta_{i}^0(r) = \beta_{i}^0 \ \text{for all } r \ge 0.
$$
This result shows that regardless of the number of factors used, the optimal regression coefficient $\beta_i$ always equals the true value $\beta_{i}^0$. Then, we have
$$
W_{it}^0(r) = \varepsilon_{it} + U_{it}(r),
$$
where
\begin{equation}\nonumber
U_{it}(r) = \lambda_{i}^{0\top} f_{t}^0 - \lambda_{i}^0(r)^\top f_{t}^0(r)
\end{equation}
represents the systematic component not captured when using only $r$ factors.

We first analyze the properties of $U_{it}(r)$. When $r \ge r^0$, we take $$
\lambda_{i}^0(r) = \begin{pmatrix} \lambda_{i}^0 \\ \mathbf{0}_{r-r^0} \end{pmatrix} \in \RR^r,\ f_{t}^0(r) = \begin{pmatrix} f_{t}^0 \\ \mathbf{0}_{r-r^0} \end{pmatrix} \in \RR^r.
$$ 
Then, we have
$$
\lambda_{i}^0(r)^\top f_{t}^0(r) = \lambda_{i}^{0\top} f_{t}^0,
$$
and therefore,
$$
U_{it}(r) = 0 \ \text{for all } i,t.
$$
When $r < r^0$, consider the rank-$r^0$ matrix $L^0 = \Lambda^0 F^{0\top}$ with entries $L_{it}^0 = \lambda_{i}^{0\top} f_{t}^0$. We approximate $L$ by a rank-$r$ matrix $L^0(r) = \Lambda^0(r) F^0(r)^\top$. By Assumption \ref{assum:factor&loading},
$$
\Sigma_F = \lim_{T \to \infty} \frac{1}{T} \sum_{t=1}^T f_{t}^0 f_{t}^{0\top} > 0, \quad
\Sigma_\Lambda = \lim_{N \to \infty} \frac{1}{N} \sum_{i=1}^N \lambda_{i}^0 \lambda_{i}^{0\top} > 0.
$$
Consider the normalized matrix $(NT)^{-1/2} L = (NT)^{-1/2} \Lambda^0 F^{0\top}$. Then, the eigenvalues of $(NT)^{-1} L L^\top$ converge to the eigenvalues of $\Sigma_\Lambda^{1/2} \Sigma_F \Sigma_\Lambda^{1/2}$. Denote these eigenvalues as $\delta_1 \ge \delta_2 \ge \cdots \ge \delta_{r^0} > 0$. By the Eckart-Young theorem,
$$
\frac{1}{NT} \sum_{i=1}^N \sum_{t=1}^T U_{it}(r)^2=\frac{1}{NT}\|L^0 - L^0(r)\|_F^2 \stackrel{p}{\to} \sum_{k=r+1}^{r^0} \delta_k:=\Delta(r).
$$
Define the following four error components,
\begin{equation}\nonumber
\begin{aligned}
A_{it}&=W_{it}^0(r) = \varepsilon_{it} + U_{it}(r)\\
B_{it}&=Z_{it}^\ast(\widehat{\alpha}_{i}) - Z_{it}^\ast(\alpha_{i}^0)\\
C_{it}&=X_{it}^\top (\widehat{\beta}_{i}(r) - \beta_{i}^0(r))\\
D_{it}&=\widehat{\lambda}_{i}(r)^\top \widehat{f}_{t}(r) -\lambda_{i}^0(r)^\top f_{t}^0(r)
\end{aligned}
\end{equation}
Then, we have
\begin{equation}\nonumber
\widehat{V}(r) = \frac{1}{NT} \sum_{i=1}^N \sum_{t=1}^T \left( A_{it} + B_{it} - C_{it} - D_{it} \right)^2.
\end{equation}
Expanding the square yields ten terms, that is,
\begin{equation}\nonumber
\begin{aligned}
\widehat{V}(r) =& \frac{1}{NT}\sum_{i=1}^N\sum_{t=1}^T A_{it}^2+ \frac{1}{NT}\sum_{i=1}^N\sum_{t=1}^T B_{it}^2 
+ \frac{1}{NT}\sum_{i=1}^N\sum_{t=1}^T C_{it}^2 
+ \frac{1}{NT}\sum_{i=1}^N\sum_{t=1}^T D_{it}^2 \\
&- \frac{2}{NT} \sum_{i=1}^N\sum_{t=1}^T A_{it} C_{it} 
- \frac{2}{NT} \sum_{i=1}^N\sum_{t=1}^T A_{it} D_{it} 
+ \frac{2}{NT} \sum_{i=1}^N\sum_{t=1}^T A_{it} B_{it}\\
&- \frac{2}{NT} \sum_{i=1}^N\sum_{t=1}^T B_{it} C_{it} 
- \frac{2}{NT} \sum_{i=1}^N\sum_{t=1}^T B_{it} D_{it} 
+ \frac{2}{NT} \sum_{i=1}^N\sum_{t=1}^T C_{it} D_{it}\\
:=&S_1+\cdots+S_{10}.
\end{aligned}
\end{equation}
For the first term $S_1$,
\begin{equation}\nonumber
S_1 = \frac{1}{NT} \sum_{i=1}^N\sum_{t=1}^T \varepsilon_{it}^2 + \frac{1}{NT} \sum_{i=1}^N\sum_{t=1}^T U_{it}(r)^2 + \frac{2}{NT} \sum_{i=1}^N\sum_{t=1}^T \varepsilon_{it} U_{it}(r).
\end{equation}
By Assumptions \ref{assum:error} and \ref{assum:weak dependence}, and based on the above analysis, we have
\begin{equation}\nonumber
\frac{1}{NT} \sum_{i=1}^N\sum_{t=1}^T \varepsilon_{it}^2 = \sigma_{\varepsilon}^2 + o_p(1),
\end{equation}
\begin{equation}\nonumber
\frac{1}{NT} \sum_{i=1}^N\sum_{t=1}^T U_{it}(r)^2 = \Delta(r) \cdot \mathbbm{1}(r < r^0) + o_p(1),
\end{equation}
and
\begin{equation}\nonumber
\frac{1}{NT} \sum_{i=1}^N\sum_{t=1}^T \varepsilon_{it} U_{it}(r) = O_p\left( \frac{1}{\sqrt{NT}} \right).
\end{equation}
Therefore,
\begin{equation}\nonumber
S_1 = \sigma_{\varepsilon}^2 + \Delta(r) \cdot \mathbbm{1}(r < r^0) + o_p(1).
\end{equation}
For second term $S_2$, by Lemma \ref{lemmaA.1}, we have
$$
Z_{it}^\ast(\widehat{\alpha}_{i}) - Z_{it}^\ast(\alpha_{i}^0)=O_p(T^{-1/2}),
$$
then,
$$
\frac{1}{NT}\sum_{i=1}^N\sum_{t=1}^T B_{it}^2=O_p(T^{-1}).
$$
For the third term $S_4$, by Assumption \ref{assum:indentification} and Theorem \ref{theo:consistency},
$$
\frac{1}{NT}\sum_{i=1}^N\sum_{t=1}^T C_{it}^2=O_p(T^{-1}).
$$
For the fourth term, decompose $D_{it}$, 
$$
D_{it}=(\widehat{\lambda}_{i}(r) - \lambda_{i}^0(r))^\top \widehat{f}_{t}(r)+\lambda_{i}^0(r)^\top (\widehat{f}_{t}(r) - f_{t}^0(r)),
$$
and by Proposition \ref{pro:B.1} and apply Cauchy-Schwarz inequality, we have
$$
\frac{1}{NT}\sum_{i=1}^N\sum_{t=1}^T D_{it}^2=O_p\left(\frac{1}{\min(N,T)}\right).
$$
For terms $S_5$, by Cauchy-Schwarz inequality, we have
$$
|S_5| \le 2 \sqrt{ \left( \frac{1}{NT} \sum_{i=1}^N\sum_{t=1}^T A_{it}^2 \right) \left( \frac{1}{NT} \sum_{i=1}^N\sum_{t=1}^T C_{it}^2 \right) } = O_p(T^{-1/2}),
$$
For terms $S_6$ and $S_7$, when $r<r_0$,
$$
|S_6| \le 2 \sqrt{ \left( \frac{1}{NT} \sum_{i=1}^N\sum_{t=1}^T A_{it}^2 \right) \left( \frac{1}{NT} \sum_{i=1}^N\sum_{t=1}^T D_{it}^2 \right) } = O_p\left( \frac{1}{\sqrt{\min(N,T)}} \right),
$$
$$
|S_7| \le 2 \sqrt{ \left( \frac{1}{NT} \sum_{i=1}^N\sum_{t=1}^T A_{it}^2 \right) \left( \frac{1}{NT} \sum_{i=1}^N\sum_{t=1}^T B_{it}^2 \right) } = O_p(T^{-1/2}).
$$
When $r\ge r^0$, $A_{it}=\varepsilon_{it}$, then by Proposition \ref{pro:B.1} and Assumption \ref{assum:weak dependence}, we have
$$
|S_6| = O_p\left( \frac{1}{\min(N,T)} \right),
$$
and by Neyman orthogonality, we have
$$
|S_7| = O_p\left( \frac{1}{T} \right).
$$
For terms $S_8$-$S_{10}$, by Cauchy-Schwarz inequality, we have
$$
|S_8| \le 2 \sqrt{ \left( \frac{1}{NT} \sum_{i=1}^N\sum_{t=1}^T B_{it}^2 \right) \left( \frac{1}{NT} \sum_{i=1}^N\sum_{t=1}^T C_{it}^2 \right) } = O_p(T^{-1}),
$$
$$
|S_9| \le 2 \sqrt{ \left( \frac{1}{NT} \sum_{i=1}^N\sum_{t=1}^T B_{it}^2 \right) \left( \frac{1}{NT} \sum_{i=1}^N\sum_{t=1}^T D_{it}^2 \right) }= O_p\left( \frac{1}{\sqrt{T \cdot \min(N,T)}} \right),
$$
$$
|S_{10}| \le 2 \sqrt{ \left( \frac{1}{NT} \sum_{i=1}^N\sum_{t=1}^T C_{it}^2 \right) \left( \frac{1}{NT} \sum_{i=1}^N\sum_{t=1}^T D_{it}^2 \right) }= O_p\left( \frac{1}{\sqrt{T \cdot \min(N,T)}} \right).
$$
Then, the dominant cross term is $S_6 = O_p(1/\sqrt{\min(N,T)})$. Based on the above analysis, when $r < r^0$, we have
\begin{equation}\nonumber
\widehat{V}(r) = \sigma_{\varepsilon}^2 + \Delta(r) + o_p(1),
\end{equation}
and when $r \ge r^0$,
\begin{equation}\nonumber
\widehat{V}(r) = \sigma_{\varepsilon}^2 + O_p\left( \frac{1}{\min(N,T)} \right).
\end{equation}
This completes the proof.
\end{proof}
\section{Proof of Theorem \ref{theo:asymptotic}}\label{secD}
\begin{proof}
We begin with the asymptotic representation derived from the proof of Theorem \ref{theo:consistency}:
\begin{equation}\nonumber
\begin{aligned}
\left( \frac{1}{T}X_i^\top M_{F^0} X_i \right)\sqrt{T}\left(\widehat{\beta}_{i}-\beta_i^0\right)
=&\frac{1}{\sqrt{T}N}\sum_{k=1}^N\left(X_i^\top M_{F^0}X_k a_{ik}\right)\sqrt{T}\left(\widehat{\beta}_k-\beta_k^0\right)\\
&+\frac{1}{\sqrt{T}} X_i^\top M_{{F}^0}\varepsilon_i - \frac{1}{\sqrt{T}N} \sum_{k=1}^{N} a_{ik} X_i^\top M_{{F}^0} \varepsilon_k+o_p(1).
\end{aligned}\end{equation}
Multiplying both sides by the inverse of $\frac{1}{T}X_i^\top M_{F^0} X_i$, we obtain
\begin{equation}\nonumber
\begin{aligned}
\sqrt{T}\left(\widehat{\beta}_{i}-\beta_i^0\right)
=&\left( \frac{1}{T}X_i^\top M_{F^0} X_i \right)^{-1}\frac{1}{\sqrt{T}N}\sum_{k=1}^N\left(X_i^\top M_{F^0}X_k a_{ik}\right)\sqrt{T}\left(\widehat{\beta}_k-\beta_k^0\right)\\
&+\left( \frac{1}{T}X_i^\top M_{F^0} X_i \right)^{-1}\left[\frac{1}{\sqrt{T}} X_i^\top M_{{F}^0}\varepsilon_i - \frac{1}{\sqrt{T}N} \sum_{k=1}^{N} a_{ik} X_i^\top M_{{F}^0} \varepsilon_k\right]+o_p(1).
\end{aligned}
\end{equation}
Now, introduce the notation:
$$
A_i=\left( \frac{1}{T}X_i^\top M_{F^0} X_i\right)^{-1},\quad H_{ik}=A_i\left(\frac{1}{T}X_i^\top M_{F^0}X_k a_{ik}\right),
$$
and
$$
U_i=\frac{1}{\sqrt{T}} X_i^\top M_{{F}^0}\varepsilon_i - \frac{1}{\sqrt{T}N} \sum_{k=1}^{N} a_{ik} X_i^\top M_{{F}^0} \varepsilon_k.
$$
Then the equation simplifies to
$$
\sqrt{T}\left(\widehat{\beta}_{i}-\beta_i^0\right)=\frac{1}{N}\sum_{k=1}^N H_{ik}\sqrt{T}\left(\widehat{\beta}_{k}-\beta_k^0\right)+A_i U_i+o_p(1).
$$
Writing this in vector form for all units, define
\begin{equation}\nonumber
\begin{gathered}
\widehat{\beta}-\beta^0=\left((\widehat{\beta}_{1}-\beta_1^0)^\top,\ldots,(\widehat{\beta}_{N}-\beta_N^0)^\top\right)^\top, \\
H=[H_{ik}],\quad A=\operatorname{diag}(A_1,\ldots,A_N),\quad U=(U_1^\top,\ldots,U_N^\top)^\top.
\end{gathered}
\end{equation}
Then we have the system
$$
\sqrt{T}\left(\widehat{\beta}-\beta^0\right)=\frac{1}{N}H\sqrt{T}(\widehat{\beta}-\beta^0)+AU+o_p(1),
$$
which rearranges to
$$
\left(I-\frac{1}{N}H\right)\sqrt{T}(\widehat{\beta}-\beta^0)=AU+o_p(1).
$$
By Assumption \ref{assum:cross-sectional dependence}, the matrix $I-\frac{1}{N}H$ is invertible with probability approaching 1, and its inverse has bounded norm:
$$
\left\|\left(I-\frac{1}{N}H\right)^{-1}\right\|=O_p(1).
$$
Therefore,
$$
\sqrt{T}\left(\widehat{\beta}-\beta^0\right)=\left(I-\frac{1}{N}H\right)^{-1}AU+o_p(1).
$$
Let $G=(I-N^{-1}H)^{-1}$. Then, for the $i$-th unit:
$$
\sqrt{T}\left(\widehat{\beta}_{i}-\beta_i^0\right)=\sum_{j=1}^N G_{ij}A_jU_j+o_p(1),
$$
where $G_{ij}$ is the $(i,j)$-th block of $G$. We now replace the random weights $G_{ij}A_j$ by their probability limits. By Assumption \ref{assum:uniform convergence}, we have
$$
A_j \xrightarrow{p} A_j^0 = \Sigma_{X_j,X_j}^{-1} = \left(\lim_{T\to \infty} E\left[\frac{1}{T}X_j^\top M_{F^0}X_j\right]\right)^{-1}.
$$
Moreover, since
$$
H_{ij} = A_i\left(\frac{1}{T}X_i^\top M_{F^0}X_j a_{ij}\right) \xrightarrow{p} \Sigma_{X_i,X_i}^{-1} \Sigma_{X_i,X_j} a_{ij}^0 = H_{ij}^0,
$$
where $a_{ij}^0 = \lambda_i^\top \Sigma_{\Lambda}^{-1} \lambda_j$, the full matrix $H$ converges in probability to $H^0 = [H_{ij}^0]$. Hence, by continuous mapping,
$$
G = \left(I - \frac{1}{N}H\right)^{-1} \xrightarrow{p} \left(I - \frac{1}{N}H^0\right)^{-1} = G^0,
$$
and similarly $G_{ij} \xrightarrow{p} G_{ij}^0$, where $G_{ij}^0$ is the $(i,j)$-th block of $G^0$.
We now have the following decomposition:
$$
\sum_{j=1}^N G_{ij}A_jU_j = \sum_{j=1}^N G_{ij}^0 A_j^0 U_j + \sum_{j=1}^N \left(G_{ij}A_j - G_{ij}^0 A_j^0\right)U_j.
$$
Let
$$
R_i = \sum_{j=1}^N \left(G_{ij}A_j - G_{ij}^0 A_j^0\right)U_j.
$$
We show that $R_i = o_p(1)$. Consider its conditional variance given $X, F^0$:
\begin{equation}\nonumber
\begin{aligned}
\mathrm{Var}(R_i \mid X, F^0) &= \sum_{j=1}^N \sum_{k=1}^N \left(G_{ij}A_j - G_{ij}^0 A_j^0\right) \mathrm{Cov}(U_j, U_k \mid X, F^0) \left(G_{ik}A_k - G_{ik}^0 A_k^0\right)^\top.
\end{aligned}
\end{equation}
Taking the norm and applying the triangle inequality, we have
\begin{equation}\nonumber
\begin{aligned}
\left\| \mathrm{Var}(R_i \mid X, F^0) \right\| &\leq \sum_{j=1}^N \sum_{k=1}^N \left\| G_{ij}A_j - G_{ij}^0 A_j^0 \right\| \cdot \left| \mathrm{Cov}(U_j, U_k \mid X, F^0) \right| \cdot \left\| G_{ik}A_k - G_{ik}^0 A_k^0 \right\| \\
&\leq \left( \max_{1 \le k \le N} \left\| G_{ik}A_k - G_{ik}^0 A_k^0 \right\| \right) \sum_{j=1}^N \left\| G_{ij}A_j - G_{ij}^0 A_j^0 \right\| \left( \sum_{k=1}^N \left| \mathrm{Cov}(U_j, U_k \mid X, F^0) \right| \right).
\end{aligned}
\end{equation}
By Assumption \ref{assum:Asymptotic Normality}, $\sum_{k=1}^N \left| \mathrm{Cov}(U_j, U_k \mid X, F^0) \right| \leq M$. By Assumption \ref{assum:uniform convergence}, we have both
$$
\max_{1 \le j \le N} \left\| G_{ij}A_j - G_{ij}^0 A_j^0 \right\| \xrightarrow{p} 0
$$
and
$$
\sum_{j=1}^N \left\| G_{ij}A_j - G_{ij}^0 A_j^0 \right\| = O_p(1).
$$
Therefore,
$$
\left\| \mathrm{Var}(R_i \mid X, F^0) \right\| \leq M \cdot o_p(1) \cdot O_p(1) = o_p(1).
$$
Since $E[R_i \mid X, F^0] = 0$, Chebyshev's inequality implies $R_i = o_p(1)$. Thus,
$$
\sqrt{T}(\widehat{\beta}_i - \beta_i^0) = \sum_{j=1}^N G_{ij}^0 A_j^0 U_j + o_p(1) = S_i + o_p(1).
$$
Let $W_{ij}^0 = G_{ij}^0 A_j^0$. We analyze the asymptotic distribution of $S_i$ conditional on $X, F^0$. Given $X, F^0$, the weights $W_{ij}^0$ are fixed. Note that $E[U_j \mid X, F^0] = 0$, so
$$
E[S_i \mid X, F^0] = 0.
$$
The conditional variance is
$$
\Omega_i \equiv \mathrm{Var}(S_i \mid X, F^0) = \sum_{j=1}^N \sum_{k=1}^N W_{ij}^0 \, \Sigma_{jk} \, (W_{ik}^0)^\top,
$$
where $\Sigma_{jk} = \lim_{T\to\infty}\mathrm{Cov}(U_j, U_k \mid X, F^0)$.
Under the joint asymptotic normality of $U = (U_1^\top, \dots, U_N^\top)^\top$ (strengthened form of Assumption \ref{assum:Asymptotic Normality}), and given the fixed weights $W_{ij}^0$, the weighted sum $S_i$ converges in distribution conditional on $X, F^0$ to a normal random vector. Since the limiting distribution does not depend on $X, F^0$, the unconditional convergence also holds:
$$
S_i \xrightarrow{d} \mathcal{N}(0, \Omega_i).
$$
Therefore, by Slutsky's theorem:
$$
\sqrt{T}\left(\widehat{\beta}_{i}-\beta_i^0\right) \xrightarrow{d} \mathcal{N}(0, \Omega_i), \quad \text{as } N, T \to \infty,
$$
where
$$
\Omega_i = \sum_{j=1}^N \sum_{k=1}^N (G_{ij}^0 A_j^0) \, \Sigma_{jk} \, (G_{ik}^0 A_k^0)^\top.
$$
This completes the proof.  
\end{proof}

\section{Proof of Proposition \ref{pro:quantile}}\label{secE}
\begin{proof}
For each cross-sectional unit $i = 1,\dots,N$, define the standardized parameter $\gamma = \Sigma_i^{1/2} (\alpha - \alpha_{i}^0) \in \mathbb{R}^{p+1}$. Let $\widehat{\gamma}_i = \Sigma_i^{1/2} (\widehat{\alpha}_{i} - \alpha_{i}^0)$. Define the reparameterized loss function
$$
\widehat{D}_i(\gamma) = \widehat{Q}_i(\alpha_{i}^0 + \Sigma_i^{-1/2} \gamma) - \widehat{Q}_i(\alpha_{i}^0), \quad 
D_i(\gamma) = E[\widehat{D}_i(\gamma)],
$$
where $\widehat{Q}_i(\alpha) = \frac{1}{T} \sum_{t=1}^T \rho_\tau(Y_{it} - X_{it}^\top \alpha)$ is the empirical quantile loss. Compute the Hessian matrix of the population loss function $Q_i(\alpha) = E[\widehat{Q}_i(\alpha)]$. The gradient is
$$
\nabla Q_i(\alpha) = E\left[ (\tau - \mathbbm{1}(Y_{it} \leq X_{it}^\top \alpha)) X_{it} \right].
$$
At the true parameter, by definition of the quantile, we have $\nabla Q_i(\alpha_{i}^0) = 0$.Then, the Hessian matrix is
$$
\nabla^2 Q_i(\alpha) = E\left[ f_{\epsilon|X}(X_{it}^\top(\alpha - \alpha_{i}^0) \mid X_{it}) X_{it} X_{it}^\top \right],
$$
where $f_{\epsilon|X}(\cdot \mid X_{it})$ is the conditional density of $\epsilon_{it} = Y_{it} - X_{it}^\top \alpha_{i}^0$. 

For any $\gamma \in \mathbb{R}^{p+1}$, consider the second-order expansion, that is
\begin{equation}\nonumber
\begin{aligned}
D_i(\gamma) &= \int_0^1 \langle \nabla Q_i(\alpha_{i}^0 + s\Sigma_i^{-1/2} \gamma) - \nabla Q_i(\alpha_{i}^0), \Sigma_i^{-1/2} \gamma \rangle ds\\
&=\int_0^1 \int_0^s \langle \gamma, \Sigma_i^{-1/2} \nabla^2 Q_i(\alpha_{i}^0 + u\Sigma_i^{-1/2} \gamma) \Sigma_i^{-1/2} \gamma \rangle du ds.
\end{aligned}
\end{equation}
Define the standardized design vector $W_{it} = \Sigma_i^{-1/2} X_{it}$. Then,
$$
D_i(\gamma) = \int_0^1 \int_0^s E\left[ f_{\epsilon|X}(u W_{it}^\top \gamma \mid X_{it}) (W_{it}^\top \gamma)^2 \right] du ds.
$$
By Assumption \ref{assum:Regularity}, $f_{\epsilon|X}(u W_{it}^\top \gamma \mid X_{it}) \geq \underline{f} - L_0 |u W_{it}^\top \gamma|$, therefore,
\begin{equation}\label{eq:E.1}
\begin{aligned}
D_i(\gamma) &\geq \int_0^1 \int_0^s \left[ \underline{f} \cdot E[(W_{it}^\top \gamma)^2] - L_0 u \cdot E[|W_{it}^\top \gamma|^3] \right] du ds \\
&= \frac{1}{2} \underline{f} \|\gamma\|_2^2 - \frac{1}{6} L_0 \cdot E[|W_{it}^\top \gamma|^3].
\end{aligned}
\end{equation}
By H\"older's inequality:
$$
E[|W_{it}^\top \gamma|^3] \leq \left( E[(W_{it}^\top \gamma)^4] \right)^{3/4}.
$$
Define $\kappa_4 = \sup_{\|u\|_2=1} E[(u^\top W_{it})^4]$. By the sub-Gaussianity in Assumption \ref{assum:Sub-Gaussianity}, there exists a constant $C > 0$ such that $\kappa_4 \leq C \upsilon_1^4$. Hence,
$$
E[|W_{it}^\top \gamma|^3] \leq \kappa_4^{3/4} \|\gamma\|_2^3 \leq C \upsilon_1^3 \|\gamma\|_2^3.
$$
Substituting into \eqref{eq:E.1} yields the population curvature lower bound
$$
D_i(\gamma) \geq \frac{1}{2} \underline{f} \|\gamma\|_2^2 - \frac{1}{6} C L_0 \upsilon_1^3 \|\gamma\|_2^3.
$$
The core is to establish a uniform bound for the empirical process $\widehat{D}_i(\gamma) - D_i(\gamma)$ over the ball $\mathbb{B}(r) = \{\gamma \in \mathbb{R}^{p+1}: \|\gamma\|_2 \leq r\}$.

Set $r_0 = 4C_0 \upsilon_1 \underline{f}^{-1} \sqrt{(p + t)/T}$. Assume the time dimension $T$ is sufficiently large such that,
$$
\underline{f}^2 > \frac{8}{3} C_0 C L_0 \upsilon_1^4 \sqrt{\frac{p + t}{T}},
$$
which is equivalent to
$$
T > \left( \frac{8 C_0 C L_0 \upsilon_1^4}{3 \underline{f}^2} \right)^2 (p + t).
$$
For any $\gamma$ satisfying $\|\gamma\|_2 = r_0$, from the Lemma \ref{lemma:E.1}, we have
\begin{equation}\nonumber
\begin{aligned}
\widehat{D}_i(\gamma) &\geq D_i(\gamma) - |\widehat{D}_i(\gamma) - D_i(\gamma)| \\
&\geq \frac{1}{2} \underline{f} r_0^2 - \frac{1}{6} C L_0 \upsilon_1^3 r_0^3 - C_0 \upsilon_1 r_0 \sqrt{\frac{p + t}{T}}.
\end{aligned}
\end{equation}
Substituting the expression for $r_0$,
\begin{equation}\nonumber
\begin{aligned}
\widehat{D}_i(\gamma) &\geq 8C_0^2 \upsilon_1^2 \underline{f}^{-1} \frac{p + t}{T} - \frac{64}{6} C_0^3 C L_0 \upsilon_1^4 \underline{f}^{-3} \left( \frac{p + t}{T} \right)^{3/2} - 4C_0^2 \upsilon_1^2 \underline{f}^{-1} \frac{p + t}{T} \\
&= 4C_0^2 \upsilon_1^2 \underline{f}^{-1} \frac{p + t}{T} \left[ 1 - \frac{8}{3} C_0 C L_0 \upsilon_1^2 \underline{f}^{-2} \sqrt{\frac{p + t}{T}} \right] \\
&> 0.
\end{aligned}
\end{equation}
Let $\widehat{\gamma}_i = \Sigma_i^{1/2} (\widehat{\alpha}_{i} - \alpha_{i}^0)$. By the definition of $\widehat{\alpha}_{i}$, we have $\widehat{D}_i(\widehat{\gamma}_i) \leq 0$. Note that $\widehat{D}_i(\cdot)$ is convex, as it is an affine transformation of the convex function $\widehat{Q}_i$. For a convex function, if the function values are positive on a sphere, then the minimum must lie inside the ball (\cite{wainwright2019high}, Lemma 9.21). Specifically, since $\widehat{D}_i(\gamma) > 0$ for all $\gamma$ with $\|\gamma\|_2 = r_0$ and $\widehat{D}_i(\widehat{\gamma}_i) \leq 0$, it must be that, $\|\widehat{\gamma}_i\|_2 \leq r_0$. Then,
$$
\|\widehat{\alpha}_{i} - \alpha_{i}^0\|_{\Sigma_i} = \|\widehat{\gamma}_i\|_2 \leq 4C_0 \upsilon_1 \underline{f}^{-1} \sqrt{\frac{p + t}{T}}.
$$
Take $t = \log(1/\delta)$. Then the above bound holds with probability at least $1 - \delta$. Define constants,
$$
C_1 = 4C_0 \upsilon_1, \quad C_2 = \left( \frac{8 C_0 C L_0 \upsilon_1^4}{3 \underline{f}^2} \right)^2,
$$
which yields the conclusion of the proposition, that is,
$$
\|\widehat{\alpha}_{i} - \alpha_{i}^0\|_{\Sigma_i} \leq C_1 \underline{f}^{-1} \sqrt{\frac{p + \log(1/\delta)}{T}},
$$
requiring $T \geq C_2 L_0^2 \underline{f}^{-4} (p + \log(1/\delta))$.
\end{proof}
\begin{lemma}\label{lemma:E.1}
Under Assumption \ref{assum:Sub-Gaussianity}, for any $r > 0$ and $t \geq 0$, there exists a constant $C_0 > 0$ such that,
$$
\sup_{\gamma \in \mathbb{B}(r)} |\widehat{D}_i(\gamma) - D_i(\gamma)| \leq C_0 \upsilon_1 r \sqrt{\frac{p + t}{T}}
$$
with probability at least $1 - e^{-t}$.
\end{lemma}
\begin{proof}
Define the function class
$$
\mathcal{F} = \left\{ f_{\gamma}(x,y) = \rho_\tau(y - x^\top(\alpha_{i}^0 + \Sigma_i^{-1/2} \gamma)) - \rho_\tau(y - x^\top \alpha_{i}^0) : \|\gamma\|_2 \leq r \right\}.
$$
For fixed $(x,y)$, the function $\gamma \mapsto f_\gamma(x,y)$ is Lipschitz continuous with respect to $x^\top \Sigma_i^{-1/2} \gamma$, with Lipschitz constant $\bar{\alpha} = \max(\tau, 1-\tau) \leq 1$. This is because the derivative of the check function $\rho_\tau$ (almost everywhere) satisfies $|\rho_\tau^\prime(u)| \leq \bar{\alpha}$.

Let $\varepsilon_1,\dots,\varepsilon_T$ be independent Rademacher random variables ($P(\varepsilon_t = \pm 1) = 1/2$), independent of the data $\{Z_{it} = (X_{it}, Y_{it})\}$. Define
$$
\Delta(r) = \sqrt{T} \sup_{\gamma \in \mathbb{B}(r)} \frac{D_i(\gamma) - \widehat{D}_i(\gamma)}{4 \upsilon_2 \bar{\alpha} r},
$$
where $\upsilon_2 = \sqrt{2} \upsilon_1$. By the symmetrization inequality, we have
$$
E e^{\lambda \Delta(r)} \leq E \exp\left\{ 2\lambda \sup_{\gamma \in \mathbb{B}(r)} \frac{1}{4 \upsilon_2 \bar{\alpha} r \sqrt{T}} \sum_{t=1}^T \varepsilon_t \cdot r(\gamma; Z_{it}) \right\},
$$
where $r(\gamma; Z_{it}) = f_\gamma(X_{it}, Y_{it})$. By the Ledoux-Talagrand contraction inequality, since $r(\cdot; Z_{it})$ is $\bar{\alpha}$-Lipschitz with respect to $X_{it}^\top \Sigma_i^{-1/2} \gamma$, we have
$$
E e^{\lambda \Delta(r)} \leq E \exp\left\{ \frac{\lambda}{2 \upsilon_2 r} \sup_{\gamma \in \mathbb{B}(r)} \frac{1}{\sqrt{T}} \sum_{t=1}^T \varepsilon_t (X_{it}^\top \Sigma_i^{-1/2} \gamma) \right\}.
$$
Denote $W_{it} = \Sigma_i^{-1/2} X_{it}$. Then,
$$
E e^{\lambda \Delta(r)} \leq E \exp\left\{ \frac{\lambda}{2 \upsilon_2} \left\| \frac{1}{\sqrt{T}} \sum_{t=1}^T \varepsilon_t W_{it} \right\|_2 \right\}.
$$

Let $\mathcal{N}$ be a $(1/2)$-covering of the unit sphere $\mathbb{S}^p = \{u \in \mathbb{R}^{p+1}: \|u\|_2 = 1\}$. By the covering lemma, there exists such a covering with $|\mathcal{N}| \leq 5^{p+1}$. For any vector $v \in \mathbb{R}^{p+1}$, $\|v\|_2 \leq 2 \max_{u \in \mathcal{N}} |u^\top v|$. Applying this bound gives,
$$
E e^{\lambda \Delta(r)} \leq E \exp\left\{ \frac{\lambda}{\upsilon_2} \max_{u \in \mathcal{N}} \left| \frac{1}{\sqrt{T}} \sum_{t=1}^T \varepsilon_t (W_{it}^\top u) \right| \right\}.
$$
By symmetry and Jensen's inequality,
$$
E e^{\lambda \Delta(r)} \leq 2 E \exp\left\{ \frac{\lambda}{\upsilon_2} \max_{u \in \mathcal{N}} \frac{1}{\sqrt{T}} \sum_{t=1}^T \varepsilon_t (W_{it}^\top u) \right\}.
$$

Fix $u \in \mathcal{N}$ and let $\xi_t = W_{it}^\top u / \upsilon_2$. By Assumption 4.3, $\xi_t$ is $1/\sqrt{2}$-sub-Gaussian, i.e., $\|\xi_t\|_{\psi_2} \leq 1/\sqrt{2}$. Hence its moment generating function satisfies
$$
E e^{\lambda \xi_t} \leq e^{\lambda^2/2}, \quad \forall \lambda \in \mathbb{R}.
$$
Since $\varepsilon_t$ is a Rademacher variable independent of $\xi_t$:
$$
E e^{\lambda \varepsilon_t \xi_t} = E \cosh(\lambda \xi_t) \leq E e^{\lambda^2 \xi_t^2/2} \leq e^{\lambda^2/2},
$$
where the last inequality follows from $E \xi_t^2 \leq 1$ and the sub-Gaussian property. For fixed $u \in \mathcal{N}$,
$$
E \exp\left\{ \frac{\lambda}{\sqrt{T}} \sum_{t=1}^T \varepsilon_t \xi_t \right\} = \prod_{t=1}^T E \exp\left\{ \frac{\lambda}{\sqrt{T}} \varepsilon_t \xi_t \right\} \leq \left( e^{\lambda^2/(2T)} \right)^T = e^{\lambda^2/2}.
$$
Now take the expectation of the maximum. By Jensen's inequality and the union bound,
\begin{equation}\nonumber
\begin{aligned}
E \exp\left\{ \frac{\lambda}{\upsilon_2} \max_{u \in \mathcal{N}} \frac{1}{\sqrt{T}} \sum_{t=1}^T \varepsilon_t (W_{it}^\top u) \right\}
&\leq \sum_{u \in \mathcal{N}} E \exp\left\{ \frac{\lambda}{\sqrt{T}} \sum_{t=1}^T \varepsilon_t \xi_t \right\} \\
&\leq |\mathcal{N}| \cdot e^{\lambda^2/2} \\
&\leq 5^{p+1} e^{\lambda^2/2}.
\end{aligned}
\end{equation}
Therefore, $E e^{\lambda \Delta(r)} \leq 2 \cdot 5^{p+1} e^{\lambda^2/2}.$

By Chernoff's bound, for any $s > 0$:
$$
P\{\Delta(r) \geq s\} \leq \exp\left[ -\sup_{\lambda \geq 0} \left\{ \lambda s - \log E e^{\lambda \Delta(r)} \right\} \right].
$$
Then, $P\{\Delta(r) \geq s\} \leq \exp\left[ -\sup_{\lambda \geq 0} \left\{ \lambda s - \frac{\lambda^2}{2} \right\} + (p+1)\log 5 + \log 2 \right]$. Optimizing the quadratic function: $\sup_{\lambda \geq 0} (\lambda s - \lambda^2/2) = s^2/2$ (achieved at $\lambda = s$). Hence, $P\{\Delta(r) \geq s\} \leq 2 \cdot 5^{p+1} e^{-s^2/2}$. Set $s^2 = 2[(p+1)\log 5 + t + \log 2]$. Then, $P\{\Delta(r) \geq s\} \leq 2 \cdot 5^{p+1} e^{-s^2/2}$.
Returning to the definition of $\Delta(r)$,
$$
\sup_{\gamma \in \mathbb{B}(r)} |\widehat{D}_i(\gamma) - D_i(\gamma)| \leq \frac{4 \upsilon_2 \bar{\alpha} r s}{\sqrt{T}}.
$$
Since $\upsilon_2 = \sqrt{2} \upsilon_1$, $\bar{\alpha} \leq 1$, and $s \asymp \sqrt{p + t}$, there exists a constant $C_0 > 0$ such that,
$$
\sup_{\gamma \in \mathbb{B}(r)} |\widehat{D}_i(\gamma) - D_i(\gamma)| \leq C_0 \upsilon_1 r \sqrt{\frac{p + t}{T}}.
$$
Taking $t = \log(1/\delta)$ gives the lemma's conclusion.
\end{proof}

\section{Proof of Theorem \ref{theo:nonasymES}}\label{secF}
\begin{proof}
By the construction of the two-step estimator, for each cross-sectional unit $i$,
$$
\widehat{\beta}_{i} = \left( X_i^\top M_{\widehat{F}} X_i \right)^{-1} X_i^\top M_{\widehat{F}} Z_i^\ast(\widehat{\alpha}_{i}).
$$
Expand $Z_i^\ast(\widehat{\alpha}_{i})$ around the true parameter,
\begin{equation}\nonumber
Z_i^\ast(\widehat{\alpha}_{i})= Z_i^\ast(\alpha_{i}^0) + \Delta_i^\alpha= X_i \beta_{i}^0 + F^0 \lambda_{i}^0 + \varepsilon_i + \Delta_i^\alpha,
\end{equation}
where the second equality uses the model specification and the definition of $\varepsilon_i$, and $\Delta_i^\alpha = Z_i^\ast(\widehat{\alpha}_{i}) - Z_i^\ast(\alpha_{i}^0) \in \mathbb{R}^T$. Substituting yields,
\begin{equation}\label{eq:F.1}
\widehat{\beta}_{i} - \beta_{i}^0 = \left( X_i^\top M_{\widehat{F}} X_i \right)^{-1} X_i^\top M_{\widehat{F}} \left[ \Delta_i^\alpha + F^0 \lambda_{i}^0 + \varepsilon_i \right].
\end{equation}

We need to control $\frac{1}{T} X_i^\top M_{\widehat{F}} F^0 \lambda_{i}^0$. By the convergence of factor estimation, there exists an invertible rotation matrix $H$ such that $\widehat{F}$ is close to $F^0 H$. Define $\Delta_F = \widehat{F} - F^0 H$, then $F^0 = (\widehat{F} - \Delta_F) H^{-1}$. Since $M_{\widehat{F}} \widehat{F} = 0$, we have $M_{\widehat{F}} F^0 = - M_{\widehat{F}} \Delta_F H^{-1}$. Therefore,
$$
\frac{1}{T} X_i^\top M_{\widehat{F}} F^0 \lambda_i^0 = -\frac{1}{T} X_i^\top M_{\widehat{F}} \Delta_F v,
$$
where $v = H^{-1} \lambda_i^0 \in \mathbb{R}^r$. By Assumption \ref{assum:factor&loading}, $\|\lambda_i^0\| \leq M_\lambda$, and $\|H^{-1}\|_{2}$ is bounded, so there exists a constant $C_v$ such that $\|v\| \leq C_v$. By the concentration inequality for factor estimation, there exists a constant $C_F > 0$ such that with probability at least $1 - \delta$,
$$
\frac{1}{T} \|\Delta_F\|_F^2 \leq C_F \left( \frac{1}{N} + \frac{1}{T} \right) \log(1/\delta).
$$
By Assumption \ref{assum:Sub-Gaussianity} and the spectral concentration inequality for sub-Gaussian matrices, there exists a constant $C_X > 0$ such that with probability at least $1 - \delta$,
$$
\frac{1}{\sqrt{T}} \|X_i\|_{2} \leq C_X \left( \sqrt{p} + \sqrt{\log(1/\delta)} \right).
$$
Since $M_{\widehat{F}} = I_T - P_{\widehat{F}}$ with $P_{\widehat{F}} = \widehat{F}\widehat{F}^\top / T$, we have
$$
\frac{1}{T} X_i^\top M_{\widehat{F}} \Delta_F v = \frac{1}{T} X_i^\top \Delta_F v - \frac{1}{T} X_i^\top P_{\widehat{F}} \Delta_F v.
$$
We control these two terms separately. By the Cauchy-Schwarz inequality and properties of the spectral norm,
$$
\left\| \frac{1}{T} X_i^\top \Delta_F v \right\|_2 \leq \frac{1}{T} \|X_i\|_{2} \|\Delta_F v\|_2 \leq \left( \frac{1}{\sqrt{T}} \|X_i\|_{2} \right) \cdot \left( \frac{1}{\sqrt{T}} \|\Delta_F\|_F \right) \cdot \|v\|_2.
$$
Thus, with probability at least $1 - \delta$, we have
\begin{equation}\nonumber
\begin{aligned}
\left\| \frac{1}{T} X_i^\top \Delta_F v \right\|_2 &\leq C_X \left( \sqrt{p} + \sqrt{\log(1/\delta)} \right) \cdot \sqrt{C_F \left( \frac{1}{N} + \frac{1}{T} \right) \log(1/\delta)} \cdot C_v\\
&\leq C \left( \frac{1}{\sqrt{N}} + \frac{1}{\sqrt{T}} \right) \cdot (p + \log(1/\delta)).
\end{aligned}
\end{equation}
Using $P_{\widehat{F}} = \widehat{F}\widehat{F}^\top / T$, we obtain
$$
\frac{1}{T} X_i^\top P_{\widehat{F}} \Delta_F v = \frac{1}{T^2} (X_i^\top \widehat{F})(\widehat{F}^\top \Delta_F v).
$$
Hence,
$$
\left\| \frac{1}{T} X_i^\top P_{\widehat{F}} \Delta_F v \right\|_2 \leq \frac{1}{T^2} \|X_i^\top \widehat{F}\|_{2} \|\widehat{F}^\top \Delta_F v\|_2.
$$
Since $\widehat{F}^\top \widehat{F} = T I_r$, we have $\|\widehat{F}\|_{2} = \sqrt{rT}$, and
$$
\|\widehat{F}^\top \Delta_F v\|_2 \leq \|\widehat{F}\|_{2} \|\Delta_F v\|_2 \leq \sqrt{rT} \cdot \|\Delta_F\|_F \|v\|_2,
$$
$$
\|X_i^\top \widehat{F}\|_{2} \leq \|X_i\|_{2} \|\widehat{F}\|_{2} = \|X_i\|_{2} \sqrt{rT}.
$$
Substituting these bounds gives
$$
\left\| \frac{1}{T} X_i^\top P_{\widehat{F}} \Delta_F v \right\|_2 \leq \frac{1}{T^2} \cdot (\|X_i\|_{2} \sqrt{rT}) \cdot (\sqrt{rT} \cdot \|\Delta_F\|_F \|v\|_2) = \frac{r}{T} \|X_i\|_{2} \|\Delta_F\|_F \|v\|_2.
$$
Thus,
$$
\left\| \frac{1}{T} X_i^\top P_{\widehat{F}} \Delta_F v \right\|_2 \leq r \left( \frac{1}{\sqrt{T}} \|X_i\|_{2} \right) \cdot \left( \frac{1}{\sqrt{T}} \|\Delta_F\|_F \right) \cdot \|v\|_2.
$$
This has the same form as the first term, with an extra factor $r$. Therefore, by the triangle inequality, with probability at least $1 - \delta$,
$$
\left\| \frac{1}{T} X_i^\top M_{\widehat{F}} F^0 \lambda_i^0 \right\|_2 \leq C^\prime \left( \frac{1}{\sqrt{T}} + \frac{1}{\sqrt{N}} \right) \cdot (p + \log(1/\delta)).
$$
Since $\|\cdot\|_{\Sigma_i^{-1}} = \|\Sigma_i^{-1/2} \cdot\|_2$ and, by Assumption 3.10, $\|\Sigma_i^{-1/2}\|_{2} \leq 1/\sqrt{\kappa_{\min}}$, we have
$$
\left\| \frac{1}{T} X_i^\top M_{\widehat{F}} F^0 \lambda_i^0 \right\|_{\Sigma_i^{-1}} \leq \frac{1}{\sqrt{\kappa_{\min}}} \left\| \frac{1}{T} X_i^\top M_{\widehat{F}} F^0 \lambda_i^0 \right\|_2.
$$
Then,
$$
\left\| \frac{1}{T} X_i^\top M_{\widehat{F}} F^0 \lambda_i^0 \right\|_{\Sigma_i^{-1}} \leq \frac{C^\prime}{\sqrt{\kappa_{\min}}} \left( \frac{1}{\sqrt{T}} + \frac{1}{\sqrt{N}} \right) \cdot (p + \log(1/\delta)).
$$
Set $C_3^{\prime} = C^{\prime} / \sqrt{\kappa_{\min}}$. The bound holds with probability at least $1 - \delta$,
\begin{equation}\label{eq:F.2}
\left\| \frac{1}{T} X_i^\top M_{\widehat{F}} F^0 \lambda_i^0 \right\|_{\Sigma_i^{-1}} \leq C_3^{\prime} \left( \frac{1}{\sqrt{T}} + \frac{1}{\sqrt{N}} \right) \cdot (p + \log(1/\delta)).
\end{equation}

Next, we need to control $\frac{1}{T} X_i^\top M_{\widehat{F}} \varepsilon_i$. By the projection property of $M_{\widehat{F}}$, $\|M_{\widehat{F}}\|_{2} = 1$, so,
$$
\left\| \frac{1}{T} X_i^\top M_{\widehat{F}} \varepsilon_i \right\|_{\Sigma_i^{-1}} \leq \left\| \frac{1}{T} W_i^\top \varepsilon_i \right\|_2.
$$
Fix $u \in \mathbb{S}^{p}$, and define $Z_u = \frac{1}{T} \sum_{t=1}^T (u^\top W_{it}) \varepsilon_{it}$. 
Conditional on $X_i$, $\{\varepsilon_{it}\}$ are zero-mean, and by Assumption \ref{assum:error}, $E[\varepsilon_{it}^2 \mid X_{it}] \leq M_\varepsilon$. From the definition of $\varepsilon_{it}$ and the model specification,
$$
\varepsilon_{it} = Z_{it}^\ast(\alpha_{i}^0) - X_{it}^\top\beta_{i}^0 - \lambda_{i}^{0\top} f_{t}^0.
$$
Recall that $Z_{it}^\ast(\alpha) = \frac{1}{\tau} Z_{it}(\alpha)$ where $Z_{it}(\alpha) = (Y_{it} - X_{it}^\top\alpha)\mathbbm{1}(Y_{it} \leq X_{it}^\top\alpha) + \tau X_{it}^\top\alpha$.
Define $\epsilon_{it,-} = (Y_{it} - X_{it}^\top\alpha_{i}^0)\mathbbm{1}(Y_{it} \leq X_{it}^\top\alpha_{i}^0)$, then
$$
Z_{it}^\ast(\alpha_{i}^0) = \frac{1}{\tau} \epsilon_{it,-} + X_{it}^\top\alpha_{i}^0.
$$
Hence,
$$
\varepsilon_{it} = \frac{1}{\tau} \epsilon_{it,-} + X_{it}^\top\alpha_{i}^0 - X_{it}^\top\beta_{i}^0 - \lambda_{i}^{0\top} f_{t}^0
= \frac{1}{\tau} \epsilon_{it,-} + X_{it}^\top(\alpha_{i}^0 - \beta_{i}^0) - \lambda_{i}^{0\top} f_{t}^0.
$$
Under the true parameters, $E[\epsilon_{it,-} \mid X_{it}] = \tau \lambda_{i}^{0\top} f_{t}^0 + \tau X_{it}^\top(\beta_{i}^0 - \alpha_{i}^0)$, 
hence $E[\varepsilon_{it} \mid X_{it}] = 0$. 
Furthermore, by Assumption \ref{assum:TailBehavior} and the Cauchy-Schwarz inequality, we have
\begin{equation}
\begin{aligned}
E[\varepsilon_{it}^2 \mid X_{it}] &\leq 3 \left( \frac{1}{\tau^2} E[\epsilon_{it,-}^2 \mid X_{it}] 
+ \|X_{it}\|^2 \|\alpha_{i}^0 - \beta_{i}^0\|^2 
+ \|\lambda_{i}^0\|^2 \|f_{t}^0\|^2 \right).
\end{aligned}
\end{equation}

Since $E[\epsilon_{it,-}^2 \mid X_{it}] \leq \overline{\sigma}^2 + (E[\epsilon_{it,-} \mid X_{it}])^2$, and the latter two terms are bounded 
(by Assumptions \ref{assum:UniformBound}), there exists $C_\tau$ such that,
$$
E[\varepsilon_{it}^2 \mid X_{it}] \leq C_\tau \overline{\sigma}^2.
$$
Now $Z_u$ is a sum of weakly dependent zero-mean random variables, with conditional variance,
$$
\mathrm{Var}(Z_u \mid X_i) \leq \frac{C_\tau \overline{\sigma}^2}{T^2} \sum_{t=1}^T (u^\top W_{it})^2 
= \frac{C_\tau \overline{\sigma}^2}{T} \cdot \frac{1}{T} \|W_i u\|_2^2.
$$
By sub-Gaussianity (Assumption \ref{assum:Sub-Gaussianity}), $\frac{1}{T} \|W_i u\|_2^2 \leq C\upsilon_1^2$ holds with high probability. 
Applying Hoeffding's inequality (for sub-Gaussian multipliers),
$$
P\left( |Z_u| \geq C_\varepsilon \overline{\sigma} \sqrt{\frac{\log(1/\delta)}{T}} \mid X_i \right) \leq \delta,
$$
where $C_\varepsilon = C_\varepsilon(\upsilon_1, \tau)$. 
Taking a union bound over an $\epsilon$-net of $\mathbb{S}^{p}$ and adjusting $\delta$ to account for the covering number, 
we obtain: there exists $C_1^\prime$ such that with probability at least $1 - \delta$,
\begin{equation}\label{eq:F.3}
\left\| \frac{1}{T} X_i^\top M_{\widehat{F}} \varepsilon_i \right\|_{\Sigma_i^{-1}} \leq C_1^\prime \overline{\sigma} \sqrt{\frac{p + \log(1/\delta)}{T}}.
\end{equation}

Decompose $\Delta_i^\alpha = D_i + S_i$, where $D_i = E[\Delta_i^\alpha \mid X_i]$ and $S_i = \Delta_i^\alpha - D_i$. By Neyman orthogonality and Assumption \ref{assum:Smoothness}, for each $t$,
\begin{equation}\nonumber
\begin{aligned}
E[Z_{it}^\ast(\alpha_{i}^0 + \delta_i) - Z_{it}^\ast(\alpha_{i}^0) \mid X_{it}] &= \frac{1}{\tau} \int_0^{\Delta_{it}} s \left[ f_{Y|X}(X_{it}^\top \alpha_{i}^0 + s \mid X_{it}) - \tau \right] ds \\
&= \frac{1}{\tau} \int_0^{\Delta_{it}} s \left[ F_{\epsilon|X}(s \mid X_{it}) - F_{\epsilon|X}(0 \mid X_{it}) \right] ds.
\end{aligned}
\end{equation}
By Assumption \ref{assum:Smoothness}, $|F_{\epsilon|X}(s \mid X_{it}) - F_{\epsilon|X}(0 \mid X_{it})| \leq \overline{f} |s|$, hence
$$
\left| E[Z_{it}^\ast(\alpha_{i}^0 + \delta_i) - Z_{it}^\ast(\alpha_{i}^0) \mid X_{it}] \right| \leq \frac{\overline{f}}{\tau} \int_0^{|\Delta_{it}|} s^2 ds = \frac{\overline{f}}{3\tau} |\Delta_{it}|^3.
$$
Since $|\Delta_{it}| = |X_{it}^\top \delta_i| \leq \|X_{it}\|_{\Sigma_i^{-1}} \|\delta_i\|_{\Sigma_i}$, and $\|X_{it}\|_{\Sigma_i^{-1}}^2 = X_{it}^\top \Sigma_i^{-1} X_{it} \leq \kappa_{\min}^{-1} \|X_{it}\|^2$, combined with the bounded moment property of $\|X_{it}\|$, we obtain, $\|D_i\| \leq C_D(\overline{f}, \tau) \|\delta_i\|_{\Sigma_i}^2.$.
Therefore,
$$
\left\| \frac{1}{T} X_i^\top M_{\widehat{F}} D_i \right\|_{\Sigma_i^{-1}} \leq \frac{1}{T} \|W_i\|_{2} \|D_i\| \leq C_D^\prime \sqrt{\frac{p + \log(1/\delta)}{T}} \|\delta_i\|_{\Sigma_i}^2.
$$
Define the stochastic process,
$$
\mathcal{R}(\delta) = \frac{1}{T} \sum_{t=1}^T \xi_{it}(\delta) W_{it}, \quad \delta \in \mathbb{R}^{p+1},
$$
where $\xi_{it}(\delta) = Z_{it}^\ast(\alpha_{i}^0 + \delta) - Z_{it}^\ast(\alpha_{i}^0) - E[\cdots \mid X_{it}]$. By construction of $Z_{it}^\ast(\cdot)$, $\xi_{it}(\delta)$ is Lipschitz in $\delta$ with Lipschitz constant satisfying
$$
|\xi_{it}(\delta_1) - \xi_{it}(\delta_2)| \leq \left(1 + \frac{1}{\tau}\right) |X_{it}^\top (\delta_1 - \delta_2)|.
$$
Perform the change of variable $\tilde{\delta} = \Sigma_i^{1/2} \delta$, so that $X_{it}^\top \delta = W_{it}^\top \tilde{\delta}$. Define the new process,
$$
\widetilde{\mathcal{R}}(\tilde{\delta}) = \frac{1}{T} \sum_{t=1}^T \xi_{it}(\Sigma_i^{-1/2} \tilde{\delta}) W_{it}.
$$
For $\tilde{\delta}_1, \tilde{\delta}_2 \in \mathbb{R}^{p+1}$,
$$
|\xi_{it}(\Sigma_i^{-1/2} \tilde{\delta}_1) - \xi_{it}(\Sigma_i^{-1/2} \tilde{\delta}_2)| \leq \left(1 + \frac{1}{\tau}\right) |W_{it}^\top (\tilde{\delta}_1 - \tilde{\delta}_2)|.
$$
Consider the ball $\mathbb{B}_2(r) = \{\tilde{\delta}: \|\tilde{\delta}\|_2 \leq r\}$. By the empirical process theory for sub-Gaussian designs (specifically applying Dudley's entropy integral), there exists a constant $C_S = C_S(\upsilon_1, \tau)$ such that,
$$
E \sup_{\tilde{\delta} \in \mathbb{B}_2(r)} \|\widetilde{\mathcal{R}}(\tilde{\delta})\|_2 \leq C_S \sqrt{\frac{p}{T}} r.
$$
Further, by the concentration version of Talagrand's contraction principle, for any $t > 0$,
$$
P\left( \sup_{\tilde{\delta} \in \mathbb{B}_2(r)} \|\widetilde{\mathcal{R}}(\tilde{\delta})\|_2 \geq C_S \sqrt{\frac{p}{T}} r + t \right) \leq 2 \exp\left( - \frac{c T t^2}{r^2} \right).
$$
Taking $r = \|\delta_i\|_{\Sigma_i}$ and $t = C_S \sqrt{\frac{\log(1/\delta)}{T}} r$, we obtain:
\[
\|S_i\|_{\Sigma_i^{-1}} = \|\widetilde{\mathcal{R}}(\Sigma_i^{1/2} \delta_i)\|_2 \leq C_S^\prime \sqrt{\frac{p + \log(1/\delta)}{T}} \|\delta_i\|_{\Sigma_i},
\]
with probability at least $1 - \delta$.
Let $C_2^\prime = \max\{C_D^\prime, C_S^\prime\}$. Then there exists a constant $C_2^\prime$ such that with probability at least $1 - 2\delta$,
\begin{equation}\label{eq:F.4}
\left\| \frac{1}{T} X_i^\top M_{\widehat{F}} \Delta_i^\alpha \right\|_{\Sigma_i^{-1}} \leq C_2^\prime \left( \sqrt{\frac{p + \log(1/\delta)}{T}} \|\delta_i\|_{\Sigma_i} + \overline{f} \|\delta_i\|_{\Sigma_i}^2 \right).
\end{equation}

On the event $E_1 \cap E_2 \cap E_3 \cap E_4$, where $E_1$ represents Lemma \ref{lemma:F.1} holds (probability $\geq 1-\delta$), $E_2$ represents factor term bound \eqref{eq:F.2} holds (probability $\geq 1-\delta$), $E_3$ represents ES error term bound \eqref{eq:F.3} holds (probability $\geq 1-\delta$) and $E_4$ represents first-stage error bound \eqref{eq:F.4} holds (probability $\geq 1-2\delta$). By the union bound, $P(E_1 \cap E_2 \cap E_3 \cap E_4) \geq 1 - 5\delta$. On this intersection event, from \eqref{eq:F.1} and Lemma \ref{lemma:F.1},
$$
\|\widehat{\beta}_{i} - \beta_{i}^0\|_{\Sigma_i} \leq \frac{2}{\kappa_{\min}} \left\| \frac{1}{T} \Sigma_i^{-1/2} X_i^\top M_{\widehat{F}} \left[ \Delta_i^\alpha + F^0 \lambda_{i}^0 + \varepsilon_i \right] \right\|_2.
$$
Applying the triangle inequality and substituting \eqref{eq:F.2} to \eqref{eq:F.4},
\begin{equation}\nonumber
\begin{aligned}
\|\widehat{\beta}_{i} - \beta_{i}^0\|_{\Sigma_i} &\leq \frac{2}{\kappa_{\min}} \Bigg[ C_1^\prime \overline{\sigma} \sqrt{\frac{p + \log(1/\delta)}{T}} \\
&\quad + C_2^\prime \left( \sqrt{\frac{p + \log(1/\delta)}{T}} \|\delta_i\|_{\Sigma_i} + \overline{f} \|\delta_i\|_{\Sigma_i}^2 \right) \\
&\quad + C_3^\prime \left( \frac{1}{\sqrt{T}} + \frac{1}{\sqrt{N}} \right)\cdot (p + \log(1/\delta)) \Bigg].
\end{aligned}
\end{equation}
Let $r_0 = \max_{1 \le i \le N} \|\delta_i\|_{\Sigma_i}$, and define
$$
C_1 = \frac{2C_1^\prime}{\kappa_{\min}}, \quad C_2 = \frac{2C_2^\prime}{\kappa_{\min}}, \quad C_3 = \frac{2C_3^\prime}{\kappa_{\min}},
$$
then,
$$
\tau\|\widehat{\beta}_{i} - \beta_{i}^0\|_{\Sigma_i} \leq C_1 \overline{\sigma} \sqrt{\frac{p + \log(1/\delta)}{T}} + C_2 \left( \sqrt{\frac{p + \log(1/\delta)}{T}} r_0 + \overline{f} r_0^2 \right) + C_3 \left( \frac{1}{\sqrt{T}} + \frac{1}{\sqrt{N}} \right)(p + \log(1/\delta)).
$$
%Finally, note that our derivation actually gives a bound for $\tau \|\widehat{\beta}_{i} - \beta_{i}^0\|_{\Sigma_i}$, because $Z_i^\ast(\cdot) = Z_i(\cdot)/\tau$. 
Therefore, the above inequality is exactly the theorem's conclusion.
\end{proof}
\begin{lemma}\label{lemma:F.1}
Under Assumptions \ref{assum:UniformBound} and \ref{assum:Sub-Gaussianity}, there exists a constant $c_0 = c_0(\upsilon_1, \kappa_{\min}) > 0$ such that when $T \geq c_0(p + \log(1/\delta))$, with probability at least $1 - 3\delta$,
$$
\lambda_{\min}\left( \frac{1}{T} X_i^\top M_{\widehat{F}} X_i \right) \geq \frac{\kappa_{\min}}{2}.
$$
\end{lemma}
\begin{proof}
%Let $A = \frac{1}{T} X_i^\top X_i$ and $B = \frac{1}{T} X_i^\top P_{\widehat{F}} X_i$, where $P_{\widehat{F}} = \widehat{F}\widehat{F}^\top/T$.For the first term $A$, by the spectral concentration inequality for sub-Gaussian matrices,$$P\left( \|A - \Sigma_i\|_{2} \geq C\upsilon_1^2 \left( \sqrt{\frac{p}{T}} + \frac{\log(1/\delta)}{\sqrt{T}} \right) \right) \leq \delta.$$When $T$ is sufficiently large, $\lambda_{\min}(A) \geq \kappa_{\min}/2$.For the second term $B$, since $P_{\widehat{F}}$ is a projection matrix of rank $r$,$$\|B\|_{2} \leq \frac{1}{T} \|X_i\|_{2}^2 \leq C\upsilon_1^2 \frac{p + \log(1/\delta)}{T},$$holds with high probability. Since $r$ is fixed and $T \geq c_0(p + \log(1/\delta))$ for sufficiently large $c_0$, we have $\|B\|_{2} \leq \kappa_{\min}/4$. By Weyl's inequality, $\lambda_{\min}(A - B) \geq \lambda_{\min}(A) - \|B\|_{2} \geq \kappa_{\min}/4$. Adjusting constants yields the conclusion.
Because $M_{\widehat{F}} = I_T - P_{\widehat{F}}$, where $P_{\widehat{F}} = \widehat{F}\widehat{F}^\top / T$ is the projection matrix onto the column space of $\widehat{F}$, then
$$
\frac{1}{T} X_i^\top M_{\widehat{F}} X_i = \frac{1}{T} X_i^\top X_i - \frac{1}{T} X_i^\top P_{\widehat{F}} X_i := A-B.
$$
We need to control terms $A$ and $B$ separately. By Assumption \ref{assum:Sub-Gaussianity}, $W_{it} = \Sigma_i^{-1/2} X_{it}$ is an isotropic sub-Gaussian random vector with sub-Gaussian norm bounded by $\upsilon_1$. According to the spectral concentration inequality for sub-Gaussian matrices, there exists an absolute constant $C_1 > 0$ such that for any $\delta \in (0,1)$,
$$
P\left[ \|A - \Sigma_i\|_2 \geq C_1 \upsilon_1^2 \left( \sqrt{\frac{p}{T}} + \sqrt{\frac{\log(1/\delta)}{T}} \right) \right] \leq \delta.
$$
Choose $T$ sufficiently large such that 
$$
C_1 \upsilon_1^2 \left( \sqrt{\frac{p}{T}} + \sqrt{\frac{\log(1/\delta)}{T}} \right) \leq \frac{\kappa_{\min}}{4},\ \text{i.e.,}\ T \geq C_1' \upsilon_1^4 \cdot \frac{p + \log(1/\delta)}{\kappa_{\min}^2}
$$
for some constant $C_1'$. On this event, by Weyl's inequality,
$$
\lambda_{\min}(A) \geq \lambda_{\min}(\Sigma_i) - \|A - \Sigma_i\|_{2} \geq \kappa_{\min} - \frac{\kappa_{\min}}{4} = \frac{3\kappa_{\min}}{4}.
$$
Denote this condition as $T \geq c_1 (p + \log(1/\delta))$, where $c_1$ depends on $\upsilon_1$ and $\kappa_{\min}$.

For term $B$,
$$
B = \frac{1}{T} X_i^\top \left( \frac{1}{T} \widehat{F}\widehat{F}^\top \right) X_i = \frac{1}{T^2} (X_i^\top \widehat{F})(X_i^\top \widehat{F})^\top.
$$
Therefore,
$$
\|B\|_{2} = \frac{1}{T^2} \|X_i^\top \widehat{F}\|_{2}^2 = \frac{1}{T^2} \lambda_{\max}\left( \widehat{F}^\top X_i X_i^\top \widehat{F} \right).
$$
By the convergence of factor estimation, there exists an invertible rotation matrix $H \in \mathbb{R}^{r \times r}$ and a constant $C_2 > 0$ such that with probability at least $1 - \delta$,
$$
\frac{1}{T} \|\widehat{F} - F^0 H\|_F^2 \leq C_2 \log(1/\delta)\left( \frac{1}{N} + \frac{1}{T} \right).
$$
Decompose $X_i^\top \widehat{F}$ as $X_i^\top \widehat{F} = X_i^\top F^0 H + X_i^\top (\widehat{F} - F^0 H)$. Then $\|X_i^\top \widehat{F}\|_{2} \leq \|X_i^\top F^0 H\|_{2} + \|X_i^\top (\widehat{F} - F^0 H)\|_{2}$. For the first term, by Assumption \ref{assum:Sub-Gaussianity} and the boundedness of $F^0$, there exists a constant $C_3 > 0$ such that with probability at least $1 - \delta$,
$$
\frac{1}{T^2} \|X_i^\top F^0\|_{2}^2 \leq C_3 \cdot \frac{p + \log(1/\delta)}{T}.
$$
For the second term, by the Cauchy–Schwarz inequality,
$$
\frac{1}{T^2} \|X_i^\top (\widehat{F} - F^0 H)\|_{2}^2 \leq \frac{1}{T} \|X_i\|_{2}^2 \cdot \frac{1}{T} \|\widehat{F} - F^0 H\|_F^2.
$$
Moreover, $\frac{1}{T} \|X_i\|_{2}^2 \leq C_4 \cdot \frac{p + \log(1/\delta)}{T}$ holds with probability at least $1 - \delta$, and $\frac{1}{T} \|\widehat{F} - F^0 H\|_F^2 \leq C_2 \left( \frac{1}{N} + \frac{1}{T} \right) \log(1/\delta)$. When $N, T$ are sufficiently large, this term is of order $O_p((p + \log(1/\delta))/T^2)$, which is negligible relative to the first term. Specifically, there exists a constant $C_5$ such that with probability at least $1 - \delta$,
$$
\|B\|_{2} \leq C_5 \cdot \frac{p + \log(1/\delta)}{T}.
$$
Take $c_2 = 4C_5 / \kappa_{\min}$, then when $T \geq c_2 (p + \log(1/\delta))$, we have $\|B\|_{2} \leq \kappa_{\min}/4$.

Set $c_0 = \max\{c_1, c_2\}$. Then when $T \geq c_0 (p + \log(1/\delta))$, with probability at least $1 - 3\delta$, we have
$$
\lambda_{\min}(A) \geq \frac{3\kappa_{\min}}{4}, \quad \|B\|_{2} \leq \frac{\kappa_{\min}}{4}.
$$
By Weyl's inequality,
$$
\lambda_{\min}\left( \frac{1}{T} X_i^\top M_{\widehat{F}} X_i \right) = \lambda_{\min}(A - B) \geq \lambda_{\min}(A) - \|B\|_{2} \geq \frac{3\kappa_{\min}}{4} - \frac{\kappa_{\min}}{4} = \frac{\kappa_{\min}}{2}.
$$

\end{proof}
\section{Proof of Theorem \ref{theo:ESnon-asy}}\label{secG}
\begin{proof}
Let $\mathcal{X} = (X, F^0)$ denote the conditioning information. From the proof of Theorem \ref{theo:asymptotic}, we have the following Bahadur representation
\begin{equation}\label{eq:G.1}
\end{equation}
$$
\sqrt{T}(\widehat{\beta}_{i} - \beta_{i}^0) = S_i^M + R_{i},
$$
where the main term is
$$
S_i^M = \sum_{j=1}^N G_{ij} A_{j} U_{j},
$$
and the remainder term $ R_{i} $ contains higher-order effects from the first-stage quantile regression estimation error and the factor estimation error. Here
$$
A_{j} = \left( \frac{1}{T} X_j^\top M_{F^0} X_j \right)^{-1}, \quad 
U_{j} = \frac{1}{\sqrt{T}} X_j^\top M_{F^0} \varepsilon_{j} - \frac{1}{\sqrt{T}N} \sum_{k=1}^N a_{jk} X_j^\top M_{F^0} \varepsilon_{k},
$$
where $ G = (I - N^{-1} H)^{-1} $ as defined in the main text, and $ \varepsilon_{j} = (\varepsilon_{j1}, \dots, \varepsilon_{jT})^\top $. Define the coefficient vectors $v_{ij}^\top = a^\top G_{ij} A_{j} \in \mathbb{R}^{1 \times p},\ j=1,\dots,N$, so that
$$
S_i^M = \sum_{j=1}^N v_{ij}^\top U_{j}.
$$

Rewrite $ S_i^M $ in a time-aggregated form,
\begin{equation}\label{eq:G.2}
S_i^M = \frac{1}{\sqrt{T}} \sum_{t=1}^T \eta_t^{(i)}, \quad 
\eta_t^{(i)} = \sum_{k=1}^N \gamma_{kt}^{(i)} \varepsilon_{kt},
\end{equation}
where the coefficients are
$$
\gamma_{kt}^{(i)} = v_{ik}^\top \tilde{X}_{kt} - \frac{1}{N} \sum_{j=1}^N v_{ij}^\top a_{jk} \tilde{X}_{jt}, \quad 
\tilde{X}_{kt} = M_{F^0} X_{kt}.
$$

Let
$$
\sigma_T^2 = \mathrm{Var}(S_i^M \mid \mathcal{X}) = \frac{1}{T} \sum_{t=1}^T \mathrm{Var}(\eta_t^{(i)} \mid \mathcal{X}).
$$
From Assumption \ref{assum:error}(i)-(ii), there exists $\underline{\sigma}^2 > 0$ such that $\mathrm{Var}(\varepsilon_{kt} \mid \mathcal{X}) \geq \underline{\sigma}^2$ almost surely. Combined with Lemma \ref{lemma:G.1},
\begin{equation}\label{eq:G.3}
\sigma_T^2 \geq \underline{\sigma}^2 \cdot \frac{1}{T} \sum_{t=1}^T \sum_{k=1}^N [\gamma_{kt}^{(i)}]^2 \geq \underline{\sigma}^2 \underline{\gamma} =: \sigma_0^2 > 0 \ \text{almost surely}.
\end{equation}

By Assumption \ref{assum:error}(iii)-(iv), the sequence $\{\varepsilon_{kt}\}$ satisfies weak dependence conditions ($\alpha$-mixing). Consequently $\{\eta_t^{(i)}\}_{t=1}^T$ is a stationary weakly dependent sequence whose mixing coefficients satisfy $\sum_{h=0}^\infty \rho(h) < \infty$. Applying the Berry–Esseen theorem for weakly dependent sequences, there exist universal constants $C_{BE}, C_{BE}^\prime$ such that
\begin{equation}\label{eq:G.4}
\sup_{t \in \mathbb{R}} \left| P(S_i^M / \sigma_T \le t \mid \mathcal{X}) - \Phi(t) \right|
\leq C_{BE} \frac{E[|\eta_1^{(i)}|^3 \mid \mathcal{X}]}{\sigma_T^3 \sqrt{T}} + C_{BE}^\prime \frac{\log T}{\sqrt{T}}.
\end{equation}

From Lemma \ref{lemma:G.1} and Assumption \ref{assum:error}(i),
$$
E[|\eta_t^{(i)}|^3 \mid \mathcal{X}] \leq C_\gamma^3 N^3 M_3 =: C_\eta.
$$
Substituting into \eqref{eq:G.4} and using \eqref{eq:G.3} yields
\begin{equation}\label{eq:G.5}
\sup_{t \in \mathbb{R}} \left| P(S_i^M / \sigma_T \le t \mid \mathcal{X}) - \Phi(t) \right|
\leq \frac{C_{BE} C_\eta}{\sigma_0^3 \sqrt{T}} + \frac{C_{BE}^\prime \log T}{\sqrt{T}} \leq C_1 \frac{\log T}{\sqrt{T}} \ \text{almost surely},
\end{equation}
where $C_1 = C_{BE} C_\eta / \sigma_0^3 + C_{BE}^\prime$.

The remainder term $R_{i}$ consists of two parts, that is, $R_{i} = R_{i}^{(QR)} + R_{i}^{(F)}$, where $R_{i}^{(QR)}$ originates from the first‑stage quantile estimation error and $R_{i}^{(F)}$ from the factor estimation error. By the Neyman orthogonality, the influence of the first‑stage error is of second order. According to Proposition \ref{pro:quantile}, there exists a constant $C_\alpha > 0$ such that for each fixed $i \in [N]$ and any $\delta \in (0,1)$, with probability at least $1 - \delta$,
$$
\|\widehat{\alpha}_i - \alpha_i^0\|_{\Sigma_i} \leq C_\alpha \sqrt{\frac{p + \log(1/\delta)}{T}}.
$$
Take $\delta = (NT)^{-1}$. Then for a single $i$, the inequality holds with probability at least $1 - (NT)^{-1}$. Taking the union bound over $i = 1, \ldots, N$, we obtain that with probability at least $1 - {T}^{-1}$, we have
$$
\max_{1 \le i \le N} \|\widehat{\alpha}_i - \alpha_i^0\|_{\Sigma_i} \leq C_\alpha \sqrt{\frac{p + \log(NT)}{T}}.
$$
Hence,
$$
\|R_i^{(QR)}\|\lesssim \frac{p + \log N + \log T}{T}.
$$
From Proposition \ref{pro:average consistency}, the factor estimation error satisfies
$$
\frac{1}{T} \|\widehat{F} - F^0 H\|_F^2 = O_p\left( \frac{1}{N} + \frac{1}{T} \right).
$$
To maintain consistency with the probability level for the first-stage error, we control the factor estimation error with probability at least $1 - T^{-1}$. A perturbation analysis of the projection matrix yields
$$
\|R_i^{(F)}\| \lesssim\sqrt{\frac{\log N + \log T}{N}} + \sqrt{\frac{\log N + \log T}{T}} .
$$
Combining the above, there exists a constant $C_R > 0$ such that with conditional probability at least $1 - T^{-1}$,
\begin{equation}\label{eq:G.6}
|a^\top R_i| \leq C_R \left( \frac{p + \log N + \log T}{T} + \sqrt{\frac{\log N + \log T}{N}} + \sqrt{\frac{\log N + \log T}{T}} \right).
\end{equation}

Because the standard normal density is bounded, $ \phi(x) \leq 1/\sqrt{2\pi} $, we have
\begin{equation}\label{eq:G.7}
\left| P(S_i^M + a^\top R_{i} \le t \mid \mathcal{X}) - P(S_i^M \le t \mid \mathcal{X}) \right|
\leq \frac{|a^\top R_{i}|}{\sqrt{2\pi} \sigma_0}.
\end{equation}

Putting together \eqref{eq:G.5}–\eqref{eq:G.7}, we obtain
\begin{equation}\label{eq:G.8}
\begin{aligned}
&\sup_{t \in \mathbb{R}} \left| P\!\left( \frac{S_i^M + a^\top R_{i}}{\sigma_T} \le t \;\middle|\; \mathcal{X} \right) - \Phi(t) \right| \\
&\leq C_1 \frac{\log T}{\sqrt{T}} + \frac{C_R}{\sqrt{2\pi} \sigma_0} \left( \frac{p + \log N + \log T}{T}+\sqrt{\frac{\log N + \log T}{N}} + \sqrt{\frac{\log N + \log T}{T}} \right) \\
&\leq C_2 \left( \frac{p+\log N +\log T}{\sqrt{T}} + \sqrt{\frac{\log N + \log T}{N}} \right),
\end{aligned}
\end{equation}
where $C_2 = \max\left\{ C_1, \frac{C_R}{\sqrt{2\pi} \sigma_0} \right\}$.

Notice that $ \sigma_T^2 = a^\top \Omega_{i} a / T $, where
$$
\Omega_{i} = \mathrm{Var}\!\left( \sqrt{T} \, a^\top (\widehat{\beta}_{i} - \beta_{i}^0) \mid \mathcal{X} \right)
$$
is the true conditional variance matrix. Replacing $\sigma_T$ in \eqref{eq:G.8} by $\sqrt{a^\top \Omega_{i} a / T}$ gives
$$
\sup_{t \in \mathbb{R}} \left| P\!\left( \frac{\sqrt{T} \, a^\top (\widehat{\beta}_{i} - \beta_{i}^0)}{\sqrt{a^\top \Omega_{i} a}} \le t \;\middle|\; \mathcal{X} \right) - \Phi(t) \right|\lesssim \frac{p + \log N + \log T}{\sqrt{T}} + \sqrt{\frac{\log N + \log T}{N}}.
$$
\end{proof}

\begin{lemma}\label{lemma:G.1}
There exist constants $ C_\gamma, \underline{\gamma} > 0 $ such that almost surely
$$
\max_{k,t} |\gamma_{kt}^{(i)}| \leq C_\gamma, \quad 
\frac{1}{T} \sum_{t=1}^T \sum_{k=1}^N [\gamma_{kt}^{(i)}]^2 \geq \underline{\gamma}.
$$
\end{lemma}
\begin{proof}
By Assumptions \ref{assum:UniformBound} and \ref{assum:Sub-Gaussianity}, there exists $ C_X < \infty $ such that $\|\tilde{X}_{kt}\| \leq C_X$ almost surely. From Assumption \ref{assum:cross-sectional dependence}, $\sum_{j=1}^N \|G_{ij}\| \leq C_G$; and by Assumption \ref{assum:eigenvalue}(i), $\|A_{j}\| \leq c_X^{-1}$. Hence $\|v_{ij}\| \leq \|a\| \|G_{ij}\| \|A_{j}\| \leq C_G c_X^{-1}$. Combined with the boundedness of $\lambda_{i}^0$ from Assumption \ref{assum:factor&loading}, the quantities $a_{jk}$ are uniformly bounded. Therefore
$$
|\gamma_{kt}^{(i)}| \leq \|v_{ik}\| \|\tilde{X}_{kt}\| + \frac{1}{N} \sum_{j=1}^N \|v_{ij}\| |a_{jk}| \|\tilde{X}_{jt}\| \leq C_G c_X^{-1} C_X (1 + C_\lambda) =: C_\gamma.
$$
The lower bound follows from Assumption \ref{assum:eigenvalue}(ii) and the non‑degeneracy of the design. Specifically, because the diagonal blocks $G_{ii}$ of $G$ are full‑rank and $A_{i}$ is positive definite, there exists $c_v > 0$ such that $\|v_{ii}\| \geq c_v$ almost surely. Using the lower bound on the minimum eigenvalue of the design matrix from Assumption \ref{assum:eigenvalue}(i), we obtain
$$
\frac{1}{T} \sum_{t=1}^T (v_{ii}^\top \tilde{X}_{it})^2 \geq c_0 c_v^2 > 0.
$$
The cross‑terms are controlled by the uniform boundedness, hence there exists $\underline{\gamma} > 0$ satisfying the requirement.
\end{proof}

\section{Additional Simulation Results}\label{sec:Additional Simulation Results}
\paragraph{Latent factors and loadings.}
The ES factors $F_t\in\mathbb{R}^{r_0}$ follow a stationary AR(1) process (componentwise) and the loadings $\lambda_i\in\mathbb{R}^{r_0}$ are drawn independently across $i$ and scaled to control factor strength. The factor-driven tail scale is generated from the interactive component $\lambda_i'F_t$ through an exponential link, $\sigma_{it}=\exp\!\Bigl(c_{\sigma}\cdot \frac{\lambda_i'F_t}{\mathrm{sd}(\lambda'F)}\Bigr)$,
with truncation to avoid extreme values. The constant $c_{\sigma}$ controls the intensity of tail comovement.

\paragraph{Cross-sectional heterogeneity.}
To reflect heterogeneous exposures to observed risk drivers, we allow slope coefficients in the ES regression to vary across units via a grouped design: units are randomly assigned to a small number of groups, and the slope vector is shifted across groups. This introduces meaningful cross-sectional dispersion in $\beta_i$ while keeping the regressor dimension fixed.

\subsection{Data-generating Scenarios}\label{subsec:Simulation Scenarios}

We consider seven economically motivated scenarios (indexed by $\texttt{scenarioID}\in\{1,\cdots,7\}$), all based on (4.1) in main text and differing only in how $\sigma_{it}$ or $\varepsilon_{it}$ are modified.

\begin{itemize}
\item \textbf{Scenario 1 (Baseline tail-factor structure).}
Tail dependence is introduced through a latent additive component in the conditional expected shortfall. 
Specifically, the ES contains a low-rank factor structure $\lambda_i'F_t$, while the conditional quantile location remains $\mu_it = X_it'\alpha_i$. This baseline design captures cross-sectional tail comovement driven by latent factors.

\item \textbf{Scenario 2 (Stronger tail-factor dependence).}
We increase the strength of the latent factor component in the ES structure. This design amplifies cross-sectional tail comovement and evaluates whether the estimator can accurately recover the factor space when the tail component dominates.

\item \textbf{Scenario 3 (Heterogeneous slopes).}
We strengthen cross-sectional heterogeneity in the slope coefficients by grouping units and shifting $\beta_i$ across groups, while maintaining factor-driven tail scale. This scenario evaluates robustness to heterogeneous covariate effects.

\item \textbf{Scenario 4 (Endogenous covariates).}
We introduce correlation between covariates and latent tail factors by letting one regressor load on the interactive component $\lambda_i'F_t$. This creates an omitted-factor-type bias for methods that do not account for latent tail components.

\item \textbf{Scenario 5 (Volatility-factor dominance).}
We increase the strength of the factor effect in $\sigma_{it}$, so that tail comovement is largely volatility-driven. This scenario stresses the setting where ignoring factors should be particularly costly for ES regression.

\item \textbf{Scenario 6 (Jump episodes).}
We add rare but large shocks to the innovation by augmenting $\varepsilon_{it}$ with a jump term that occurs with small probability. This generates episodic systemic stress while preserving the basic location--scale structure.

\item \textbf{Scenario 7 (Asymmetric tails).}
We modify the innovation distribution by introducing an additional asymmetric component that thickens the lower tail. This design captures asymmetric downside risk beyond symmetric heavy-tailed innovations.
\end{itemize}

Across these scenarios, ES regression is intentionally misspecified whenever latent tail factors drive $\sigma_{it}$, whereas ESFM explicitly models and estimates the latent tail component. The simulations therefore provide a direct check of the central mechanism of ESFM: incorporating tail factors improves the accuracy of $\beta_i$ estimation and yields reliable recovery of the factor space in finite samples.

\subsection{Simulation Results}\label{subsec:Simulation Results}
The main text results assume the number of factors is known. Here, we examine the performance of the factor selection procedure. To assess the finite-sample behavior of the estimator, we report the Monte Carlo mean of the estimated factor number $\hat{r}_{\tau}$ under different sample sizes and tail probability levels. For each simulation design, the factor number is estimated using the information criterion described in Section \ref{subsec:determine}. The true number of factors in the data-generating process is fixed at $r_\tau=2$, while the cross-sectional dimension $N$, the time dimension $T$, and the tail probability level $\tau$ vary across designs.
\begin{table}[htbp]
  \centering
  \caption{Monte Carlo Mean of the Estimated Number of Factors. {\footnotesize{\textit{Notes}. The table reports the Monte Carlo mean of the estimated factor number $\hat r$ across different simulation settings. 
The true number of factors in the data-generating process is $r=2$. The table reports the Monte Carlo mean of the estimated factor number $\hat r$ across different simulation settings. The true number of factors in the data-generating process is $r=2$.}}}
  \label{tab:Error for number of factors}

  \setlength{\tabcolsep}{4pt}      
  \renewcommand{\arraystretch}{0.95} 
  \small                  

  \begin{adjustbox}{max totalsize={\textwidth}{0.85\textheight},center}
  \begin{tabular}{lccccccc}
    \toprule
          & Scenario 1 & Scenario 2 & Scenario 3 & Scenario 4 & Scenario 5 & Scenario 6 & Scenario 7 \\
    \midrule
    \multicolumn{8}{c}{Panel A: $\tau$=0.10} \\
    \midrule
    \multicolumn{8}{c}{$N=100$} \\
    \midrule
    $T=100$ & 2.1300  & 2.2500  & 2.1200  & 2.2400  & 2.2300  & 2.2200  & 2.2800  \\
    $T=200$ & 2.1100  & 2.1400  & 2.1000  & 2.1900  & 2.1600  & 2.1600  & 2.1600  \\
    $T=300$ & 2.0600  & 2.0500  & 2.0400  & 2.0900  & 2.0100  & 2.0600  & 2.1300  \\
    \midrule
    \multicolumn{8}{c}{$N=200$} \\
    \midrule
    $T=100$ & 2.1100  & 2.1600  & 2.1100  & 2.3500  & 2.1800  & 2.1800  & 2.1600  \\
    $T=200$ & 2.1000  & 2.1400  & 2.1000  & 2.1200  & 2.1100  & 2.1000  & 2.1400  \\
    $T=300$ & 2.0700  & 2.0500  & 2.0500  & 2.0300  & 2.0800  & 2.0500  & 2.0600  \\
    \midrule
    \multicolumn{8}{c}{$N=300$} \\
    \midrule
    $T=100$ & 2.1000  & 2.1500  & 2.1000  & 2.1000  & 2.1000  & 2.1300  & 2.1500  \\
    $T=200$ & 2.0900  & 2.1000  & 2.0700  & 2.0600  & 2.0300  & 1.9500  & 2.1100  \\
    $T=300$ & 2.0500  & 2.0400  & 2.0600  & 1.9800  & 1.9700  & 1.9600  & 2.0400  \\
    \midrule
    \multicolumn{8}{c}{Panel B: $\tau$=0.05} \\
    \midrule
    \multicolumn{8}{c}{$N=100$} \\
    \midrule
    $T=100$ & 2.0900  & 2.0900  & 2.0900  & 2.1500  & 2.1500  & 2.2000  & 2.2000  \\
    $T=200$ & 2.0500  & 2.0400  & 2.0600  & 2.1300  & 2.1500  & 2.1500  & 2.1900  \\
    $T=300$ & 2.0000  & 2.0300  & 2.0500  & 2.0100  & 2.0500  & 2.1300  & 2.0600  \\
    \midrule
    \multicolumn{8}{c}{$N=200$} \\
    \midrule
    $T=100$ & 2.1500  & 2.1800  & 2.1900  & 2.1400  & 2.1300  & 2.1800  & 2.1900  \\
    $T=200$ & 2.1100  & 2.1200  & 2.0900  & 2.0600  & 2.0800  & 2.0900  & 2.0600  \\
    $T=300$ & 2.0900  & 2.0800  & 2.0200  & 2.0300  & 2.0500  & 2.0500  & 2.0200  \\
    \midrule
    \multicolumn{8}{c}{$N=300$} \\
    \midrule
    $T=100$ & 2.0900  & 2.1000  & 1.9100  & 1.9200  & 1.9200  & 1.9100  & 1.9000  \\
    $T=200$ & 2.0500  & 2.0900  & 1.9300  & 1.9400  & 1.9600  & 1.9500  & 1.9500  \\
    $T=300$ & 2.0800  & 2.0600  & 1.9500  & 1.9600  & 2.0200  & 1.9800  & 1.9900  \\
    \midrule
    \multicolumn{8}{c}{Panel C: $\tau$=0.01} \\
    \midrule
    \multicolumn{8}{c}{$N=100$} \\
    \midrule
    $T=100$ & 1.9400  & 2.1000  & 1.9800  & 2.0600  & 1.9500  & 1.9400  & 1.9400  \\
    $T=200$ & 1.9600  & 2.0800  & 1.9600  & 2.0500  & 1.9600  & 1.9600  & 1.9500  \\
    $T=300$ & 2.0100  & 2.0200  & 2.0000  & 2.0100  & 1.9900  & 1.9800  & 1.9900  \\
    \midrule
    \multicolumn{8}{c}{$N=200$} \\
    \midrule
    $T=100$ & 1.9600  & 1.9600  & 1.9800  & 1.9600  & 1.9800  & 2.0400  & 2.0500  \\
    $T=200$ & 1.9800  & 1.9600  & 2.0200  & 1.9900  & 1.9900  & 2.0300  & 2.0200  \\
    $T=300$ & 1.9900  & 1.9900  & 2.0100  & 2.0000  & 2.0000  & 2.0100  & 2.0000  \\
    \midrule
    \multicolumn{8}{c}{$N=300$} \\
    \midrule
    $T=100$ & 1.9500  & 2.0400  & 1.9500  & 2.0300  & 1.9600  & 1.9600  & 2.0300  \\
    $T=200$ & 1.9900  & 1.9000  & 1.9800  & 1.9000  & 2.0100  & 2.0000  & 1.9900  \\
    $T=300$ & 2.0000  & 2.0000  & 2.0000  & 2.0000  & 2.0000  & 2.0000  & 2.0000  \\
    \bottomrule
  \end{tabular}
  \end{adjustbox}
\end{table}

Table \ref{tab:Error for number of factors} summarizes the Monte Carlo averages of $\hat{r}_\tau$ across the considered simulation settings. Overall, the estimated factor number is close to the true value $r_\tau=2$ across all designs, with deviations diminishing as the sample size increases. In particular, when both $N$ and $T$ are moderately large, the average estimate concentrates tightly around 2, indicating reliable identification of the latent factor structure. The results are largely stable across different tail probability levels, suggesting that the factor-number selection procedure is not sensitive to the degree of tail sparsity.

\section{Additional Empirical Results}\label{sec:Additional Empirical Results}

\subsection{Estimation Results}\label{subsec:Estimation Results}
\begin{figure}[htb]
	\vspace{0pt}
	\centering
	\includegraphics[scale=0.45]{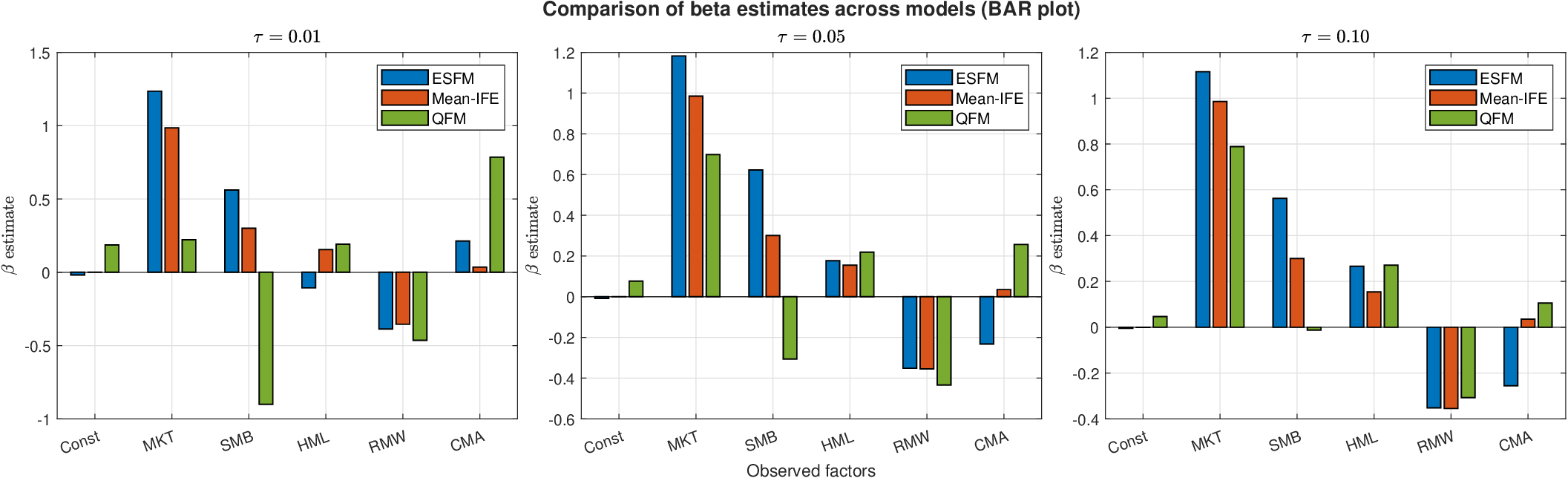}
	\caption{\footnotesize Estimated coefficients ($\beta$) on observable risk factors. \textit{Notes}. The figure reports the estimated coefficients on observable factors (MKT, SMB, HML, RMW, CMA, and a constant) from three models: ESFM, the mean factor model (Mean-IFE), and the quantile factor model (QFM). The estimates are obtained using EW observable factor returns, and results are shown for three tail levels ($\tau = 0.01, 0.05, 0.10$). Each panel corresponds to a different $\tau$, and bars represent the average coefficients across assets.}\label{fig:beta_EW}
\end{figure}

\begin{figure}[htb]
	\vspace{0pt}
	\centering
	\includegraphics[scale=0.59]{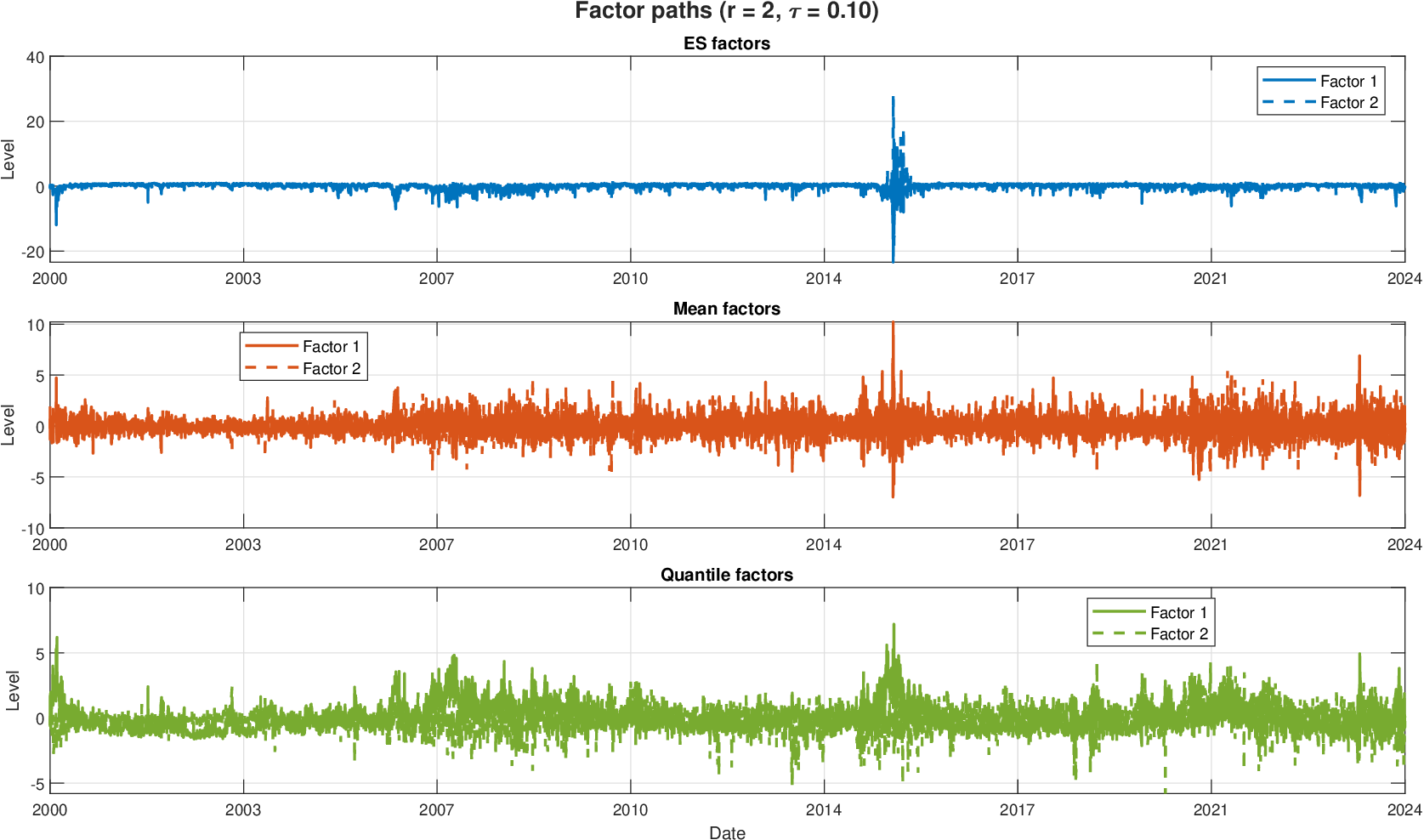}
	\caption{\footnotesize Estimated factor paths across models ($r=2,\tau=0.10$). \textit{Notes}. The figure plots the estimated latent factor paths from three models: ESFM (top panel), the mean factor model (middle panel), and the quantile factor model (bottom panel). Each model extracts two factors ($r = 2$) based on equal-weighted Fama–French five factors. The quantile and ES factors are constructed using tail information at level $\tau = 0.10$.}\label{fig:factor_VW_0p10}
\end{figure}

The additional results reported in Figures \ref{fig:beta_EW}–\ref{fig:factor_VW_0p10} provide robustness checks under alternative specifications, including different tail levels and weighting schemes.

Figure \ref{fig:beta_EW} presents the corresponding coefficient estimates across models. The overall pattern remains stable relative to the main text: the ESFM continues to produce larger and more pronounced loadings across most observable factors, while the quantile-based model exhibits weaker and less stable estimates, particularly in the lower tail. The relative ordering across models is largely preserved across $\tau$ levels.

Figures \ref{fig:factor_VW_0p01}–\ref{fig:factor_VW_0p05} report the factor paths under value-weighted specifications for $tau = 0.01$ and $\tau = 0.05$, respectively, while Figures \ref{fig:factor_VW_0p01}–\ref{fig:factor_VW_0p10} present the corresponding results under equal-weighted specifications. Several features are worth noting. First, the ES factors consistently display episodic spikes aligned with periods of market stress, whereas the mean factors remain comparatively smooth and the quantile factors exhibit persistent but less differentiated fluctuations. Second, these patterns are robust across weighting schemes, suggesting that the tail factor structure is not driven by portfolio construction choices. Third, as $\tau$ increases (Figures \ref{fig:factor_VW_0p05}–\ref{fig:factor_VW_0p10}), the distinction between ES and quantile factors becomes less pronounced, consistent with the interpretation that tail-specific information diminishes away from the extreme left tail.

\begin{figure}[htb]
	\vspace{0pt}
	\centering
	\includegraphics[scale=0.59]{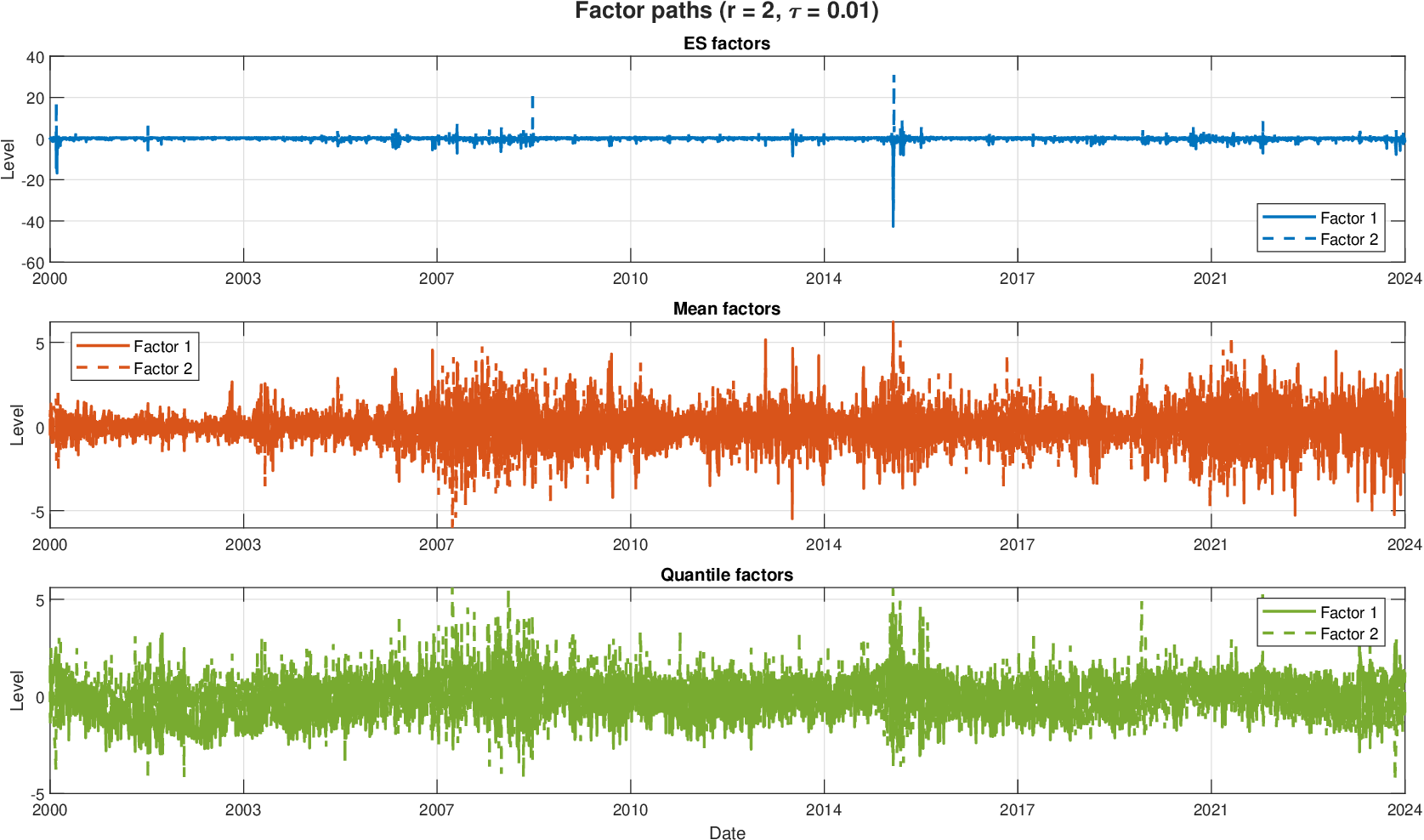}
	\caption{\footnotesize Estimated factor paths across models ($r=2,\tau=0.01$). \textit{Notes}. The figure plots the estimated latent factor paths from three models: ESFM (top panel), the mean factor model (middle panel), and the quantile factor model (bottom panel). Each model extracts two factors ($r = 2$) based on value-weighted Fama–French five factors. The quantile and ES factors are constructed using tail information at level $\tau = 0.01$.}\label{fig:factor_VW_0p01}
\end{figure}

\begin{figure}[htb]
	\vspace{0pt}
	\centering
	\includegraphics[scale=0.59]{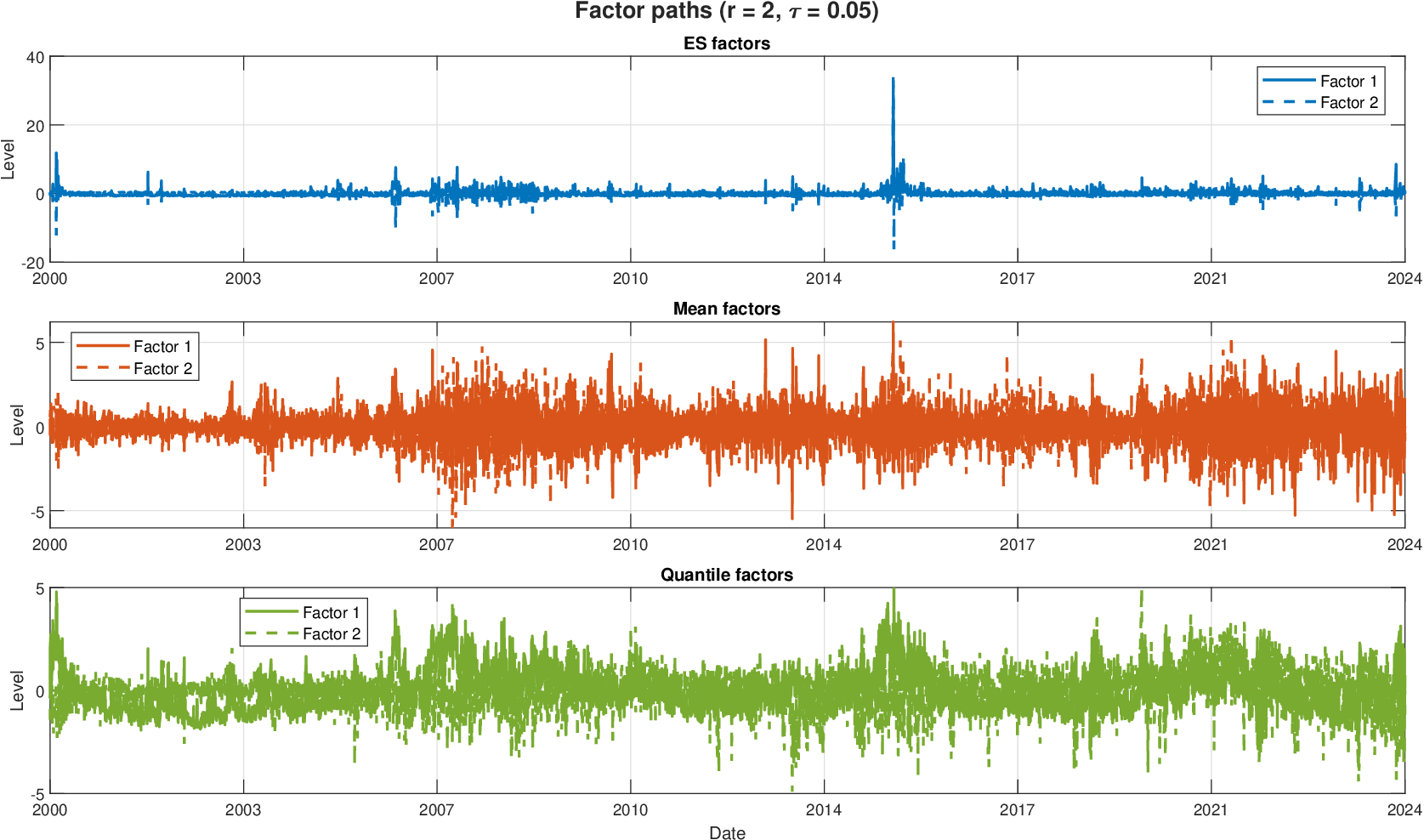}
	\caption{\footnotesize Estimated factor paths across models ($r=2,\tau=0.05$). \textit{Notes}. The figure plots the estimated latent factor paths from three models: ESFM (top panel), the mean factor model (middle panel), and the quantile factor model (bottom panel). Each model extracts two factors ($r = 2$) based on value-weighted Fama–French five factors. The quantile and ES factors are constructed using tail information at level $\tau = 0.05$.}\label{fig:factor_VW_0p05}
\end{figure}

\begin{figure}[htb]
	\vspace{0pt}
	\centering
	\includegraphics[scale=0.59]{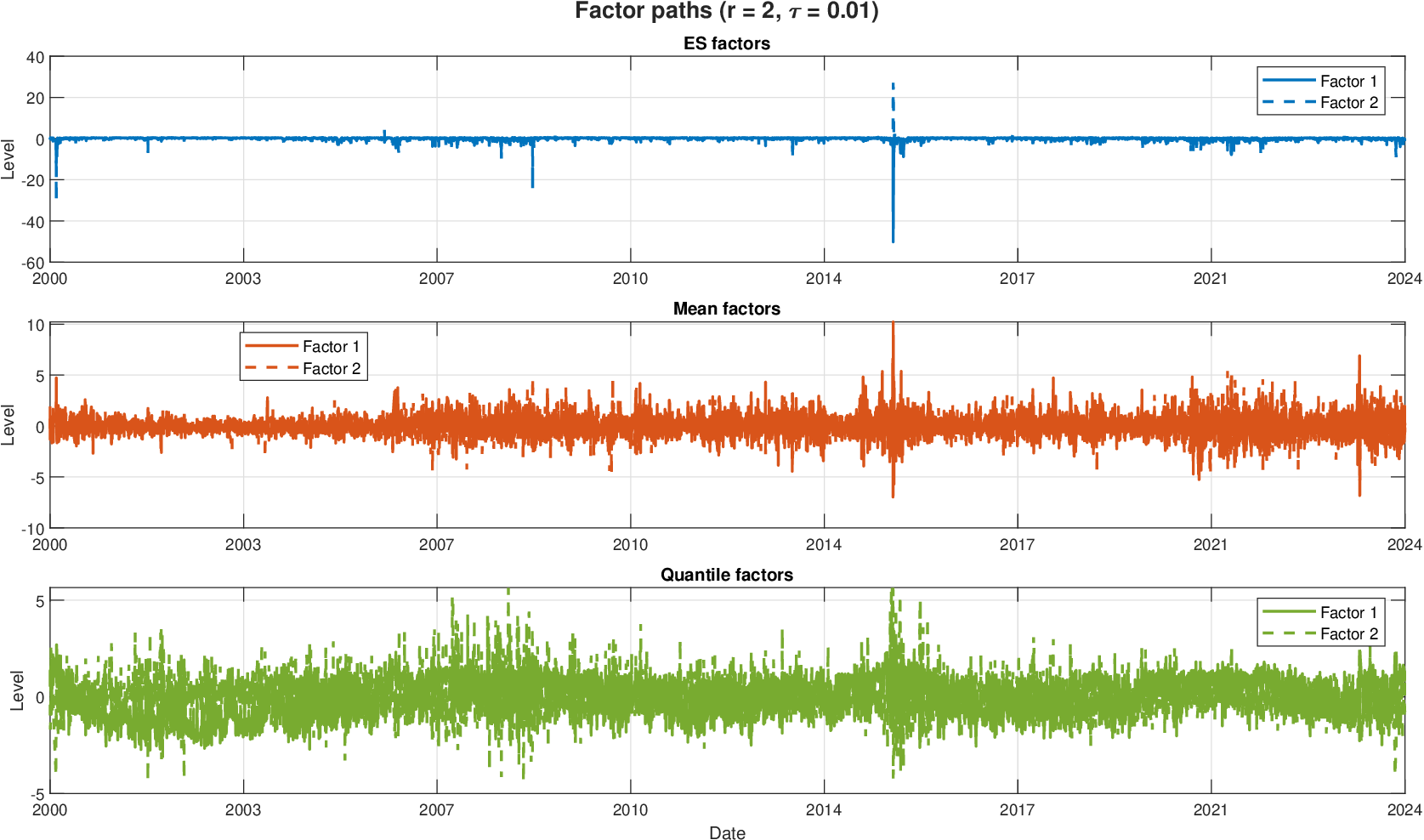}
	\caption{\footnotesize Estimated factor paths across models ($r=2,\tau=0.01$). \textit{Notes}. The figure plots the estimated latent factor paths from three models: ESFM (top panel), the mean factor model (middle panel), and the quantile factor model (bottom panel). Each model extracts two factors ($r = 2$) based on equal-weighted Fama–French five factors. The quantile and ES factors are constructed using tail information at level $\tau = 0.01$.}\label{fig:factor_VW_0p01}
\end{figure}

\begin{figure}[htb]
	\vspace{0pt}
	\centering
	\includegraphics[scale=0.59]{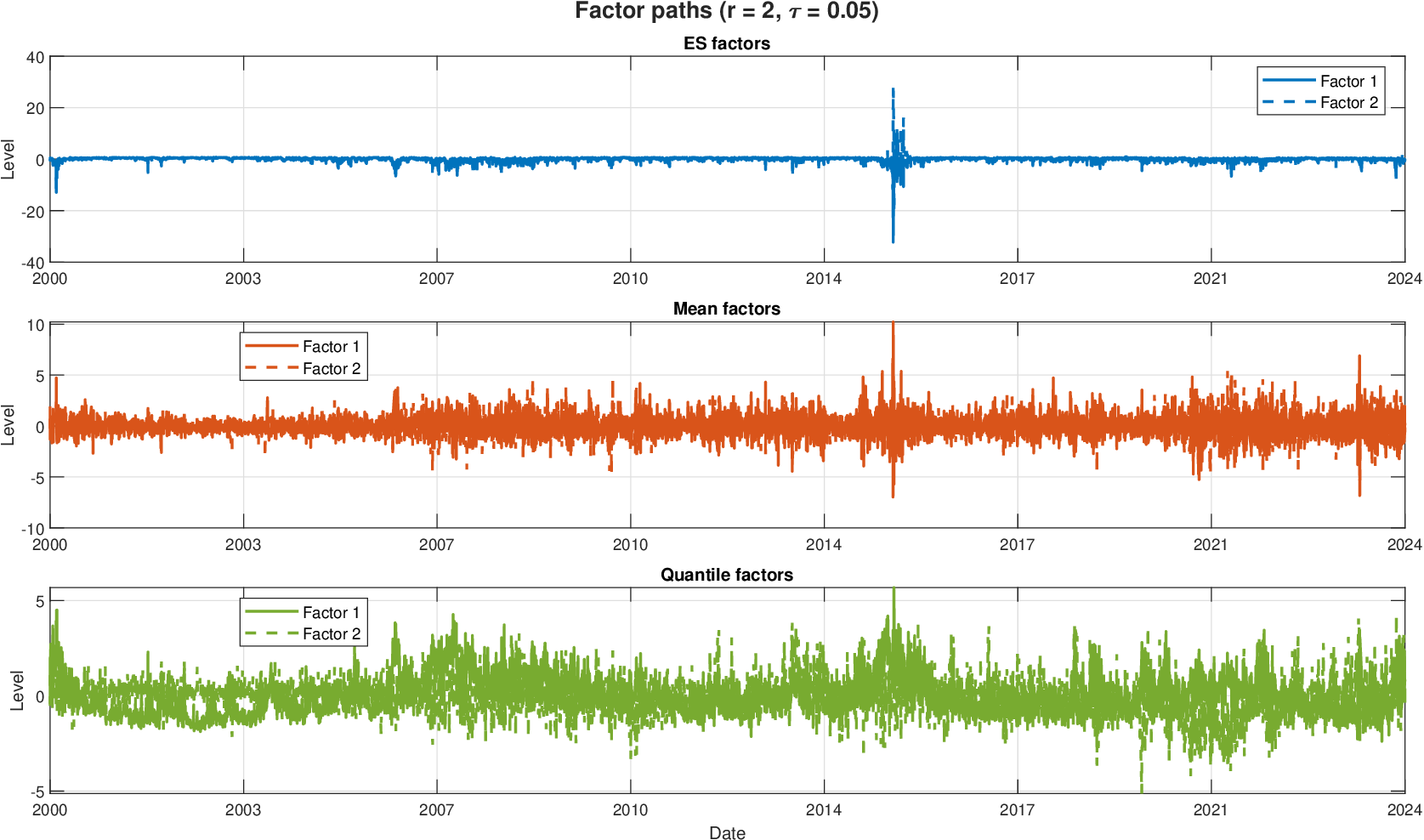}
	\caption{\footnotesize Estimated factor paths across models ($r=2,\tau=0.05$). \textit{Notes}. The figure plots the estimated latent factor paths from three models: ESFM (top panel), the mean factor model (middle panel), and the quantile factor model (bottom panel). Each model extracts two factors ($r = 2$) based on equal-weighted Fama–French five factors. The quantile and ES factors are constructed using tail information at level $\tau = 0.05$.}\label{fig:factor_VW_0p05}
\end{figure}

Overall, the appendix results confirm that the main empirical findings are not sensitive to alternative specifications. In particular, the ES factors continue to isolate tail risk components that are not captured by mean or quantile factor models, reinforcing the empirical relevance of the proposed framework.

\subsection{Asset Pricing}
\begin{figure}[htb]
	\vspace{0pt}
	\centering
	\includegraphics[scale=0.45]{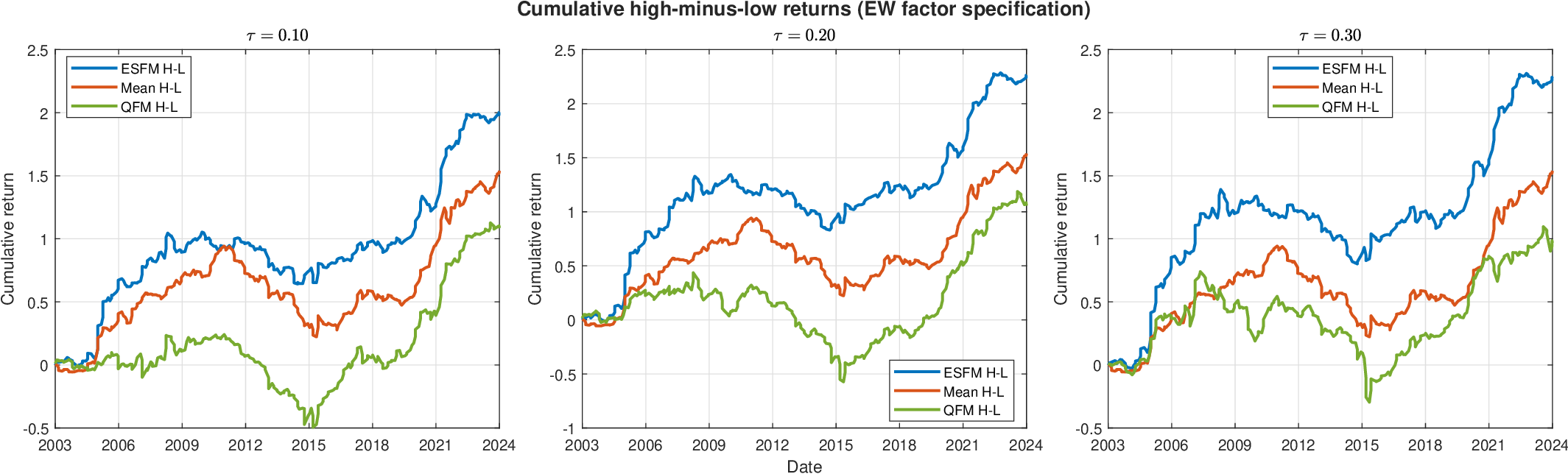}
	\caption{\footnotesize Cumulative high-minus-low returns across models. \textit{Notes}. The figure plots cumulative returns of high-minus-low (H–L) portfolios formed on factor exposures estimated from three models: ESFM, the mean factor model (Mean), and the quantile factor model (QFM). Stocks are sorted into five portfolios based on estimated exposures, and the H–L portfolio is constructed as the difference between the highest- and lowest-exposure portfolios. Portfolio returns are computed using equal-weighted (EW) returns, while the underlying Fama–French five factors are constructed using EW returns. Each panel corresponds to a different tail level ($\tau=0.10,0.20,0.30$).}\label{fig:cumu_returns_EW}
\end{figure}

As a robustness check, we re-compute the factor-mimicking portfolios using equal-weighted (EW) returns for the underlying Fama–French factors. Figure \ref{fig:cumu_returns_EW} plots the cumulative high-minus-low returns, and Table \ref{tab:Alphas in EW} reports the corresponding average returns and pricing alphas.

The overall patterns remain closely aligned with the results in main text. In Figure \ref{fig:cumu_returns_EW}, the ESFM-based portfolios continue to deliver the strongest and most persistent H–L performance across all tail levels, while the spreads implied by the mean and quantile factor models are noticeably weaker. This ranking is particularly clear at higher tail levels, where the ESFM spreads widen further relative to the alternatives.

\begin{table}[htbp]
\centering
\caption{Portfolio returns and alphas sorted on factor exposures across models. {\footnotesize{\textit{Notes}. The table reports annualized portfolio returns and alphas (in \%) sorted on factor exposures estimated from different models (ESFM, Mean, and QFM) across three tail levels ($\tau=0.10,0.20,0.30$). We report estimated intercept (alphas) from regressing the returns on various sets of asset pricing factors: market (CAPM), three factor factors of \cite{fama1993common} (FF3), and five factors of \cite{fama2015five}. Stocks are sorted into No.G $=5$ or $10$ portfolios based on estimated exposures, and portfolio returns are computed using equal-weighted (EW) scheme. The underlying Fama–French
three and five factors are constructed using EW returns. Reported numbers correspond to time-series averages of one-period-ahead returns, with Newey-West $t$-statistics (six lags) shown in parentheses. The ``High–Low'' spread captures the return difference between the highest- and lowest-exposure portfolios (the average is placed in column ``Average'').}}}
\label{tab:Alphas in EW}
\setlength{\tabcolsep}{2.5pt}       % Reduce column spacing
\renewcommand{\arraystretch}{0.95} % Reduce row spacing
\small                       % Font size
\begin{adjustbox}{max totalsize={\textwidth}{0.85\textheight},center}
\begin{tabular}{rrccccrccccrcccc}
    \toprule
          &       & \multicolumn{4}{c}{$\tau=0.10$} &       & \multicolumn{4}{c}{$\tau=0.20$} &       & \multicolumn{4}{c}{$\tau=0.30$} \\
\cmidrule{3-6}\cmidrule{8-11}\cmidrule{13-16}      No.G    &       & Average & CAPM  & FF3   & FF5   &       & Average & CAPM  & FF3   & FF5   &       & Average & CAPM  & FF3   & FF5 \\
    \midrule
    \multicolumn{1}{l}{5} & \multicolumn{1}{l}{ESFM} & 9.16\% & 10.18\% & 11.58\% & 11.32\% &       & 10.33\% & 11.45\% & 13.41\% & 12.17\% &       & 10.37\% & 11.58\% & 14.02\% & 12.38\% \\
          &       & (2.86) & (3.32) & (3.96) & (3.62) &       & (2.93) & (3.35) & (4.16) & (3.68) &       & (2.80) & (3.22) & (4.27) & (3.74) \\
    \multicolumn{1}{l}{5} & \multicolumn{1}{l}{Mean} & 7.02\% & 7.41\% & 7.82\% & 8.05\% &       & 7.02\% & 7.41\% & 7.82\% & 8.05\% &       & 7.02\% & 7.41\% & 7.82\% & 8.05\% \\
          &       & (2.43) & (2.55) & (2.83) & (2.58) &       & (2.43) & (2.55) & (2.83) & (2.58) &       & (2.43) & (2.55) & (2.83) & (2.58) \\
    \multicolumn{1}{l}{5} & \multicolumn{1}{l}{QFM} & 4.95\% & 5.94\% & 4.94\% & 5.08\% &       & 4.82\% & 5.73\% & 6.93\% & 6.05\% &       & 4.42\% & 5.07\% & 8.62\% & 4.66\% \\
          &       & (1.66) & (2.04) & (1.72) & (1.72) &       & (1.46) & (1.76) & (2.33) & (1.91) &       & (1.21) & (1.35) & (2.67) & (1.48) \\
    \multicolumn{1}{l}{10} & \multicolumn{1}{l}{ESFM} & 10.75\% & 11.99\% & 13.66\% & 13.48\% &       & 12.73\% & 14.15\% & 16.68\% & 15.80\% &       & 12.79\% & 14.34\% & 17.65\% & 16.06\% \\
          &       & (2.71) & (3.18) & (3.85) & (3.45) &       & (3.04) & (3.56) & (4.51) & (4.03) &       & (2.95) & (3.46) & (4.65) & (4.07) \\
    \multicolumn{1}{l}{10} & \multicolumn{1}{l}{Mean} & 9.92\% & 10.43\% & 10.96\% & 11.50\% &       & 9.92\% & 10.43\% & 10.96\% & 11.50\% &       & 9.92\% & 10.43\% & 10.96\% & 11.50\% \\
          &       & (2.76) & (2.87) & (2.99) & (2.78) &       & (2.76) & (2.87) & (2.99) & (2.78) &       & (2.76) & (2.87) & (2.99) & (2.78) \\
    \multicolumn{1}{l}{10} & \multicolumn{1}{l}{QFM} & 7.56\% & 8.77\% & 7.58\% & 7.28\% &       & 6.10\% & 7.22\% & 8.97\% & 7.47\% &       & 5.20\% & 5.85\% & 10.16\% & 5.92\% \\
          &       & (2.01) & (2.41) & (2.05) & (1.89) &       & (1.60) & (1.92) & (2.53) & (1.98) &       & (1.32) & (1.46) & (2.91) & (1.71) \\
    \bottomrule
    \end{tabular}
\end{adjustbox}
\end{table}

Table \ref{tab:Alphas in EW} provides complementary evidence from the alpha perspective. The ESFM portfolios generate the largest return spreads and corresponding alphas across specifications, and these alphas remain economically large after controlling for CAPM, FF3, and FF5 factors constructed using EW returns. In contrast, the pricing performance of the mean and quantile-based portfolios is substantially weaker and less stable.

Taken together, these results confirm that our main findings are not driven by the weighting scheme used to construct the benchmark factors. The ESFM-based tail factors continue to exhibit strong and robust pricing ability under alternative implementations.
\bibliographystyle{chicago}
\bibliography{ref}